\documentclass[12pt,a4paper]{article}
\pdfoutput=1
\usepackage{jheppub}
\usepackage[utf8]{inputenc}
\usepackage[T1]{fontenc}

\usepackage{tikz, pgf}  
\usetikzlibrary{patterns}
\usetikzlibrary{positioning}
\usepackage{amsmath,physics,float}  
\usepackage{amssymb}  
\usepackage{latexsym}
\usepackage{booktabs}
\usepackage{multirow}
\usepackage{amsfonts}
\usepackage{braket}
\usepackage{mathrsfs}

\usepackage{subfig} 
\usepackage{empheq}

\usepackage[shortlabels]{enumitem}

\usepackage{pgfplots}
\pgfplotsset{compat=1.13}
\usepackage{natbib} 
\usepackage{comment}

\usepackage{bigints}
\definecolor{ashgrey}{rgb}{0.6, 0.5, 0.5}
\definecolor{ashgrey2}{rgb}{0.7, 0.73, 0.71}

\newcommand{\bg}{\begin{equation}}
\newcommand{\nd}{\end{equation}}
\newcommand{\beq}{\begin{equation}}
\newcommand{\eeq}{\end{equation}} 
\newcommand{\ef}{\rm eff}

\newcommand{\del}{\partial}
\newcommand{\lc}{\left(}
\newcommand{\rc}{\right)}

\newcommand{\ca}{\bf}

\begin{document}

\title{What if string theory has a de Sitter excited state?}

\author{Joydeep Chakravarty and}
\emailAdd{joydeep.chakravarty@mail.mcgill.ca}

\author{Keshav Dasgupta}
\emailAdd{keshav@hep.physics.mcgill.ca}

\affiliation{Department of Physics, McGill University,\\
3600 Rue University, Montreal, H3A 2T8, QC Canada.}

\begin{abstract}
{ We propose precise effective field theory criteria to obtain a four-dimensional de Sitter space within M-theory. To this effect, starting with the state space described by the action of metric perturbations, fluxes etc over the supersymmetric Minkowski vacuum in eleven-dimensions,  we discuss the most general low energy effective action in terms of the eleven-dimensional fields including non-perturbative and non-local terms. Given this, our criteria to obtain a valid four-dimensional de Sitter solution at far IR involve satisfying the Schwinger-Dyson equations of the associated path integral, as well as obeying positivity constraints on the dual IIA string coupling and its time derivative. For excited states, the Schwinger-Dyson equations imply an effective emergent potential different from the original potential. We show that while vacuum solutions and arbitrary coherent states fail to satisfy these criteria, a specific class of excited states called the Glauber-Sudarshan states obey them. Using the resurgent structure of observables computed using the path integral over the Glauber-Sudarshan states, four-dimensional de Sitter in the flat slicing can be constructed using a Glauber-Sudarshan state in M-theory. 

Among other novel results, we discuss the smallness of the positive cosmological constant, including the curious case where the cosmological constant is very slowly varying with time. We also discuss the resolution of identity with the Glauber-Sudarshan states, generation and the convergence properties of the non-perturbative and the non-local effects, the problems with the static patch and other related topics. We analyze briefly the issues related to the compatibility of the Wilsonian effective action with Borel resummations and discuss how they influence the effective field theory description in a four-dimensional de Sitter space.}
\end{abstract}

\maketitle


\section{Introduction  \label{secintro}}

An important question in theoretical physics is to understand whether our four dimensional observable universe can be described using a UV complete description. Over the last few decades string theory/M-theory has provided penetrating insights into important questions in quantum gravity. The theory is consistent in ten/eleven dimensions, and we obtain a four dimensional universe upon compactification. In this light, we can ask whether upon string compactification, is it possible to obtain a four dimensional universe with a positive cosmological constant or not.

The Gibbons-Maldacena-Nunez no-go theorem \cite{GMN} has played a guiding light in searching for such solutions. By circumventing the no-go theorem, various works have provided a seemingly positive answer to the above question by constructing meta-stable de Sitter like vacua in string theory which involve lifting from a scale-separated AdS space \cite{kklt}. However it has been pointed out that there can be various issues with such meta-stable realization of de Sitter vacua \cite{whatif, issues, dologoto, swampland}.  

In our work, we attempt to address the topic by focussing on a rather simple question: can the issues associated with the construction of a vacuum de Sitter solution be circumvented if we search for a de Sitter {\it excited state} instead? Regarding this, a very conservative criterion would be to reproduce the quantum analogue of the equations of motion such as the Einstein equations, {\it i.e.} the Schwinger-Dyson equations, for the on-shell degrees of freedom describing four-dimensional de Sitter spacetime. 

The goal of our present work is two-fold. The \textit{first objective} is to lay down precise effective field theory criteria to obtain four dimensional de Sitter spacetime in type IIB string theory starting with M-theory, and compactifying over an internal eight-manifold. We lay down our criteria in terms of the Schwinger-Dyson equations from the associated path integral description. As we show, the criteria are readily satisfied by a certain class of excited states called Glauber-Sudarshan states, but not by the vacuum or by arbitrary excited states. The \textit{second objective} is to precisely understand the implications of the observables computed over the Glauber-Sudarshan states. In order to overcome the Gibbons-Maldacena-Nunez no-go theorem we need to take non-perturbative contributions to such observables, which requires an analysis of the resurgent structure of the associated path integral.

\vskip.1in

\noindent \textbf{A reading guide:} For the readers' convenience, the key results of the paper are summarized in \S \ref{subb1} as well as in {\bf Table \ref{Tablee1}}. Following that a partial list of new results derived in this paper is listed in \S \ref{newresults}. The outline of the paper appears in section \ref{outline} with appropriate links. 
We pedagogically discuss the Glauber-Sudarshan state in M-theory in \S \ref{sec2.2.2} and \S \ref{sec2.2.3}. The rest of the work deals with understanding the consequences, building upto the EFT analysis in \S \ref{sec7}.
In the conclusion section \ref{sec8.1}, we have a slightly non-technical longer discussion in a question and answer format which summarizes the results from this paper exhaustively without going into technical details.
In {\bf Table \ref{firzacut3}} we compare results from a vacuum configuration and a GS state.

\subsection{Brief summary of the key results}
\label{subb1}

\begin{table}[h]
\begin{center}
  \begin{tabular}{ |c | c|} 
  \hline
  
  &\bf{Four-dimensional de Sitter in flat slicing}
  \\
  &\bf{within the framework of M-theory}
  \\
  \hline
 
   \bf{Metric} & $ds^2 = g_s^{-8/3}\left(-dt^2 + dx_1^2 + dx_2^2\right) + g_s^{-2/3}{\rm H}^2(y) g_{mn}(y) dy^m dy^n + g_s^{4/3} \delta_{ab} dw^a dw^b$, \\
   (M-theory) & on ${\bf R}^{2,1}\times {\cal M}_6 \times \dfrac{\mathbb{T}^2}{{\cal G}}$ (see \eqref{duipasea2})\\
    & $g_s$ is the Type IIA string coupling. \\

    &Compactification to 4d dS: Take $g_s \to 0$, and T-duality along $x_3$.  \\
  \hline
  \bf{Path Integral}& Total action $\hat{\bf S}_{\rm tot}\lc {\bf g}_{\rm AB}, {\bf C}_{\rm ABD} \rc$ includes all possible perturbative terms,  \\
  & with specific non-perturbative and non-local contributions, as given in \eqref{kaittami4}.\\
  &Path integral description utilizes the action $\hat{\bf S}_{\rm tot}\lc {\bf g}_{\rm AB}, {\bf C}_{\rm ABD} \rc$ from \eqref{katuli} \\
  & with the supersymmetric vacuum $\ket{\Omega}$ defined over ${\bf R}^{2,1}\times {\cal M}_6 \times \dfrac{\mathbb{T}^2}{{\cal G}}$ (see \eqref{dunbar}). \\
  \hline
   \bf{EFT Criteria} & {\bf 1.} Restriction of time dependence of Type IIA string coupling:\\
    on action &  $g_s(t) \ll 1, \, \, \, \, $ and $ \, \, \,\, \dfrac{\partial g_s(t)}{ \partial t} \propto g_s^{k} \, \, (k>0)$, (see \eqref{atryan})\\
    for four-dim dS & {\bf 2.} Satisfying the Schwinger Dyson equation for 11-dim fields: \\
     & $\dfrac{\delta \hat{\bf S}_{\rm tot}\lc \langle{ {\bf g}_{\rm AB}}\rangle_\sigma, \langle{ {\bf C}_{\rm ABD}}\rangle_\sigma \rc}{ \delta\langle{ {\bf g}_{\rm AB}}\rangle_\sigma}  =   \dfrac{\delta \hat{\bf S}_{\rm tot}\lc \langle{ {\bf g}_{\rm AB}}\rangle_\sigma, \langle{ {\bf C}_{\rm ABD}}\rangle_\sigma \rc}{ \delta\langle{{\bf C}_{\rm ABD}}\rangle_\sigma} =0$, (see \eqref{trancom}) \\
  \hline
  \bf{Solutions} & The vacuum $\ket{\Omega}$ itself does not satisfy EFT criteria.\\
 & Glauber-Sudarshan states satisfy the above EFT criteria, defined as \\
  & $\vert\sigma\rangle = \mathbb{D}\lc \sigma, {\bf g}_{\rm AB}, {\bf C}_{\rm ABD}\rc \ket{\Omega} $ \\
  & where $\mathbb{D}(\sigma, {\bf g}_{\rm AB}, {\bf C}_{\rm ABD})$ is a {\it displacement} operator for the GS state given in \eqref{28rms}.\\
  & The IIB 4d metric appears from $\langle{\bf g}_{\mu \nu}\rangle_\sigma={\langle \sigma |{\bf g}_{\mu \nu}|\sigma \rangle/ \langle\sigma\vert\sigma\rangle}$, {\it i.e.} \eqref{ripley}, as \eqref{engrose}. \\  
  \hline
\end{tabular} 
  \caption{\label{Tablee1} An overview of the EFT criteria. Here, in our EFT criteria, for representation purposes we have only included the bosonic sector.}
\end{center}
\end{table}

 To pose the effective field theory (EFT) criteria, we work with M-theory on a supersymmetric background with the topology ${\bf R}^{2,1}\times {\cal M}_6 \times {\mathbb{T}^2\over {\cal G}}$, where ${\cal M}_6 $ denotes a six-dimensional  internal manifold while ${\cal G}$ is an orbifold action without a fixed point. Here as conventional, one direction of $\mathbb{T}^2$ denotes the M-theory circle, while a T-duality along the other direction combined with ${\bf R}^{2,1}$ gives a four dimensional universe upon compactification in type IIB theory. Alternatively, using this setup, we can simply attempt to lift a four-dimensional de Sitter space in type IIB in the flat slicing to a eleven-dimensional configuration. To systematically do this, we need to carefully specify the dependence of the eleven-dimensional metric on the dual type IIA string coupling $g_s$, which is a function of the flat-slicing (conformal) temporal coordinate as does the volume of ${\cal M}_6 $. 

 We write down the most general perturbative action ${\bf Q}_{\rm pert}({\bf \Xi})$ in \eqref{QT3} at far IR using the set of on-shell fields $\{{\bf \Xi}\}$, where the set goes over the G-flux ${\bf G}_{\rm ABCD}$ and the metric ${\bf g}_{\rm AB}$ components with $({\rm A, B}) \in {\bf R}^{2,1}\times {\cal M}_6 \times {\mathbb{T}^2\over {\cal G}}$. (We will leave the fermionic degrees of freedom for later work.) Here ${\bf Q}_{\rm pert}({\bf \Xi})$ incorporates all possible higher derivative terms corresponding to the bosonic fields. The coefficients of the higher derivative terms 
 systematically carry information about the full UV theory in the sense of an exact renormalization group (ERG) approach.

However, from the Gibbons-Maldacena-Nunez theorem, a simple scaling analysis tells us that accounting solely from perturbative corrections does not allow us to create a de Sitter solution. Consequently, we must take into account the full action $\hat{\bf S}_{\rm tot} ({\bf \Xi})$, in \eqref{kaittami4} which incorporates the perturbative series  ${\bf Q}_{\rm pert}({\bf \Xi})$ from \eqref{QT3} in the full non-perturbative and non-local interactions. An example of a non-perturbative contribution is given by wrapping five-brane instantons on the manifold $\mathcal{M}_6$. 

The generic interacting theory on the supersymmetric backgound on ${\bf R}^{2,1}\times {\cal M}_6 \times {\mathbb{T}^2\over {\cal G}}$ provides us with an interacting vacuum state $\ket{\Omega}$. A state in our theory, called the Glauber-Sudarshan state and denoted by $\ket{\sigma}$, is obtained by acting with operator insertions over the interacting vacuuum. More precisely, $\ket{\sigma}= {\mathbb D}_{\sigma}({\bf \Xi})\ket{\Omega}$, where ${\mathbb D}_{\sigma}({\bf \Xi})$ denotes the displacement operator with a specific insertions of the on-shell fields from the set $\{{\bf \Xi}\}$ as in \eqref{28rms}.

With all this in place, we now outline the key results related to the first objective, which are also summarized in {\bf Table \ref{Tablee1}}. In order to obtain a consistent four-dimensional description, the EFT criteria involve satisfying the Schwinger-Dyson equations for the on-shell fields, along with imposing  positivity conditions on Type IIA string coupling $g_s$. The EFT criteria are not satisfied by the vacuum state or by a generic coherent state. However we show that a very specific class of excited states labelled by Glauber-Sudarshan (GS) states satisfy the aforementioned criteria. 

We now describe some key results obtained within our second objective, i.e. implications of the Glauber-Sudarshan states. Over a Glauber-Sudarshan state, we show that the series expansion of perturbative action leads to an asymptotic series with a specific factorial growth. In literature, this specific form of the asymptotic series is known as the Gevrey series. A Borel resummation of the Gevrey series leads to a resurgent trans-series from where one could predict the non-perturbative terms in the effective action. For the Glauber-Sudarshan states, the non-perturbative term in $\hat{\bf S}_{\rm tot} ({\bf \Xi})$ can be thus be recovered this way leading us to the final form as given in \eqref{kaittami4}. This allows us to determine the complete non-perturbative, as well as the non-local, contributions to the Schwinger-Dyson equations as shown in \eqref{tranishonal3}. The non-local terms, which appear from integrating out the massless off-shell degrees of freedom are also shown to take a trans-series form.
 Lastly, the Borel resummation of the Gevrey series can be used to understand the smallness of the positive cosmological constant, including the curious case of a very slowly varying cosmological constant. The latter also leads to a consistent Glauber-Sudarshan state which we illustrate here in some details.


\subsection{What are the new results in this paper? \label{newresults}}

Apart from the key results discussed above, we now provide a quick glance at the other results in our work. These are important towards concretely establishing the central results in {\bf Table \ref{Tablee1}} as well as opening up avenues for further research.

\vskip.1in

\noindent $\bullet$ A new form for the displacement operator which is taken here to be complex instead of real.

\vskip.1in

\noindent $\bullet$ Resolution of identity with the Glauber-Sudarshan states.

\vskip.1in

\noindent $\bullet$  The factorial growths of the nodal diagrams and the corresponding amplitudes.

\vskip.1in

\noindent $\bullet$ Possible reasons for the Gevrey growth, and subsequent Borel resummation.

\vskip.1in

\noindent $\bullet$ Generation of non-local quantum terms by integrating out the massless off-shell states.

\vskip.1in

\noindent $\bullet$  Generation of the non-perturbative quantum effects by summing over the instanton saddles.

\vskip.1in

\noindent $\bullet$ Convergence properties of all the quantum terms in the Borel resummed effective action.

\vskip.1in

\noindent $\bullet$ Elucidating the differences between the Glauber-Sudarshan states and the coherent states.

\vskip.1in

\noindent $\bullet$ Reason for the smallness of the positive cosmological constant.

\vskip.1in

\noindent $\bullet$ Constructing a Glauber-Sudarshan state associated to a slowly varying cosmological constant.

\vskip.1in

\noindent $\bullet$ Proof and the derivation of the equations of motion as Schwinger-Dyson equations.

\vskip.1in

\noindent $\bullet$ Computing the emergent potential and satisfying the Effective Field Theory (EFT) criteria.

\vskip.1in

\noindent $\bullet$ The problems and the deceptive dynamics with the usage of the de Sitter static patch.

\vskip.1in

\noindent $\bullet$ Studying the compatibility of the Wilsonian effective action with Borel resummed action.

\subsection{Organization of the paper and summary \label{outline}}

The paper is organized in the following way. In \S \ref{sec2} we analyze the reasons why would a four-dimensional de Sitter spacetime {\it cannot} exist as a vacuum solution in string theory. Our analysis is in type IIB but the problems are generic. In \S \ref{sec2.1} we study this from a M-theory uplift, with special emphasis on the vacuum EOMs in \S \ref{analysis} and including the quantum corrections in \S \ref{sec2.1.3}. In \S \ref{sec2.1.4} we discuss how {\it narrowly} a de Sitter spacetime misses being a vacuum solution. This section partially constitutes a review of the failure of a  de Sitter vacua in string theory as a follow-up to \cite{whatif}, but put in a different light.

Having discussed the failure, our aim in \S \ref{sec3} is to realize this as a Glauber-Sudarshan state in string theory. We start in section \ref{sec3.1} by pointing out issues in the static patch, where one might have thought of realizing a de Sitter spacetime, by including the problems and the subtleties coming from the Wilsonian integrating-out procedure itself. The way out is to realize de Sitter spacetime as an excited states, and in \S \ref{sec2.2.2} we point out why we couldn't realize it as a {\it coherent} state, but only as a Glauber-Sudarshan state. In \S \ref{sec2.2.3} we quantify the displacement operator and using this show how the de Sitter metric appears from a path-integral formulation as an {\it emergent} quantity.

The path-integral formulation has its own set of subtleties which we point out in \S \ref{patho}. In \S \ref{sec3.4} we argue that the path-integral \eqref{mmtarfox} can be rewritten as a binomial expansion, using which we show in \S \ref{sec4.2} how one may determine the nested structure and how one may actually compute it. The binomial expansion form also helps us to see the factorial growth much more clearly which we discuss in \S \ref{factoria}. In fact using this we can compute the one-point functions in a more efficient way. The special asymptotic behavior appearing from the path-integral leads to the so-called Gevrey series on which we can do the Borel resummation. This is detailed in \S \ref{sec3.7}, wherein we also discuss the origin of the non-perturbative effects. 

The Borel resummation actually leads, for a certain well-defined toy set-up with scalar degrees of freedom, to an actual determination of the cosmological constant. This and other details are the subject of \S \ref{smallcc}. In \S \ref{sec3.8} we combine the ideas from the binomial expansion and the Borel resummation to provide a {\it closed form} expression for the four-dimensional cosmological constant in 
\eqref{tinmey}. This was already argued in \cite{borel} to be positive definite, but it wasn't clear from there that it could be made {\it small}. In \S \ref{sec3.9} we discuss procedures how to make the four-dimensional cosmological constant small. Our procedures include introducing new computational tools called the {\it Borel Boxes}, as well as finding ways to allow for convergent summing over the Borel Boxes, which we describe in some details in \S \ref{sec3.9}. The final expression after summing over a subset of Borel Boxes is given by \eqref{palazzome}, which does show considerable decrease if we carefully take the degeneracies into account. However the analysis is technically challenging and is not clear whether the decrease could be as small as 
$10^{-120} {\rm M}_p^2$. We leave a detailed study on this for an upcoming work \cite{ccpaper}. Finally in \S \ref{sec5.3} we study the curious case of the cosmological constant slowly varying with time. 
In the limit where the variation is very slow, we argue that this may be easily accommodated in our set-up by constructing a new Glauber-Sudarshan state. It is possible that this might have some connection to the recent DESI BAO result \cite{desibao}, but we do not make any detailed elaboration here and leave this for future work.

The path-integral formulation in \S \ref{patho} provides an emergent metric configuration, but doesn't provide a detailed connection to other emergent degrees of freedom like fluxes and fermionic condensates. Such a connection appears from the Schwinger-Dyson equations which may also be derived from the path-integral. Our aim in \S \ref{sec4} is the derivation of the Schwinger-Dyson equations and study the consequences. These equations are specifically designed to provide the dynamics of the emergent degrees of freedom, but are in general harder to implement in real practice. This improves drastically once we determine the resolution of identity for the Glauber-Sudarshan states as we show in \S \ref{sec4.1}. The resolution of identity allows us to rewrite the Schwinger-Dyson equations in a more efficient way which, in turn, leads us to determine the exact reason for the failure of the vacuum solution from \S \ref{sec2}. In addition to that, it provides a way to incorporate the non-local and the non-perturbative effects much more clearly. We discuss this in \S \ref{sec6.2.2} wherein we show two things: one, how the Borel resummation of the Gevrey series introduces the non-perturbative effects and two, how by integrating out the massless off-shell states introduces the non-local effects. For the latter, to get a convergent form, one needs a somewhat more detailed mathematical analysis which we elaborate in \S \ref{mathy}. The final convergent forms for both the non-perturbative and the non-local effects are useful but raises a different question related to the compatibility of the Wilsonian effective action with Borel resummation. We discuss this briefly in \S \ref{sec4.20}, but leave a more detailed study for future work. Finally in \S \ref{sec6.3} we show that the Schwinger-Dyson equations lead to two set of equations in the presence of Faddeev-Popov ghosts and gauge fixing terms. One set of equation, given by the first equation in \eqref{eventhorizon}, is somewhat similar to the EOMs that we studied in \S \ref{sec2} in the sense that they are almost the same EOMs except the fields are replaced by their expectation values. However the second set of equations are new. These set of equations, coming from the second equation in \eqref{eventhorizon} and the two other equations from \eqref{necampbel}, signify the {\it deviations} from the vacuum configuration.  

One consequence of our study of path-integrals, Borel resummations and the Schwinger-Dyson equations is the construction of the two EFT criteria given by \eqref{atryan}, that we discuss in \S \ref{sec7}. Another consequence of our analysis, and especially from the study of the Schwinger-Dyson equations, is the construction of an {\it emergent} potential which is different from the vacuum potential. We study this in \S \ref{sec7.1} and show that, using this emergent potential, the EOMs for the expectation values of the on-shell degrees of freedom take a simpler form as in \eqref{alto2mei}. Moreover, since the expectation values are never at the minima of the potential $-$ plus the potential being an emergent one $-$ they are not subjected to the EFT criteria of \cite{swampland}. The first EFT criterion of \eqref{atryan} then restricts the dynamics to lie within the temporal domain of \eqref{tcc}, and in \S \ref{sec7.2} we argue that this criterion is violated in the static patch leading to non-convergence of the instanton series. The second criterion of \eqref{atryan} is deeply rooted in the quantum series expressed using  \eqref{QT3}, and is determined from actual computations of wrapped five-brane instantons. \S \ref{sec7.3} is dedicated to the study of these instantonic effects. We specifically take only five-brane instantons to illustrate this, dedicating \S \ref{sec7.3.1} and \S \ref{sec7.3.2} to the BBS instanons \cite{bbs} and \S \ref{sec7.3.3} for the KKLT instantons \cite{kklt}. The study of the BBS instantons in \S \ref{sec7.3.1} however reveals the incompleteness of the non-local interactions that we had derived in \S \ref{sec6.2.2}, which we rectify in \S \ref{sec7.3.2}. Once all corrections are made, the effective low energy action takes the form as in \eqref{kaittami4}, and the corresponding Schwinger-Dyson equations become \eqref{tranishonal3}. 
In \S \ref{sec7.3.3} we study the KKLT instantons and discuss the special case of a temporally varying cosmological constant $-$ that we studied in \S \ref{sec5.3} $-$ and show how our Schwinger-Dyson equations may easily accommodate such a scenario. 

In \S \ref{sec8.1}, we take the opportunity to give a detailed non-technical summary of our work in a question-and-answer format. There is a dual motive for such a format: not only is our aim to answer questions justifying the validity of the Glauber-Sudarshan states in providing a positive cosmological constant background, but also to summarize some of the earlier related works that are sometimes difficult to access due to their overly technical natures. We conclude with a technical summary in \S \ref{sec8} and point out some of the future directions and open questions. In Appendix \ref{append} we list various coordinates patches that one puts to study four-dimensional de Sitter spacetime.

\section{de Sitter space as a vacuum solution in string theory? \label{sec2}}

To see why we would like to view four-dimensional de Sitter space-time as a coherent state, or more appropriately as a Glauber-Sudarshan state\footnote{The distinction between coherent states and Glauber-Sudarshan states will be spelled out in section \ref{sec2.2.2}.}, we need to take a step back and start by asking some basic questions. With this in mind, let us take the following four-dimensional action:
\bg\label{einstein1}
{\rm S} = \int d^4x \sqrt{-{\bf g}_4}\left[\frac{1}{16 \pi G_{\rm N}}\left({\bf R}_4 - \frac{\Lambda}{2}\right) + {1\over 4g_{\rm YM}^2} {\bf F} \wedge \ast{\bf F} + .... \right], \nd
where $G_{\rm N}$ is the four-dimensional Newton's constant, $g_{\rm YM}$ is the gauge coupling for gauge field ${\bf F}$, $\Lambda$ is the positive cosmological constant, and the dotted terms denote higher order interactions. 

If \eqref{einstein1} is considered, then there is {\it nothing} to show! The theory admits, in four-dimension, a positive cosmological constant solution which we can identify with the de Sitter space-time. The issue here is the middle term that is proportional to $\sqrt{-{\bf g}_4} \Lambda$: string theory or M-theory {\it does not} come equipped with a term like this! The question then is the following:
Can we show that this term comes out from dimensional reduction of a ten or eleven dimensional action? In other words, can the following action in (say) eleven-dimensions:
\bg\label{einstein2}
{\rm S} = {\rm M}_p^{9}\int d^{11}x \sqrt{-{\bf g}_{11}}\left({\bf R}_{11} + {\bf G}_4 \wedge \ast {\bf G}_4\right) + {\rm M}_p^9\int {\bf C}_3 \wedge {\bf G}_4 \wedge {\bf G}_4 + {{\rm M}_p^3}\int {\bf C}_3 \wedge \mathbb{X}_8 + ..., \nd
reproduce the action \eqref{einstein1} by some dimensional reduction to four-dimensions? Here we denote the three-form flux as ${\bf C}_3$ with ${\bf G}_4 = d{\bf C}_3 + ...$, and $\mathbb{X}_8$ is a fourth order curvature polynomial. Note that \eqref{einstein2} has no scale other than ${\rm M}_p$ and the {\it size} of the internal manifold; and therefore both $G_{\rm N}$ and $g^2_{\rm YM}$ in \eqref{einstein1} should come only from the two aforementioned scales.  

The challenge is to reproduce the four-dimensional cosmological constant term. Clearly since $\Lambda$ does not explicitly show up in \eqref{einstein2}, it must appear from the flux and the dotted terms in \eqref{einstein2} which we, for our purpose, package them as the energy-momentum tensor ${\bf T}_{\rm MN}$. The condition to get a positive cosmological constant solution is simply the statement that \cite{GMN}:
\bg\label{cc}
\mathbb{T}^\mu_\mu ~ > ~ {4\over 5} \mathbb{T}^m_m, \nd
where $x^\mu$ is the coordinate of four-dimensional space-time and $y^m$ is the coordinate of the internal seven-dimensional compact manifold. If \eqref{cc} is satisfied then we can get a four-dimensional de Sitter space-time. Unfortunately, as shown in \cite{GMN}, this simple condition {\it cannot} be satisfied by any combination of fluxes, branes, anti-branes, and orientifold planes. Somewhat surprisingly, even perturbative quantum correction (to any order) fails to satisfy \eqref{cc} \cite{GMN}. Why is this the case? Where do all the energies from fluxes, branes, planes etc. go, and why do they fail to give positive energy to the space-time? A more puzzling question is regarding the perturbative corrections. Why aren't they effective in lifting the energy of space-time to some positive value? However even if perturbative corrections were effective, why is the cosmological constant, which measures the {\it amount} of positive energy, so small? 

We have already raised too many questions, so in the following we will try to answer at least a subset of them. The first one, related to the energies of the classical sources like fluxes, branes and planes, is relatively easy. The answer lies in the strange property of gravity itself: the back-reactions of the classical sources not only curve the background, but also create {\it negative} potentials. Thus all the positive energies from the classical sources are basically nullified by the negative potentials created by the back-reactions of them on space-time. This is also the reason why negative cosmological constant solutions are so ubiquitous in string theory: it doesn't require much effort to create an effective negative potential using classical sources. 

This means, to create an effective positive energy we require some kind of {\it sources} that do not back-react so strongly on the space-time and yet still provide enough positive energies to make the net effect positive. Clearly since classical sources cannot do this, the sole burden now falls on quantum corrections. Herein, and as alluded to earlier, lies a surprise: the perturbative quantum corrections fail to do this. Why is this the case?
Could non-perturbative corrections save the day? If yes, does this mean that de Sitter space can exist as a vacuum solution in the presence of non-perturbative corrections?

To start answering the aforementioned questions, let us first quantify some details. Instead of finding a four-dimensional de Sitter space in M-theory, we will look for such a space-time in type IIB. One specific ans\"atze for the metric configuration is the following:
\bg\label{metansatze}
ds^2 = {1\over {\rm H}^2(y)\Lambda \vert t\vert^2}(-dt^2 + dx_1^2 + dx_2^2 + dx_3^2) + {\rm H}^2(y) g_{mn}(y) dy^m dy^n, \nd
which would give us a four-dimensional de Sitter space-time in a flat-slicing with the conformal time $t$ ranging as $-\infty < t < 0$ as shown in {\bf figure \ref{planar}}; and
where ${\rm H}(y)$ is the warp-factor with $g_{mn}(y)$ being the metric on the six-dimensional internal space that is generically a non-K\"ahler manifold (which may or may not be complex). One could however raise a question on the form of the metric \eqref{metansatze} itself: could we even express the background metric in the above form? The answer is more clear from a M-theory uplift, where we will show that a description like \eqref{metansatze} is only valid in a limited temporal domain.

\subsection{The M-theory uplift of the IIB background \label{sec2.1}}

\begin{figure}[h]
    \centering
\begin{tikzpicture}[scale=1.5]
    \draw (5,5) -- (5,0) -- (0,5) -- (5,5);
    \draw[blue] (0,5) .. controls (2.7,3.7) .. (5,3.3);
    \draw[dashed] (0,5) .. controls (3.5,2.5) .. (5,2);
    \draw[fill] (3,5)  node[above] { $t =0$ };
    \draw[fill] (3,2.9) node [rotate=-25,above] {$t_c = -\frac{1}{\sqrt{\Lambda}}$ };
    \draw[fill] (3.4,3.7) node [rotate=-10,above] {$t = $ constant};
    \draw[fill] (2.5,2.5) node [rotate=-45,below] {$t = -\infty$};
\end{tikzpicture} 
    \label{potential}
 \caption{Flat slicing, with a constant Poincare time slice denoted in blue and $t$ being the conformal time. Here $t_c$ denotes the TCC cutoff and the relevant part of the diagram is the interval \eqref{tcc}.}
    \label{planar}
\end{figure}
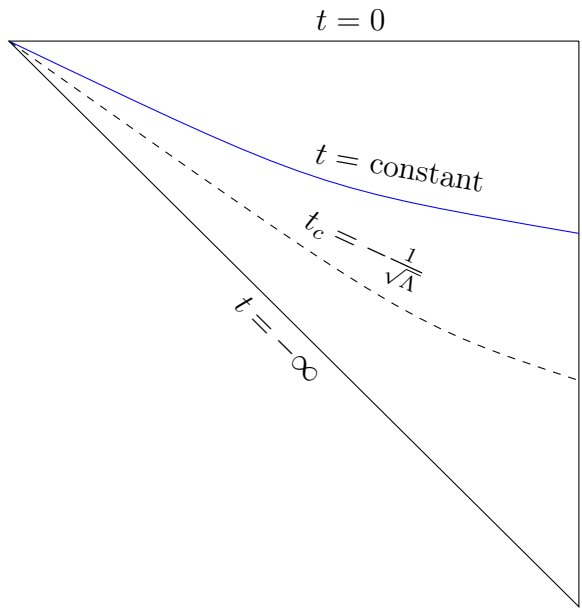

As mentioned above, somewhat surprisingly such a configuration is dual to M-theory on a compact eight-manifold\footnote{And not a compact seven-manifold!}, which is a torus fibered over a six-dimensional base, implying that we can continue using the action \eqref{einstein2} \cite{DRS}. The step to reach M-theory is to first T-dualize along the spatial $x_3$ direction of \eqref{metansatze}, and then uplift the resulting configuration to eleven-dimensions. T-dualizing along $x_3$ leads to a time-dependent type IIA coupling $g_s \equiv {\rm H}(y) \sqrt{\Lambda}\vert t \vert$. The coupling ${g_s\over {\rm H}(y)}$ remains smaller than 1, {\it i.e.} ${g_s \over {\rm H}(y)} < 1$, in the interval:
\bg\label{tcc}
-{1\over \sqrt{\Lambda}} ~ < ~  t ~ < ~ 0, \nd
which is surprisingly the trans-Planckian bound advocated in \cite{tcc}! Thus the de Sitter space-time is automatically defined within the allowed temporal domain governed by the so-called TCC (see also {\bf figure \ref{planar}}). One might however ask why is this the case. For example, lifting to M-theory means we are automatically going to strong type IIA coupling, so shouldn't this be an issue here?  

The answer is actually no because of the contributions from the non-perturbative and the non-local counter-terms. In fact, and as mentioned earlier, it is the non-perturbative and the non-local quantum terms that are actually responsible in generating the kind of background we want. (The perturbative terms are just red-herrings in the problem.) Both these quantum terms go as ${\rm exp}\left(-{{\rm M}_p^2\over g_s^2}\right)$ and therefore if we keep $g_s$ strong then it will be difficult to allow for a convergent series of these quantum effects. This means, while M-theory is defined at strong type IIA coupling, we want to be in the {\it opposite} limit, {\it i.e.} in the limit when ${g_s\over {\rm H}(y)} < 1$. 
(A more elaborate analysis of the aforementioned statement has appeared in \cite{desitter2, coherbeta, coherbeta2} which the readers may look up for more details.)

Before moving further, let us quickly clarify one question that may arise at this stage: why go to M-theory? Can we not do the analysis directly in the type IIB set-up? The answer is unfortunately no because the type IIB background with the metric \eqref{metansatze} is actually at the constant coupling limit of F-theory \cite{senkd} and therefore the IIB coupling $g_s^{(b)} = 1$ where even S-duality does not help. Going away from this point would switch on time-dependent dilaton and axion (because of the dynamical seven-branes) making the analysis even harder to track. (Additionally, we really do not have a well defined Lagrangian description for the type IIB case.) In fact we will soon see that keeping the internal manifold and the fluxes time-independent will become problematic. There will also more severe problems associated with the very {\it existence} of the vacuum solution \eqref{metansatze} itself that we will encounter soon. Such a conclusion will be borne out from a careful analysis of the system, but for the time-being we shall suffice with the  metric ans\"atze \eqref{metansatze}.

Assuming this to be the case, we can now lift\footnote{The {\it lifting} from type IIB involves at least one T-duality, so one needs to take into account all the $\alpha'$ corrections to the Buscher's T-duality rules. This is surprisingly tractable as shown in the upcoming work \cite{hete8}, but we will avoid these subtle nuances here. \label{petbeta}} the metric \eqref{metansatze} to eleven dimensions by using the type IIA string coupling $g_s \equiv {\rm H}(y) \sqrt{\Lambda} \vert t \vert$ in the following way:
\bg\label{mmetric}
ds^2 = g_s^{-8/3}\left(-dt^2 + dx_1^2 + dx_2^2\right) + g_s^{-2/3}{\rm H}^2(y) g_{mn}(y) dy^m dy^n + g_s^{4/3} \delta_{ab} dw^a dw^b, \nd
where $(w^a, w^b) = (x_3, x_{11}) \in {\mathbb{T}^2\over {\cal G}}$ with ${\cal G}$ being an orbifold action without a fixed point; and $(y^m, y^n) \in {\cal M}_6$ such that the internal eight-manifold is locally ${\cal M}_6 \times {\mathbb{T}^2\over {\cal G}}$. The warp-factor ${\rm H}(y)$ is assumed to be smooth over the six-manifold ${\cal M}_6$. Question is whether \eqref{mmetric}, or \eqref{metansatze}, solves the background EOMs? 

When we talk of EOMs in supergravity we assume that we are at low energies where all massive stringy and KK degrees of freedom have been integrated away, and the theory is completely specified by a small subset of known low energy degrees of freedom. In other words we want the EOMs of a low energy Wilsonian effective action whose solution is \eqref{mmetric}. Note that we have made at least two crucial assumptions:

\vskip.1in

\noindent $\bullet$ Existence of a well-defined Wilsonian effective action for the finite set of low energy degrees of freedom, and

\vskip.1in

\noindent $\bullet$ Existence of a supergravity limit where we can consistently truncate any quantum series to a finite subset.

\subsection{Analysis of the EOMs and consistency conditions \label{analysis}}

\noindent We shall see below that both the above assumptions are problematic, and in particular it looks unlikely for the second assumption to be true. The reasoning is simple: even if we are at weak type IIA coupling, the curvature tensors can become very strong in the temporal domain \eqref{tcc} and specifically when we approach late time where $g_s \to 0$. There is fortunately a resolution to the above conundrum of strong curvature \cite{hete8}, but even then the first problem could actually rule out our whole procedure itself! Unfortunately however these are not the only issues. There are other difficulties that we will face soon when we start writing the EOMs themselves, {\it assuming of course that it is allowed to do so}. 

We will then start with a {\it narrow minded} approach wherein we will assume that the aforementioned two points are true: namely, there exists a Wilsonian effective action that allows for a well-defined truncated supergravity limit. Can \eqref{mmetric}, or \eqref{metansatze}, be a solution to such a system? 

To see this we will have to write all possible EOMs allowed by the system from say eleven-dimensional point of view. This is a formidable exercise, but can be done with some efforts. We have already proposed a metric ans\"atze in \eqref{mmetric} with dominant $g_s$ behavior, and we take the following ans\"atze for the G-flux components:
\bg\label{gfluxes}
{\bf G}_{\rm MNPQ}({\bf x}, y; g_s) = \sum_{k = 0}^\infty {\cal G}^{(k)}_{\rm MNPQ}(x, y) \left({g_s\over {\rm H}(y)}\right)^{2k\over 3}, \nd
where $k \in {\mathbb{Z}\over 2}$, $({\rm M, N, P, Q}) \in {\bf R}^{2, 1} \times {\cal M}_6 \times {\mathbb{T}^2\over {\cal G}}$, ${\bf x} \in {\bf R}^2, y \in {\cal M}_6$ and $t = t(g_s)$ is the conformal time. Notice that, according to \eqref{gfluxes}, $k = 0$ constitute the time-independent components of the G-flux, and $k > 0$ are the sub-dominant contributions that control the temporal behavior of the G-flux components\footnote{The actual ans\"atze that works for all string theories descending from M-theory is a bit more involved as discussed in \cite{hete8}. But for the present exposition, these technicalities are not needed and therefore may be avoided. Interested readers may look up our upcoming paper \cite{hete8}.}. Plugging \eqref{mmetric} and \eqref{gfluxes} in the EOMs lead to a complicated set of coupled equations as shown in  
\cite{desitter2, coherbeta, coherbeta2} which can nevertheless be combined together to extract the following consistency condition:
\bg\label{consistency}
\begin{split}
& ~12\Lambda \int d^6y \sqrt{{\bf g}_6(y)}~ {\rm H}^8(y) + {5\over 8} \int d^6 y \sqrt{{\bf g}_6(y)}~{\cal G}^{(3/2)}_{mnab}(y) {\cal G}^{(3/2)mnab}(y)\\   
+ & ~2\kappa^2 {\rm T}_2\left(n_b + \overline{n}_b\right)
+ {1\over 2} \int d^6 y
\sqrt{{\bf g}_6(y)} ~{\rm H}^4(y) \Big(\left[\mathbb{C}^a_a(y)\right]^{(0)} + 
\left[\mathbb{C}^m_m(y)\right]^{(0)} - 2{\rm H}^4(y)\left[\mathbb{C}^i_i(y)\right]^{(0)}\Big)= 0,
\end{split} \nd
where $\Lambda$ is the four-dimensional cosmological constant, ${\rm T}_2$ is the tension of the 2-brane and the anti-2-brane (their number being $n_b$ and $\overline{n}_b$ respectively), $\kappa$ is a constant, ${\rm H}(y)$ is the {\it smooth} warp-factor appearing in \eqref{mmetric}, 
$\left[\mathbb{C}^{\rm M}_{\rm M}(y)\right]^{(k)}$ is the trace of the quantum term\footnote{The notation differs a bit from \cite{desitter2} because we will define the quantum series in a slightly different way to avoid repeating details from \cite{desitter2}.} (see \eqref{QT}), and the indices are raised or lowered by the {\it unwarped} metric components ({\it i.e.} the $g_s$ independent parts of the metric components).

The first term in \eqref{consistency} is a consequence of the backreactions from fluxes, branes and O-planes and is positive definite. The second and the third terms, which are also positive-definite, are from the G-fluxes and from branes and anti-branes (both integer and fractional) respectively, and the last term is from the quantum effects. Some of the results mentioned above can now be easily quantified from \eqref{consistency}.

\vskip.1in

\noindent $\bullet$ In the absence of quantum terms, there is no solution in the system for $\Lambda > 0$. If $\Lambda < 0$, solutions can easily exist. This shows that branes, anti-branes, O-planes and fluxes are simply red-herrings in the problem, {\it i.e.} they {\it cannot} give rise to positive $\Lambda$. 

\vskip.1in

\noindent $\bullet$ The time-independent part of the G-flux, namely ${\cal G}^{(0)}_{\rm MNPQ}$, does not enter \eqref{consistency}. In fact one can easily argue that the time-independent part of the G-flux components {\it cannot} solve the background EOMs. This issue does not go away if the use static-patch because of other factors plaguing the physics in a 
static-patch as we shall discuss later.

\vskip.1in

\noindent $\bullet$ Solution with $\Lambda > 0$ would appear to exist only in the presence of quantum terms and in particular only if the quantum terms satisfy the following inequality:
\bg\label{inequality}
\int d^6y \sqrt{{\bf g}_6(y)}~ {\rm H}^8(y)~\left[\mathbb{C}^i_i(y)\right]^{(0)} ~
> ~ {1\over 2} \int d^6y \sqrt{{\bf g}_6(y)}~ {\rm H}^4(y)\Big(\left[\mathbb{C}^a_a(y)\right]^{(0)} + \left[\mathbb{C}^m_m(y)\right]^{(0)}\Big), \nd
where $x^i \in {\bf R}^2, w^a \in {\mathbb{T}^2\over {\cal G}}$ and 
$y^m \in {\cal M}_6$. The condition \eqref{inequality}, while encouragingly reminiscent of the bound \eqref{cc} (but now completely in the language of quantum terms), actually creates more puzzles than resolving them. The first is the issue of hierarchy, which amounts to saying as to {\it how many} quantum terms should we take to justify \eqref{inequality}; and second is the unclear nature of \eqref{inequality}, which amounts to asking whether only perturbative, non-perturbative or non-local quantum terms, or a combination of all of them, would be required to maintain the inequality {\it in the temporal domain $-{1\over \sqrt{\Lambda}} < t < 0$}.

The last few points raised above are important and therefore let us clarify them in the following. From the discussion it is clear that only the quantum terms can possibly allow a $\Lambda > 0$ solution to exist in the system, but this immediately raises the following subtlety in 
\eqref{consistency}: the first three set of terms therein are classical, whereas the last set of terms are quantum. Thus somehow we are balancing the classical terms with the quantum terms, so we have to carefully specify what \eqref{consistency} entails. The quantum terms $\mathbb{Q}_{\rm T}$ are typically classified  as a series in
${g_s^a\over {\rm M}_p^b}$ perturbatively, or as a series in 
${\rm exp}\left(-{{\rm M}_p^b\over g_s^a}\right)$ non-perturbatively, 
so the contributions that actually enter \eqref{consistency} are the traces of the corresponding energy-momentum tensors {\it i.e.} 

{\footnotesize
\bg\label{QT}
\mathbb{C}^{\rm P}_{\rm M}(y, g_s) = {g}^{\rm PN}(y) {\del \mathbb{Q}_{\rm T}(y, [g_s])\over \del {\bf g}^{\rm MN}(y, g_s)} \equiv \sum_{k = 0}^\infty \sum\limits_{{}_{\{l_i\},\{n_j\}}}\left[\mathbb{C}^{\rm P}_{\rm M}(y)\right]^{(k, \{l_i\}, \{n_j\})}\left({g_s\over {\rm H}(y)}\right)^{\theta(\{l_i\},\{n_j\}) - \bar{a}^{\rm P}_{\rm M} + {2k\over 3}}, \nd}
where ${\bf g}_{\rm MN}(y, g_s)$ is the warped and $g_{\rm MN}(y)$ is the un-warped ({\it i.e.} $g_s$ independent part of the) metric components from \eqref{mmetric}; $\theta(\{l_i\}, \{n_j\})$ is the quantum scaling to be defined momentarily, and $\bar{a}^{\rm P}_{\rm M} \in {\mathbb{Z}\over 3}$ such that:
\bg\label{katya}
\left(\bar{a}^0_0, ~\bar{a}^i_i, ~\bar{a}^m_m, ~\bar{a}^a_a\right) = \left({8\over 3}, ~{8\over 3}, ~{2\over 3}, ~-{4\over 3}\right), \nd
which are clearly related to the $g_s$ exponents in \eqref{mmetric}. This is not surprising because the definition \eqref{QT} involves derivative with respect to the {\it warped} metric. Putting everything together, this means that the only way \eqref{consistency} would make sense is if the 
$g_s$ and ${\rm M}_p$ scalings of all terms match with each other. The $g_s$ and ${\rm M}_p$ scalings of the classical terms are easy to compute 
using \eqref{mmetric} and \eqref{gfluxes}, so the whole burden now lies in matching them to the $g_s$ and ${\rm M}_p$ scalings of the quantum pieces. The question now is following. How do we express the quantum series $\mathbb{Q}_{\rm T}$ succinctly to accomplish the aforementioned job? \textcolor{blue}{In the following sub-section we will analyze this in some details, but due to the technical nature of the content the readers, who would like to see the final answers, could skip it on their first reading}.

\subsection{Analysis of the quantum series and the Exact RG constraints \label{sec2.1.3}}

\begin{figure}[h]
    \centering
 \begin{tikzpicture}
\draw[thick,->] (0,-1) -- + (0,5.5) ;
\draw (-0.25,-0.9) --  (0.25,-0.9);
\node (x) at (-0.25,-0.9) [label=left:$k_{\text{IR}}$] {};
\node (x) at (0.25, -0.9) [label=right: IR cutoff] {};
\draw (-0.25,0) --  (0.25,0);
\node (x) at (-0.25,0) [label=left:$\mu$] {};
\node (x) at (0.25,0) [label=right:LHC cutoff] {};
\draw (-0.25,2) --  (0.25,2);
\node (x) at (-0.25,2) [label=left:$\hat{\mu}$] {};
\node (x) at (0.25,2) [label=right:KK scale] {};
\draw (-0.25,3) --  (0.25,3);
\node (x) at (-0.25,3) [label=left:${\rm M}_s$] {};
\node (x) at (0.25,3) [label=right:String scale] {};
\draw (-0.25,4) --  (0.25,4);
\node (x) at (-0.25,4) [label=left:${\rm M}_p$] {};
\node (x) at (0.25,4) [label=right:Planck scale] {};
\draw[pattern={north east lines},pattern color=ashgrey]
    (-0.20,-0.89) rectangle +(0.4,0.88);
    \end{tikzpicture} 
    \caption{Energy scales in the problem.}
    \label{scales}
\end{figure}
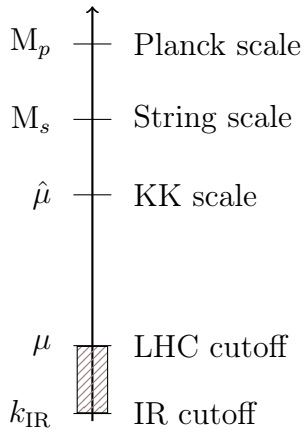

The quantum series $\mathbb{Q}_{\rm T}$ that we want to write should be expressed in terms of the {\it on-shell} degrees of freedom so as to appear in the EOMs, and therefore consequently in the constraint \eqref{consistency}. This means all the massless off-shell degrees of freedom will have to be integrated out giving rise to the {\it non-local} quantum terms. These are exactly the non-local counter-terms discussed in details in \cite{desitter2, coherbeta, coherbeta2}. (For a simpler derivation of this the readers may refer to section \ref{sec4.20} and our upcoming paper \cite{hete8}.) In fact, when we talk of the on-shell degrees of freedom, we have inherently assumed that we have some knowledge of the allowed degrees of freedom (DOFs) in M-theory. Since the UV behavior of M-theory is completely unknown, all we can say with certainty are the DOFs at the energy scale much below the KK scale $\hat\mu$. The various energy scales are shown in {\bf figure \ref{scales}}.

This means, as long as $k_{\rm IR} < k < \mu << \hat\mu$, where 
$k_{\rm IR}$ is the IR cut-off, we can confidently ascertain the DOFs. Such a scenario calls for a Wilsonian effective action (or more appropriately an action constructed from an Exact Renormalization Group procedure), wherein all the massive stringy, massless off-shell, as well as the KK DOFs have been integrated away, including the massive stringy, as well as the massless off-shell, DOFs on the branes. Thus we keep only the light states\footnote{We take the LHC cut-off as the scale of relevance once we integrate out all the massive stringy states as well as the KK states. However this could be a concern if some of the massive states become massless in the moduli space of the theory. Fortunately in a time-independent space-time with stabilized moduli this issue does not happen and therefore we can fix the LHC cut-off scale (denoted by $\mu$) as the only high-energy scale up to which the dynamics are governed by very light states. Once we take massive states like the standard model DOFs, the masses appear from the Higgs mechanism, so the massive stringy or the KK scales again play no part in the analysis and $\mu$ continues to be a good scale for us. \label{referee1}} in the energy regime $k_{\rm IR} < k < \mu << \hat\mu$. To provide a finer reiteration of the aforementioned claims: our analysis with the quantum series relies on two very important conditions.

\vskip.1in

\noindent {\bf 1}. Knowledge of the precise set of degrees of freedom in the energy scale $k$ where $k_{\rm IR} < k < \mu << \hat\mu$ with $\mu$ being the typical energy scale in a scattering experiment and $\hat\mu$ being the KK scale, and 

\vskip.1in

\noindent {\bf 2}. Existence of an Exact Renormalization Group effective action constructed from integrating out {\it all} the massive, as well as the massless off-shell, degrees of freedom.

\vskip.1in

\noindent Once the aforementioned two conditions are satisfied we can easily quantify the quantum series $\mathbb{Q}_{\rm T}$. The Exact Renormalization Group procedure now instructs us to keep all possible powers of the curvatures and the G-flux components, including all possible derivatives, to construct the perturbative series. The non-perturbative and the non-local terms may then be constructed by exponentiating these terms in appropriate ways as shown in \cite{desitter2, coherbeta, coherbeta2}. Putting everything together it is not too hard to see that generically there can at most be 60 on-shell curvature tensors and at most 40 on-shell G-flux components. (The fermionic extension can be easily inserted in, but we will not do so here. Interested readers may look up \cite{hete8, bergshoeff}.) This would mean that a typical quantum series may be presented succinctly using these on-shell tensors as the following series:

{\scriptsize
\bg\label{QT2}
\mathbb{Q}_{\rm T}({\bf x}, y, w, [g_s]) = \sum_{\{l_i\}, \{n_j\}} {{\cal C}_{\{l_i\}, \{n_j\}} \over {\rm M}_p^{\sigma_{nl}}} \left[{\bf g}^{-1}\right] \prod_{j = 0}^3 \left[\partial\right]^{n_j} \prod_{k = 1}^{60} \left({\bf R}_{\rm A_k B_k C_k D_k}\right)^{l_k} \prod_{p = 61}^{100} \left({\bf G}_{\rm A_p B_p C_p D_p}\right)^{l_p} + {\cal O}\left({\rm exp}(-{\rm M}_p^{\sigma'})\right), \nd}
where ${\cal C}_{\{l_i\}, \{n_j\}}$ are constants for a given values in the set $(\{l_i\}, \{n_j\})$; $(l_i, n_j) \in (+\mathbb{Z}, +\mathbb{Z})$;
$(n_0, n_1, n_2, n_3)$ are the number of derivatives along temporal, ${\bf R}^2, {\cal M}_6$ and ${\mathbb{T}^2\over {\cal G}}$ respectively; and 
$\sigma_{nl} \equiv \sum\limits_{j = 0}^3 n_j + 2\sum\limits_{k = 1}^{60} l_k + \sum\limits_{p = 61}^{100} l_p$ with $\sigma'$ being the scaling of a similar series \cite{desitter2, coherbeta, coherbeta2}. All the curvature and the flux tensors, including the inverse metric components are expressed in terms of their warped form, and therefore include the $g_s$ factors. The presence of 
$[g_s]$ in $\mathbb{Q}_{\rm T}({\bf x}, y, w, [g_s])$ signifies this, including $({\bf x}, y, w) \in ({\bf R}^2, {\cal M}_6, {\mathbb{T}^2\over {\cal G}})$ dependence generically. 

Despite the complicated nature of \eqref{QT2} one can, with some effort, work out the precise $g_s$ scalings at any orders in $(\{l_i\}, \{n_j\})$. This have been demonstrated in \cite{desitter2} for simpler cases and in \cite{coherbeta, coherbeta2} for more complicated cases. We will not repeat any of the details here $-$ the readers may look up \cite{desitter2, coherbeta, coherbeta2} $-$ and only present the form of the $g_s$ scalings of \eqref{QT2} in the following way:
\bg\label{gsscale}
\mathbb{Q}_{\rm T}({\bf x}, y, w, [g_s]) = \sum_{k = 0}^\infty \sum_{\{l_i\}, \{n_j\}} {{\cal C}_{\{l_i\}, \{n_j\}}\over {\rm M}_p^{\sigma_{nl}}}
\left({g_s\over {\rm H}(y){\rm H}_o({\bf x})}\right)^{\theta(\{l_i\}, \{n_j\}) + {2k\over 3}} + {\cal O}\left({\rm exp}(-{\rm M}_p^{\sigma'})\right), \nd
where ${\rm H}_o({\bf x})$ is a new warp-factor that only appears if we demand a dependence on the spatial coordinate ${\bf x} \in {\bf R}^2$. The $g_s$ scaling of the non-perturbative and the non-local terms, shown here simply as ${\cal O}\left({\rm exp}(-{\rm M}_p^{\sigma'})\right)$ is quite non-trivial and has been worked out in \cite{coherbeta, coherbeta2}. Taking  derivatives of this term with respect to the warped metric components, and then making it traceless using the unwarped metric components, introduce the factors $\bar{a}^{\rm P}_{\rm M}$ from \eqref{katya} in \eqref{QT}. The EOM constraint then tells us that at the lowest order, {\it i.e.} for $k = 0$ in \eqref{gsscale}, the following identification may be done:
\bg\label{identification}
\theta(\{l_i\}, \{n_j\}) - \bar{a}^{\rm M}_{\rm M} = \zeta^{\rm M}_{\rm M}, ~~~ \sigma_{nl} = \sigma_{\rm classical}, ~~~ \left[\mathbb{C}^{\rm M}_{\rm M}(y)\right]^{(0)} \equiv \sum\limits_{{}_{\{l_i\},\{n_j\}}}\left[\mathbb{C}^{\rm P}_{\rm M}(y)\right]^{(0, \{l_i\}, \{n_j\})}, \nd
where the repeated indices are not summed over and $\sigma_{\rm classical}$ is the ${\rm M}_p$ scaling of the classical pieces (we are ignoring $\sigma'$ to maintain the simplicity of the ensuing analysis). The set $(\{l_i\}, \{n_j\})$ is then the number of integer solutions to
$\theta(\{l_i\}, \{n_j\}) - \bar{a}^{\rm M}_{\rm M} = \zeta^{\rm M}_{\rm M}$, with $\zeta^{\rm M}_{\rm M}$ taking the following values:
\bg\label{zetavalve}
\zeta^{\rm M}_{\rm M} \equiv (\zeta^0_0, ~ \zeta^i_i, ~ \zeta^m_m, ~ \zeta^a_a) = (-2,~ -2, ~0, ~ +2), \nd
which reproduces the following two choices for $\theta(\{l_i\}m \{n_j\})$, namely, $\theta(\{l_i\}, \{n_j\}) = {2\over 3}$ perturbatvely and $\theta(\{l_i\}, \{n_j\}) = {8\over 3}$ non-perturbatively. One may then combine \eqref{identification} with \eqref{zetavalve} to see whether they may provide possible ways to realize the consistency conditions in \eqref{consistency} and \eqref{inequality}.

\subsection{An {\it almost} possible way of attaining a de Sitter vacuum \label{sec2.1.4}}

The readers who have skipped section \ref{sec2.1.3} may note that 
the identifications \eqref{identification} and \eqref{zetavalve} are the only results we need to resolve the classical-quantum issue that we raised in the paragraph above \eqref{QT}. The classical terms, which are the Einstein tensors ${\bf G}_{\mu\nu} \equiv {\bf R}_{\mu\nu} - {1\over 2} {\bf g}_{\mu\nu} {\bf R}$ for $(\mu, \nu) \in {\bf R}^{2, 1}$, scale in the following way:
\bg\label{kerrussel}
\left({g_s\over {\rm H}(y) {\rm H}_o({\bf x})}\right)^{-2} \times {1\over {\rm M}_p^{\sigma_{\rm classical}}}, \nd
and therefore, from \eqref{identification} and \eqref{zetavalve}, we look for $\theta(\{l_i\}, \{n_j\})$ in the quantum terms that would scale in exactly the same way. To facilitate the discussion, let us take two case: one, with time-independent G-fluxes and the other with time-dependent G-fluxes. With the first case, the $g_s$ scalings of various terms entering the EOMs along ${\bf R}^{2, 1}$ directions are given in the second column of {\bf Table \ref{firoksut}}. We see that all other ingredients would scale as \eqref{kerrussel} except the fluxes. Unfortunately however the three set of quantum terms: perturbative, non-perturbative and the non-local ones, the scaling behavior imply:
\bg\label{wendicasin}
\theta_1 - \theta_2 = {2\over 3}, ~~~~ {\rm and}~~~~ \theta_1 - \theta_2 = {8\over 3}\nd
where with $\theta_i$, for $i = 1, 2$, the former denote the $g_s$ scalings of the perturbative quantum series in \eqref{QT2} and the latter for the non-perturbative and the non-local quantum series (which we don't discuss here. See \cite{coherbeta} and \cite{coherbeta2} for details). The precise values of $\theta_i$, while can be seen from \cite{desitter2, coherbeta, coherbeta2} or easily worked out directly from \eqref{QT2}, are not important. What is important is the relative {\it minus} sign, which implies an infinite number of solutions with no $g_s$ hierarchy. Due to the subtlety with localized fluxes, the ${\rm M}_p$ hierarchy is also removed (see \cite{desitter2}), meaning that to order ${g_s^{-2}\over {\rm M}_p^{\sigma_{\rm classical}}}$, there are an infinite number of quantum terms with no apparent $g_s$ or ${\rm M}_p$ hierarchies, signalling a breakdown of EFT. Similar breakdown happens for both the non-perturbative and the non-local terms, the latter appearing from integrating out the massless off-shell states. 

This is a serious problem, but
what if we try to ignore the quantum terms altogether and solve the space-time EOMs using only the branes and the anti-branes? From the second column, since these states scale exactly as \eqref{kerrussel}, it appears that we should be able to make use of this. Plugging it in the EOMs, the result we get is the following:
\bg\label{venetiankalume}
\Lambda = -{1\over 6{\rm V}_6} {\rm T}_2 \kappa^2(n_b + \bar{n}_b), \nd
where $(\kappa^2, {\rm T}_2)$ are the data related to the branes; ${\rm V}_6$ is the volume of the six-dimensional base of the internal eight-manifold; and $(n_b, \bar{n}_b)$ are the number of the $-$ integer and fractional $-$ M2 and the $\overline{\rm M2}$ branes respectively. The cosmological constant $\Lambda$ from \eqref{venetiankalume} is {\it negative}! This means in the dual type IIB side, we are dealing with the following configuration:
\bg\label{TIgrandluxcafe}
{1\over \Lambda t^2}\left(-dt^2 + dx_1^2 + dx_2^2 + dx_3^2\right)~ \xrightarrow[x_3 \to i\eta]{t \to iz}~ {1\over \vert\Lambda\vert z^2}\left(-d\eta^2 + dx_1^2 + dx_2^2 + dz^2\right), \nd
which would generically be a non-supersymmetric ${\bf AdS}_4$ space-time in the Poincare patch. Including the internal space, we expect the configuration to be of the form ${\bf AdS}_4 \times {\bf M}_6$, with ${\rm V}_6$ giving the volume of ${\bf M}_6$. Because of this volume factor, it is a {\it scale-separated} AdS space. Unfortunately due to the ill-behaved quantum terms, such a configuration is most likely in the swampland. (See \cite{maxim} for more details on the scale-separated AdS spaces using the language of \eqref{QT2}.) This means branes or the anti-branes, including the IIB O-planes\footnote{The information of O-planes is encoded in the geometry of the eight-manifold in M-theory \cite{DRS}.}, can at best give a non-supersymmetric ${\bf AdS}_4$ spacetime but never a de Sitter space. 

The story appears to improve rather dramatically once we take time-dependent G-fluxes and slowly moving branes. (The latter is not important, at least for solving the EOMs, but is crucial for flux quantizations and anomaly cancellations. We will however avoid these discussion for the time being.) The $g_s$ scalings of various contributing factors are shown in the third column of {\bf Table \ref{firoksut}}. The crucial change from \eqref{wendicasin} is that $\theta_1 \to \theta_1'$ and $\theta_2 \to -\theta_2'$, such that:
\bg\label{elenacasin}
\theta_1' + \theta_2' \equiv \theta(\{l_i\}, \{n_j\}) = {2\over 3}, ~~~~ {\rm and} ~~~~ \theta_1' + \theta_2' \equiv \theta(\{l_i\}, \{n_j\}) = {8\over 3}, \nd
for the perturbative and the non-perturbative (as well as the non-local) quantum series respectively; and $\theta(\{l_i\}, \{n_j\})$ is defined earlier in section \ref{sec2.1.3}. 
Unfortunately however the perturbative solution with $\theta(\{l_i\}, \{n_j\}) = {2\over 3}$ {\it does not} satisfy either \eqref{consistency} or \eqref{inequality}, which amounts to saying that the perturbative quantum terms, just like the classical terms, are red herrings in the problem. This however doesn't mean that the perturbative terms are completely useless. They do contribute at higher orders in $g_s$, a fact that may be verified by working out the EOMs in the presence of \eqref{QT2}. At the lowest order in $g_s$, it is only the non-perturbative and non-local quantum terms that can save the day. These terms come from the BBS-like instanton effects \cite{bbs}, which are basically instantonic five-branes wrapped around the six-manifold ${\cal M}_6$. These non-perturbative (and non-local) effects are slightly more complicated to compute because of the temporal behavior of the six-manifold ${\cal M}_6$ as seen from \eqref{mmetric}, but they can be done (see details in \cite{coherbeta, coherbeta2}). The result precisely gives $\theta(\{l_i\}, \{n_j\}) = {8\over 3}$, and plugging the values of $(l_i, n_j)$ in \eqref{QT2} provides us with quantum terms that are quartic in curvature tensors (amongst other higher order mixed terms). The back-reaction of these quartic terms shows that \eqref{mmetric} solves the EOMs provided we modify the internal metric $g_s^{-2/3} g_{mn}(y)$ of ${\cal M}_6$ by adding sub-dominant contributions as $g_s^{-2/3} g_{mn}(y) \to \sum\limits_{k = 0}^\infty g_{mn}^{(k)}(y) g_s^{-2/3 + 2k/3}$ with $k \in {\mathbb{Z}\over 2}$, which in-turn will make the internal metric \eqref{metansatze} in the IIB side 
time-dependent\footnote{To keep the four-dimensional Newton's constant time-independent in type IIB, a simple way would be to split ${\cal M}_6$ to ${\cal M}_4 \times {\cal M}_2$ and define the sub-dominant temporal dependence as multiplicative factors, say for example ${\rm F}_1(t)$ and ${\rm F}_2(t)$ respectively. Making ${\rm F}_1^2(t) F_2(t) = 1$ would keep the four-dimensional Newton's constant time-independent. Happily, such a choice also solves the EOMs as demonstrated in \cite{desitter2}.}.  The inclusion of quartic corrections  was anticipated for sometime (see \cite{issues}), and here we see why this may be the case. Four further encouraging results appear simultaneously:

\vskip.1in

\noindent {\bf 1}. The {\it non-locality} factors in the  non-local quantum terms  are integrated away to give rise to local interactions which contribute to 
$\theta(\{l_i\}, \{n_j\}) = {8\over 3}$, thus behaving very similar to the BBS instantons.

\vskip.1in

\noindent {\bf 2}. The fermionic terms, coming from the Rarita-Schwinger fermions in M-theory, contribute only as polynomial powers of the {\it condensates} of lower dimensional Dirac fermions thus creating the anticipated non-perturbative effects.

\vskip.1in

\begin{table}[tb]  
 \begin{center}
\renewcommand{\arraystretch}{1.5}
{\begin{tabular}{|c|c|c|}
\hline
Contributions to the EOMs & \multicolumn{2}{c|}{$\hat{g}_s \equiv {g_s\over {\rm H}(y){\rm H}_o({\bf x})}$ scaling}\\
\cline{2-3}
${\bf G}_{\mu\nu} = {1\over 4\pi{\rm G}_{\rm N}}\sum\limits_{i = 1}^6\mathbb{T}^{[i]}_{\mu\nu}$& Time-independent fluxes & Time-dependent fluxes \\
\hline
${\bf G}_{\mu\nu} \equiv {\bf R}_{\mu\nu} - {1\over 2}{\bf g}_{\mu\nu} {\bf R}$ & ${1\over \hat{g}_s^2}$ & ${1\over \hat{g}_s^2}$  \\
\hline
$\mathbb{T}_{\mu\nu}^{[1]}= \mathbb{T}_{\mu\nu}^{[{\rm fluxes}]}$ & ${1\over \hat{g}_s^4}$ & ${1\over \hat{g}_s^2}$\\
\hline
$\mathbb{T}_{\mu\nu}^{[2]}= \mathbb{T}_{\mu\nu}^{[{\rm M2}_{({\rm i+f})}]}$ & ${1\over \hat{g}_s^2}$ & ${1\over \hat{g}_s^2}$\\
\hline
$\mathbb{T}_{\mu\nu}^{[3]}= \mathbb{T}_{\mu\nu}^{[\overline{\rm M2}_{({\rm i+f})}]}$ & ${1\over \hat{g}_s^2}$ & ${1\over \hat{g}_s^2}$\\
\hline
$\mathbb{T}_{\mu\nu}^{[4]}= \mathbb{T}_{\mu\nu}^{[{\rm perturbative}]}$ & $\hat{g}_s^{\theta_1-\theta_2 - {8\over 3}}$ & $\hat{g}_s^{\theta'_1 + \theta'_2 - {8\over 3}}$\\
\hline
$\mathbb{T}_{\mu\nu}^{[5]}= \mathbb{T}_{\mu\nu}^{[{\rm non-perturbative}]}$ & ${\hat{g}_s^{\theta_1-\theta_2 - {8\over 3}}\over \hat{g}_s^2}~{\rm exp}\left(-{\hat{g}_s^{\theta_1 - \theta_2}\over \hat{g}_s^2}\right)$ & ${\hat{g}_s^{\theta'_1+\theta'_2 - {8\over 3}}\over \hat{g}_s^2}~{\rm exp}\left(-{\hat{g}_s^{\theta'_1 + \theta'_2}\over \hat{g}_s^2}\right)$\\
\hline
$\mathbb{T}_{\mu\nu}^{[6]}= \mathbb{T}_{\mu\nu}^{[{\rm non-local}]}$ & ${\hat{g}_s^{\theta_1-\theta_2 - {8\over 3}}\over \hat{g}_s^2}~{\rm exp}\left(-{\hat{g}_s^{\theta_1 - \theta_2}\over \hat{g}_s^2}\right)$ & ${\hat{g}_s^{\theta'_1+\theta'_2 - {8\over 3}}\over \hat{g}_s^2}~{\rm exp}\left(-{\hat{g}_s^{\theta'_1 + \theta'_2}\over \hat{g}_s^2}\right)$\\
\hline
\end{tabular}}
\renewcommand{\arraystretch}{1}
\end{center}
 \caption[]{The $g_s$ scalings of various terms appearing the EOMs along the $2+1$ space-time directions. The second column depicts the scalings with time-independent fluxes and the third column depicts the same with time-dependent fluxes. ${\rm M2}_{({\rm i+f})}$ denotes both integer and fractional M2 branes. $\theta_i$ and $\theta'_i$, for $i = 1, 2$, denote the $g_s$ scaling for the quantum series \eqref{QT2} for the time-independent and time-dependent fluxes respectively. Details about what these $g_s$ scalings imply are described in the text.} 
  \label{firoksut}
 \end{table}

\noindent ${\bf 3.}$ The expression for the four-dimensional cosmological constant changes from being purely negative in \eqref{venetiankalume}, to now taking the following form (see the fourth paper in \cite{GMN}; and \cite{desitter2}):
\bg\label{binionethiopia}
\Lambda = {1\over 12 {\rm V}_6} ~\vert \mathbb{T}\vert^{[{\rm NP, NL}]} - {1\over 6{\rm V}_6} {\rm T}_2\kappa^2(n_b + \bar{n}_b) - {5\over 384{\rm V}_6} \vert {\rm G}\vert^2, \nd
where $\vert\mathbb{T}\vert^{[{\rm NP, NL}]}$ involve integrals of the traces of the unwarped, {\it i.e.} $g_s$ independent, non-perturbative and non-local energy-momentum tensors along the ${\bf R}^{2, 1}$ directions. $\vert{\rm G}\vert^2$ is also defined in a similar vein. For both cases, the integrals are carried over the unwarped six-dimensional base (meaning that they involve the unwarped six-dimensional metric components). 

\vskip.1in

\noindent ${\bf 4.}$ The order by order study in $g_s$ is important. To the lowest order in $g_s$ the expression for cosmological constant is 
\eqref{binionethiopia}. At the next order in $g_s$, new flux components (recall the ${2k\over 3}$ factor in \eqref{gfluxes}), as well as new quantum terms {\it et cetera} appear. They are all balanced against each other at every order in $g_s$ in such a way that the lowest order result, for example \eqref{binionethiopia}, does not change.  

\vskip.1in

\noindent The last two points confirm what we have been saying all along: the branes, anti-branes, O-planes and the perturbative quantum corrections are all red-herrings in the problem. Positive cosmological constant can only come from the non-perturbative and the non-local quantum terms, and only if they have the requisite $g_s$ and ${\rm M}_p$ hierarchies. However \eqref{binionethiopia} is more delicate: it appears to balance the quantum terms, {\it i.e.} the non-perturbative and the non-local quantum terms, against the classical terms, {\it i.e.} the branes, fluxes and the O-planes. This would have been impossible if the quantum terms were vanishing for small $g_s$. Fortunately, while the usual instanton like effects go as ${\rm exp}\left(-{1\over g_s^2}\right)$, thus vanishing for $g_s \to 0$, here the effects go as  ${\rm exp}\left(-{{g}_s^{\theta'_1 + \theta'_2}\over {g}_s^2}\right) = {\rm exp}\left(-g_s^{2/3}\right)$, which do not vanish for $g_s \to 0$. This may be easily confirmed by plugging in the second relation from \eqref{elenacasin} in the non-perturbative and the non-local results from the third column of {\bf Table \ref{firoksut}}. Interestingly, using the first relation from \eqref{elenacasin}, the instanton like effects would have vanished for small $g_s$, thus confirming once again that the first relation in \eqref{elenacasin} cannot help in constructing a positive cosmological constant solution in the type IIB side. 

Despite this, the temporal $-$ albeit sub-dominant $-$ dependence of the internal metric in the type IIB side is a slight matter of concern now. Questions can be raised regarding Bianchi identities, flux quantizations and anomaly cancellation. However it has been shown in \cite{desitter2, coherbeta, coherbeta2} that all the three criteria {\it can} be satisfied order-by-order in $g_s$. For example, taking derivatives of \eqref{QT2} with respect to the ${\bf C}_{\rm MNP}$ fields provide corrections to the EOMs of the three-form fields in the presence of all order quantum terms. Once we restrict the flux components along the internal directions and integrate the equations over the eight-manifold, we get the anomaly cancellation formula to any order in $g_s$. Similarly taking the derivatives of \eqref{QT2} with respect to the dual six-form fields (not field strengths!), provide the Bianchi identities to any order in $g_s$. Restricting over four-cycles (considered as boundaries of  five-cycles), provide the necessary flux quantization conditions, again to any order in $g_s$. These have been discussed in \cite{desitter2, coherbeta, coherbeta2} for type IIB theory and to some extent in the upcoming work \cite{hete8} for the $SO(32)$ and the ${\rm E}_8 \times {\rm E}_8$ theories suggesting the possibility of {\it almost} attaining a de Sitter vacuum solution. What goes wrong, or more appropriately, where is the error in the above analysis?

\section{Glauber-Sudarshan states and the failure of vacuum solutions \label{sec3}}

Our above analysis may provide a false hope to the readers that a four-dimensional de Sitter space-time can exist as a vacuum solution in string theory. Unfortunately this is not the case, despite all the apparent  successes mentioned in the previous section. Where did we go wrong? The mistake lies in the two assumptions that we made earlier, namely the existence of an Exact Renormalization Group effective action, and the existence of a supergravity description at the energy scale $k_{\rm IR} < k < \mu << \hat\mu$. Let us clarify this by first recalling some, often misunderstood, facts about supergravity and string theory.

\vskip.1in

\noindent $\bullet$ String theory is {\it not} supergravity, not even approximately. When a string is quantized over a flat {\it time-independent}
background, the spectrum consists of massless states and a tower of massive states appearing from the scale ${\rm M}_s$ and above (as depicted in {\bf figure \ref{scales}}).

\vskip.1in

\noindent $\bullet$ Supergravity generically\footnote{But not always! Type IIB string theory is an example.} provides a Lagrangian description of these massless states and are represented by some consistent truncation procedure that keeps the marginal and relevant interactions between these states. Most irrelevant interactions (with respect to the string or the Planck scale) take us away from the supergravity description.

\vskip.1in

\noindent $\bullet$ The energy scale is very important, as we have no idea how string theory behaves in the far UV ({\it i.e.} at very short distances) and what degrees of freedom control the far UV dynamics, although it is presumed that the UV behavior is {\it well-defined} without any pathologies. The point is that, no matter what degrees of freedom exist in the far UV, an Exact Renormalization Group (ERG) procedure will tell us that once we integrate from $\Lambda_{\rm UV}$ till $\mu$ (see {\bf figure \ref{scales}}) all the information of the UV can be encoded in the {\it coefficients} of the infinite towers of marginal, relevant and irrelevant interactions of the massless states!

\vskip.1in

\noindent $\bullet$ Over a curved but still {\it time-independent} background, although we do not know how to quantize a string, it is believed that the supergravity interactions involving the {\it same} massless states continue to provide the dynamics of the theory. Unfortunately the irrelevant interactions are not well known and are difficult to derive in general. Even going to the energy scale $\hat\mu$ that introduces the KK states (see {\bf figure \ref{scales}}) becomes hard to deal within the supergravity approximation. Thus the safest energy scale is the one where $k_{\rm IR} < k < \mu << \hat\mu$ wherein we could take all order perturbative interactions involving only the {\it massless} states and venture beyond the supergravity approximation consistently.

\vskip.1in

\noindent $\bullet$ String perturbation theory is {\it not} convergent but is rather asymptotic, which is good because it tells us how to introduce non-perturbative effects, like D-branes, instantons, world-sheet instantons et cetera in a natural and consistent way. The original discovery of the D-branes successfully used this asymptotic behavior to argue for the existence of states whose mass go as inverse the string coupling instead of inverse square of the string coupling \cite{shenker}. These states themselves carry towers of massless and massive degrees of freedom on their world-volumes and within the energy range $k_{\rm IR} < k < \mu << \hat\mu$ one could express the complete UV dynamics from an infinite towers of interactions involving the world-volume massless states following ERG. This perturbative series is again {\it not} convergent but is asymptotic which tells us what the non-perturbative contributions are in the aforementioned energy range. 

\vskip.1in

\noindent $\bullet$ Supersymmetry is also important. The world-sheet GSO projection typically eliminates the tachyon and preserves certain amount of 
space-time supersymmetry. However there is a misconception that breaking supersymmetry, spontaneously or explicitly, always guarantees a positive energy. This is definitely untrue. For example anti-brane breaks supersymmetry spontaneously, but {\it does not} change the cosmological constant of the theory. Similarly non self-dual four-form flux in M-theory breaks supersymmetry but keeps the cosmological constant unchanged. In fact no classical sources can break suspersymmetry and create positive cosmological constant without violating the no-go theorems \cite{GMN}. The connection between supersymmetry breaking and the sign of the cosmological constant is more subtle. For example zero and negative cosmological constant spacetimes may or may not be supersymmetric, but a positive cosmological constant spacetime is always non-supersymmetric\footnote{As an example taking ${\rm AdS}_4$ in the Poincare patch and changing the coordinates $z \to it$ and $t \to iz$, where $t$ is the conformal time, would appear to give a {\it supersymmetric} ${\rm dS}_4$. (This is sometime also termed as a dS slicing of an AdS space.) However a careful study of the fluxes supporting ${\rm AdS}_4$ will tell us that the fluxes now become imaginary, thus ruling out this oft-used trick to generate ${\rm dS}_4$.}. Our problem here is with the latter: can we generate a four-dimensional positive cosmological constant, and therefore non-supersymmetric, space-time as a {\it vacuum} solution from string theory? The answer will turn out to be {\it no}.

\subsection{Why doesn't de Sitter spacetime exist as a vacuum solution? \label{sec3.1}}

\noindent Our little discussion above concluded with the claim that four-dimensional de Sitter spacetime cannot exist as a vacuum solution in string theory.
Why is this the case? The reason is that all of the above facts change when we have a {\it time-dependent} background: an example being four-dimensional de Sitter space-time in the flat-slicing as shown in {\bf figure \ref{planar}}. The reasoning is simple, and can be illustrated directly in four-dimensional QFT. Consider a QFT defined over a four-dimensional de Sitter background in a flat-slicing. (The slicing will not be important in the ensuing discussion and we will make this clear soon. Meanwhile we will continue with the flat-slicing because of its technical simplicity.) From our modern understanding of QFT, we know that there is no distinction between {\it renormalizable} or {\it non-renormalizable} theories: ``non-renormalizable$"$ theory simply means that, beyond some scale, the theory would need new degrees of freedom for it to make sense. An example is Einstein gravity itself. Traditionally it was classified as a ``non-renormalizable$"$ theory, but now we know that beyond the string scale, the theory gets a tower of increasingly massive degrees of freedom which would render {\it all} UV amplitudes finite. Alternatively, a theory at a given scale may be defined by integrating out exactly those massive degrees of freedom. Once we do that, we should not think of extending the theory beyond that specific scale. Failing to maintain this would result in all kind of wrong conclusions and nomenclatures\footnote{Despite the simplicity of the above argument, it has unfortunately been constantly misused and mis-interpreted in the literature. We want to affirm the readers early on that we will always be in the energy range $k_{\rm IR} < k < \mu << \hat\mu$ where no massive degrees of freedom will enter and the dynamics will be completely governed by the light degrees of freedom, albeit beyond the supergravity description. See also footnote \ref{referee1}.}. 

\begin{figure}[h]
    \centering
    \begin{tikzpicture}[scale=0.5]
    \draw[thick] (0,8.0) -- (0,0)        (8,0) -- (8,8.0) (0,8.0) -- (8,0) (0,0) --(8,8) (0,8) -- (8,8) (0,0) --(8,0) ;   
    \draw[fill] (6,4)  node[right] {  R };
   \draw[fill] (2,4)  node[left] {  L };
   \draw[fill] (4,6)  node[above] {  F };
   \draw[fill] (4,2)  node[below] {  P };
    \end{tikzpicture}
    \qquad \quad
    \begin{tikzpicture}[scale=0.5]
\draw[thick] (0,6.5) -- (0,-0.5)        (3,-0.5) -- (3,6.5)  ;
\draw[blue] (0,1) .. controls (0.5,2) .. (1,1) .. controls (1.5,0) .. (2.0,1) .. controls  (2.5, 2) .. (3,1) ; 
\draw[blue] (0,3) .. controls (0.5,4) .. (1,3) .. controls (1.5,2) .. (2.0,3) .. controls  (2.5, 4) .. (3,3) ; 
\draw[blue] (0,5) .. controls (0.5,6) .. (1,5) .. controls (1.5,4) .. (2.0,5) .. controls  (2.5, 6) .. (3,5) ; 

\draw[fill] (3,-0.5)  node[right] {  IR };
\draw[fill] (3,6.5)  node[right] {  UV };
\end{tikzpicture} 
 \begin{tikzpicture}[scale=0.5]
\draw[thick] (0,6.5) -- (0,-0.5)        (3,-0.5) -- (3,6.5)  ;
\draw[blue] (0,1) .. controls (0.5,2) .. (1,1) .. controls (1.5,0) .. (2.0,1) .. controls  (2.5, 2) .. (3,1) ; 
\draw[blue] (0,3) .. controls (0.5,4) .. (1,3) .. controls (1.5,2) .. (2.0,3) .. controls  (2.5, 4) .. (3,3) ; 
\draw[blue] (0,5) .. controls (0.5,6) .. (1,5) .. controls (1.5,4) .. (2.0,5) .. controls  (2.5, 6) .. (3,5) ; 
\draw[blue] (3,5) .. controls (3.5, 4) .. (4,5) .. controls (4.5, 6) .. (5,5);
\draw[blue, dashed] (5,5) .. controls (5.5, 4) .. (6,5) ;
\draw[blue] (0,5) .. controls (-0.5, 4) .. (-1,5) .. controls (-1.5, 6) .. (-2,5);
\draw[blue, dashed] (-2,5) .. controls (-2.5, 4) .. (-3,5) ;
\draw[purple] (3,3) .. controls (3.75, 2) .. (4.5,3) .. controls (5.25, 4) .. (6,3);
\draw[purple, dashed] (6,3) .. controls (6.75, 2) .. (7.5,3) ;
\draw[purple] (0,3) .. controls (-0.75, 2) .. (-1.5,3) .. controls (-2.25, 4) .. (-3.0,3);
\draw[purple, dashed] (-3.0,3) .. controls (-3.75, 2) .. (-4.5,3) ;
\draw[red] (3,1) .. controls (4, 0) .. (5,1) .. controls (6, 2) .. (7,1);
\draw[red, dashed] (7,1) .. controls (8, 0) .. (9,1) ;
\draw[red] (0,1) .. controls (-1, 0) .. (-2,1) .. controls (-3, 2) .. (-4,1);
\draw[red, dashed] (-4,1) .. controls (-5, 0) .. (-6,1) ;
\draw[fill] (3,-0.5)  node[right] {  IR };
\draw[fill] (3,6.5)  node[right] {  UV };
\end{tikzpicture} 
\caption{\textcolor{blue}{Left}: Penrose diagram for a static patch of a four-dimensional de Sitter space-time. \textcolor{blue}{Middle}: The region inside the black lines indicate the static patch, where a mode can possess the same frequency. \textcolor{blue}{Right}: Similar configuration as the middle where, inside a static patch, a mode can possess the same frequency, while it changes outside the static patch. The colour scheme outside the static patch indicates slow red-shifting of the modes as we go from the UV to the IR. The underlying temporal coordinate used here is the one from the flat-slicing.}
    \label{staticpatch3}
\end{figure}
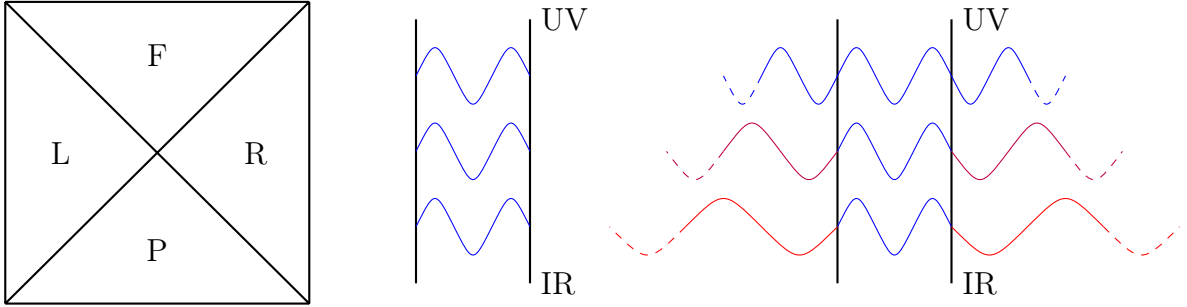

The above conclusion relies on an important fact: our ability to integrate out the UV degrees of freedom. {\it What if we could not do this}? Such a scenario could arise if there is no well-defined UV description, or if the frequencies of the fluctuations are changing with respect to time. The former could be ruled out because, despite our ignorance of the UV degrees of freedom and the UV dynamics, the short-distance behavior in string theory is well-defined with no pathologies whatsoever as mentioned earlier. This means, the only way a Wilsonian integrating out procedure would fail if the frequencies are themselves changing with respect to time. Unfortunately this is exactly what happens when we study fluctuations over a de Sitter background: the frequencies, and therefore the corresponding energies, are constantly red-shifted. This immediately implies a failure of the integrating-out procedure because if the energies of the modes are changing with respect to time, there is no meaning of integrating out the {\it high energy modes}. A common misconception at this point is to view the de Sitter space-time in a {\it static patch} and declare it as a way out of this problem. Unfortunately this argument doesn't work. Part of the reason being the highly misleading dynamics in a static patch. To see this, consider the two scenarios presented in the middle and right of {\bf figure \ref{staticpatch3}}, the first one corresponds to the behavior of a mode {\it inside} a static patch, 
and the second one corresponds to the behavior of the same mode when seen {\it outside} the static patch.

The regions inside the black lines in both the figures correspond to the static patch. An observer inside the static patch would see the UV frequencies (drawn in blue). From UV to IR, the frequencies would appear to {\it not} change with respect to time $-$ using the temporal coordinate of the flat-slicing $-$ and therefore the observer would attempt to integrate them out to write an EFT in the IR. However from {\bf figure \ref{staticpatch3}} we see that this would be a mistake because the frequencies do change {\it outside} the static patch, so at any given energy scale the frequencies are actually constantly changing {\it even inside the static patch}! Going to other coordinate patches related to the static patch would fail to alleviate the 
problem. This and other related issues fall under the so-called trans-Planckian problems in de Sitter space-time \cite{tcc}.

Our above argument should convince the readers that allowing a de Sitter as a vacuum solution in string theory spoils the two essential properties that we need to allow for a low energy description, namely the integrating out mechanism following ERG, and the existence of a supergravity description. In addition to that, since the fluctuations are related to the zero modes of the closed string, the quantization of the string becomes a problem and it is not clear how the spectra of the string would look like. This issue would extend to the open string quantization also, leading to the same issue with the world-volume spectra on branes. Non-existence of supersymmetry is another problem: the bulk zero point energies do not cancel, and therefore lead to the cosmological constant problem that we barely avoided using supergravity descriptions with time-independent backgrounds. Question is, are there ways out of this conundrum? In the following we tabulate three possible ways.

\vskip.1in

\noindent ${\bf 1}.$ Discard completely the possibility of de Sitter as a vacuum solution and replace it by quintessence, or string gas cosmology or other alternatives.

\vskip.1in

\noindent ${\bf 2.}$ Allow the existence of a de Sitter vacuum solution but replace Wilsonian Effective Field Theories by {\it Open} Quantum Field Theories.

\vskip.1in

\noindent ${\bf 3.}$ Discard completely a de Sitter vacuum solution and replace it by an excited state in the theory over a {\it supersymmetric} Minkowski vacuum.

\vskip.1in

\noindent The {\bf first} one is an interesting alternative. While quintessence has it's own share of problems (see for example \cite{donof}), string gas cosmology \cite{robert1} {\it can} serve as a viable alternative: it is based on string theory and there are recent developments connecting it to M(atrix) models \cite{suddhomatrix} which appear to be very promising. It will be interesting to see how ERG ideas play out in this setting and how EFTs are constructed. We will have more to say on this a little later. 

The {\bf second} one is more subtle. Open QFTs \cite{vernon} are traditionally described by isolating relevant degrees of freedom from an ``environment$"$ which allows them to gain or lose energies to the environment. Since neither energy is conserved, nor an EFT description exists, the dynamics of the theory typically follow some Markovian process which may be quantified in certain settings (see for example \cite{suddhomarkov} and references therein). The problem with this picture appears when we try to use it in string theory: in a UV complete theory, like string theory or M-theory, there appears to be no clear demarcation between the relevant degrees of freedom and the environment, and therefore a concrete realization over a temporally varying background is equally hard. So far there has not been much progress implementing this idea in string theory, although some recent works in field theories have shed some interesting light on the de Sitter problem \cite{burgess}.

This brings us to the {\bf third} one: realizing a de Sitter space-time as an excited state over a supersymmetric Minkowski background. Our bet is on this. Three immediate encouraging  results are as follows.

\vskip.1in

\noindent ${\bf 1.}$ Being supersymmetric Minkowski, or more appropriately being a warped supersymmetric Minkowski with a compact non-K\"ahler internal space, the zero point energies automatically cancel, and so does the Dine-Seiberg runaway \cite{dineseiberg} problem. The latter being due to the stabilized moduli.

\vskip.1in

\noindent ${\bf 2.}$ Being time-independent background, a Wilsonian EFT {\it can} be constructed. The fluctuations over such a vacuum are time-independent and, although our knowledge of the short-distance physics is negligible, ERG techniques can be implemented leading to a well-defined EFT in the energy scale $k_{\rm IR} < k < \mu << \hat\mu$, where $\hat\mu$ is quantified by the KK modes from the non-K\"ahler internal space.

\vskip.1in

\noindent ${\bf 3.}$ Being an excited state over a supersymmetric Minkowski, the supersymmetry breaking required for a de Sitter space can be realized {\it spontaneously}, meaning that the state breaks supersymmetry whereas the vacuum doesn't. Since the zero point energies continue to cancel, the positive cosmological constant would come from those quantum corrections responsible for violating the no-go theorems \cite{GMN}. 

\subsection{Coherent states versus the Glauber-Sudarshan states \label{sec2.2.2}}

\begin{figure}[h]
    \centering
    \begin{tikzpicture}[scale=0.7]
    \draw (0,5) .. controls (0.3,3) .. (1,1) .. controls (1.5,0) .. (2.0,1) .. controls  (2.5, 2) .. (3,1) .. controls (3.5,0) ..(4.0,1) .. controls (4.5,2.5).. (5,5) ;  
    \node [right] at (8.95,2.65) {$ \longrightarrow$}; 
    \end{tikzpicture} 
\qquad  \qquad
\begin{tikzpicture}[scale=0.7]
        \draw (0,5) -- (0,0)  (2,5) -- (2,0) (0,0) -- (2,0) (0,0.1) -- (2,0.1); 
        \filldraw (1, 0.5) circle(1pt) (1,1.0) circle(1pt) (1, 1.5) circle(1pt) ; 
\end{tikzpicture}

\medskip 
\medskip

\begin{tikzpicture}[scale=0.7]
\draw (0,5) .. controls (0.3,3) .. (1,1) .. controls (1.5,0) .. (2.0,1) .. controls  (2.5, 2) .. (3,1) .. controls (3.5,0) ..(4.0,1) .. controls (4.5,2).. (5,1) ; 
\filldraw[black] (5.2,1) circle (1pt)   (5.6,1) circle (1pt)   (5.9,1) circle (1pt);
\draw (9,5) .. controls (8.7,3) .. (8,1) .. controls (7.5,0) .. (7.0,1) .. controls  (6.5, 2) .. (6,1);
 \node [right] at (10.25,2.65) {$ \longrightarrow \qquad $}; 
\end{tikzpicture}
\begin{tikzpicture}[scale=0.7]
        \draw (0,5) -- (0,0)  (2,5) -- (2,0) (0,0) -- (2,0) (0,1) -- (2,1) (0,2.5) -- (2,2.5) (0,3.5) -- (2,3.5); 
        \fill[ashgrey2] (1.98,2.3) rectangle (0.0,3.51);
        \fill[ashgrey] (1.98,0.99) rectangle (0.0,0.01);
        \filldraw (1, 4.5) circle(1pt) (1,4.0) circle(1pt) (1, 5.0) circle(1pt) ; 
\end{tikzpicture}
\caption{Top diagrams indicate the lowest two states for the quartic potential, i.e. $\ket{0}_S$ and $\ket{0}_A$, with the true vacuum being the symmetric wavefunction $\ket{0}_S$. Bottom diagrams indicate that for a multi-valley potential the energy levels split into different bands, with the bottom of the brown energy band being the true vacuum $\ket{0}_S$ corresponding to the most symmetric wavefunction.}
    \label{pot1}
\end{figure}
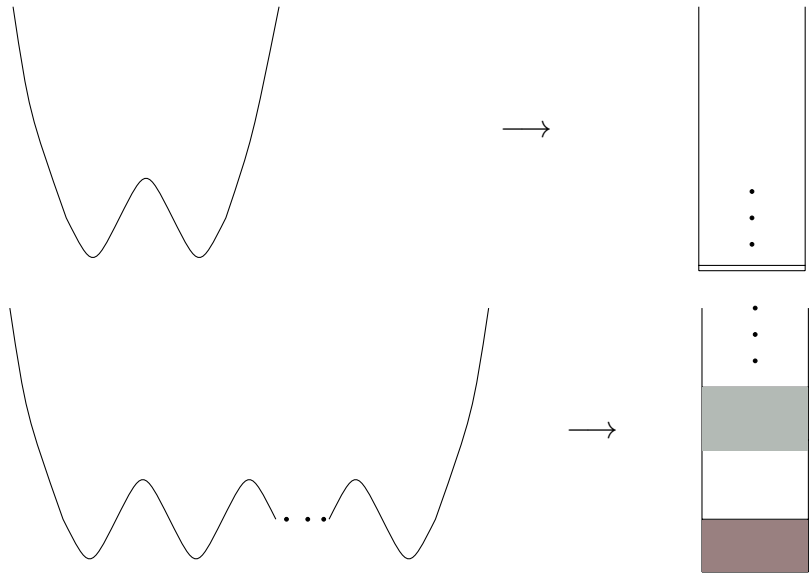

\noindent With the aforementioned encouragements, we now ask what kind of excited state over a Minkowski minimum\footnote{By this we mean {\it local minima} unless mentioned otherwise.} should we look for to reproduce a de Sitter space-time once we take an expectation value of the metric operator over this state. Since the space-time is almost ``classical$"$, as a first trial we should look for a quantum state that is closest to classical trajectories\footnote{Such a line of thought will eventually lead us to a dead-end, but we will push on.}.  An immediate answer would be a {\it coherent state} which is the closest analog to a classical trajectory from quantum theory. Coherent states are usually described as displacement operators acting on free Gaussian vacua, which may alternatively be represented by specific linear combinations of the number operators associated with the quantization of harmonic oscillators. These harmonic oscillators clearly arise from quantizing a {\it free} field theory over a Minkowski minima. Once we have multiple minima (with identical values), as shown in {\bf figure \ref{pot1}}, the coherent states are not as simple as the one known for a single minimum and are typically represented in quantum mechanics by the so-called Gazeau-Klauder coherent states \cite{klauder}:
\bg\label{dadari}
\vert\sigma\rangle \equiv \vert {\rm J}, \gamma\rangle = {1\over \sqrt{\rm N(J)}} \sum_{n \ge 0} {{\rm J}^{n/2}\over \sqrt{\rho_n}} ~e^{-i{\rm E}_n \gamma} \vert n \rangle, \nd
where ${\rm E}_n$ are the eigenvalues of a Hamiltonian ${\rm H}$, such that 
${\rm H} \vert n \rangle = \omega {\rm E}_n \vert n \rangle$, $\gamma \in \mathbb{R}, 0 \le {\rm J} \in \mathbb{R}, \rho_n$ is an arbitrary function of $n$ with $\rho_0 = 1$, such that when ${\rm E}_n = n$ and $\rho_n = n!$, we reproduce the well-known coherent states for the case with a single minimum. Additionally demanding $\langle\sigma\vert\sigma\rangle = 1$ implies that ${\rm N(J)} \equiv \sum\limits_{n \ge 0}{{\rm J}^n\over \rho_n}$ be a convergent series within an appropriate radius of convergence.
Within this radius of convergence if the moment problem allows a positive 
solution for the moment, then there exists a resolution of the identity (see \cite{klauder} for details).

\begin{figure}
    \centering
    \includegraphics[scale=0.7]{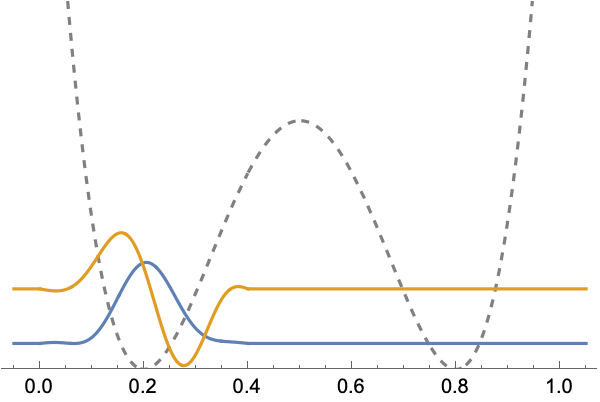} \, \, \,
    \includegraphics[scale=0.75]{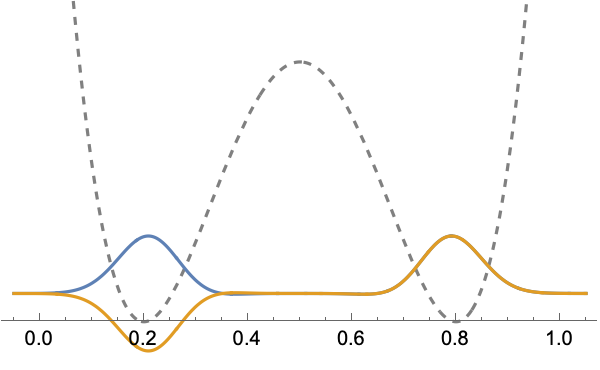}
    \caption{\textcolor{blue}{Left}: excitations over the left potential well in the example a quartic potential. \textcolor{blue}{Right}: The symmetric and the antisymmetric wavefunctions corresponding to the true ground state and first excitation of the double well problem.}
    \label{doublewell}
\end{figure}

The coherent states for the multiple minima case, or more appropriately $-$ since they have the same values $-$ multiple wells' case, are not necessarily the mimimum uncertainity states although their temporal evolutions follow quantum approximations to classical motions. As an example consider the double well case as depicted by the first figure in {\bf figure \ref{pot1}}, or by a more elaborate version in {\bf figure \ref{doublewell}}. The symmetric wave-function will have the lowest energy, with the energy of the anti-symmmetric one being slightly more than the symmetric one. One can construct a coherent state following \eqref{dadari} and its temporal evolution follows the sequence of graphs plotted in figure 6 of \cite{novaes}. From the figure one may see the appearance of Schr\"odinger cat-like states. For more complicated wells, the system forms energy bands with the lowest energy states at the bottom of the grey band as depicted in {\bf figure \ref{pot1}}. Coherent states are difficult to construct for such cases, although some numerical works have been done. 

Few issues prohibit us from using the aforementioned coherent states as representations of the excited states over a Minkowski vacuum in string theory. First, the coherent state construction in \eqref{dadari} is mostly in quantum mechanics and its extension to quantum field theory is non-trivial. Secondly, interacting field theories are an infinite collection of {\it interacting} harmonic oscillators which would make the construction of coherent states pretty hard. One might think of alleviating the second problem by going to weakly coupled scenarios where typically we would expect almost decoupled oscillators. {\it Unfortunately, this is exactly what we cannot do.} Thus the three main issues that actually prohibit us from using the standard notion of the coherent states in the low energy effective field theory description of string or M-theory are as follows.

\vskip.1in

\noindent ${\bf 1.}$ Type IIB theory in which we want to realize a coherent state over a supersymmetric Minkowski vacuum is at a constant coupling point of F-theory \cite{senkd}. This means that the IIB coupling $g^{(b)}_s \equiv 1$, with the moduli stabilized at vanishing axio-dilaton. Zero type IIB coupling would correspond to infinitely negative value of the background dilaton. We would however like to maintain $g^{(b)}_s \equiv 1$ even for the {\it emergent} de Sitter background to avoid dynamical seven-branes et cetera (the latter also being states in the theory \cite{dileep}). 

\vskip.1in

\noindent ${\bf 2.}$ Once we have a non-trivial coupling, there are no {\it free} vacua $\{\vert 0 \rangle\}$ and one has to replace them by an {\it interacting vacuum} $\vert\Omega\rangle$. However now we have no idea how to construct the equivalent of higher states $\{\vert n \rangle\}$ in the interacting system! In QFT we avoid this delicate problem by using an $i\epsilon$ prescription and by computing the correlation functions by tilting the temporal intervals along the slightly imaginary direction, {\it i.e.} using ${\rm T} \to \pm\infty(1 - i\epsilon)$
prescription.

\vskip.1in

\noindent ${\bf 3.}$ Coherent states are supposed to mimic classical trajectories or classical configurations. However de Sitter space-time is {\it not} actually a classical configuration as non-perturbative effects are needed to stabilize it within the temporal domain \eqref{tcc}. Thus whatever excited states we construct should have the relevant set of quantum terms somehow {\it embedded} in them. 

\vskip.1in

\begin{figure}[h]   
    \centering
\begin{tikzpicture}[scale=1.0]
    \draw (0,5) .. controls (0.5,2.5) .. (1,1) .. controls (1.5,0) .. (1.8,1) .. controls  (2.5, 4) .. (3.2,1) .. controls (3.5,0) ..(4.0,1) .. controls (4.5,2.5).. (5,5) ; 
     \draw[magenta] (0,1) ..controls (0.7,1) .. (0.9,1.3) .. controls (1.3,2.0) .. (1.9,1.3) ..controls (2.2,1.0).. (2.5,0.95) ..controls (2.8,1.0) .. (3.1,1.3) .. controls (3.7,2.0) .. (4.1,1.3) ..controls (4.3,1.0).. (5,1.0);
    \draw[blue] (0,1.5) ..controls (0.7,1.5) .. (0.9,1.8) .. controls (1.3,2.5) .. (1.8,1.9) ..controls (2.2,1.5).. (2.5,1.5) ..controls (2.8,1.5) .. (3.05,1.2) .. controls (3.5,0.6) .. (4.0,1.2) ..controls (4.3,1.5).. (5,1.5);
\end{tikzpicture} 
\qquad \qquad \qquad 
\begin{tikzpicture}[scale=1.0]
    \draw (0,5) .. controls (0.5,2.5) .. (1,1) .. controls (1.5,0) .. (1.8,1) .. controls  (2.5, 4) .. (3.0,3) .. controls (3.5,2) ..(4.0,3) .. controls (4.7,4.5).. (4.9,5) ; 
     \draw[blue] (0,1) ..controls (0.7,1) .. (0.9,1.3) .. controls (1.3,2.0) .. (1.9,1.3) ..controls (2.2,1.0).. (2.5,0.95) --(5,1.0);
\end{tikzpicture} 
 \caption{\textcolor{blue}{Left}: Out of the two wave-functions for the double-well potential  the symmetric one (drawn in red) represents the lowest energy state $\vert 0 \rangle$. The Glauber-Sudarshan state will be related to the {\it interacting} vacuum for the full potential and not the lowest energy state $\vert 0 \rangle$, {\it i.e.} 
 $\vert 0 \rangle \rightarrow \vert\Omega\rangle$. However using \eqref{saw1} we could express it around any one of the (Minkowski) minima in the full double well potential. \textcolor{blue}{Right}: For more general potential with unequal minima,  the Glauber-Sudarshan state will again be related to the interacting vacuum $\vert\Omega\rangle$ associated with the full potential. Again for computational purpose using \eqref{saw1} one could use the localized wave-function near the lowest energy, {\it i.e.} the Minkowski minimum, to realize the state. However the second minimum, associated with positive energy, should be removed by quantum corrections so that only the Minkowski minimum survives with no potential conflict with EFT considerations.}
\label{multsaddle} 
\end{figure}
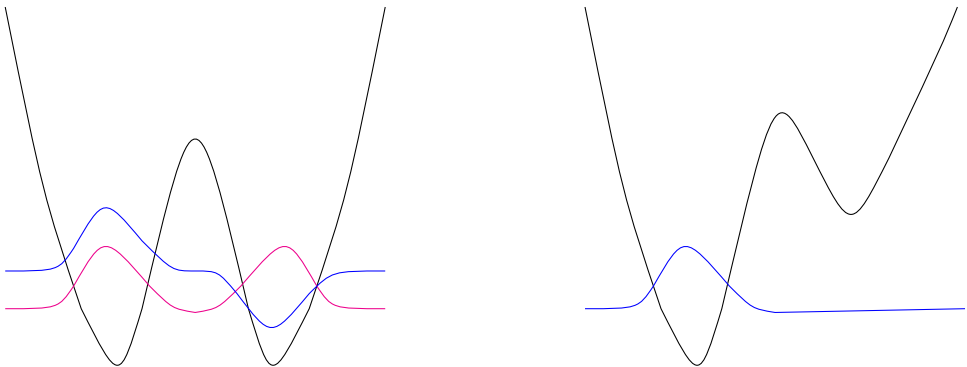

\noindent The second and the third problems raised above are fatal: they clearly tell us that we cannot use the standard notion of coherent state from say \eqref{dadari}. What can we use, now that we only have the interacting vacuum $\vert\Omega\rangle$ to our disposal? Our proposal 
is to use the Glauber-Sudarshan state $-$ named after \cite{GSstate} $-$ which is constructed by displacing the interacting vacuum in the following way \cite{desitter2, coherbeta, coherbeta2}:
\bg\label{saw1}
\vert\sigma\rangle = \mathbb{D}(\sigma) \vert\Omega\rangle \propto \mathbb{D}(\sigma) \lim_{{\rm T}\to \infty(1-i\epsilon)} {\rm exp}\left(-i {\bf H} {\rm T}\right) \vert 0\rangle_{\rm min}, \nd
where $\mathbb{D}(\sigma)$ is the {\it displacement} operator that is not unitary; $\vert 0 \rangle_{\rm min}$ is the free vacuum associated to any chosen Minkowski minimum\footnote{To see the second equality, express $\vert 0 \rangle_{\rm min}$ in terms of the linear combination of the eigenstates of the total Hamiltonian ${\bf H}$, in the following way:
\bg \vert 0 \rangle_{\rm min} = c_o \vert\Omega\rangle + \sum_{n = 1}^\infty c_n\vert n \rangle, \nonumber \nd
where $c_o = \langle\Omega\vert 0\rangle_{\rm min}$. Acting with 
$e^{-i{\bf H}{\rm T}}$ on $\vert 0\rangle_{\rm min}$, with the assumption that there is an energy gap between the interacting vacuum $\vert\Omega\rangle$ and the higher states $\vert n\rangle$, and then taking the limit of ${\rm T} \to \infty(1-i\epsilon)$, decouples all the higher states $\vert n\rangle$ and gives us:
\bg \vert\Omega\rangle = \lim_{{\rm T} \to \infty(1-i\epsilon)} {e^{-i({\bf H} - {\rm E}_o){\rm T}}\vert 0\rangle_{\rm min}\over \langle \Omega \vert 0\rangle_{\rm min}}, \nonumber \nd
implying that the proportionality constant in \eqref{saw1} is 
${e^{-i{\rm E}_o {\rm T}}\over \langle\Omega\vert 0\rangle_{\rm min}}$. Plugging the second equality in the definition of $\langle\varphi\rangle_\sigma = {\langle\sigma\vert \varphi\vert \sigma\rangle\over \langle\sigma\vert \sigma\rangle}$ then restricts the path integral over the Minkowski minimum. \label{saddledistance}}; and ${\bf H}$ is the total Hamiltonian that contains the full non-perturbative and non-local contributions. The latter will be described in details in section \ref{sec4.2}. By construction it is {\it not} a coherent state per se, but mimics somewhat the construction of a coherent state for a single well which is a displacement of the free vacuum. (As an example, a coherent state $\vert \sigma\rangle_{\rm c}$ can be described as $\vert\sigma\rangle_{\rm c} = \mathbb{D}_{\rm c}(\sigma) \vert 0 \rangle_{\rm min}$, where $\mathbb{D}_{\rm c}(\sigma)$ is the displacement operator, or more appropriately as in \eqref{dadari}.) However the definition \eqref{saw1} remains unchanged even if we have multiple minima: the first equality in \eqref{saw1} is universal, whereas the second equality depends on which Minkowski minimum we choose. As an example consider the two potentials depicted in {\bf figure \ref{multsaddle}}. The Glauber-Sudarshan state is the displacement of the interacting vacua associated with the minimum energy state for the {\it full} interacting potential. For computational purpose we could, using \eqref{saw1}, describe it as {\it localized around a Minkowski minimum} in each of the two configurations respectively by choosing $\vert 0 \rangle_{\rm min}$ appropriately. For the potential on the left of {\bf figure \ref{multsaddle}} the states could be localized around any one of the two minima. For the potential on the right of {\bf figure \ref{multsaddle}}, the second minima associated with the positive potential should be removed by quantum corrections in the theory, and therefore only the Minkowski (and AdS) minima remain. Glauber-Sudarshan states are then defined, using \eqref{saw1}, over the Minkowski $-$ and {\it not} the AdS $-$ minima. Such states clearly answer all the objections raised above. 

\vskip.1in

\noindent ${\bf 1.}$ By construction \eqref{saw1} is not classical as $\vert\Omega\rangle$ is far from classical. Because of this non-classical nature, $\vert\sigma\rangle$ somehow {\it contains} all order quantum corrections which are in fact necessary to overcome the no-go theorems.

\vskip.1in

\noindent ${\bf 2.}$ Since the Glauber-Sudarshan states can be localized around the Minkowski minimum of the potential, but are not related to free vacuum associated with the potential, they cannot be generated by simple Bogoliubov transformations. The corresponding EOMs satisfied by these states will consequently be more involved, albeit tractable quantitatively.

\vskip.1in

\noindent ${\bf 3.}$ The definition of the Glauber-Sudarshan state is universal and works for all types of potentials irrespective of their natures. All we need are the interacting vacua expressed using the Minkowski minima (as in \eqref{saw1}) for each cases, and by appropriately displacing them we can construct the required states. Even if the potential has, in addition, a lowest energy AdS minimum (or a set of AdS minima), the Glauber-Sudarshan states are still constructed only over the Minkowski minimum\footnote{This raises two questions. \textcolor{blue}{One}, what if the potential only has AdS minima? And \textcolor{blue}{two}, how far the AdS minima should be from the Minkowski minima in a given potential? The answer to the first question, albeit an unlikely scenario in string theory, is that for such cases construction of the Glauber-Sudarshan states would be difficult. These states are not constructed over the AdS minimum as the boundary condition would not match. One possibility from our earlier discussion would be that the quantum corrections uplift some of the AdS minima to Minkowski minima (and nothing beyond that to avoid the aforementioned EFT issues!). Glauber-Sudarshan states may then be constructed over any one of such Minkowski vacuum using \eqref{saw1}. However if the potential also has Minkowski minima this uplift might be harder because the Minkowski minima {\it cannot} be uplifted further. For the second question, we expect the AdS minima to be far away in the field space, or be separated by a large potential wall, from the Minkowski minima to reduce the tunnelling effects.}.

\vskip.1in

\noindent The above arguments suggest that the Glauber-Sudarshan states, and {\it not} the coherent states, are responsible in creating a four-dimension de Sitter space-time with positive cosmological constant (see also \cite{dvali}). However one could raise the following questions. How do we quantify the displacement operator $\mathbb{D}(\sigma)$, and especially the value of $\sigma$? How do the EOMs show up in this new viewpoint? How is the positive energy realized, or how are the no-go theorems avoided using the Glauber-Sudarshan states? How are the other constraints like anomaly cancellation, flux quantizations et cetera in M-theory are taken care of now? How do these states satisfy the Effective Field Theory criteria? In the following sections we will answer most of these questions but leave the details of the EFT criteria, and how these states satisfy them, for section \ref{sec7}. 

\subsection{Quantifying the displacement operator and the emergent metric \label{sec2.2.3}}

Our first attempt would be to quantify the displacement operator $\mathbb{D}(\sigma)$. For the standard coherent state, the displacement operator is unitary and may be expressed by exponentiating the creation operator $a_k^\dagger$ associated with the momentum $k$ (typically in $3+1$ dimensions, although this definition could be extended to higher dimensions). Going beyond the single minumum to the multiple minima case, the coherent state takes the form \eqref{dadari}, which may not have a simple representation using a displacement operator. Additionally, and as emphasized earlier, the low energy effective field theory in type IIB theory is at order 1 coupling, so clearly the theory is interacting. For such a case, as mentioned earlier, we talk about Glauber-Sudarshan state \eqref{saw1} (instead of the coherent state) where the displacement operator isn't necessarily unitary, nor it may be represented by exponentiating a creation operator because defining a creation operator itself is complicated in an interacting theory. Nevertheless one could define an effective annihilation operator $a_{\rm eff}(k)$, constructed from mixing the creation and the annihilation operators for all $k$ as described in \cite{coherbeta}, that would annihilate these states:
\bg\label{virper}
a_{\rm eff}(k) \vert\sigma(k)\rangle = 0, \nd
where $\vert\sigma(k)\rangle$ is the Glauber-Sudarshan state for a given mode $k$ over the supersymmetric Minkowski background. Unfortunately this definition is not very illuminating, partly because of the non $c$-number nature of the commutator of $a_{\rm eff}(k)$ and 
$a_{\rm eff}^\dagger(k)$, and partly because of the absence of a simple description of $\mathbb{D}(\sigma)$, where $\sigma = \{\sigma(k)\} ~\forall~k$, in terms of $a_{\rm eff}(k)$ and 
$a_{\rm eff}^\dagger(k)$. How do we then quantify $\mathbb{D}(\sigma)$?

The answer actually appears from a slightly different consideration. To see this let us restrict ourselves to only a set of scalar fields $\varphi_1, \varphi_2, \varphi_3$ and $\varphi_4$ which form the representative samples of the spacetime metric in ${\bf R}^{2, 1}$, the internal metric in 
${\cal M}_4 \times {\cal M}_2 \times {\mathbb{T}^2\over {\cal G}}$, the three-form flux components and the fermionic condensates respectively. Such a choice has a hidden agenda: we can avoid inserting ghosts in the effective action, although we will show a way to insert them in the generic case. One might however get worried about representing tensor and scalar components by purely scalar fields $\varphi_i$. Notwithstanding the fact that this is just a toy model, in string field theory we do use something like a scalar field $\Phi$ to represent the level of the string field (see for example the recent work \cite{sft}). We are not using string field theory, so the choice of scalar fields should be viewed as an exercise in simplicity\footnote{Our work, as discussed in \cite{borel, ccpaper} has other parallels to string field theory, but we will not dwell on this aspect here anymore.}. The quantity $\sigma$, that enters in the definition of the displacement operator $\mathbb{D}(\sigma)$, then may be quantified as:
\bg\label{patarket}
\sigma = \{\sigma_i\} = \left(\alpha_{1\mu\nu}, \alpha_{2mn}, \beta_{\rm ABC}, \gamma\right) ~ \to ~ \left(\alpha_1, \alpha_2, \beta, \gamma\right), \nd
where $(\mu, \nu) \in {\bf R}^{2, 1}, ~(m, n) \in {\cal M}_4 \times {\cal M}_2 \times {\mathbb{T}^2\over {\cal G}}, ~({\rm A, B})\in {\bf R}^{2, 1} \times{\cal M}_4 \times {\cal M}_2 \times {\mathbb{T}^2\over {\cal G}}$ with $\gamma$ related to the condensate $\overline\Psi^{\rm A} \Psi_{\rm A}$. Since we are taking scalar degrees of freedom, the tensor indices are not important and therefore $\sigma_i$ may be related directly to $(\alpha_1, \alpha_2, \beta, \gamma)$ as shown on the right of \eqref{patarket}. A possible choice of the displacement operator then takes the following form:
\bg\label{28rms}
\log~\mathbb{D}(\sigma) = \int_{k_{\rm IR}}^\mu \prod_{j = 1}^4 d^{11}k_j \sum_{\{n_i\},\{m_i\}, \{p_i\}} z_{n_1m_1p_1...n_4m_4p_4}~\prod_{i = 1}^4 k_i^{2n_i}\sigma_i^{(m_i, p_i)}(k_i) \widetilde\varphi_i^{\ast m_i}(k_i), \nd
which is expressed completely in the language of the Fourier components of the field fluctuations over a solitonic background, and $z_{n_1m_1p_1...n_4m_4p_4}$ are dimensionful\footnote{All quantities are measured with respect to ${\rm M}_p$, although we could also choose the KK scale $\hat\mu$. For the latter there would be additional suppression factors proportional to powers of ${\hat\mu\over {\rm M}_p}$. See \cite{borel} for more details.} constants with $n_i \ge 0, m_i \ge 1, p_i \ge 1$. The eleven-dimension integral is henceforth done over 
the Minkowski spacetime, unless mentioned otherwise, with a metric $(\eta_{\mu\nu}, \delta_{mn}, \delta_{ab})$ for $(\mu, \nu) \in {\bf R}^{2, 1}, (m, n) \in {\cal M}_6, (a, b) \in {\mathbb{T}^2\over {\cal G}}$.
The form of the displacement operator \eqref{28rms} differs from the one we considered in \cite{borel} in the sense that it is complex and more elaborate, although one could make it even more generic by viewing $\widetilde\varphi_i^m(k_i)$ to be the first term in the binomial expansion of $\varphi_i^m(x, y)$ in Fourier modes. However no matter how generic the displacement operator is, it is clear that we need to impose the following constraints on the $z_{n_1m_1p_1...n_4m_4p_4}$ coefficients:
\bg\label{dotperv}
z_{n_12p_10..0} ~ = ~ z_{0..n_22p_20..0} ~ = ~ z_{0..n_32p_3000} ~ = ~ z_{0..n_41p_4} ~ = ~ 0, \nd
to avoid generating massive states in the theory\footnote{There could be potential {\it ghosts} for $n_i > 0$ and $m_i = 2, ~\forall p_i \ge 1$, which will require more careful analysis. The constraint \eqref{dotperv} eliminates them here, but generically ghosts could in principle appear once we take the actual tensors into account. We will discuss this later.}. The other quantity defined in \eqref{28rms} is $\sigma_i^{(m_i, p_i)}$ such that $\sigma_i^{(1, 1)}(k_i) \equiv \sigma_i(k_i)$. Without loss of generality we can keep $z_{0110..0} = z_{0000110..0} = .. = {1\over 2}$. Note that for $m_i > 2$, \eqref{28rms} introduces polynomial interactions in the theory whereas for $(m_1, m_2, m_3, m_4)$ taking values $(1, 0, 0, 0), (0, 1, 0, 0), (0, 0, 1, 0)$ and $(0, 0, 0, 1)$, \eqref{28rms} simply shifts the kinetic terms once we insert $\mathbb{D}(\sigma)$ inside a path-integral. To see how this works out precisely, let us consider an example where we want to compute the expectation value of $\varphi_1$ over the Glauber-Sudarshan state $\vert\sigma\rangle$ in the following way:
\bg\label{mmtarfox}
\langle {\hat\varphi}_1\rangle_\sigma \equiv {\langle \sigma\vert {\hat \varphi}_1\vert \sigma\rangle \over \langle\sigma \vert \sigma\rangle} 
= {\int \prod\limits_{i = 1}^4 {\cal D}\varphi_i~ e^{-{\bf S}_{\rm tot}}~\varphi_1(x, y) ~\mathbb{D}^\dagger(\sigma) \mathbb{D}(\sigma) \over 
\int \prod\limits_{i = 1}^4 {\cal D}\varphi_i~ e^{-{\bf S}_{\rm tot}}~\mathbb{D}^\dagger(\sigma) \mathbb{D}(\sigma)}, \nd 
where $\hat\varphi_1$ is the operator associated to the field $\varphi_1$, and ${\bf S}_{\rm tot}$ is the total action that involves the kinetic terms of the four fields as well as their all-order {\it perturbative} interactions. Other possible ingredients in ${\bf S}_{\rm tot}$, as well as the reason for choosing only the {\it perturbative} interactions, will be elaborated soon\footnote{The $i\epsilon$ prescription appears in the usual way from the path integral \eqref{mmtarfox} over the Minkowski background. \label{ieps}}. {\it Note that the integrations are done over the ``on-shell'' degrees of freedom, which means that we have given small masses to the ``off-shell'' degrees of freedom to avoid generating non-local interactions}. As such they do not appear in the energy range $k_{\rm IR} < k < \mu$. A more careful analysis will be presented in section \ref{sec4.2}, wherein we will analyze the consequence of the non-local interactions. Coming back, the form of $\mathbb{D}(\sigma)$ in \eqref{28rms} implies that it is complex, but $\mathbb{D}^\dagger(\sigma) \ne \mathbb{D}^{-1}(\sigma)$, and therefore inserting this in the path integral \eqref{mmtarfox} and imposing \eqref{dotperv}, gives us the following three results.

\vskip.1in

\noindent ${\bf 1.}$ The kinetic terms of the scalar fields, namely 
$k_i^2 \vert\widetilde\varphi_i(k_i)\vert^2$, shift by the $\sigma_i(k_i)$ factors from the $m_i = p_i = 1, n_i = 0$ terms of \eqref{28rms} in the following way:
\bg\label{monwhip}
k_i^2 \vert \widetilde\varphi_i(k_i)\vert^2 ~ \rightarrow ~ k_i^2 \Big\vert \widetilde\varphi_i(k_i) - {\sigma_i(k_i)\over k_i^2}\Big\vert^2, \nd
which is exactly what we require to keep the one-point functions in \eqref{mmtarfox} non-zero. The solitonic part will not change the above decomposition. This is then the meaning of the {\it displacement} in the displacement operator $\mathbb{D}(\sigma)$. For $m_i = 1$ but $p_i > 1$ and $n_i > 0$, there will be additional contribution to 
\eqref{monwhip} which may be related to the $\alpha'$ corrections to the T-duality rules that we alluded to in footnote \ref{petbeta} (see the upcoming paper \cite{hete8} for more details on this).

\vskip.1in

\noindent {\bf 2.} The displacement operator presented in \eqref{28rms} is {\it complex}, which implies that the any other Glauber-Sudarshan state $\vert\sigma'\rangle$ is not orthogonal\footnote{Unfortunately even for a real choice of $\mathbb{D}(\sigma)$, as in \cite{borel}, orthogonality cannot be guaranteed.} to $\vert\sigma\rangle$, {\it i.e.} 
$\langle\sigma'\vert\sigma\rangle$ is 
non-zero even for large difference between $\sigma'$ and $\sigma$. Because of this the resolution of identity is harder to find here. Nevertheless let us define the following operator:
\bg\label{metmedicmey}
{\cal O} = \sum_{\sigma, \sigma^\ast} ~{\underset{k}{\bigotimes}~\vert\sigma(k)\rangle \langle\sigma(k)\vert\over \prod\limits_k~\langle\sigma(k)\vert \sigma(k)\rangle}, \nd
where the operation $\underset{k}{\bigotimes}$ collects all the kets with different momenta on one side and the bras, again with different momenta, on the other side without contractions; and $\prod\limits_k$ is the usual product over $k$. For any arbitrary state $\vert\sigma'\rangle$, one can see that ${\cal O}\vert\sigma'\rangle \propto \vert\sigma'\rangle$, because 
when ${\rm Re}~\sigma \ne {\rm Re}~\sigma'$, the exponential factors in $\mathbb{D}(\sigma)$ make sure that the corresponding path-integral is suppressed by:
\bg\label{wombet}
{\rm exp}\left(-{\vert\sigma(k) -\sigma'(k)\vert^2\over 4k^2} + {\vert\sigma(k) + \sigma'(k)\vert^2\over 4k^2} - {\vert\sigma(k)\vert^2\over k^2} + {\sigma^\ast(k)\sigma'(k) - \sigma(k)\sigma^{'\ast}(k)\over 2k^2}\right),
\nd
where for both ${\rm Re}~\sigma(k) > {\rm Re}~\sigma'(k)$ and ${\rm Re}~\sigma(k) < {\rm Re}~\sigma'(k)$ this becomes arbitrarily small provided ${\rm Im}~\sigma > {\rm Im}~\sigma'$; and for $\sigma(k) = \sigma'(k)$ this becomes identity\footnote{Alternatively one could think of transforming $\sigma(k) \to - \sigma(k) \equiv \hat\sigma(k)$, Then $\forall ~{\rm Re}~\sigma'(k) > 0$ and ${\rm Re}~\sigma'(k) \ne {\rm Re}~\hat\sigma(k)$ orthogonality can be ensured. Unfortunately, when $\sigma'(k) = + \hat\sigma(k)$, \eqref{wombet} is non-zero but not an identity anymore.}. Thus 
it is the closest we can come to decoupling these states despite the absence of a clear notion of {\it orthogonality}. Therefore ${\cal O}$ does almost behave like an identity operator for all practical purposes.
(We will provide a better construction in section \ref{sec4.1} where we will see that we need to add another operator to \eqref{metmedicmey} to convert it into an identity.)

\vskip.1in

\noindent ${\bf 3.}$ The remaining terms in the displacement operator \eqref{28rms} for $m_i \ge 3$ only changes the coefficients of the interaction terms in the total action ${\bf S}_{\rm tot} \equiv {\bf S}_{\rm kin} + {\bf S}_{\rm int}$. If we denote the interaction term 
${\bf S}_{\rm int}$ as a function of the four fields $\varphi_i$ and the coefficients $\{g_l\}$ of the irrelevant perturbative interactions, then in the presence of \eqref{28rms}, we expect:
\bg\label{indydial}
{\bf S}_{\rm int}(\varphi_1,..,\varphi_4; \{g_l\}) ~ \rightarrow ~ 
{\bf S}_{\rm int}(\varphi_1,..,\varphi_4; \{g_l\} + \{z\}), \nd
where $\{z\}$ denotes the set of the coefficients $z_{n_1m_1...m_4p_4}$, 
$\forall (n_i, m_i, p_i)$. Thus in the presence of \eqref{28rms}, the action that appears in the path-integral \eqref{mmtarfox} could become slightly {\it different} from the expected Wilsonian effective action, or more appropriately, the action after the ERG operation. Note that we have only considered perturbative pieces in both ${\bf S}_{\rm int}$ and the displacement operator $\mathbb{D}(\sigma)$ from \eqref{28rms}. An alternative choice of the displacement operator that shifts $\varphi_i$ in the interaction term ${\bf S}_{\rm int}$ as:
\bg\label{noharamcam}
{\bf S}_{\rm int}(\varphi_1,...,\varphi_4; \{g_l\}) ~ \rightarrow ~ 
{\bf S}_{\rm int}(\varphi_1 - \sigma_1, ..., \varphi_4 - \sigma_4; \{g_l\}), \nd
{\it i.e.} not as in \eqref{indydial} but as global change like \eqref{monwhip}, does not help because taking an expectation value \eqref{mmtarfox} only gives the tree-level result that reduces $\sigma_i$ to 0. This is a self-consistency test otherwise the result would have been in tension with the no-go theorems \cite{GMN}.

\vskip.1in

\noindent With the above inputs we are ready to interpret \eqref{mmtarfox} as a way to extract the form of the background using path-integrals. The background data, like metric and flux components as well as the fermionic condensates, may now be viewed as expectation values or as {\it emergent} quantities. As an example, the space-time metric (albeit from M-theory point of view) now takes the form:

{\footnotesize
\bg\label{russmameye}
\langle \hat{\bf g}_{\mu\nu}\rangle_{\sigma} \equiv {\langle\sigma\vert \hat{\bf g}_{\mu\nu}\vert\sigma\rangle\over \langle\sigma\vert\sigma\rangle} = {\int \{{\cal D}{g}_{\rm AB}\}\{ {\cal D}{\rm C}_{\rm ABD}\}\{ {\cal D}{\Psi}_{\rm A}\}  \{ {\cal D}\overline{\Psi}_{\rm A}\}\{ {\cal D}\Upsilon_g\}~e^{-{\bf S}_{\rm tot}}~ \mathbb{D}^\dagger(\sigma)
{g}_{\mu\nu}(x, y)
\mathbb{D}(\sigma) \over 
\int \{{\cal D}{g}_{\rm AB}\}\{ {\cal D}{\rm C}_{\rm ABD}\}\{ {\cal D}{\Psi}_{\rm A}\} \{ {\cal D}\overline{\Psi}_{\rm A}\}\{ {\cal D}\Upsilon_g\}~e^{-{\bf S}_{\rm tot}} ~\mathbb{D}^\dagger(\sigma) 
\mathbb{D}(\sigma)},
\nd}
where $\Upsilon_g$ is the ghost sector; and $\mathbb{D}(\sigma)$ now differs from \eqref{28rms} by the replacement of the scalar fields with the tensor fields related to the metric and the flux components and with the fermionic condensates (the latter is to avoid Grassmanian integrals). The total action is also different because now it involves, in addition to ${\bf S}_{\rm kin}$ and ${\bf S}_{\rm int}$, the topological interactions ${\bf S}_{\rm top}$, gauge fixing terms ${\bf S}_{\rm gf}$, as well as ghost interactions ${\bf S}_{\rm ghost}$. Two immediate questions arise:

\vskip.1in

\noindent ${\bf 1.}$ What equations of motion do the expectation values satisfy? Can they be derived in the generic set-up?

\vskip.1in

\noindent ${\bf 2.}$ Is it possible to compute explicitly the path-integrals in say \eqref{mmtarfox} or \eqref{russmameye} even if we don't know the exact coefficients of the irrelevant operators? 

\vskip.1in

\noindent In the following sections we will argue that both the above questions {\it can} be answered satisfactorily at least in the toy model with scalar fields. With tensors and fermions, complications as well as subtleties arise due to the formidable nature of ${\bf S}_{\rm tot}$ from the presence of ghosts and gauge-fixing terms. In this paper we will spare the readers, as well as ourselves, from these challenges and only consider the theory with four scalar fields. This is because, once we know how to tackle the path integral in \eqref{mmtarfox}, the generic case \eqref{russmameye} will not be very hard. Therefore, in the following, we will start by answering the second question first and then derive the corresponding EOMs in section \ref{sec4}.

\section{Path integral, factorial growth and Borel resumming a Gevrey series \label{patho}}

The path integral in \eqref{mmtarfox} basically computes the one-point function of the scalar field $\varphi_1$. In the absence of the displacement operator $\mathbb{D}(\sigma)$ from \eqref{28rms}, one point functions would typically vanish, and the only non-zero answer from \eqref{mmtarfox} would be the background value (which is the warped Minkowski background here). Thus to get a non-zero answer we need \eqref{monwhip}, but also the full perturbative interactions ${\bf S}_{\rm int}$. Why is that, and what about the non-perturbative interactions? These and other questions will be discussed in this section, and in particular we will be able to quantify precisely how the path-integral computation leads to a Gevrey growth. Previously such a claim was presented without a formal proof in \cite{borel}. Here we will fill in the gaps. In the process we will also show that the path integral computation follows a simple binomial expansion.

\subsection{Organizing the path integral \eqref{mmtarfox} using binomial expansions \label{sec3.4}}

Let us start with the perturbative interactions.
The presence of these interactions is easy to see: it appears directly from the presence of the interacting vacua $\vert\Omega\rangle$ in the definition of the Glauber-Sudarshan state \eqref{saw1}. The shift in \eqref{monwhip} appears from the non-unitary nature of the displacement operator when we define $\mathbb{D}(\sigma) \mathbb{D}^\dagger(\sigma)$, or more appropriately $\mathbb{D}(\sigma, \varphi_i(x, y)) \mathbb{D}^\dagger(\sigma, \varphi_i(x, y))$, at a coincident point. Clearly this is very different from what we would have expected using coherent states, and therefore this gives yet another motivation to use the Glauber-Sudarshan states instead of the coherent states\footnote{One could however ask the following question: what would happen if we take a unitary $\mathbb{D}(\sigma, \varphi_i)$ but define the product $\mathbb{D}(\sigma, \varphi_i(x_1, y_1)) \mathbb{D}^\dagger(\sigma, \varphi_i(x_2, y_2))$ at two different points? The answer could be provided in multiple ways, but the simplest one is as follows. Because of the presence of a {\it free} vacuum in the definition of the displacement operator for a coherent state, the path integral structure, if any, will unfortunately only have the tree-level pieces but no interaction terms. This will clash with the no-go theorems \cite{GMN}, plus when we make $x_1 \to x_2$, the system will have no solution other than the Minkowski one.}. Using the Glauber-Sudarshan states, somewhat surprisingly, the non-perturbative effects are not that hard to see because they depend on factorial growths of the amplitudes at a given order. The factorial growths are of the {\it Gevrey} kind \cite{gevrey, borelborel, gevrey2, dorigoni, ecalle, borel}, so what we want here is to Borel resum a Gevrey series. The net result of such a summation is the appearance of a condensed form for the expression of $\langle\varphi_i\rangle_\sigma$. From here the {\it non-perturbative} effects could be discerned.

The aforementioned story is non-trivial, so we need to carefully
address the path integral in \eqref{mmtarfox}. Fortunately this has already been done in great details in \cite{borel}, so we will present and elaborate on some salient features that were not discussed in \cite{borel}. One of the key question is of course the reason behind the Gevrey growth, and how to perform the Borel resummation of a Gevrey series. To answer this, we need to address an even more basic question: how to compute the path integral in \eqref{mmtarfox} efficiently?

Before starting the path integral computation we should note that, because of the shift in the kinetic term from \eqref{monwhip}, one-point functions would be non-zero and therefore this cannot be the usual analysis that we perform for the Feynman path integral. Two key differences are:

\vskip.1in

\noindent ${\bf 1.}$ The path integral in \eqref{mmtarfox} can be represented by the so-called {\it nodal diagrams} that form a bigger class of diagrams than the Feynman diagrams. 

\vskip.1in

\noindent ${\bf 2.}$ Feynman diagrams require the presence of complex fields otherwise certain non-trivial (and necessary) cancellations cannot be performed. Nodal diagrams work for both real and complex fields. With the latter, certain restrictions appear, but they do not alter much of the physics.

\vskip.1in

\noindent Both the above points have been described in \cite{borel}, which the readers may look up for details. Here we want to emphasize the fact that the easiest way to do the path integral in \eqref{mmtarfox} is to break up the momenta from $k_{\rm IR} < k < \mu$ into small and equal pieces so as to convert any integrals into summation via the standard formula:
\bg\label{fleabag}
\int_a^b dx~f(x) = \lim_{{\rm N} \to \infty} {b-a\over {\rm N}} \sum_{j = 0}^{\rm N} f\left(a + j\left({b-a\over {\rm N}}\right)\right). \nd
Since our integrals will be over eleven-dimensional momenta, that factor of ${b-a\over {\rm N}}$ may be converted to the inverse volume, {\it i.e.} ${1\over {\rm V}}$, of the eleven-dimensional space. This is related to the IR cut-off $k_{\rm IR}$ that we continue to impose here. With all these in places, the path integral structure in \eqref{mmtarfox} may be represented by eq. (3.4) in \cite{borel}. The path integral has the following structural breakdown.

\vskip.1in

\noindent ${\bf 1.}$ The measures of all the fields in terms of their real and complex Fourier components. In \cite{borel} we took only three scalar fields, but here we take four.

\vskip.1in

\noindent ${\bf 2.}$ The displacement operator $\mathbb{D}^\dagger(\sigma) \mathbb{D}(\sigma)$. In \cite{borel} we took a real displacement operator. Here we take a complex one \eqref{28rms} although 
$\mathbb{D}^\dagger(\sigma) \mathbb{D}(\sigma)$ doesn't change from what we had in \cite{borel}.

\vskip.1in

\noindent ${\bf 3.}$ The kinetic terms of all the fields written in terms of their Fourier components. In \cite{borel} we took a Lorentzian formalism, here we take an Euclidean approach.

\vskip.1in

\noindent ${\bf 4.}$ The interactions terms in terms of powers and derivatives of the four fields. Here we take one specific set of interactions, and we will discuss the generic case a bit later. 

\vskip.1in

\noindent ${\bf 5.}$ Momentum conservation directly imposed on the Fourier components of the fields. As in \cite{borel}, we can impose momentum conservation on one Fourier component of the $\varphi_4$ field.

\vskip.1in

\noindent The important parts in the above breakdown of the path integral structure are in points 2, 3 and 4. Combining 2 and 3, gives us \eqref{monwhip}, which is what we term as the shift in the vacua. This shift, as mentioned earlier, is one of the important aspects of our construction as it keeps the one-point functions non-zero. Point 3 is the interaction term, which may be given the following binomial expansion:
\bg\label{phobeindy}
\partial^n \varphi^p_1(x, y) ~ \to ~ {\cal Z}_{\varphi_1} \equiv {1\over {\rm V}^p p!} \Big(\sum_{i = 1}^p k'_i\Big)^n \bigodot_{k'_i\to k_i}\Big(\widetilde\varphi_1(k_1) + \widetilde\varphi_1(k_2) + \widetilde\varphi_1(k_3) + ..... + 
\widetilde\varphi_1(k_{{\rm N}_\mu})\Big)^p, \nd
with $n > 0$ because we chose to integrate out the off-shell degrees of freedom by giving them a small mass (see discussion after \eqref{mmtarfox}).
The operation $\underset{k'_i \to k_i}{\bigodot}$ chooses the appropriate set of ({\it i.e.} $p$ number of) $k'_i$ momenta that would tally with the $p$ product of the Fourier components; and ${\rm N}_\mu$ is the number of division of the momenta $k$ between $k_{\rm IR} \equiv k_1$ and $\mu << \hat\mu$.
This is most clearly presented as products of the Fourier components, as we did in eq. (3.4) of \cite{borel}, but the binomial form is easier to handle. The most dominant terms in the binomial expansion go as products of $p$ number of Fourier components with all momenta {\it unequal}. They all scale as $p!$, which are cancelled by the ${1\over p!}$ factor in \eqref{phobeindy}. If we denote the dominant contribution as ${\cal A}(k_i, k_{i+1},.., k_{i+p})$
via an ordered set of $p$ momenta, then \eqref{phobeindy} may be rewritten as:
\bg\label{antikytheria}
\begin{split}
{\rm V}^p {\cal Z}_{\varphi_1}= & ~{\cal A}(k_1, k_2, .., k_p) \Big(\sum_{i = 1}^p k_i\Big)^n + {\cal A}(k_{p+1}, k_{p+2}, .., k_{2p}) \Big(\sum_{i = p+1}^{2p} k_i\Big)^n + .....
\\
+ & ~{1\over (p-1)!} {\cal A}(k''_1, k''_1, .., k''_1, k''_2) \left((p-1)k''_1+k''_2\right)^n  + {1\over p!} 
{\cal A}(k'_1, k'_1,.., k'_1) \left(pk'_1\right)^n,
\end{split}
\nd
where we see that the sub-dominant contributions are being suppressed by 
$(p-n)!$ with $n$ being the order of the binomial expansion. The total number of terms  would depend on ${\rm N}_\mu$, {\it i.e.} on how many finer pieces are we dividing the momenta $k_{\rm IR} < k < \mu$. For the present case it should be ${\rm N}_\mu^p$, and clearly ${\rm N}_\mu >> 1$. This means that the number of interaction terms from the $\varphi_1$ sector may be simply be arranged as the following series:
\bg\label{dialdest}
{\cal Z}_{\varphi_1}= {1\over {\rm V}^p} \Big({\cal A}_1 + {\cal A}_2 + {\cal A}_3 + {\cal A}_4 + ..... + {\cal A}_{{\rm N}_\mu^p-1} + {\cal A}_{{\rm N}_\mu^p}\Big), \nd
where ${\cal A}_1 = {\cal A}(k_1,..,k_p)(k_1+..+ k_p)^n = \widetilde\varphi_1(k_1)..\widetilde\varphi_1(k_p)(k_1+..+k_p)^n$, and so on in the same order as in \eqref{antikytheria}. The above is of course only the $\varphi_1$ sector as mentioned, so we will have similar binomial series for $\varphi_2, \varphi_3$ and $\varphi_4$. However momentum conservation will put some restrictions in the $\varphi_4$ sector\footnote{The analysis is independent of how we assign the momentum conservation rule, as is shown in \cite{borel} and \cite{ccpaper}. This is also quite easy to see from our above analysis, so will leave it for our diligent readers as an exercise.} because (a) not all terms of the Fourier components for $\varphi_4$ are independent, and (b) the total momentum expansion for the $\varphi_4$ sector would now simply be a linear combination of the momenta for the other three sectors. These subtleties need to be carefully taken care of, which we did in \cite{borel}, but here we will not discuss this further. Instead, we just rearrange the path integral in \eqref{mmtarfox}, using the aforementioned information, in the following suggestive way:
\bg\label{mcahini}
{\rm Num}~\langle\hat\varphi_1\rangle_\sigma = \int \prod_{i = 1}^4 [{\cal D}\widetilde\varphi_i]~e^{-{\bf S}_{\rm kin} + \log\left[\mathbb{D}^\dagger(\sigma)\mathbb{D}(\sigma)\right]} \Big(1 +g 
\overbracket[1pt][7pt]{{\cal Z}_{\varphi_1}{\cal Z}_{\varphi_2} {\cal Z}_{\varphi_3} \hat{\cal Z}_{\varphi_4} + .....\Big) \cdot {1\over {\rm V}}\sum_{j= 1}^{{\rm N}_\mu}\widetilde\varphi_1}(\hat{k}_j)\psi_{\hat{k}_j}(z) \nd
where Num denotes the numerator of the path integral \eqref{mmtarfox}, the over-bracket denotes the contraction of the $\widetilde\varphi_1(\hat{k}_j)$ field with one of the field in ${\cal Z}_{\varphi_1}$, and $\psi_{\hat{k}_j}(z)$ for $z \equiv (x, y)$ denotes the wave-function over the Minkowski background. The momentum conservation condition is imposed on the $\varphi_4$ field components and it is reflected by $\hat{\cal Z}_{\varphi_4}$ in \eqref{mcahini}. $\mathbb{D}(\sigma)$ is the displacement operator from \eqref{28rms}, $g$ is the coupling constant expressed in inverse powers of ${\rm M}_p$, and the dotted terms are the higher order interactions. Two questions arise:

\vskip.1in

\noindent ${\bf 1.}$ Can we compute the over-bracket contraction explicitly?

\vskip.1in

\noindent ${\bf 2.}$ Can we go to all orders in the perturbative expansion and compute the over-bracket contraction explicitly? 

\vskip.1in 

\noindent The answer to both the questions are in the affirmative, but we will have to develop the story a little bit more before we can get to the final answer. This is what we turn to next.

\subsection{Computing the path integral \eqref{mmtarfox} and nested integrals \label{sec4.2}}

The numerator of the path integral \eqref{mmtarfox} is as shown in \eqref{mcahini}, and the denominator will be similar but without the source term. The shift in the vacuum is denoted by the $\log\left[\mathbb{D}^\dagger(\sigma) \mathbb{D}(\sigma)\right]$ term, and this results to at least \eqref{monwhip}, but could in-principle also shift the coupling constant $g$ (see earlier discussion). The computation of the over-bracket contraction is quite a challenging exercise, as evident from the diagramatic study in \cite{borel}, so we will not pursue this here. Instead we will give a toy computation to motivate the actual analysis. Let us then start with the following {\it interacting} part of the path integral:

{\footnotesize
\bg\label{ivleague}
{\cal W}_\phi = 
{1\over p!{\rm V}^p}\int d\phi_1 d\phi_2 d\phi_3 .... e^{-b_1\left(\phi_1 - {\beta_1\over b_1}\right)^2} e^{-b_2\left(\phi_2 - {\beta_2\over b_2}\right)^2} e^{-b_3\left(\phi_3 - {\beta_3\over b_3}\right)^2}...
(k''_1 + k''_2 +..+k''_p)^n \bigodot_{k''\to k'}\left(\phi_{1'} + \phi_{2'} + ....\right)^p,
\nd}
where $\phi_i = \phi(k_i)$, $\phi_{i'} = \phi(k'_i), b_i = b(k_i), \beta_i = \beta(k_i)$ and the operation $\underset{k''_i \to k'_i}{\bigodot}$, as in \eqref{phobeindy}, chooses the appropriate set ({\it i.e.} $p$ number) of $k''_i$ momenta that would tally with the $p$ product of the field components $\phi(k'_i)$. In the end all set of momenta, $k_i, k'_i$ and $k''_i$, needs to be matched properly to extract the value of the path integral. For example if we consider $p-1$ fields to have the same momenta and one field with a different moments, we are looking at the following interaction in the path integral \eqref{ivleague}:
\bg\label{ivile1}
{1\over (p-1)!{\rm V}^p} ~\phi^{p-1}_{i'} \phi_{j''} \Big((p-1)k'_i + k''_j\Big)^n, \nd
for all values of $i$ and $j$. Looking at \eqref{ivile1} one might get worried about a term linear in $\phi$, but due to the shifted vacua in \eqref{ivleague}, such a term is non-zero. A simple way to then perform the path integral is to first fix a value of $i$, and then sum over all values of $j$. After which we repeat the process with a different value of $i$, and then sum over all the corresponding $j$, and so on. The result may be summarized in the following way:

{\footnotesize
\bg\label{blamortimer}
\begin{aligned}
{\cal W}^{(1)}_\phi =~ &{1 \over {\rm V}^p}\sum_{r, l} {\bf A}^{(1)}_{rl} \left({\beta_1\over b_1}\right)^{p-1-r} k_1^l\left({\beta_2\over b_2}~k_2^{n-l} + {\beta_3\over b_3}~k_3^{n-l} + 
{\beta_4\over b_4}~k_4^{n-l} + {\beta_5\over b_5}~k_5^{n-l} + 
{\beta_6\over b_6}~k_6^{n-l} + ... +{\beta_{{\rm N}_\mu}\over b_{{\rm N}_\mu}}~k_{{\rm N}_\mu}^{n-l}\right)\\
+ ~&{1 \over {\rm V}^p}\sum_{r, l} {\bf A}^{(2)}_{rl} \left({\beta_2\over b_2}\right)^{p-1-r} k_2^l\left({\beta_1\over b_1}~k_1^{n-l} + {\beta_3\over b_3}~k_3^{n-l} + 
{\beta_4\over b_4}~k_4^{n-l} + {\beta_5\over b_5}~k_5^{n-l} + 
{\beta_6\over b_6}~k_6^{n-l} + ...+{\beta_{{\rm N}_\mu}\over b_{{\rm N}_\mu}}~k_{{\rm N}_\mu}^{n-l}\right)\\
+ ~&{1 \over {\rm V}^p}\sum_{r, l} {\bf A}^{(3)}_{rl} \left({\beta_3\over b_3}\right)^{p-1-r} k_3^l\left({\beta_1\over b_1}~k_1^{n-l} + {\beta_2\over b_2}~k_2^{n-l} + 
{\beta_4\over b_4}~k_4^{n-l} + {\beta_5\over b_5}~k_5^{n-l} + 
{\beta_6\over b_6}~k_6^{n-l} + ...+{\beta_{{\rm N}_\mu}\over b_{{\rm N}_\mu}}~k_{{\rm N}_\mu}^{n-l}\right)\\
+ ~& {1 \over {\rm V}^p}\sum_{r, l} {\bf A}^{(4)}_{rl} \left({\beta_4\over b_4}\right)^{p-1-r} k_4^l\left({\beta_1\over b_1}~k_1^{n-l} + {\beta_2\over b_2}~k_2^{n-l} + 
{\beta_3\over b_3}~k_3^{n-l} + {\beta_5\over b_5}~k_5^{n-l} + 
{\beta_6\over b_6}~k_6^{n-l} + ...+{\beta_{{\rm N}_\mu}\over b_{{\rm N}_\mu}}~k_{{\rm N}_\mu}^{n-l}\right)\\
& ~~~~~~~~~~~~~~~~~~~~~~~~~~~~~~~~~~~~~~~~~~~~~~~ .........~~~~~~~~~~~~~~~~~~ ......... \\
+ ~& {1 \over {\rm V}^p}\sum_{r, l} {\bf A}^{(m)}_{rl} \left({\beta_m\over b_m}\right)^{p-1-r} k_m^l\left({\beta_1\over b_1}~k_1^{n-l} + {\beta_2\over b_2}~k_2^{n-l} + 
{\beta_3\over b_3}~k_3^{n-l} + {\beta_5\over b_5}~k_5^{n-l} + 
{\beta_6\over b_6}~k_6^{n-l} + ...+{\beta_{q}\over b_{q}}~k_{q}^{n-l}\right)
\end{aligned}
\nd}
where $m \equiv {\rm N}_\mu$ and $q \equiv {\rm N}_\mu - 1$, which are decided from the way we divided the momentum $k$, lying between $k_{\rm IR}$ and $\mu$, into ${\rm N}_\mu$ equal pieces. The other quantities, namely ${\bf A}^{(i)}_{rl} = {\bf A}^{(i)}_{rl}[b_i]$, are defined in the following way:
\bg\label{ghumaypasa}
{\bf A}^{(s)}_{rl}[b_s] = { (p-1)^{l-1}\over (p-2)!}\left(\begin{matrix} p-1 \cr r\end{matrix}\right) \left(\begin{matrix} n \cr l\end{matrix}\right) \int_0^\infty dy ~e^{-b_s y^2}~y^r, \nd
where the $s$ dependence comes from $b_s$ in the Gaussian integrals, and we have integrated the fields from 0 to $\infty$, instead of $-\infty$ to $+\infty$. Thus depending on even or odd $r$, the coefficients will be slightly different, although powers of $b_i$ will remain unaltered. The set of terms in \eqref{blamortimer} has a hidden nested structure which is easy to see. Moving {\it diagonally} and summing over the terms in the upper triangle leads to one nested integral structure. In the lower triangle, if we move from bottom to top we can see another nested integral structure. These may be combined to give us:
\bg\label{relcha}
\begin{aligned}
{\cal W}^{(1)}_\phi = ~ &{1\over {\rm V}^{p-2}} \sum_{r, l}\int_{k_{\rm IR}}^\mu d^{11}k ~\left({\beta(k)\over b(k)}\right)^{p-1-r} k^l \int_k^\mu d^{11}k'~{\bf A}_{rl}[b(k), b(k')]~{k^{'(n-l)}\beta(k')\over b(k')}\\
 + ~ &{1\over {\rm V}^{p-2}} \sum_{r, l}\int_{k_{\rm IR}}^\mu d^{11}k ~{k^{(n-l)}\beta(k)\over b(k)}
\int_k^\mu d^{11}k'~{\bf A}_{rl}[b(k), b(k')] \left({\beta(k')\over b(k')}\right)^{p-1-r} k'^l,
\end{aligned}
\nd
showing that the amplitude coming from an interaction like \eqref{ivile1} is suppressed by ${\rm V}^{p-2}$. But there is more. Since two triangles can form a square, the two nested integrals may be combined under one roof to give us the final amplitude:

\bg\label{lafyr}
\begin{aligned}
{\cal W}^{(1)}_\phi \equiv ~ & {\frac{1}{{\rm V}^p} \sum_{i, j} \quad
\begin{tikzpicture}[thick, baseline={(0, 0cm)},
main/.style = {draw, circle, fill=black, minimum size=12pt},
dot/.style={inner sep=0pt,fill=black, circle, minimum size=4pt},
dots/.style={inner sep=0pt, fill=black, circle, minimum size=2pt}]
  \node[main] (a) at (-3.2, 0) {};
  \node[dot] (1) at (-1.7, 0) {};
  \node[dot] (2) at (0, 0) {};
  \node[main] (3) at (60:1.5) {};
  \node[main] (4) at (30:1.5) {};
  \node[main] (5) at (0:1.5) {};
  \node[main] (6) at (-30:1.5) {};
  \node[main] (7) at(-65:1.5) {};
  \node[dots] (d1) at (-40:1.3) {};
  \node[dots] (d2) at (-50:1.3) {};
  \node[dots] (d3) at (-60:1.3) {};
\draw (a) -- node[midway, above right, pos=0.2] {$k$} (1);
\draw (1) -- node[midway, above right, pos=0.01] {$(p-1)k'$} (2);
\draw (2) -- node[midway, left, pos=0.7] {$k'$} (3);
\draw (2) -- node[midway, left, pos=0.8] {$k'$} (4);
\draw (2) -- node[midway, above right, sloped, pos=0.7] {$k'$} (5);
\draw (2) -- node[midway, right, pos=0.7] {$k'$} (6);
\draw (2) -- node[midway, right, pos=0.7] {$k'$} (7);
\end{tikzpicture}}\\
= ~ & {1\over {\rm V}^{p-2}} \sum_{r, l}\int_{k_{\rm IR}}^\mu d^{11}k ~{k^{(n-l)}\beta(k)\over b(k)}
\int_{k_{\rm IR}}^\mu d^{11}k'~{\bf A}_{rl}[b(k), b(k')] \left({\beta(k')\over b(k')}\right)^{p-1-r} k'^l,
\end{aligned}
\nd
which is a general result and continues to hold even if we have multiple distribution of the momenta. The diagrammatic representation that we present in \eqref{lafyr} is an example of a {\it nodal diagram}, and is a very convenient computational tool for the one-point functions in a path integral like \eqref{mmtarfox}. The suppression factor of ${\rm V}^{p-2}$ tells us that the most dominant diagram is for the case when all the $p$ field components in \eqref{ivleague} have different momenta. The amplitude for this case will simply be products of integrals with integrands proportional to ${\beta(k)\over b(k)}$ along with powers of momenta coming from the $n$-th power of the momentum sum in \eqref{ivleague}. All integrals are computed from $k_{\rm IR}$ to $\mu$ since all nested pieces add up to this.

The readers may have easily guessed that the above computation is related to the $\varphi_2$ or $\varphi_3$ fields. For $\varphi_4$ additional complications come from momentum conservation which we will not discuss here. (The factor $b(k)$ appearing here and in \eqref{lafyr} is precisely the massless propagator with $\beta$ being the shift for the $\varphi_3$ field in \eqref{patarket}.) For the $\varphi_1$ fields, as evident from \eqref{mcahini}, the analysis is slightly different due to the presence of a source. The most dominant diagram with the source is clearly given by the following contraction:
\bg\label{julbinoc}
\underbrace{\widetilde\varphi_1(k_1)~ \widetilde\varphi_1(k_2) ~\widetilde\varphi_1(k_3)~... ~\widetilde\varphi_1(k_{p-1})~\overbracket[1pt][7pt]{\widetilde\varphi_1(k)~\widetilde\varphi_1}}_{p~{\rm fields} ~+ ~1~{\rm source}}(k), \nd
which is suppressed by ${\rm V}$. The completely un-suppressed diagram will be when the source itself has a different momentum compared to the $p$ fields. Unfortunately such a diagram is projected out by the 
denominator of the path integral \eqref{mmtarfox}. The story is similar to the vacuum bubble case in field theory, which we shall leave it for our diligent readers to work out. The amplitude for the contraction in \eqref{julbinoc} is now easy to work out and it takes the form:

{\footnotesize
\bg\label{shubarsh}
{\cal A}_p \equiv \sum_{\{r_i\}}  \prod_{j = 1}^{p-1} \int_{k_{\rm IR}}^\mu d^{11}k_j~k_j^{r_j}~{\alpha(k_j)\over a(k_j)} \cdot {1\over {\rm V}}\int_{k_{\rm IR}}^\mu d^{11}k~{\bf A}_{r_1r_2...r_{p-1}r_p}[\{a(k_j)\}, a(k)]~
k^{r_p}\left({\alpha^2(k)\over a^2(k)} + {1\over 2a(k)}\right)\psi_k(x, y), \nd}
where ${\bf A}_{r_1r_2...r_{p-1}r_p}[\{a(k_j)\}, a(k)]$ is similar to the 
structure in \eqref{ghumaypasa} but involve multiple Gaussian integrals and more complicated permutation factors. The explicit form is not important, so we will not work it out here. The integrands are defined in terms of the propagators $a(k_j), a(k)$ as well as the shift parameters $\alpha(k_j), \alpha(k)$ which may be related to $\alpha_1$ or $\alpha_2$ factors from \eqref{patarket}. 
Note also the presence of the volume suppression factor for the integral involving source wave-function $\psi_k(x, y)$. This is a consequence of the {\it contraction} in \eqref{julbinoc} with the source term. Since both $\alpha(k)$ and $a(k)$ have hidden volume factors, the ${\alpha^2(k)\over a^2(k)}$ term becomes sub-dominant compared to the ${1\over 2a(k)}$ term. The latter is related to the Feynman propagator (but since we are taking real fields, it is not exactly a two-point function).

All these that we said above is interesting, but is only the result to first order in $g$. Our aim is to extend it to all orders in $g$ and see how the diagrams grow. We plan to answer the following two questions.

\vskip.1in

\noindent ${\bf 1.}$ Can we consistently argue that the nodal diagrams show factorial growths even if we don't know the exact coefficients of the higher order terms?

\vskip.1in

\noindent ${\bf 2.}$ Is it possible to sum up all the {\it dominant} nodal diagrams to all perturbative orders in $g$ in a self-consistent way? Will such summation reveal the non-perturbative structure underlying the system?

\vskip.1in

\noindent In the following section we will answer the first question, and leave the details of the second question for the subsequent section.

\subsection{Factorial growths of the nodal diagrams and 1-point functions
\label{factoria}}

To study the growth of the dominant nodal diagrams we go back to the product structure in the path integral \eqref{mcahini}. There are four key factors given by ${\cal Z}_{\varphi_i}$ with $i = 1,.., 4$. Let us start with ${\cal Z}_{\varphi_1}$ associated with the $\varphi_1$ field components. This appears from the binomial expansion in \eqref{phobeindy}, and may be arranged as a series in \eqref{dialdest}. The most dominant terms, which are products of $p$ fields, appear when all the $p$ momenta are different. Such terms scale as $p!$, but since we have a ${1\over p!}$, they scale as 1. When we raise the series in \eqref{dialdest} to ${\rm N}$-th power, the dominant contributions, which are again terms with unequal momenta, would naturally scale as ${\rm N}!$. Due to the magic of the binomial expansion, the dominant scaling is always ${\rm N}!$ irrespective of the number of terms in the expansion\footnote{For example, if we have $(a + b + c + d)^4$, or $(a + b + c + d + e)^4$, or $(a + b + c + d + e + ...)^4$, the dominant terms, which are here with different alphabets, always scale as $4!$.}. This should partially explain the ${\rm N}!$ growths at order ${\rm N}$, but the story is a bit more subtle because there are many $-$ depending on what value we assign to ${\rm N}_\mu$ $-$ terms with the dominant contributions. Does this effect the growth of the dominant nodal diagrams as we are eventually {\it adding} up these contributions? 

The answer actually appears from the nested structure that we discussed earlier. Here however, since all fields have different momenta, the nested structure is pretty involved, but a simple diagrammatic approach can shed some light on it. Recall that now we are looking at contractions of the following form:
\bg\label{julbinoc2}
\underbrace{{\cal Z}^{{\rm N}-1}_{\varphi_1}~\widetilde\varphi_1(k_2)..\widetilde\varphi_1(k_{({\rm N} -1)p + 1})~ \widetilde\varphi_1(k_{({\rm N}-1)p + 2}) ~\widetilde\varphi_1(k_{({\rm N}-1)p + 3})~... ~\widetilde\varphi_1(k_{{\rm N}p})~\overbracket[1pt][7pt]{\widetilde\varphi_1(k_1)~\widetilde\varphi_1}}_{{\rm N}p~{\rm fields} ~+ ~1~{\rm source}}(k_1), \nd
with ${\rm N}p-1$ fields having different momenta from the source, which in turn is taken to have a momentum $k_1$. The nested structure appears from choosing the momentum $k_1$ and arranging the remaining ${\rm N}p-1$ fields with different (and hence unequal) momenta. This is depicted by {\bf figure \ref{lenmoon}}
\begin{figure}
    \centering
 \begin{tikzpicture}[baseline={(0, 0)}, thick,
main/.style = {draw, circle, fill=black, minimum size=4pt},
dot/.style={inner sep=0pt,fill=black, circle, minimum size=2pt},
dots/.style={inner sep=0pt,fill=black, circle, minimum size=1pt}, 
ball/.style={ball  color=white, circle,  minimum size=15pt}] 
\node[] (a) at (-1,-1.1) {$\widetilde\varphi_1\overbracket[1pt][7pt]{(k_1)~\widetilde\varphi_1}(k_1)$}; 
\node[dots] (a1) at (1,-1.1) {};

\node[dot] (0) at (7.5, 1) {};
\node[dot] (0) at (7.5, 0.5) {};
\node[dot] (0) at (7.5, 0) {};
\node[dot] (0) at (7.5, -0.4) {};

\node[dot] (0) at (7.5, -3) {};
\node[dot] (0) at (7.5, -3.5) {};
\node[dot] (0) at (7.5, -4) {};

\draw [red](a) -- (a1);

\node [dots] (a2) at (1,-0.5) {};
\node [dots] (a3) at (1,2) {};
\node [dots] (a4) at (1,-2.7) {};
\node [dots] (a5) at (1,-4.5) {};

\node (a21) at (2,-0.5) {$\widetilde\varphi_1(k_3)$};
\node (a31) at (2,2) {$\widetilde\varphi_1(k_2)$};
\node (a41) at (2,-2.7) {$\textcolor{red}{\widetilde\varphi_1(k_4)}$};
\node (a51) at (2,-4.5) {$\widetilde\varphi_1(k_{{\rm N}_\mu})$};

\draw[->] (a2) -- (a21);
\draw[->] (a3) -- (a31);
\draw[red][->] (a4) -- (a41);
\draw[->] (a5) -- (a51);
\draw [red](a4) -- (a1);
\draw (a1) -- (a3);
\draw [dashed] (a4) -- (a5);

\node [dots] (b) at (3, 2) {};
\node [dots] (b1) at (3, 2.5) {};
\node [dots] (b2) at (3, 3) {};
\node [dots] (b3) at (3, 1) {};

\node (b0) at (4, 2) {$\widetilde\varphi_1(k_5)$};
\node (b11) at (4, 2.5) {$\widetilde\varphi_1(k_4)$};
\node (b21) at (4, 3) {$\widetilde\varphi_1(k_3)$};
\node (b31) at (4.4, 1) {$\widetilde\varphi_1(k_{{\rm N}_\mu})$};

\draw (a31) -- (b);
\draw[->] (b) -- (b0);
\draw[->] (b1) -- (b11);
\draw[->] (b2) -- (b21);
\draw[->] (b3) -- (b31);
\draw (b2) -- (b);
\draw[dashed] (b) -- (b3);

\node [dots] (c) at (3, -2.7) {};
\node [dots] (c0) at (3, -1) {};
\node [dots] (c1) at (3, -2) {};
\node [dots] (c2) at (3, -2.7) {};
\node [dots] (c3) at (3, -3.5) {};

\node (c01) at (4, -1) {$\widetilde\varphi_1(k_5)$};
\node (c11) at (4, -2) {$\textcolor{red}{\widetilde\varphi_1(k_3)}$};
\node (c21) at (4, -2.7) {$\widetilde\varphi_1(k_2)$};
\node (c31) at (4.4, -3.5) {$\widetilde\varphi_1(k_{{\rm N}_\mu})$};

\draw [red](a41) -- (c);
\draw[->] (c0) -- (c01);
\draw[red][->] (c1) -- (c11);
\draw[->] (c2) -- (c21);
\draw[->] (c3) -- (c31);
\draw (c0) -- (c1);
\draw [red](c1) -- (c2);
\draw[dashed] (c2) -- (c3);

\node [dots] (d) at (3, -4.5) {};
\node [dots] (d0) at (3, -4) {};
\node [dots] (d1) at (3, -5) {};
\node [dots] (d2) at (3, -5.5) {};

\node (d01) at (4, -4) {$\widetilde\varphi_1(k_2)$};
\node (d11) at (4, -5) {$\widetilde\varphi_1(k_3)$};
\node (d21) at (4.4, -5.5) {$\widetilde\varphi_1(k_{{\rm N}_\mu-1})$};

\draw (a51) -- (d);
\draw[->] (d0) -- (d01);
\draw[->] (d1) -- (d11);
\draw[->] (d2) -- (d21);
\draw (d1) -- (d0);
\draw[dashed] (d0) -- (d2);

\node [dots] (e) at (6, 2.5) {};
\node [dots] (e1) at (6, 3) {};
\node [dots] (e2) at (6, 2.3) {};
\node [dots] (e3) at (6, 1.5) {};

\node (e11) at (7.8, 3) {$\widetilde\varphi_1(k_3)$};
\node (e21) at (7.8, 2.3) {$\widetilde\varphi_1(k_5)$};
\node (e31) at (7.8, 1.5) {$\widetilde\varphi_1(k_{{\rm N}_\mu})$};

\draw (b11) -- (e);
\draw[->] (e1) -- (e11);
\draw[->] (e2) -- (e21);
\draw[->] (e3) -- (e31);
\draw (e1) -- (e2);
\draw[dashed] (e2) -- (e3);

\node [dots] (f) at (6, -2) {};
\node [dots] (f1) at (6, -1) {};
\node [dots]  (f2) at (6, -1.5) {};
\node[dots] (f3) at (6, -2.5) {};

\node (f11) at (7.8, -1) {$\widetilde\varphi_1(k_2)$};
\node (f21) at (7.8, -1.5) {$\textcolor{red}{\widetilde\varphi_1(k_5)}$};
\node (f31) at (8, -2.5) {$\widetilde\varphi_1(k_{{\rm N}_\mu})$};

\draw [red](c11) -- (f);
\draw[->] (f1) -- (f11);
\draw[red][->] (f2) -- (f21);
\draw[->] (f3) -- (f31);
\draw (f1) -- (f2);
\draw[dashed] (f2) -- (f3);

\node [dots] (g) at (6, -5) {};
\node [dots] (g1) at (6, -4.5) {};
\node [dots] (g2) at (6, -5.5) {};
\node [dots] (g3) at (6, -6.4) {};

\node (g11) at (7.8, -4.5) {$\widetilde\varphi_1(k_2)$};
\node (g21) at (7.8, -5.5) {$\widetilde\varphi_1(k_4)$};
\node (g31) at (8.2, -6.4) {$\widetilde\varphi_1(k_{{\rm N}_\mu-1})$};

\draw (d11) -- (g);
\draw[->] (g1) -- (g11);
\draw[->] (g2) -- (g21);
\draw[->] (g3) -- (g31);
\draw (g1) -- (g2);
\draw[dashed] (g2) -- (g3);

\node (A1) at (9.5, 3.3) {};
\node (B1) at (9.5, 1) {};
\usetikzlibrary{snakes} (A1) -- (B1);


\node (A2) at (9.5, -0.7) {};
\node (B2) at (9.5, -2.9) {};
\usetikzlibrary{snakes}(A2) -- (B2);


\node (A3) at (9.7, -4.1) {};
\node (B3) at (9.7, -6.8) {};
\usetikzlibrary{snakes} (A3) -- (B3);

\end{tikzpicture}
    \caption{Diagrammatic representation of the contraction from \eqref{vaccmey} depicted by a path in \textcolor{red}{red} here. One can similarly work out other paths in the above diagram from \eqref{julbinoc2}.}
    \label{lenmoon}
\end{figure}
where we see that the first row has all momenta lying between $k_2$ till 
$k_{{\rm N}_\mu}$. The second row is more involved but the above diagrammatic representation clearly depicts the proliferation of the chain. The latter continues till ${\rm N}p-1$ steps (the {\bf figure \ref{lenmoon}} only shows the first three steps in the chain). Tracing one such path, shown in \textcolor{red}{red}, say we find the following string of operators:
\bg\label{vaccmey}
\widetilde\varphi_1(k_1)~ \widetilde\varphi_1(k_1)~ \widetilde\varphi_1(k_4)~ \widetilde\varphi_1(k_3)~ \widetilde\varphi_1(k_5)~...~ \widetilde\varphi_1(k_{{\rm N}_\mu}), \nd
which, from our binomial expansion, generically scale as $\hat{p}$ where $\hat{p} < {\rm N}!$. (In other words, keeping $\widetilde\varphi_1\overbracket[1pt][7pt]{(k_1)~\widetilde\varphi_1}(k_1)$ fixed, the total number of paths with the same set of field contents in any arbitrary order scales as $\hat{p} < {\rm N}!$.) In fact, there are infinite paths (labelled by $l$) connecting the operators given in the ${\rm N}_\mu$ rows in the above diagram $-$ but differing from \eqref{vaccmey} $-$ with each scaling as $p_l < {\rm N}!$. The total amplitude is given by summing over all these infinite string of operators, {\it i.e.} summing over $\widetilde\varphi_1\overbracket[1pt][7pt]{(k_j)~\widetilde\varphi_1}(k_j)$ for $1 \le j \le {\rm N}_\mu$ with $j \in \mathbb{Z}$, which would basically appear as nested integrals. After the dust settles, the final answer of the total amplitude is not very different from what we had in \eqref{shubarsh}, albeit a bit more involved, and takes the form:

{\footnotesize
\bg\label{shubarsh02}
{\cal A}_{{\rm N}p} \equiv \sum_{\{r_i\}} \prod_{j = 1}^{{\rm N}p-1} \int_{k_{\rm IR}}^\mu d^{11}k_j~k_j^{r_j}~{\alpha(k_j)\over a(k_j)} \cdot {1\over {\rm V}}\int_{k_{\rm IR}}^\mu d^{11}k~{\bf A}_{r_1r_2...r_{{\rm N}p-1}r_{{\rm N}p}}[\{a(k_j)\}, a(k)]~k^{r_{{\rm N}p}}\left({\alpha^2(k)\over a^2(k)} + {1\over 2a(k)}\right)\psi_k(x, y), \nd}
which, once carefully observed, reveals the full structure presented in the diagram above. The coefficients ${\bf A}_{r_1r_2...r_{{\rm N}p-1}r_{{\rm N}p}}[\{a(k_i)\}, a(k)]$ are more complicated than \eqref{ghumaypasa} but, since their explicit forms are not necessary here, we will not work them out and leave them again for our diligent readers. From the arguments presented earlier, the whole amplitude scales as ${\rm N}!$, which should point towards the factorial growth of the one-point functions.

There are however couple of subtleties here that need clarification. First, if we compute the magnitude of the amplitude \eqref{shubarsh02}, {\it i.e.} $\vert{\cal A}_{{\rm N}p}\vert$, then the result may appear to scale as:
\bg\label{voyter}
\vert{\cal A}_{{\rm N}p}\vert = {\rm N}! f(k_{\rm IR}, \mu, p, {\rm N}, {\rm V}; x, y), \nd 
where $f(k_{\rm IR}, \mu, p, {\rm N}, {\rm V}; x, y)$ is a function that can be worked out from \eqref{shubarsh02}.
Note the appearance of ${\rm N}$ in two places, one as an overall factorial ${\rm N}!$, and the other in the function $f(k_{\rm IR}, \mu, p, {\rm N}, {\rm V}; x, y)$. The latter is typically in the polynomial form, whereas the former is in the factorial form which is what we meant when we used the term ``factorial growth''. The polynomial growth is also very important and we will discuss it's consequence in the next section.

Secondly, \eqref{voyter} cannot quite be the full answer because this ignores the power $p$ of $\varphi_1^p$ at the tree-level. To see what exactly replaces the ${\rm N}!$ in \eqref{voyter}, let us quantify the interactions in our model with four fields $(\varphi_1, .., \varphi_4)$: 
\bg\label{katumara}
{\bf S}_{\rm int} = \int d^{11}x ~{c_o\over {\rm M}_p^{n+m+l+h}}~\partial^n\varphi_1^p ~\partial^m\varphi_2^q ~\partial^l\varphi_3^r ~\partial^h\varphi_4^s, \nd
where $(n, m, l, h) \in +2\mathbb{Z}, c_o$ is yet another unknown coefficient, and we will define $g$, the coupling constant, as $g \equiv {c_o\over {\rm M}_p^{n+m+l+h}}$. As usual, the fields are taken to be dimensionless, and the Minkowski space has a metric $(\eta_{\mu\nu}, \delta_{mn}, \delta_{ab})$ for $(\mu, \nu) \in {\bf R}^{2, 1}, (m, n) \in {\cal M}_6, (a, b) \in {\mathbb{T}^2\over {\cal G}}$. The growth ${\rm N}!$ in \eqref{voyter} should be replaced as:
\bg\label{bartsfo}
\begin{split}
{\rm N}! ~ & \rightarrow ~ {(p{\rm N})!(q{\rm N})!(r{\rm N})!((s-1){\rm N})!\over \left(p!q!r!(s-1)!\right)^{\rm N}} \\
~& = \left[{p^p q^q r^r (s-1)^{s-1}\over p! q! r! (s-1)!}\right]^{\rm N} 
\left({\rm N}!\right)^{p+q+r + s-1}\bigg\vert_{{\rm N} >> 1} \equiv 
{\cal A}_1^{\rm N} \left({\rm N}!\right)^{\alpha \check{s}\left(1 + {1\over\alpha} - {1\over \alpha\check{s}}\right)},
\\
\end{split}
\nd
where ${\cal A}_1 ={p^p q^q r^r (s-1)^{s-1}\over p! q! r! (s-1)!}$, $\alpha +1$ is the total number of fields such that $p + q + r + s \equiv (\alpha + 1) \check{s}$. The $s-1$ factor is taken to imply the momentum conservation in the $\varphi_4$ sector where one of the Fourier components
of the $s$ fields becomes redundant. And in going from first line to the second one we have used $(p{\rm N})! \propto ({\rm N}!)^p$, where the proportionality factor\footnote{Let $\check{p}$ be the proportionality factor. Taking log on both sides shows that the RHS is $\log~\check{p} + p{\rm N}\log~{\rm N} - p{\rm N}$ for ${\rm N} >> 1$. This equals LHS for $\check{p} = p^{p{\rm N}}$. \label{berkeley}} is $p^{p{\rm N}}$ for ${\rm N} >> 1$, whereas for small ${\rm N}$, the proportionality factor is again some polynomial in $p$. 

There are still a few details that need to be considered. 
First, due to ${1\over {\rm N}!}$ suppression coming from the exponential\footnote{Recall that the path integral has $e^{-{\bf S}_{\rm int}}$, so expanding this gives us the ${1\over {\rm N}!}$ suppression at order ${\rm N}$. \label{pathos}}, the growth that we saw above should get further modified to $\left({\rm N}!\right)^{\alpha \check{s}\left(1 + {1\over\alpha} - {2\over \alpha\check{s}}\right)}$. 
Secondly, there is a mixing between the fields.  However considering the fact that our theory has more than one fields, {\it i.e.} say $\alpha + 1$ fields, and since both $\alpha > 1$ and $\check{s} > 1$, we can ignore the 
${1\over \alpha}$ and ${2\over \alpha \check{s}}$ in the factorial growth above. This means that, everything considered including the fact that 
for every field components we can draw the nested interactions as in the diagram above, the theory should have at least $\left({\rm N}!\right)^{\alpha\check{s}}$ growth\footnote{It is interesting to compare the result with the factorial growth in the standard $\lambda\varphi^4$ theory in the absence of the displacement operator $\mathbb{D}(\sigma)$. As before, let $\widetilde\varphi(k_i)$ be the Fourier components, and we are now looking at binomial expansion of the following form:
\bg\label{lidia}
{\cal Z}_{\varphi} \equiv \sum_{{\rm N} = 0}^\infty{\lambda^{\rm N}\over {\rm V}^{4{\rm N}}(4!)^{\rm N}{\rm N}!}
\left(\sum_{i = 1}^{{\rm N}_\mu} \vert\widetilde\varphi(k_i)\vert^2\right)^{2{\rm N}}, \nonumber \nd
where the modulus sign takes care of the momentum conservation. The dominant term in the binomial expansion now grows as ${(2{\rm N})!\over {\rm N}!} \approx {\rm N}!$, which is the factorial growth expected from the usual Feynman diagrams. Once we introduce the displacement operator $\mathbb{D}(\sigma)$ the story changes quite a bit as discussed here.}.

This is {\it not} exactly true because of yet another interesting subtlety. As an example consider the two sectors $\varphi_2$ and $\varphi_3$ that do not directly interact with the source. For simplicity we will take cubic interactions between the two set of fields, {\it i.e.} $g^3\varphi_2^3(x, y) \varphi_3^3(x, y)$, which is basically ${\rm N} = 3$ with $n = m = l = h = 0$ and $p = q = r = s = 1$ in \eqref{katumara}, but we shall ignore the $\varphi_1$ and $\varphi_4$ sectors for simplicity, as one can absorb the ${1\over {\rm N}!}$ suppression and the other can absorb the momentum conservation. In terms of Fourier modes this may be represented as:

{\footnotesize
\bg\label{millerelo}
{g^3\over {\rm V}^6}\left(\sum_{i = 1}^{{\rm N}_\mu}\widetilde\varphi_2(k_i)\psi_{k_i}(x, y)\right)^3 \left(\sum_{j = 1}^{{\rm N}_\mu}\widetilde\varphi_3(k_j)\psi_{k_j}(x, y)\right)^3
~\to ~ \begin{matrix} \textcolor{blue}{\widetilde\varphi_2(k_1)} & \widetilde\varphi_2(k_1) & 
\widetilde\varphi_2(k_1) & \widetilde\varphi_3(k_1) & \widetilde\varphi_3(k_1) &\textcolor{red}{\widetilde\varphi_3(k_1)} \cr
~~ & ~~ & ~~ & ~~ & ~~ & ~~ \cr
\widetilde\varphi_2(k_2) & \textcolor{blue}{\widetilde\varphi_2(k_2)} & 
\widetilde\varphi_2(k_2) & \widetilde\varphi_3(k_2) & 
\textcolor{red}{\widetilde\varphi_3(k_2)} & \widetilde\varphi_3(k_2) \cr
~~ & ~~ & ~~ & ~~ & ~~ & ~~ \cr
\widetilde\varphi_2(k_3) & \widetilde\varphi_2(k_3) & 
\textcolor{blue}{\widetilde\varphi_2(k_3)} &\textcolor{red}{\widetilde\varphi_3(k_3)} & \widetilde\varphi_3(k_3) & \widetilde\varphi_3(k_3) \cr
~~ & ~~ & ~~ & ~~ & ~~ & ~~ \cr
\widetilde\varphi_2(k_4) & \widetilde\varphi_2(k_4) & 
\textcolor{red}{\widetilde\varphi_2(k_4)} &\textcolor{blue}{\widetilde\varphi_3(k_4)} & \widetilde\varphi_3(k_4) & \widetilde\varphi_3(k_4) \cr
~~ & ~~ & ~~ & ~~ & ~~ & ~~ \cr
\textcolor{red}{\widetilde\varphi_2(k_5)} & \widetilde\varphi_2(k_5) & 
\widetilde\varphi_2(k_5) & {\widetilde\varphi_3(k_5)} & 
\textcolor{blue}{\widetilde\varphi_3(k_5)} & \widetilde\varphi_3(k_5) \cr
~~ & ~~ & ~~ & ~~ & ~~ & ~~ \cr
\widetilde\varphi_2(k_6) &\textcolor{red}{\widetilde\varphi_2(k_6)} & 
\widetilde\varphi_2(k_6) & \widetilde\varphi_3(k_6) & \widetilde\varphi_3(k_6) &\textcolor{blue}{\widetilde\varphi_3(k_6)} \cr
.. & .. & ~~ & .. & ~~ & .. \cr
.. & .. & ~~ & .. & ~~ & .. \cr
\widetilde\varphi_2(k_{{\rm N}_\mu}) & \widetilde\varphi_2(k_{{\rm N}_\mu}) & 
\widetilde\varphi_2(k_{{\rm N}_\mu}) & \widetilde\varphi_3(k_{{\rm N}_\mu}) & \widetilde\varphi_3(k_{{\rm N}_\mu}) & \widetilde\varphi_3(k_{{\rm N}_\mu}) \cr
\end{matrix}
\nd}
where $\psi_{k_i}(x, y)$ are the wave-functions over the Minkowski vacuum, and on the RHS we represent the field components in six rows related to the ${\rm N} = 3$ case. The wave-functions will simply give momentum conservation in the $\varphi_4$ sector so we ignore them on the RHS. Following two paths, one in \textcolor{blue}{blue} and the other in \textcolor{red}{red}, provides the two strings of operators:
\bg\label{secretlive}
\begin{split}
{\cal B}_1 = ~&\textcolor{blue}{\widetilde\varphi_2(k_1) ~\widetilde\varphi_2(k_2) ~\widetilde\varphi_2(k_3) ~\widetilde\varphi_3(k_4)~\widetilde\varphi_3(k_5)~
\widetilde\varphi_3(k_6)}\\
{\cal B}_2 = ~& \textcolor{red}{\widetilde\varphi_2(k_5) ~\widetilde\varphi_2(k_6) ~\widetilde\varphi_2(k_4) ~\widetilde\varphi_3(k_3)~ \widetilde\varphi_3(k_2)~
\widetilde\varphi_3(k_1)}\\
\end{split}
\nd
which have the same distribution of momenta, albeit with different field contents. The dominant terms that do not have volume suppressions grow as $3!$ for the blue line and $3!$ for the red line, after we add up all the contributions. So the naive growth is $(3!)^2$, appearing from binomial product\footnote{From above, the general growth with everything considered is proportional to $\left({\rm N}!\right)^{\alpha \check{s}\left(1 + {1\over\alpha} - {2\over \alpha\check{s}}\right)}$. Taking ${\rm N} = 3, \alpha = 3$ and $\check{s} = 1$ we get $(3!)^2$.}. On the other, if we look at the distribution of momenta, then there are in fact $(2.3)! = 6! = 720$ terms contributing instead of $(3!)^2 = 36$ terms. This enhancement is due to mixing of momenta over the two set of field components. Once we take a more general interaction of the form ${g\over q! r!} \varphi_2^q(x, y) \varphi_3^r(x, y)$, with a coupling $g$, then the amplitude to order ${\rm N}$ becomes:

{\footnotesize
\bg\label{shubarsh03}
{\cal A}_{{\rm N}(q+r)} \equiv g^{\rm N}\sum_{\{r_i\}}  \prod_{j = 1}^{{\rm N}q} \int_{k_{\rm IR}}^\mu d^{11}k_j~k_j^{r_j}~{\alpha_2(k_j)\over a(k_j)} 
\prod_{l = {\rm N}q+1}^{{\rm N}(q+r)} \int_{k_{\rm IR}}^\mu d^{11}k_l~k_l^{r_l}~{\beta(k_l)\over b(k_l)}
{\bf A}_{r_1r_2...r_{{\rm N}q}...r_{{\rm N}(q+r)}}[\{a(k_j)\}, \{b(k_l)\}],
\nd}
where $(\alpha_2, \beta)$ correspond to the notations given in \eqref{patarket}. As expected, \eqref{shubarsh03} is almost a product of ${\rm N}(q+r)$ integrals with ${\rm N}q$ integrals coming from the $\alpha_2$ Glauber-Sudarshan states and ${\rm N}r$ integrals coming from $\beta$ Glauber-Sudarshan states. If $\hat{f}^{1/2}_{\rm max}(k_{\rm IR}, \mu; q, r)$ denotes the maximum value of {\it one} such integral piece, then the value of the amplitude $\vert{\cal A}_{{\rm N}(q+r)}\vert$ is bounded from above by (see also \eqref{lyttoncasi} for a more correct way to implement the bound):
\bg\label{stoneje}
\vert{\cal A}_{{\rm N}(q+r)}\vert ~ \le ~ (2{\rm N})! ~g^{\rm N}~\vert \hat{f}_{\rm max}(k_{\rm IR}, \mu; q, r)\vert^{{\rm N}}, \nd
which may be compared to the $({\rm N}!)^2$ growth for the amplitude \eqref{shubarsh02} as depicted in \eqref{bartsfo}. (Note the absence of $(x, y)$ in the definition of $\hat{f}_{\rm max}$.) The power of $1/2$ in $\hat{f}^{1/2}_{\rm max}$ suggests that the maximum value could also come from a {\it geometric mean} of the maximum values from the $(\alpha_2, \beta)$ sectors. Ultimately however, the precise value of $\hat{f}_{\rm max}$ is not very important for us here, although what is important here is the upper bound in \eqref{stoneje}. Our above analysis suggests that, given say $\alpha + 1$ set of interacting fields, the factorial growth of the amplitude of any one-point function in the shifted vacua is generically {\it bigger} than $({\rm N}!)^{\alpha\check{s}}$, and it can enhance {\it at most} to:
\bg\label{bribechmey}
\check{p}\cdot \left({\rm N}!\right)^{\alpha\check{s}} ~ = ~ \left(\alpha \check{s} {\rm N}\right)!, \nd
where $\check{p}$ is the possible number of permutations of fields and momenta, and $\alpha >> 1$ (see for example \eqref{secretlive}). For small values of ${\rm N}$, it is not too hard to show that $\check{p} \in +\mathbb{Z}$. When ${\rm N}$ is large, one can use the Sterling's approximation to show that $\check{p} = (\alpha\check{s})^{\alpha\check{s}{\rm N}} \in +\mathbb{Z}$, as $\alpha\check{s} \in +\mathbb{Z}$ for $\alpha >> 1$ (see footnote \ref{berkeley}). The relation \eqref{bribechmey} then is a simpler way of understanding how in a mixed system with $\alpha+1$ fields with $\alpha >> 1$ the system could show the so-called {\it Gevrey} growth\footnote{A more conservative approach would be to predict the growth to be $\left(\alpha {\bf s} {\rm N}\right)!$ where  
${1\over \alpha} \le {\bf s} \le \check{s}$. The lower limit would be the ${\rm N}!$ growth whereas the upper limit would be the maximum allowed $\left(\alpha \check{s} {\rm N}\right)!$ growth from \eqref{bribechmey}. \label{referee2}} of {\it at most} $(\alpha\check{s}{\rm N})!$. 

The above conclusion is not the end of the story, rather the beginning of a new direction where the Gevrey growth should point towards the appearance of non-perturbative effects hitherto ignored in our set-up. In fact the factorial growth provides a deeper reason why the system has non-perturbative effects at all. Question is, can we quantify this by Borel resumming the Gevrey series? This is what we turn to next.

\subsection{Gevrey series, Borel resummation and non-perturbative effects
\label{sec3.7}}

The $({\rm N}!)^{\alpha\check{s}}$ growth that we discussed above for $\alpha + 1$ fields with $\alpha >> 1$ generically leads to the so-called {\it Gevrey series} that has been discussed in the literature in much details \cite{gevrey}. We will not do the same here and instead refer the readers to section 4.3.1 of our recent work \cite{borel}. Interestingly, the following observation due to \'Ecalle \cite{ecalle}:
\vskip.1in

\noindent \textcolor{blue}{The Gevrey order of the asymptotic series ${\rm F}(g) = 
\sum\limits_{{\rm N} = 0}^\infty ({\rm N}!)^l g^{\rm N}$ where $l \in \mathbb{Z}$ and $g << 1$, is the same as that of another asymptotic series 
${\rm G}(g) = \sum\limits_{{\rm N} = 0}^\infty (l{\rm N})! g^{\rm N}$, thus reducing everything to Gevrey-1.}

\vskip.1in

\noindent which efficiently leads to the so-called Borel-\'Ecalle resummation, is pretty useful for us because of \eqref{bribechmey}. In fact the in-built enhancement of \eqref{bribechmey} suggests that it is no longer necessary to apply the {\it acceleration operators} of \'Ecalle \cite{ecalle} to go from Gevrey-$\alpha\check{s}$ to Gevrey-1. All we need is an upper bound of the form \eqref{stoneje} $-$ but now extended to $\alpha+1$ fields $-$ to perform the Borel resummation of a Gevrey series efficiently.

Unfortunately the actual computation is easier said than done, as evident from section 4.3.2 of our earlier work \cite{borel} and the upcoming work \cite{ccpaper}. Some of the subtleties that we actually face in our computations are as follows.

\vskip.1in

\noindent ${\bf 1.}$ The Borel resummation arising from the path integral computation isn't the only resummation. There are additional off-shell pieces, that remain invisible in the path integral, and arise from the background time-independent Minkowski space-time. These additional contributions are necessary to extract correct answers for the expectation values.

\vskip.1in

\noindent ${\bf 2.}$ The singularities on the Borel plane lead to certain principal value computations that are non-trivial to perform. In fact the exact values of these integrals are not known, but numerical results can be found (see our upcoming paper \cite{ccpaper} for details). Pin-pointing these numerical values are important to get an estimate on the four-dimensional cosmological constant.

\vskip.1in

\noindent ${\bf 3.}$ The Borel resummation presented in \cite{borel} considers only one specific type of interactions. Introducing other interactions in the theory lead to the so called {\it Borel boxes}\footnote{They are useful technical machineries that help to Borel re-sum not only the dominant amplitudes, but also all the sub-dominant ones. We will discuss this briefly in section \ref{sec3.9}. Unfortunately a detailed exposition of this is beyond the scope of the paper and the readers may refer to our upcoming work \cite{ccpaper}. \label{bbox}}, whose convergence properties are important for finiteness. This additional ``resummation'' of Borel resummed series make the analysis highly non-trivial (see our upcoming paper \cite{ccpaper} for further details).

\vskip.1in

\noindent ${\bf 4.}$ The permutation factors ${\bf A}_{r_1r_2...r_{{\rm N}p-1}r_{{\rm N}p}..}$ appearing in \eqref{shubarsh02} and \eqref{shubarsh03} create additional hurdles because they not only mix up the momenta from various sectors in a non-trivial way, but also sum over the various choices of permutations. The way out of this technical complication, as analyzed in section \ref{sec3.8} and in our upcoming work \cite{ccpaper}, is to impose {\it bounds} on the binomial growths of the series in appropriate ways. 

\vskip.1in

\noindent ${\bf 5.}$ The integral \eqref{shubarsh02} with the source term integrated in, and on which we base our Borel resummations, works for ${\rm N} \ge 1$, but {\it does not} work for ${\rm N} = 0$. This unfortunate hurdle leads to further constraints in the system that has it's root in the no-go condition \cite{GMN}. This subtlety does not show up in other sectors.

\vskip.1in

\noindent Clearly a detailed study of any of the aforementioned points is beyond the scope of this paper. Fortunately, these subtleties will not impede our progress in understanding the underlying picture, although we will take some time to elaborate the last point, namely the case for ${\rm N} \ge 1$ and ${\rm N} = 0$.  Taking the displacement operator to be \eqref{28rms}, the expectation value $\langle\varphi_1\rangle_\sigma$ from \eqref{mmtarfox} takes the following form at the ${\rm N}$-th order with $\alpha >> 1$ (see also \eqref{lyttoncasi} and \eqref{lyttoncasime}):
\bg\label{runikate}
\vert{\cal A}_{\rm N}\vert \le \left(\alpha\check{s}{\rm N}\right)! g^{\rm N}~\vert \check{f}_{\rm max}(k_{\rm IR}, \mu; p, q, r, s-1)\vert^{\rm N}\cdot {1\over {\rm V}} 
\int_{k_{\rm IR}}^\mu d^{11}k ~k^{\rho^\ast(k)}\left({\alpha_1^2(k)\over a^2(k)} + {1\over 2a(k)}\right)\psi_k(x, y), \nd
which is gathered from the two amplitudes \eqref{shubarsh02} and \eqref{shubarsh03} discussed earlier; and we have taken the maximum allowed growth from \eqref{bribechmey} (see also footnote \ref{referee2}). To avoid clutter, we can henceforth replace $\alpha\check{s}$, or more appropriately $\alpha\check{s}\left(1 + {1\over \alpha} - {2\over \alpha\check{s}}\right)$ by $\bar\alpha$.
Both $\check{f}_{\rm max}$ and $k^{\rho^\ast(k)}$ appear from imposing bounds on the permutations of the momenta (see section \ref{sec3.8} as well as \cite{borel} and \cite{ccpaper} for more details). The parameter $s-1$ signifies the momentum conservation as one of the degrees of freedom becomes redundant. The inverse volume ${1\over {\rm V}}$ is the consequence of the contraction \eqref{julbinoc2}, and in fact this is the main cause of the disparity between ${\rm N} \ge 1$ and ${\rm N} = 0$. This disparity may be quantified by the following integral function:
\bg\label{marasis}
\mathbb{T}(k_{\rm IR}, \mu; \rho, {\rm V}; x, y) \equiv \int_{k_{\rm IR}}^\mu d^{11}k 
\left[{\alpha_1(k)\over a(k)} - {k^{\rho^\ast(k)}\over {\rm V}}\left({\alpha_1^2(k)\over a^2(k)} + {1\over 2a(k)}\right) + {\cal O}\left({1\over {\rm V}^2}\right)\right]\psi_k(x, y), \nd
whose {\it vanishing} will imply no generation of de Sitter at the tree level. This is of course a consequence of the no-go theorem \cite{GMN}, and here we see how it may be implemented in the language of the Glauber-Sudarshan states. The equation $\mathbb{T} = 0$ leads to the determination of $\alpha_1(k)$ in terms of $\rho^\ast(k)$ as:
\bg\label{robinright}
\alpha_1(k) = {1\over 2} k^{\rho^\ast(k)} + {\cal O}\left({1\over {\rm V}}\right), \nd
which appears from the fact that both $\alpha_1(k)$ and $a(k)$ have inverse volume suppressions from imposing \eqref{fleabag}. Note that \eqref{robinright} does not fix the functional form for $\alpha_1(k)$ because we haven't yet determined $\rho^\ast(k)$. Question is, how to find the functional form for $\rho^\ast(k)$?

The answer is a bit non-trivial because of the appearance of additional off-shell pieces that we mentioned as point 1 earlier. There is also the sum over ${\rm N}$ which we need to perform first to conclude anything about these additional off-shell pieces. Fortunately, another advantage of the vanishing of $\mathbb{T}$ is our ability to extend ${\rm N}$ in \eqref{runikate} to ${\rm N} \ge 0$ now. What does that buy us?

Extending ${\rm N}$ in \eqref{runikate} to include ${\rm N} = 0$ means that we can perform the Borel-\'Ecalle re-summation of the series in 
\eqref{runikate}. This summation will automatically provide the full functional form for $\langle\varphi_1\rangle_\sigma$ in \eqref{mmtarfox} in the following way (\textcolor{red}{henceforth we do not put hats to denote operators}):
\bg\label{rubiem}
\langle\varphi_1\rangle_\sigma = \sum_{{\rm N} = 0}^\infty \vert {\cal A}_{\rm N}\vert = 
{1\over 2g^{1/\bar\alpha}}\left[\int_0^\infty d{\rm S} ~{\rm exp}\left(-{{\rm S}\over g^{1/\bar\alpha}}\right) {1\over 1 - \check{f}_{\rm max} {\rm S}^{\bar\alpha}}\right]_{\rm P.V} \int_{k_{\rm IR}}^\mu d^{11}k ~k^{\rho^\ast(k) - 2} ~\psi_k(x, y), \nd
where the subscript P.V means that we have to take the principal value of the integral over ${\rm S}$, and $\psi_k(x, y)$ is the wave-function over the Minkowski space-time. The total number of fields is $\alpha + 1$, but what enters \eqref{rubiem} is $\bar\alpha$ which is defined earlier as: 
\bg\label{kitkatmey}
\bar\alpha \equiv \alpha\check{s}\left(1 + {1\over \alpha} - {2\over \alpha\check{s}}\right) = p+q+r + s - 2 ~ \in ~ \mathbb{Z}, \nd
to avoid clutter.
The result \eqref{rubiem} is an {\it exact} answer got by summing over all perturbative orders in ${\rm N}$ {\it despite our ignorance of the value of $c_o$ in \eqref{katumara}, or of the functional form for $\check{f}_{\rm max}$}. The functional form for $\check{f}_{\rm max}$ provides the necessary poles in the Borel plane, but the actual principal value integral is quite non-trivial. We will not attempt this here but direct the readers to \cite{borel, ccpaper} for more details. In list 2 earlier we pointed out precisely this difficulty. Nevertheless looking at the principal value integral the readers may easily notice the non-perturbative contributions going as ${\rm exp}\left(-{n\over g^{1/\bar\alpha}}\right)$. Two questions arise.

\vskip.1in

\noindent ${\bf 1.}$ Where are these non-perturbative effects coming from? And what does the infinite sum of the non-perturbative effects mean here? Can this be quantified?

\vskip.1in

\noindent ${\bf 2.}$ Can we determine the functional value of $\rho^\ast(k)$ now that we have performed the Borel-\'Ecalle resummation? How does the additional off-shell terms manifest from the functional form for $\rho^\ast(k)$? Can this be quantified?

\vskip.1in

\noindent The answer to the first question is relatively easier: the form of the non-perturbative corrections clearly signify the presence of instantons. There is however a difference appearing from the fact that these instantons are both real and complex ones. To see how this comes about let us go back to the two potentials in {\bf figure \ref{multsaddle}}. We can make the following observations.

\vskip.1in

\noindent ${\bf 1.}$ If we consider the symmetric wave-function on the left\footnote{We will always be extending the wave-function $\Psi_0 = {\rm exp}\left[\int d^{11}k ~\log\langle{\bf f}_k\vert 0\rangle\right]$ to 
$\Psi_\Omega = {\rm exp}\left[\int d^{11}k ~\log\langle{\bf f}_k\vert \Omega\rangle\right]$, where $\vert{\bf f}_k\rangle$ are states in the configuration space in all the cases studied here unless mentioned otherwise.}, then the expectation value of $\varphi_1$ will only get contributions from the complex turning points, {\it i.e.} only from the complex instantons. 

\vskip.1in

\noindent ${\bf 2.}$ If we consider a wave-function that is localized over one minima, which is basically the linear combination of the two wave-functions (shown in \textcolor{red}{red} and \textcolor{blue}{blue}), then the contributions to the expectation value will come from {\it both} real and the complex instantons. 

\vskip.1in

\noindent ${\bf 3.}$ For the potential shown on the right, if we consider the localized wave-function over the left minimum\footnote{Recall from our earlier discussion that the positive energy minimum does not survive under quantum corrections.}, then the expectation value of $\varphi_1$ will get contributions from both real and complex instantons. The main difficulty in this case is the explicit realization of the instantons (even the real ones are harder to determine quantitatively).  

\vskip.1in

\noindent Considering the aforementioned three points, and the result in \eqref{rubiem}, then justifies what we said earlier, namely that the Glauber-Sudarshan state appears from the interacting vacuum $\vert\Omega\rangle$ associated with the full potential in the system. Alternatively, restricting ourselves to Minkowski minimum (using \eqref{saw1}), the presence of the neighboring vacua manifest themselves as the real and complex instantons in the computation of $\langle\varphi_1\rangle_\sigma$. The powerful resurgence techniques, lending eventually to the Borel-\'Ecalle resummations, succinctly capture this. More so, if we compare the {\it tree} level result with the one coming from summing over all instantons, we get:
\bg\label{recelcards}
\begin{split}
&\langle\varphi_1\rangle_{\sigma, {\rm tree}} = \int_{k_{\rm IR}}^\mu d^{11} k ~{\alpha_1(k)\over a(k)}~\psi_k(x, y)\\
&\langle\varphi_1\rangle_{\sigma, {\rm inst}} = {1\over g^{1/\bar\alpha}}\left[\int_0^\infty d{\rm S} ~{\rm exp}\left(-{{\rm S}\over g^{1/\bar\alpha}}\right) {1\over 1 - \check{f}_{\rm max} {\rm S}^{\bar\alpha}}\right]_{\rm P.V} \int_{k_{\rm IR}}^\mu d^{11}k ~{\alpha_1(k)\over a(k)}~\psi_k(x, y)\\
\end{split}
\nd
which are exactly the same upto an overall renormalization factor. This shows that the resurgent sum over all instantons, both real and complex, simply provides an overall effect like a wave-function renormalization! 

Of course the above conclusion is based on (a) scalar field components, and (b) on a bound like \eqref{runikate}, but we expect that the presence of actual metric (or other M-theory) degrees of the freedom should not change the main conclusion drastically. In the absence of a bound like \eqref{runikate}, the permutation factors ${\bf A}_{r_1r_2...r_{{\rm N}p}}$
would make the analysis a bit non-trivial. We will not discuss this here but leave it for \cite{ccpaper}. Note that the overall renormalization factor cannot be absorbed in the definition of $\alpha_1(k)$ because doing so will ruin the delicate vanishing of $\mathbb{T}$ in \eqref{marasis}. Thus the overall factor in \eqref{recelcards} is non-trivial and we will identify this with the inverse of the dimensionless cosmological constant. Before that however we need to determine $\rho^\ast(k)$ in \eqref{robinright}. These two points will be elaborated in the following section. \textcolor{blue}{Needless to say, due to the technical natures of the following section, {\it i.e.} section \ref{smallcc}, the readers not acquainted with Borel resummation and Gevrey series may choose to skip it on their first reading and jump directly to section \ref{sec4}}.

\section{Borel resummation and the smallness of the cosmological constant \label{smallcc}}
The asymptotic nature of the expectation value for $\varphi_1$ is the key reason why such a series may be Borel resummed to take the form as in the second relation in \eqref{recelcards}. From here one may extract the form of the four-dimensional cosmological constant, as we shall show here in details, but before that let us discuss a bit on the concept of the cosmological constant itself. We will start by first ruling out a few oft-wrongly-introduced ingredients, namely, fluxes, branes, anti-branes, O-planes and perturbative quantum effects. These have no effects on the four-dimensional positive cosmological constant. Another popular view is that the four-dimensional cosmological constant should come from summing up the zero point energies of the bosonic and the fermionic degrees of freedom. This is {\it not} true despite the popularity of such an idea. The reasoning is simple. Typically the sum of the zero point energies in a non-supersymmetric theory blows up, but one can regularize it in a Lorentz invariant way (see for example \cite{akhmedov}) and get a finite answer from it. The problem is $-$ taking for example a massive scalar field of mass $m$ $-$ that the result is typically $10^{120}$ times {\it bigger} than the expected answer (which is $70\%$ of the critical density $\rho_c$ of the universe\footnote{$\rho_c \approx 4.3 \times 10^{-47}~({\rm GeV})^4$.}). For massless scalar field the result vanishes\footnote{To see the behavior for a massive and a massless scalar field, we have to resort to a Lorentz-invariant framework. The zero point energy can be expressed completely in terms of a Lorentz invariant Euclidean integral in the following way:
\bg\label{tagindigo}
\rho \equiv  {1\over 2}\int_0^{\Lambda} {d^4k_{\rm E}\over (2\pi)^4}{\rm log}\left(1 + {m^2\over k^2_{\rm E}}\right)
 \approx  {1\over 64\pi^2}\left[\left(\Lambda^4 + m^4\right)\log ~ {m^2\over \lambda^2} + m^2\Lambda^2\right], \nonumber \nd
which is exactly the one-loop contribution to the effective potential of a scalar field of mass $m$ (the tree level contribution is the standard mass term ${1\over 2} m^2 \phi^2$), and shows {\it quartic} divergence. Unfortunately this result does not satisfy 
$\rho = -p$, with $p$ being the pressure in three dimensions. A more careful approach, outlined in \cite{akhmedov} shows that:
\bg\label{duitagindigo}
\rho = {1\over 64\pi^2}\left[m^2\Lambda^2 - m^4\log\left(1 + {\Lambda^2\over m^2}\right)\right], \nonumber \nd
and is therefore {\it quadratically} divergent. This divergence is unaffected by the space-time curvature. From the absence of a quartic divergence it appears that a free massless scalar field will have {\it zero} contribution to the vacuum energy \cite{akhmedov}.}. Thus in either case the vacuum energy doesn't appear to provide the small cosmological constant that we seek. It could be possible that once we take fermionic fields then the zero point energies could be tuned to be arbitrarily small. Unfortunately this delicate balance has not been demonstrated so far because it requires special choices for the masses of the bosonic and the fermionic fields. We will avoid this line of investigation and for the analysis that we perform here, we take a supersymmetric Minkowski background $-$ where the vacuum energy vanishes $-$ and claim that the cosmological constant appears from the non-perturbative contributions via Borel resumming a Gevrey series resulting from an expectation value over a Glauber-Sudarshan state. An explicit demonstration of the above statement is the content of the following two subsections.

\subsection{Determining the cosmological constant using Borel resummation  \label{sec3.8}}

\begin{figure}
    \centering
    \includegraphics[scale=0.8]{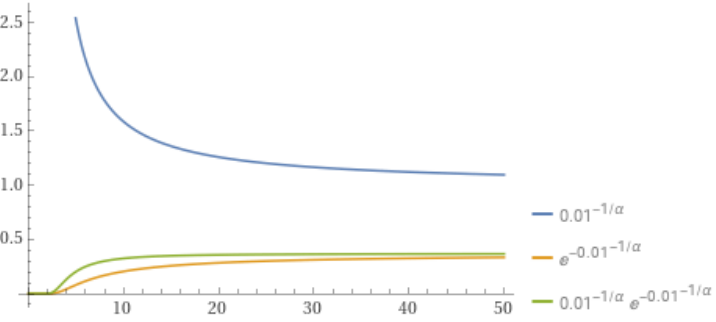} 
    \caption{Plot of three functions $f_1(\alpha) = {1\over g^{1/\alpha}}$ in \textcolor{blue}{blue}; $f_2(\alpha) = {\rm exp}\left(-{1\over g^{1/\alpha}}\right)$  in \textcolor{orange}{orange} and 
    $f_3(\alpha) ={1\over g^{1/\alpha}}{\rm exp}\left(-{1\over g^{1/\alpha}}\right)$ in \textcolor{green}{green} for $g = 0.01$ and $\alpha$ identified to $\bar\alpha$ from \eqref{kitkatmey}. We see that as $\alpha$ increases both the \textcolor{orange}{orange} and the \textcolor{green}{green} curves increase and saturate to ${1\over e}$ as $\alpha \to \infty$.}
    \label{e1overg}
\end{figure}

The short discussion outlined above suggests that the concept of a four-dimensional positive cosmological constant is a subtle one. However
before we study the cosmological constant $-$ and here we will take the dimensionless one $-$ we need to find the functional form for $\rho^\ast(k)$ appearing in \eqref{robinright} and \eqref{rubiem}. There is also another issue coming from the dependence of $\bar\alpha$ in \eqref{kitkatmey} on the {\it powers} of the field contents $(p, q, r, s)$. The worrisome feature, as evident from {\bf figure \ref{e1overg}}, is that both the instanton terms: ${\rm exp}\left(-{1\over g^{1/\bar\alpha}}\right)$ and ${1\over g^{1/\bar\alpha}}{\rm exp}\left(-{1\over g^{1/\bar\alpha}}\right)$ {\it increase} as $\bar\alpha$ increases, and saturate to $e^{-1}$ when $\bar\alpha \to \infty$. Since $\bar\alpha$ is related to the number, as well as powers, of fields, the increase in {\bf figure \ref{e1overg}} implies that we cannot consistently truncate the interaction Lagrangian \eqref{katumara} to any finite order. 

\begin{figure}
    \centering
    \includegraphics[scale=0.8]{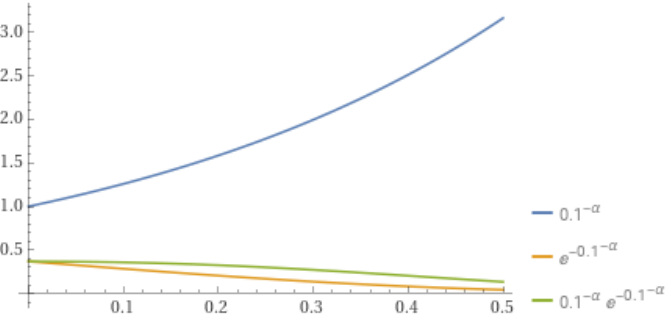} 
    \caption{Plot of three functions $f_1(\alpha) = {1\over g^{\alpha}}$ in \textcolor{blue}{blue}; $f_2(\alpha) = {\rm exp}\left(-{1\over g^{\alpha}}\right)$  in \textcolor{orange}{orange} and 
    $f_3(\alpha) ={1\over g^{\alpha}}{\rm exp}\left(-{1\over g^{\alpha}}\right)$ in \textcolor{green}{green} for $g = 0.1$ and $\alpha$ identified to $\check\alpha$ from \eqref{halfprice}. We see that as $\alpha$ increases both the \textcolor{orange}{orange} and the \textcolor{green}{green} curves decrease and go to zero as $\alpha \to \infty$.}
    \label{plotmp}
\end{figure}

String theory saves this situation in a rather interesting way. The perturbative interactions in the low energy effective action of string theory may be classified as a series in powers of curvature, fluxes and Rarita-Schwinger fermions that involve at least one derivatives of the metric, and the three-form fluxes. For fermions, we can allow interactions of the form $\left(\overline\Psi^{\rm M} \Psi_{\rm M}\right)^{s'+2}$ or of the form $\left(\overline\Psi_{\rm M} \Gamma^{\rm MP} \Psi_{\rm P}\right)^{s'+2}$ for $s' \ge 0$ alongwith multiple derivatives as $\partial^h \varphi_4^s$ in \eqref{katumara}\footnote{Recall that $\varphi_4$ is related to the Rarita-Schwinger fermionic condensate. See \cite{hete8} for more details on the fermionic Lagrangian.}. So far we have been taking $\varphi_4$ dimensionless, but the right thing would be to give some dimension to $\varphi_4$ but keep other $\varphi_i$ dimensionless. Alternatively we can impose the constraint that $n \ge p, m \ge q, l \ge r$ and $h \ge s$ in \eqref{katumara} to form a consistent picture from string theory. Taking $\vert c_o\vert \equiv 1$ (or absorbing it in the definition of ${\rm M}_p$), we get:
\bg\label{halfprice}
g^{1/\bar\alpha} = {{\rm M}_p^{-{n+m+l+h\over p+q+r+s-2}}} \equiv {\rm M}_p^{-\check\alpha} \equiv {\check{g}}^{\check\alpha}, \nd
where $\check\alpha= {n+m+l+h\over p+q+r+s-2}> 1$ and $\check{g} = {1\over {\rm M}_p}$. The scenario is plotted in {\bf figure \ref{plotmp}}. We see that the behavior now is the exact opposite of what we had in {\bf figure \ref{e1overg}}: the non-perturbative effects {\it decrease} as the number of fields or their powers are increased. This means that interactions like \eqref{katumara} are amenable to truncations and the non-perturbative series coming from higher order interactions could be made convergent. What does it buy us?

To see the advantage of the aforementioned scaling behavior, we should go back where we left off, namely, determine the functional form for 
$\rho^\ast(k)$. Our main aim is to reproduce a metric configuration of the form \eqref{mmetric}, although we have so far only used scalar fields. The reasoning is as described earlier: to avoid inserting Faddeev-Popov ghosts, gauge fixing terms {\it et cetera}. On the other hand, if we {\it know} how to approach the problem with actual metric operators, then at least we expect $\langle {\bf g}_{\mu\nu}\rangle_\sigma = g_s^{-8/3}$ from \eqref{mmetric} in M-theory. With this in mind, one possibility is the following\footnote{In deriving \eqref{traderjoeme} we have swept a subtlety under the rug. The momentum integrals are bounded by $k_{\rm IR} < k < \mu$, and as such we have to use the Riemann-Lebesgue lemma to do the integrals. Implementing this leads to equating certain series with Hermite polynomials, which eventually gives us to the result in \eqref{traderjoeme}. We point the readers to section 4.2 of \cite{borel} for further details on this.}:
\bg\label{traderjoeme}
\rho^\ast(k) =  \rho^\ast({\bf k}, k_0) = {11\over 3} - \lim_{\omega \to 0} \left[{{{\bf k}^2} + \omega^2~{\rm log}\left(\pi\vert\omega\vert\right)\over \omega^2~{\rm log}~k}\right] - {{\rm log}~{\bf k}^9\over {\rm log}~k} + {{\rm log}~2 \over {\rm log}~k}, \nd
where $k^2 = {\bf k}^2 + k_0^2$ and we have taken all momenta to be dimensionless {\it i.e.} all momenta are measured with respect to ${\rm M}_p$. Since $g_s = {\rm H}(y) \sqrt{\Lambda} \vert t \vert$, and we are taking ${\rm H}(y) = 1$ for simplicity, the {\it emergent} metric along ${\bf R}^{2, 1}$ becomes ${1\over \Lambda^{4/3} \vert t\vert^{8/3}}$. Since we have identified the dimensionless temporal coordinate $({\rm M}_p\vert t\vert)^{-8/3}$ from $\rho^\ast({\bf k}, k_0)$ in \eqref{traderjoeme}, the four-dimensional cosmological constant may then be identified from \eqref{rubiem} as:
\bg\label{tinmey}
\Lambda =  {{\rm M}_p^2 \over {1\over g^{3/4\bar\alpha}}\left[\int_0^\infty d{\rm S} ~{\rm exp}\left(-{{\rm S}\over g^{1/\bar\alpha}}\right) {1\over 1 - \check{f}_{\rm max} {\rm S}^{\bar\alpha}}\right]^{3/4}_{\rm P.V}}, \nd
where $\bar\alpha$ and $\check{f}_{\rm max}$ are defined in \eqref{kitkatmey} and \eqref{runikate} respectively; and P.V is the principal value of the integral. Somewhat remarkably this is an exact closed form expression for the cosmological constant, albeit in a set-up with scalar degrees of freedom. Looking at the integral structure of the cosmological constant, let us make the following observations on it's form.

\vskip.1in

\noindent ${\bf 1.}$ The principal value integral appearing in \eqref{tinmey} is {\it positive} definite no matter what sign of $\check{f}_{\rm max}$ we choose. A detailed proof of this statement appears in section 4.5 of \cite{borel} which the interested readers may look up.

\vskip.1in

\noindent ${\bf 2.}$ The principal value computation relies on the number of roots in the Borel plane (here it is a line stretching from ${\rm S} = 0$ to ${\rm S} = \infty$). The number of roots are the number of solutions to the equation ${\rm S}^{\bar\alpha} - {1\over \check{f}_{\rm max}} = 0$. Since $\bar\alpha$, defined in \eqref{kitkatmey}, can either be an even or an odd integer, there can only be {\it one} pole on the positive real axis of the Borel plane for any choice of positive $\check{f}_{\rm max}$. Computing the integral in the near neighborhood of this pole provides it's principal value.  

\vskip.1in

\noindent ${\bf 3.}$ In the limit $g \to 0$, or ${\rm M}_p \to \infty$, which would be in some sense the classical limit, the principal value integral becomes 1. Thus $\Lambda$ in \eqref{tinmey} is still too big to be match up with the actual four-dimensional cosmological constant which is supposedly $10^{-120} {\rm M}_p^2$. 

\vskip.1in

\noindent ${\bf 4.}$ The closed form expression of the cosmological constant in \eqref{tinmey} is clearly constructed from summing over all the real and the complex instantons in the system. Thus neither the zero point energies $-$ which are in fact cancelled here due to the underlying supersymmetry $-$ nor the perturbative and the classical terms have any contributions in the construction as alluded to earlier. 

\vskip.1in

\noindent Despite the success of the aforementioned analysis, and the appearance of an exact closed-form expression for the four-dimensional cosmological constant, two questions arise.

\vskip.1in

\noindent ${\bf 1.}$ The cosmological constant is derived from an expression like \eqref{rubiem} by identifying the Fourier integral to the dimensionless temporal coordinate $({\rm M}_p\vert t\vert)^{-8/3}$, and the instanton sum to $\left({\Lambda\over {\rm M}_p^2}\right)^{-4/3}$. This clearly leads to an ambiguity for the choice of the cosmological constant. How do we then justify an expression like \eqref{tinmey}?

\vskip.1in

\noindent ${\bf 2.}$ The instanton sum, or more appropriately the instanton integral, goes to 1 in the limit $g \to 0$. This leads to a value of the cosmological constant that is at least $10^{120}$ times bigger than the actual answer. Even if we are not near $g \to 0$ point, keeping $g< 1$ the instanton integral is never big enough to compensate for such a large discrepancy.  Why is the cosmological constant so small, and how can our analysis reproduce such a small number?

\vskip.1in

\noindent The first question can be answered by matching the boundary conditions. Recall that the Glauber-Sudarshan state description remains a valid one in the temporal domain \eqref{tcc}. At $t = -{1\over \sqrt{\Lambda}}$, $g_s = 1$, and the metric will resemble the Minkowski metric. This happens for the following choice of $\rho^\ast({\bf k}, k_0)$:
\bg\label{chipotelme}
\rho^\ast({\bf k}, k_0) = 2 - \lim_{\omega, \omega' \to 0} \left[{\omega'^2{\bf k}^2 + \omega^2 k_0^2 + (\omega\omega')^2~{\rm log}\left({\pi^2\vert\omega\vert\vert\omega'\vert} \right) \over (\omega\omega')^2~{\rm log}~k}\right] - {{\rm log}~{\bf k}^9\over {\rm log}~k} + {{\rm log}(-2\Lambda^{4/3}) \over {\rm log}~k}, \nd
where in the presence of \eqref{traderjoeme} we can make $\omega' = \omega$. From \eqref{chipotelme}, the only value of $\Lambda$ that satisfies the required boundary condition is \eqref{tinmey}, thus fixing the form for the four-dimensional cosmological constant. Additionally, \eqref{chipotelme} provides precisely the extra off-shell contribution that was alluded to earlier, and is required to get the background de Sitter configuration correctly. 

The second question requires us to find the reason for the smallness of the four-dimensional cosmological constant. This is what we turn to next.

\subsection{Why is the four-dimensional positive cosmological constant so small? \label{sec3.9}}

\begin{table}[tb]  
 \begin{center}
\renewcommand{\arraystretch}{1.5}
\resizebox{\textwidth}{!}{\begin{tabular}{|c||c||c||c||c||c||c||c||c||c||c||c|| c|}
\hline ${\rm N}$  & 0 & 1 & 2 & 3 & 4 & 5 & 6 &..& 10 & .. & 13 & $j$ \\ \hline\hline
0 & ${\bf A}_{{\rm eff}, 10}^0 k^{\nu_{10}}$  & ${\bf A}_{{\rm eff}, 11}^0 k^{\nu_{11}}$
& ${\bf A}_{{\rm eff}, 12}^0 k^{\nu_{12}}$ & ${\bf A}_{{\bf eff}, 13}^0 k^{\nu_{13}}$ & & & & & & & & \\ \hline
1 & ${\bf A}_{{\rm eff}, 10}^1 k^{\nu_{10}}$  & ${\bf A}_{\ef, 11}^1 k^{\nu_{11}}$
& ${\bf A}_{\ef, 12}^1 k^{\nu_{12}}$ & ${\bf A}_{\ef, 13}^1 k^{\nu_{13}}$ & & & & & & & & \\ \hline
2 & ${\bf A}_{\ef, 10}^2 k^{\nu_{10}}$  & ${\bf A}_{\ef, 11}^2 k^{\nu_{11}}$
& ${\bf A}_{\ef, 12}^2 k^{\nu_{12}}$ & ${\bf A}_{\ef, 13}^2 k^{\nu_{13}}$ & 
${\bf A}_{\ef, 14}^2 k^{\nu_{14}}$ & ${\bf A}_{\ef, 15}^2 k^{\nu_{15}}$ &$ {\bf A}_{\ef, 16}^2 k^{\nu_{16}}$ & & & & & \\ \hline
3 & ${\bf A}_{\ef, 10}^3 k^{\nu_{10}}$  & ${\bf A}_{\ef, 11}^3 k^{\nu_{11}}$
& ${\ca A}_{\ef, 12}^3 k^{\nu_{12}}$ & ${\ca A}_{\ef, 13}^3 k^{\nu_{13}}$ & 
${\ca A}_{\ef, 14}^3 k^{\nu_{14}}$ & ${\ca A}_{\ef, 15}^3 k^{\nu_{15}}$ & ${\ca A}_{\ef, 16}^3 k^{\nu_{16}}$ & .. & ${\ca A}_{\ef, 1, 10}^3 k^{\nu_{1, 10}}$& & & \\ \hline
4 & ${\ca A}_{\ef, 10}^4 k^{\nu_{10}}$  & ${\ca A}_{\ef, 11}^4 k^{\nu_{11}}$
& ${\ca A}_{\ef, 12}^4 k^{\nu_{12}}$ & ${\ca A}_{\ef, 13}^4 k^{\nu_{13}}$ & 
${\ca A}_{\ef, 14}^4 k^{\nu_{14}}$ & ${\ca A}_{\ef, 15}^4 k^{\nu_{15}}$ & ${\ca A}_{\ef, 16}^4 k^{\nu_{16}}$ & .. & ${\ca A}_{\ef, 1, 10}^4 k^{\nu_{1, 10}}$ & .. 
& ${\ca A}_{\ef, 1, 13}^4 k^{\nu_{1, 13}}$ &
\\ \hline
..& .. & .. & ..&..&..&..&..&..&..& .. & .. & ..\\ \hline
${\rm N}$ & ${\ca A}_{\ef, 10}^{\rm N} k^{\nu_{10}}$  & ${\ca A}_{\ef, 11}^{\rm N} k^{\nu_{11}}$
& ${\ca A}_{\ef, 12}^{\rm N} k^{\nu_{12}}$ & ${\ca A}_{\ef, 13}^{\rm N} k^{\nu_{13}}$ & 
${\ca A}_{\ef, 14}^{\rm N} k^{\nu_{14}}$ & ${\ca A}_{\ef, 15}^{\rm N} k^{\nu_{15}}$ & ${\ca A}_{\ef, 16}^{\rm N} k^{\nu_{16}}$ & .. & ${\ca A}_{\ef, 1, 10}^{\rm N} k^{\nu_{1, 10}}$ & .. 
& ${\ca A}_{\ef, 1, 13}^{\rm N} k^{\nu_{1, 13}}$ & ..
\\ \hline
\end{tabular}}
\renewcommand{\arraystretch}{1}
\end{center}
 \caption[]{Proliferation of the terms (first four are explicitly shown) contributing to expectation value of $\varphi_1$ with $p = 7$. Notice the columns are completely filled till $j = 3$, beyond which there are gaps in the representation of ${\ca A}_{\ef, 1j}$ for $j > 3$. To preserve the brevity, we have avoided showing the factors ${\cal A}_1^{\rm N}$ and $\Delta(j)$.} 
  \label{fiirrzacutt}
 \end{table}

Finding the reason for the smallness of the four-dimensional cosmological constant is more complicated and we will provide a detailed answer in our upcoming paper \cite{ccpaper}. Here we will simply sketch the main points. As before we will continue with the scalar fields and start with the action \eqref{katumara}. Let us also concentrate only on the sector with $\varphi_1$ field components. The dominant contribution has already been given in \eqref{julbinoc}, and we represent it here as:
\bg\label{julbinoc3}
\big[\underbrace{1,~ 1,~ 1,~ 1, ....,~ 1}_{p-1~{\rm fields}}, ~2\big]_{\varphi_1} ~ \equiv ~
\underbrace{\widetilde\varphi_1(k_1)~ \widetilde\varphi_1(k_2) ~\widetilde\varphi_1(k_3)~... ~\widetilde\varphi_1(k_{p-1})~\overbracket[1pt][7pt]{\widetilde\varphi_1(k)~\widetilde\varphi_1}}_{p~{\rm fields} ~+ ~1~{\rm source}}(k), \nd
where such a term grows as $p!$, but since the suppression factor is also $p!$, the growth is 1. The factor of ``1'' appearing within bracket on the LHS of \eqref{julbinoc3} represents the number of fields with a given momentum that is unequal to the other $p-1$ choices of momenta. This means ``2'' should correspond to two field components having the same momenta, and therefore {\it contracted} according to the rules mentioned earlier. In this language, the amplitude coming from an action like \eqref{katumara}, at first order in $g$, can be expressed as:
\bg\label{plearnthai}
{\bf A} \equiv\sum_{j = 0}^{\lfloor {p -1\over 2}\rfloor}{\bf A}_{1j} = {g\over p!}\sum_{j = 0}^{\lfloor {p -1\over 2}\rfloor} \big[1, ~1, ~ 1,...,~1, ~\underbrace{2,~ 2,~ 2, ....,~ 2}_j, ~2\big]_{\varphi_1} ~ + ~ {\rm permutations}\, ,\nd
which can be easily inferred from the binomial expansion discussed in section \ref{sec3.4}, with the subscript $j$ referring to the number and distributions of ``2'' in the amplitude, and $\lfloor x\rfloor$ denoting greatest integer less than or equal to $x$. Such an amplitude has a piece that is not suppressed by the volume factor ${\rm V}$ and therefore is {\it not} subdominant and should contribute to the cosmological constant that we computed earlier with only ${\bf A}_{10}$, {\it i.e.} with $j = 0$ in \eqref{plearnthai}, part of the amplitude. How does this reduce the value of $\Lambda$ in \eqref{tinmey}?

To see, we will have to go to the ${\rm N}$-th order in the amplitude. This is quite a non-trivial exercise so, as mentioned earlier, we will only sketch the main points. Again, details may be inferred from \cite{ccpaper}. At the ${\rm N}$-th order the amplitude becomes:
\bg\label{moebooks}
{\bf A}^{\rm N} \equiv\sum_{j = 0}^{\lfloor {{\rm N}p -1\over 2}\rfloor}\left({\bf A}^{\rm N}\right)_{1j} = {g^{\rm N}\over (p!)^{\rm N}}\sum_{j = 0}^{\lfloor {{\rm N}p -1\over 2}\rfloor} \big[\underbrace{1, ~1, ~ 1,...,~1}_{{\rm N}p - 2j -1}, ~\underbrace{2,~ 2,~ 2, ....,~ 2}_j, ~2\big]_{\varphi_1} ~ + ~ {\rm permutations}\, ,\nd
which works well when $j = 0$ because $\left({\bf A}^{\rm N}\right)_{10} = \left({\bf A}_{10}\right)^{\rm N}$, and therefore Borel resummation over ${\rm N}$ can be performed. This is of course what we did earlier to get \eqref{tinmey}. For $j > 0$, this is clearly not the case, which is first of the many difficulties we face once we go to more complicated nodal diagrams. There is also the denominator of the path-integral that should depend on $j$, plus we haven't shown the source momentum explicitly in \eqref{moebooks}. In fact now we expect:

{\scriptsize
\bg\label{crossover}
{{\rm Num}\over {\rm Den}} \equiv {{\cal N}\over {\cal D}} =  {\big[1\big]_{\varphi_1} ~+~  \int d^{11} k\sum\limits_{{\rm N} = 1}^\infty {g^{\rm N}\over (p!)^{\rm N}}\sum\limits_{j = 0}^{\lfloor {{\rm N}p -1\over 2}\rfloor} \sum\limits_{m_{1j}({\rm N})}\big[\underbrace{1, ~1, ~ 1,...,~1}_{{\rm N}p - 2j -1}, ~\underbrace{2,~ 2,~ 2, ....,~ 2}_j, ~2\big]_{\varphi_1}\bigodot k^{m_{1j}({\rm N})} ~ + ~ {\rm permutations} \over 
1 ~+~ \sum\limits_{{\rm N} = 1}^\infty {g^{\rm N}\over (p!)^{\rm N}}\sum\limits_{j = 0}^{\lfloor {{\rm N}p\over 2}\rfloor} \big[\underbrace{1, ~1, ~ 1,...,~1}_{{\rm N}p - 2j}, ~\underbrace{2,~ 2,~ 2, ....,~ 2}_j\big]_{\varphi_1} ~ + ~ {\rm permutations}}, \nd}
showing clearly the difference between the integrals on the numerator and the denominator, with $\bigodot$ identifying the source momentum with $k$ expressed using the permutation indices $m_{1j}({\rm N})$. One may then sum over the permutation indices $m_{1j}({\rm N})$, and then integrate over the source momentum over the IR regime $k_{\rm IR} < k < \mu$. 
These momentum factors,  
 stemming from $\partial^n$ in \eqref{katumara}, complicate the source term integral as evident from the appearance of 
$k^{\rho^\ast(k)}$ in \eqref{runikate} for the case with $j = 0$. We can generalize this by imposing the following bound on the ${\rm N}$-th order amplitude:
\bg\label{lyttoncasi}
\sum_{m_{1j}({\rm N})} \left({\bf A}^{\rm N}\right)_{1j}(m_{ij}({\rm N})) 
{k^{m_{1j}({\rm N})}\over {\cal D}} ~ \le ~ {(p{\rm N})!\over 2^{j'}(p!)^{\rm N}} ~\left({\bf A}_{{\rm eff}, 1j'}\right)^{\rm N} k^{\nu_{1j'}(k)}, \nd
for all values of ${\rm N}$ and ${\cal D} \equiv {\rm Den}$ in \eqref{crossover}. Here we have expressed the binomial factor at ${\rm N}$-th order by $m_{1j}({\rm N}) \in \mathbb{Z}_+$, and therefore the sum is over $m_{1j}({\rm N})$ with $k$ being the source momentum. Clearly $j' \le j$, and for $j = 0$, $j' = 0$ with $\nu_{10}(k) \equiv \rho^\ast(k)$. 
(The inequality in \eqref{lyttoncasi} is necessary because the denominator ${\cal D}$ is in general not normalized to identity.) We can see that the bound \eqref{lyttoncasi} is useful in multiple ways. 

\vskip.1in

\noindent ${\bf 1.}$ It has managed to control the binomial growth in a specific way. We no longer have to worry about the various binomial factors emerging from the powers of the derivatives or momenta, in the language of the Fourier components. 

\vskip.1in

\noindent ${\bf 2.}$ The sector associated with $\varphi_2, .., \varphi_4$ can be thought of as been included in ${\bf A}_{{\rm eff}, 1j'}$, provided we replace 
${(p{\rm N})!\over (p!)^{\rm N}}$ by the term replacing ${\rm N}!$ in the first line of \eqref{bartsfo}. The bound \eqref{lyttoncasi} can then be rewritten to take the following form:
\bg\label{lyttoncasime}
\left({\cal A}^{\rm N}\right)_{1j} ~ \equiv ~ \sum_{m_{1j}({\rm N})} \left({\bf A}^{\rm N}\right)_{1j}(m_{ij}({\rm N})) 
{k^{m_{1j}({\rm N})}\over {\cal D}} ~ \le ~ \Delta(j'){\cal A}_1^{\rm N} \left(\bar\alpha {\rm N}\right)! ~\left({\bf A}_{{\rm eff}, 1j'}\right)^{\rm N} k^{\nu_{1j'}(k)}, \nd
where ${\cal A}_1$ is defined on the second line of \eqref{bartsfo}, $\Delta(j')$ is the degeneracy\footnote{We can generalize this to 
$\Delta(j', \bar{p})$ for a given value of $j'$ and $\bar{p} \equiv (p, q, r, s)$, but will not do so here.} for a given value of $j'$ (which is a generalization of $2^{-j'}$ degeneracy we saw in \eqref{lyttoncasi}), 
${\cal D}$ is the denominator from \eqref{crossover}, and $\bar\alpha$ is as in \eqref{kitkatmey}. 
For $j = j' = 0$ this is exactly what gave us the required Gevrey growth, and now we see that we can extend it to include $j > 0$ case too. 

\vskip.1in

\noindent ${\bf 3.}$ It has also resolved the conundrum associated with raising the amplitude \eqref{plearnthai} to ${\rm N}$-th power in \eqref{moebooks}. Now instead of looking for $1j$-th component of ${\bf A}^{\rm N}$, we can look at ${\rm N}$-th power of ${\bf A}_{{\rm eff}, 1j}$. This is now in the same league as $({\bf A}_{10})^{\rm N}$, implying that the Borel resummation of the corresponding Gevrey series may be performed efficiently.

\vskip.1in

\noindent To proceed with the a bound like \eqref{lyttoncasime}, it will be instructive to give one concrete example. Let us consider $p = 7$ in \eqref{katumara}, which would correspond to an interaction of the form 
$\partial^n\varphi_1^7$, where from our earlier considerations, $n \ge 7$. In the language of ${\ca A}_{\ef, 1j}$ we will have components ${\ca A}_{\ef, 10}$ till ${\ca A}_{\ef, 13}$, so that the source can attach appropriately to the last node. In other words the bound \eqref{lyttoncasime} may alternatively be expressed as:

{\scriptsize
\bg\label{telegraphmey}
\begin{split}
& \sum_{j = 0}^3 \left({\cal A}^1\right)_{1j} ~\le ~ {\cal A}_1^1 \left(1\bar\alpha\right)! \left(\Delta(0){\ca A}^1_{\ef, 10} k^{\nu_{10}} + \Delta(1){\ca A}^1_{\ef, 11}k^{\nu_{11}} + \Delta(2) {\ca A}^1_{\ef, 12}k^{\nu_{12}} + \Delta(3){\ca A}^1_{\ef, 13} k^{\nu_{13}}\right)\\
& \sum_{j = 0}^6 \left({\cal A}^2\right)_{1j} ~ \le ~ {\cal A}_1^2 \left(2\bar\alpha\right)! \left(\Delta(0){\ca A}^2_{\ef, 10} k^{\nu_{10}} + \Delta(1){\ca A}^2_{\ef, 11} k^{\nu_{11}} + \Delta(2){\ca A}^2_{\ef, 12} k^{\nu_{12}} + \Delta(3){\ca A}^2_{\ef, 13} k^{\nu_{13}} + ... + \Delta(6) {\ca A}^2_{\ef, 16} k^{\nu_{16}}\right)\\
& \sum_{j = 0}^{10} \left({\cal A}^3\right)_{1j} ~ \le ~ {\cal A}_1^3 \left(3\bar\alpha\right)! \left(\Delta(0){\ca A}^3_{\ef, 10} k^{\nu_{10}} + \Delta(1) {\ca A}^3_{\ef, 11} k^{\nu_{11}} + \Delta(2) {\ca A}^3_{\ef, 12} k^{\nu_{12}} + \Delta(3){\ca A}^3_{\ef, 13} k^{\nu_{13}} + ... + \Delta(10){\ca A}^3_{\ef, 1, 10} k^{\nu_{1, 10}}\right)\\
\end{split}
\nd}
where we show the series upto ${\rm N} = 3$. In writing \eqref{telegraphmey} we have assumed $j' = j$ which, while a little simplistic, works for the example here. (As mentioned earlier, typically $j' < j$ and becomes equal for $j = 0$. We will however not dwell on these subtle nuances to avoid further complicating the story. Detailed study of this will appear in \cite{ccpaper}, and we direct the readers to that reference.) With this in mind, at ${\rm N}$-th order, we expect:

{\scriptsize
\bg\label{telephula}
\sum_{j = 0}^{\lfloor{p{\rm N}-1\over 2}\rfloor} \left({\cal A}^{\rm N}\right)_{1j} ~ \le ~ {\cal A}_1^{\rm N} \left({\rm N}\bar\alpha\right)! \Delta(j')\bigodot_{j'\to j}\left({\ca A}^{\rm N}_{\ef, 10} k^{\nu_{10}} + {\ca A}^{\rm N}_{\ef, 11} k^{\nu_{11}} + {\ca A}^{\rm N}_{\ef, 12} k^{\nu_{12}} + {\ca A}^{\rm N}_{\ef, 13} k^{\nu_{13}} + ... + {\ca A}^{\rm N}_{\ef, 1, \lfloor{p{\rm N}-1\over 2}\rfloor} k^{\nu_{1, \lfloor{p{\rm N}-1\over 2}\rfloor}}\right),
\nd}
with $p = 7$, which clearly depicts both the factorial and the polynomial growths. (The operation $\underset{j' \to j}{\bigodot}$, used earlier in \eqref{phobeindy}, replaces the $j'$ in $\Delta(j')$ with the corresponding value of $j$ from $\nu_{1j}$.) 
The situation is shown in {\bf Table \ref{fiirrzacutt}}, where we see that up to $j = 3$ the columns are all completely filled. Beyond $j = 3$, the representation of ${\ca A}_{\ef, 1j}$ will have {\it gaps} coming from \eqref{telegraphmey} and \eqref{telephula}. These gaps ruin the nice summations that we can have for all the columns beyond $j = 3$. The summing over a column in {\bf Table \ref{fiirrzacutt}}, due to the corresponding factorial growth, is precisely the Borel resummation. For the $j = 0$ column, the answer is what we had in \eqref{rubiem}, which led to the expression for $\Lambda$ in \eqref{tinmey}. Now that there are gaps in the columns in {\bf Table \ref{fiirrzacutt}}, summation over all the columns $\lfloor {p-1\over 2}\rfloor < j < \infty$ with $j \in \mathbb{Z}_+$ cannot be done. How should we take care of this subtlety?

\begin{table}[tb]  
 \begin{center}
\renewcommand{\arraystretch}{1.5}
\resizebox{\textwidth}{!}{\begin{tabular}{|c||c||c||c||c||c||c||c||c||c||c||c|| c|}
\hline ${\rm N}$  & 0 & 1 & 2 & 3 & 4 & 5 & 6 &..& 10 & .. & 13 & $j$ \\ \hline\hline
0 & ${\ca A}_{\ef, 10}^0 k^{\nu_{10}}$  & ${\ca A}_{\ef, 11}^0 k^{\nu_{11}}$
& ${\ca A}_{\ef, 12}^0 k^{\nu_{12}}$ & ${\ca A}_{\ef, 13}^0 k^{\nu_{13}}$ & 
${\textcolor{red}{{\ca A}_{\ef, 14}^{0} k^{\nu_{14}}}}$ & ${\textcolor{red}{{\ca A}_{\ef, 15}^{0} k^{\nu_{15}}}}$ & ${\textcolor{red}{{\ca A}_{\ef, 16}^{0} k^{\nu_{16}}}}$ & {\textcolor{red}{..}} & ${\textcolor{red}{{\ca A}_{\ef, 1, 10}^{0} k^{\nu_{1, 10}}}}$ & {\textcolor{red}{..}} 
& ${\textcolor{red}{{\ca A}_{\ef, 1, 13}^{0} k^{\nu_{1, 13}}}}$ & {\textcolor{red}{..}} \\ \hline
1 & ${\ca A}_{\ef, 10}^1 k^{\nu_{10}}$  & ${\ca A}_{\ef, 11}^1 k^{\nu_{11}}$
& ${\ca A}_{\ef, 12}^1 k^{\nu_{12}}$ & ${\ca A}_{\ef, 13}^1 k^{\nu_{13}}$ & 
${\textcolor{red}{{\ca A}_{\ef, 14}^{1} k^{\nu_{14}}}}$ & ${\textcolor{red}{{\ca A}_{\ef, 15}^{1} k^{\nu_{15}}}}$ & ${\textcolor{red}{{\ca A}_{\ef, 16}^{1} k^{\nu_{16}}}}$ & {\textcolor{red}{..}} & ${\textcolor{red}{{\ca A}_{\ef, 1, 10}^{1} k^{\nu_{1, 10}}}}$ & {\textcolor{red}{..}} 
& ${\textcolor{red}{{\ca A}_{\ef, 1, 13}^{1} k^{\nu_{1, 13}}}}$ & {\textcolor{red}{..}}\\ \hline
2 & ${\ca A}_{\ef, 10}^2 k^{\nu_{10}}$  & ${\ca A}_{\ef, 11}^2 k^{\nu_{11}}$
& ${\ca A}_{\ef, 12}^2 k^{\nu_{12}}$ & ${\ca A}_{\ef, 13}^2 k^{\nu_{13}}$ & 
${\ca A}_{\ef, 14}^2 k^{\nu_{14}}$ & ${\ca A}_{\ef, 15}^2 k^{\nu_{15}}$ &$ {\ca A}_{\ef, 16}^2 k^{\nu_{16}}$ & 
{\textcolor{red}{..}} & ${\textcolor{red}{{\ca A}_{\ef, 1, 10}^{2} k^{\nu_{1, 10}}}}$ & {\textcolor{red}{..}} 
& ${\textcolor{red}{{\ca A}_{\ef, 1, 13}^{2} k^{\nu_{1, 13}}}}$ & {\textcolor{red}{..}}\\ \hline
3 & ${\ca A}_{\ef, 10}^3 k^{\nu_{10}}$  & ${\ca A}_{\ef, 11}^3 k^{\nu_{11}}$
& ${\ca A}_{\ef, 12}^3 k^{\nu_{12}}$ & ${\ca A}_{\ef, 13}^3 k^{\nu_{13}}$ & 
${\ca A}_{\ef, 14}^3 k^{\nu_{14}}$ & ${\ca A}_{\ef, 15}^3 k^{\nu_{15}}$ & ${\ca A}_{\ef, 16}^3 k^{\nu_{16}}$ & .. & ${\ca A}_{\ef, 1, 10}^3 k^{\nu_{1, 10}}$& 
{\textcolor{red}{..}} 
& ${\textcolor{red}{{\ca A}_{\ef, 1, 13}^{3} k^{\nu_{1, 13}}}}$ & {\textcolor{red}{..}}\\ \hline
4 & ${\ca A}_{\ef, 10}^4 k^{\nu_{10}}$  & ${\ca A}_{\ef, 11}^4 k^{\nu_{11}}$
& ${\ca A}_{\ef, 12}^4 k^{\nu_{12}}$ & ${\ca A}_{\ef, 13}^4 k^{\nu_{13}}$ & 
${\ca A}_{\ef, 14}^4 k^{\nu_{14}}$ & ${\ca A}_{\ef, 15}^4 k^{\nu_{15}}$ & ${\ca A}_{\ef, 16}^4 k^{\nu_{16}}$ & .. & ${\ca A}_{\ef, 1, 10}^4 k^{\nu_{1, 10}}$ & .. 
& ${\ca A}_{\ef, 1, 13}^4 k^{\nu_{1, 13}}$ & {\textcolor{red}{..}}
\\ \hline
..& .. & .. & ..&..&..&..&..&..&..& .. & .. & ..\\ \hline
${\rm N}$ & ${\ca A}_{\ef, 10}^{\rm N} k^{\nu_{10}}$  & ${\ca A}_{\ef, 11}^{\rm N} k^{\nu_{11}}$
& ${\ca A}_{\ef, 12}^{\rm N} k^{\nu_{12}}$ & ${\ca A}_{\ef, 13}^{\rm N} k^{\nu_{13}}$ & 
${\ca A}_{\ef, 14}^{\rm N} k^{\nu_{14}}$ & ${\ca A}_{\ef, 15}^{\rm N} k^{\nu_{15}}$ & ${\ca A}_{\ef, 16}^{\rm N} k^{\nu_{16}}$ & .. & ${\ca A}_{\ef, 1, 10}^{\rm N} k^{\nu_{1, 10}}$ & .. 
& ${\ca A}_{\ef, 1, 13}^{\rm N} k^{\nu_{1, 13}}$ & ..
\\ \hline
\end{tabular}}
\renewcommand{\arraystretch}{1}
\end{center}
 \caption[]{We can add new terms to fill in the gaps left in {\bf Table \ref{fiirrzacutt}}. These are shown in \textcolor{red}{red}. One may now Borel resum each of the series given in the columns, and then sum over the Borel resummed values. The convergence property of the second summation, as well as how to remove the added contributions, will be described in the text.} 
  \label{fiirzacutt2}
 \end{table}

The answer is as given in  {\bf Table \ref{fiirzacutt2}}: we simply add the missing terms by hand! These are represented in \textcolor{red}{red}. Now we can easily Borel resum over all the columns from $0 \le j < \infty$. The final answer should be a summation over all the Borel resummed values. This {\it double} summation procedure is what will effectively reduce the value of the cosmological constant. However, two immediate questions arise.

\vskip.1in

\noindent ${\bf 1.}$ How do we make sense of {\it adding} new terms to {\bf Table \ref{fiirzacutt2}}? 

\vskip.1in

\noindent ${\bf 2.}$ How do we know that the summation of the Borel resummed values is {\it convergent}? 

\vskip.1in

\noindent Both the above questions are non-trivial, but can be answered by following a simple procedure. The procedure is to {\it remove} the added contribution to any column after we Borel resum the corresponding series. As an example consider the fourth column in {\bf Table \ref{fiirzacutt2}}. The first two elements are added by hand, and they could contribute a small percentage to the Borel resummed series in that column. Similarly, the contribution in the tenth column would again be a small percentage to the Borel resummed series from that column, but clearly the percentage contribution is bigger than the percentage contribution in the fourth column. As $j$ increases, the percentage contribution also increases, but not linearly. Instead the contribution takes the following form:
\begin{equation}
\label{resorworlme}
\sum_{j = 0}^\infty \sum_{{\rm N} = 0}^\infty{\cal A}_1^{\rm N} \left(\bar\alpha {\rm N}\right)! ~\Delta(j)~\left({\bf A}_{{\rm eff}, 1j}\right)^{\rm N} k^{\nu_{1j}(k)}~ 
\equiv \def\first{(0,0) --  (2.5,0) -- (2.5,1.5) -- (0,1.5) -- cycle}
        \def\second{
  (0.8, 1.5) .. controls (1.2, 0.4)  .. (2.5, 0) -- (0, 0) -- (0, 1.5) -- (0.8, 1.5)}
  \begin{tikzpicture} [baseline={(0, 0.6)}]
    \fill[red!60] \first;
    \begin{scope}
      \clip \first;
      \fill[blue!50] \second;
    \end{scope}
   \draw[color=black] \first;
   \draw \second;
   \node at (0.3, 1.8) {$j = \lfloor{\frac{p - 1}{2}}\rfloor$};
  \end{tikzpicture}
\end{equation}
where the RHS is a symbolic way of representing the Borel resummed columns in {\bf Table \ref{fiirzacutt2}}. The sum over ${\rm N}$ is the {\it vertical} Borel resummation of the columns, and the sum over $j$ is the second {\it horizontal} summation of the rows that we do after Borel resumming the columns. We will denote the picture on the RHS as {\it Borel box}, a term alluded to earlier at the beginning of section \ref{sec3.7} and in footnote \ref{bbox}. The coloring scheme inside the box denotes the contributions from the $p = 7$ interaction (in \textcolor{blue}{blue}), and the contributions added by hand (in \textcolor{red}{red}).

The summation on the LHS of \eqref{resorworlme} does not show the full story. In particular it misses the contribution from the Glauber-Sudarshan state, as well as the procedure to {\it remove} the contributions in 
\textcolor{red}{red} appearing on the RHS of \eqref{resorworlme}. Both of these may be accommodated in the summation scheme in the following suggestive way:

{\footnotesize
\bg\label{rodlalipa}
\begin{split}
\langle\varphi_1\rangle_{\sigma}  &\le  \mathbb{T} +  {1\over {\rm V}} \int_{k_{\rm IR}}^\mu d^{11}k \Bigg(1 + \sum_{j = 1}^{\left\lfloor{p-1\over 2}\right\rfloor} \Delta(j) + 
\sum_{j = 1 + \left\lfloor{p-1\over 2}\right\rfloor}^\infty {\rm exp}\left(-\vert{\rm G}(j, p)\vert\right)~\Delta(j)\Bigg) \\
&\times  {1\over g^{1/\bar\alpha}}\left[\int_0^\infty d{\rm S}~
{\rm exp}\left(-{{\rm S}\over g^{1/\bar\alpha}}\right) {k^{\nu_{1j}}\over 1 - {\rm S}^{\bar\alpha}{\cal A}_1{\bf A}_{{\rm eff}, 1j}}\right]_{\rm P. V} \left({\alpha_1^2(k)\over a^2(k)} + {1\over 2a(k)}\right)~ \psi_k(x, y) 
+ {\cal O}\left({1\over {\rm V}^2}\right),\\
\end{split}
\nd}
where $\Delta(j)$ is the degeneracy of various terms in the binomial expansion for a given value of $j$ and $p$ from \eqref{lyttoncasime}. These may be easily worked out by simply looking at the growth of the terms (which are the ones accompanying the factorial and the polynomial growths). The important term however is the exponentially decaying factor ${\rm exp}\left(-\vert{\rm G}(j, p)\vert\right)$, controlled by a monotonically increasing function ${\rm G}(j, p)$. For large $j$, ${\rm G}(j, p)$ would be large and consequently it would reduce the contributions from Borel resummation in the column $j$. When $j \to \infty$, ${\rm G}(j, p) \to \infty$, and therefore there would be no contribution\footnote{We expect ${\rm G}(j, p)$ to be independent of $p$ and only depend on the profile separating the \textcolor{blue}{blue} and the \textcolor{red}{red} regimes in the Borel box of \eqref{resorworlme}. However we keep the $p$ dependence for genericity.}. This is of course what we expect from the color scheme on the RHS of \eqref{resorworlme}.

The result \eqref{rodlalipa} is still not ready for synthesis. In particular we are lacking a precise way to define $\mathbb{T}$, the tree-level contribution. Such a definition should simplify the second line involving $k$ dependence to bring it in a more manageable form. This step however is more non-trivial, because the tree-level result is a bit unwieldy and takes the following form:
\bg\label{parismysteri}
\mathbb{T} \equiv  \int_{k_{\rm IR}}^\mu d^{11} k\Bigg[{\overline\alpha_1(k)\over a(k)} - \Bigg(1 + \sum_{j = 1}^{\left\lfloor{p-1\over 2}\right\rfloor} \Delta(j) \Bigg)
{k^{\nu_{1j}}\over {\rm V}}\left({\overline{\alpha}_1^2(k)\over a^2(k)} + {1\over 2a(k)}\right)\Bigg]~\psi_k(x, y),
\nd
which may be compared to \eqref{marasis}. Note that the summation over $j$ in \eqref{parismysteri} does not extend to $j \to \infty$. It is easy to see from {\bf Table \ref{fiirzacutt2}} why this is the case. We need the quadratic pieces to efficiently perform the Borel resummation only in the range $0 \le j \le \lfloor{p-1\over 2}\rfloor$ and therefore they appear with {\it minus} signs in the tree-level result \eqref{parismysteri}. For $1 + \lfloor{p-1\over 2}\rfloor \le j < \infty$, this is no longer a requirement as we are anyway adding the ${\rm N}= 0$ pieces by hand. While this helps us to write $\mathbb{T}$ efficiently, it also implies that
due to the $j$ dependence in $\Delta(j)$ and $\nu_{1j}$, one cannot express the quadratic piece of $\alpha_1(k)$ in terms of the linear piece so efficiently. In \eqref{marasis} we managed this by resorting to $\rho^\ast(k)$ in \eqref{traderjoeme}. Can we do this here? 

Unfortunately the story is more complicated and the simplification resulting from \eqref{marasis} cannot be repeated here. But all is not lost, because we can adapt a different strategy. The strategy would be to look for the {\it maximum} value of $\nu_{1j}$ to get a bound on the quadratic piece from the vanishing of $\mathbb{T}$ in \eqref{parismysteri}. In other words, we define:
\bg\label{TIsuite}
\nu^\ast(k) \equiv {\rm max}\left(\nu_{1j}, \forall j\right) = 
{\rm max}\left(\nu_{10},~ \nu_{11}, ~\nu_{12}, ..., ~\nu_{1\lfloor{p-1\over 2}\rfloor}, ..., ~\nu_{1\infty}\right), \nd
which makes $k^{\nu^\ast(k)}$ the least dominant one because the dimensionless $k$ satisfy $k < 1$ in the range 
$k_{\rm IR} < k < \mu < 1$ as they are all measured with respect to ${\rm M}_p$ (see \cite{coherbeta}, \cite{coherbeta2}, \cite{borel} for a discussion on the scales appearing in the problem). This means the bound coming from the vanishing of $\mathbb{T}$ in \eqref{parismysteri} will be the following:

{\footnotesize
\bg\label{vananoutside}
\int_{k_{\rm IR}}^\mu d^{11} k~
{k^{\nu^\ast(k)}\over {\rm V}}\left({\alpha_1^2(k)\over a^2(k)} + {1\over 2a(k)}\right)~\psi_k(x, y) ~ \le ~  
\Bigg({1 + \sum\limits_{j = 1}^{\left\lfloor{p-1\over 2}\right\rfloor} \Delta(j)}\Bigg)^{-1}
\int_{k_{\rm IR}}^\mu d^{11} k ~{\alpha_1(k)\over a(k)}~\psi_k(x, y), \nd}
which works perfectly because of our choice of $\nu^\ast(k)$ from \eqref{TIsuite}. The bound on the expectation value, {\it i.e.} $\langle\varphi_1\rangle_\sigma$ in \eqref{rodlalipa}, remains unaffected by the choice of $\nu^\ast(k)$, and therefore plugging \eqref{vananoutside} in \eqref{rodlalipa} leads to the following alternative expression:

{\footnotesize
\bg\label{rodlalipa2}
\begin{split}
\langle\varphi_1\rangle_{\sigma}  & ~ \le ~ \Bigg(1 + \sum\limits_{j = 1}^{\left\lfloor{p-1\over 2}\right\rfloor} \Delta(j) + 
\sum\limits_{j = 1 + \left\lfloor{p-1\over 2}\right\rfloor}^\infty {\rm exp}\left(-\vert{\rm G}(j, p)\vert\right)~\Delta(j)\Bigg) \Bigg( 
1 + \sum\limits_{j = 1}^{\left\lfloor{p-1\over 2}\right\rfloor} \Delta(j)\Bigg)^{-1}\\
&~ \times ~  {1\over g^{1/\bar\alpha}}\left[\int_0^\infty d{\rm S}~
{\rm exp}\left(-{{\rm S}\over g^{1/\bar\alpha}}\right) {1 \over 1 - {\rm S}^{\bar\alpha}{\cal A}_1{\bf A}_{{\rm eff}, 1j}}\right]_{\rm P. V} 
~ \int_{k_{\rm IR}}^\mu d^{11}k ~ ~{\alpha_1(k)\over a(k)}~\psi_k(x, y) 
+ {\cal O}\left({1\over {\rm V}^2}\right),\\
\end{split}
\nd}
which not only provides the bound for the expectation value, but also cleanly separates $j$ dependent summation from the $k$ dependent integral. One should also compare \eqref{rodlalipa2} with \eqref{rubiem}, the latter appearing from the $j = 0$ sector of full construction as seen from {\bf Table \ref{fiirzacutt2}}. Using similar arguments from the boundary condition $-$ which led us to a closed form expression for the four-dimensional cosmological constant in \eqref{tinmey} $-$ now leads to the following modified expression for the cosmological constant:

{\scriptsize
\bg\label{palazzome}
\Lambda^{4/3} ~ = ~ {{\rm M}_p^{8/3}\Bigg( 
1 + \sum\limits_{j = 1}^{\left\lfloor{p-1\over 2}\right\rfloor} \Delta(j)\Bigg) \over  {1\over g^{1/\bar\alpha}}\Bigg(1 + \sum\limits_{j = 1}^{\left\lfloor{p-1\over 2}\right\rfloor} \Delta(j) + 
\sum\limits_{j = 1 + \left\lfloor{p-1\over 2}\right\rfloor}^\infty {\rm exp}\left(-\vert{\rm G}(j, p)\vert\right)~\Delta(j)\Bigg)  \Bigg[\int_0^\infty d{\rm S}~
{\rm exp}\left(-{{\rm S}\over g^{1/\bar\alpha}}\right) {1 \over 1 - {\rm S}^{\bar\alpha}{\cal A}_1{\bf A}_{{\rm eff}, 1j}}\Bigg]}_{{}^{{}^{\rm P. V}}} <~{\rm M}_p^{8/3}, \nd}
which is {\it smaller} than the value of the cosmological constant we got in \eqref{tinmey}! This reduction in the value  appears from $j > 0$ contributions that were ignored in our earlier computations. Note that if ${\rm G}(j, p) = 0$, then the convergence of the series appearing in the denominator of \eqref{palazzome} may not be an issue. For example if $\Delta(j) = {1\over 2^j}$, and since:
\bg\label{bbdrem}
{1 + \sum\limits_{j = 1}^{\left\lfloor{p-1\over 2}\right\rfloor} \Delta(j) \over 1 + \sum\limits_{j = 1}^\infty \Delta(j)} ~ = ~ 1- \left({1\over 2}\right)^{1 + \lfloor{p-1\over 2}\rfloor} ~ < ~ 1, \nd
the cosmological constant is effectively reduced by that factor. Of course,
from the analysis that we presented above, ${\rm G}(j, p)$ {\it cannot} vanish which could in fact be easily justified from {\bf Table \ref{fiirzacutt2}} or from the coloring scheme of the Borel box in \eqref{resorworlme}. But even with non-zero ${\rm G}(j, p)$, the value of $\Lambda$ would be smaller than what we got from \eqref{tinmey}. Despite the aforementioned success, the closed form expression in \eqref{palazzome} is still {\it not} the full answer as we have chosen a specific value of $p = 7$ in \eqref{katumara}. To get the full answer, we will have to add in the contributions coming from all possible values of $p$ $-$ including all possible values for $(q, r, s)$ that we have ignored so far $-$ in \eqref{katumara}. This is a formidable exercise which will take us beyond the scope of this paper. We will then suffice ourselves by making the following observations.

\vskip.1in

\noindent ${\bf 1.}$ For $p$ powers of the field with $n$ derivatives, we have seen earlier that $n \ge p$. To avoid generating non-localities, we can choose a scheme to shut off infinite derivatives, or infinite powers of the field. This will suggest that $p$, and in turn $n$, can be very large but not infinite.

\vskip.1in

\noindent ${\bf 2.}$ Each of the $p$ powers would generate a Borel box using the analysis that we presented earlier. Including $(q, r, s)$ will increase the number of Borel boxes considerably.
Thus we will have a large, but not infinite, collection of Borel boxes contributing to the path integral analysis for the expectation value $\langle\varphi_1\rangle_\sigma$. 

\vskip.1in

\noindent ${\bf 3.}$ The expression for the four-dimensional cosmological constant $\Lambda$ from all the Borel boxes will take a form like \eqref{palazzome}, except that the denominator will be much bigger than the numerator $-$ but again not infinite. This will effectively lower the cosmological constant quite a bit from what we have in \eqref{palazzome}.

\vskip.1in

\noindent Our above considerations provide a strong reason why the four-dimensional cosmological constant $\Lambda$ can be so small, despite using only scalar degrees of freedom to mimic gravitational and other on-shell degrees of freedom. However the actual demonstration of the aforementioned three points is a pretty non-trivial exercise, which we shall present in our upcoming work \cite{ccpaper}. 


\subsection{What if the four-dimensional cosmological constant slowly varies with time? \label{sec5.3}}

Our analysis in the precious section shows the possibility that the four-dimensional cosmological constant could in principle be made small. One could also entertain the following possibility: what if the small and positive cosmological constant also varies slowly with respect to time? This is a curious scenario \cite{desibao}, but before we start we should point out that even with this set-up we could not realize such a scenario as a vacuum solution. The reasoning remains the same as before related to the problems with Wilsonian integration, quantization of strings over a temporally varying background and other similar issues. In fact the situation may be more severe now because of temporal variation of the cosmological constant $\Lambda$, albeit very slowly. In literature, these kind of cosmologies are termed as the quintessence cosmologies and their realization in string theory so far has not been very successful. In the following we will argue that a class of such models may be realized from our Glauber-Sudarshan state construction. 

We will start by first specifying the emergent metric components from the eleven-dimensional set-up\footnote{It might be useful for readers, who are not acquainted with the Schwinger-Dyson equations or their applications to the Glauber-Sudarshan states, to skip this section on their first reading and come back only after learning the derivation in \eqref{tranishonal3} from an effective action \eqref{kaittami4}.} with the topology of the internal eight-manifold to be ${\cal M}_4 \times {\cal M}_2 \times {\mathbb{T}^2\over {\cal G}}$ such that $y^m \in {\cal M}_4$ and $y^\rho \in {\cal M}_2$. These metric components, as defined earlier in 
say \eqref{russmameye}, as well as the G-fluxes\footnote{We are ignoring the fermionic terms to keep the analysis simple. A more detailed study will be presented in \cite{hete8} including their contributions to the Schwinger-Dyson equations.} are expressed as expectation values over the Glauber-Sudarshan state and for the present case we will specify them in the following way:
\bg\label{ripley}
\begin{split}
& \langle {\bf g}_{11, 11}\rangle_\sigma \equiv {\bf g}_{11, 11}({\bf x}, y; g_s) = {\bf \widetilde{g}}_{11, 11}({\bf x}, y) \left({g_s\over {\rm H}(y){\rm H}_o({\bf x})}\right)^{4\over 3}\\
& \langle {\bf g}_{33}\rangle_\sigma \equiv {\bf g}_{33}({\bf x}, y; g_s) = \sum_{k = 0}^\infty \widetilde{h}_k \delta_{33} \left({g_s\over {\rm H}(y){\rm H}_o({\bf x})}\right)^{{4\over 3} + {2k \vert\widetilde{f}_k(t)\vert\over 3}}\\
&\langle {\bf g}_{\mu\nu}\rangle_\sigma \equiv {\bf g}_{\mu\nu}({\bf x}, y; g_s) = \sum_{k = 0}^\infty h_k \eta_{\mu\nu}\left({g_s\over {\rm H}(y){\rm H}_o({\bf x})}\right)^{-{8\over 3} + {2k\vert f_k(t)\vert\over 3}}\\
& \langle{\bf g}_{\kappa\rho}\rangle_\sigma \equiv {\bf g}_{\kappa\rho}({\bf x}, y; g_s) = 
\sum_{k = 0}^\infty {\bf g}^{(k)}_{\kappa\rho}({\bf x}, y) \left({g_s\over {\rm H}(y){\rm H}_o({\bf x})}\right)^{-{2\over 3} + \beta(t) + {2k\vert\beta_k(t) \vert\over 3}}\\ 
& \langle{\bf g}_{mn}\rangle_\sigma \equiv {\bf g}_{mn}({\bf x}, y; g_s) = 
\sum_{k = 0}^\infty {\bf g}^{(k)}_{mn}({\bf x}, y) \left({g_s\over {\rm H}(y){\rm H}_o({\bf x})}\right)^{-{2\over 3} + \alpha(t) + {2k\vert\alpha_k(t)\vert\over 3}}\\ 
& \langle {\bf G}_{\rm ABCD}\rangle_\sigma \equiv {\bf G}_{\rm ABCD}({\bf x}, y; g_s) = \sum_{k = 0}^\infty {\bf G}^{(k)}_{{\rm ABCD}}({\bf x}, y) \left({g_s\over {\rm H}(y){\rm H}_o({\bf x})}\right)^{l_{\rm AB}^{\rm CD} + {2k\vert g_k(t)\vert\over 3}}\\ 
\end{split}
 \nd
where $({\rm A, B}) \in {\bf R}^{2, 1} \times {\cal M}_4 \times {\cal M}_2 \times {\mathbb{T}^2\over {\cal G}}$, $l_{\rm AB}^{\rm CD}$ is the dominant scaling for the emergent flux components; and 
$(\widetilde{f}_k(t), f_k(t), \alpha(t), \alpha_k(t), \beta(t), \beta_k(t), g_k(t))$ are the set of functions that are arranged so that they would solve the Schwinger-Dyson equations \eqref{tranishonal3} as well as equations coming from duality constraints.
The latter would typically involve dualities from M-theory to type IIB as well as the constraint to keep the four-dimensional Newton's constant in the type IIB side time-independent. Note that $g_s$ continues to be the dual type IIA coupling, as evidenced from the form of $\langle {\bf g}_{11, 11}\rangle_\sigma$, but its connection to the conformal temporal coordinate $t$ needs to be determined. We will derive this soon, but before we do so we should point out that, to maintain the isometry in the type IIB side, $\widetilde{f}_k(t)$ and $f_k(t)$ cannot be independent of each other. In fact they are related by the following polynomial equation: 
\bg\label{marge}
\sum_{k, l} h_k \widetilde{h}_l \left({g_s\over {\rm H}(y) {\rm H}_o({\bf x})}\right)^{{2\over 3}(k\vert\check{f}_k(t)\vert + l\vert\hat{f}_l(t)\vert)} = 1, \nd
where $\check{f}_k(t) = f_k(t) + {\cal O}({\rm M}_p, g_s)$ and 
$\hat{f}_l(t) = \widetilde{f}_l(t) + {\cal O}({\rm M}_p, g_s)$ with the corrections coming from the duality rules. Note the presence of modulus signs on $\hat{f}_l(t)$ in \eqref{marge} and $\widetilde{f}_k(t)$ in \eqref{ripley}. This ensures the sub-dominant behavior and subsequent duality to type IIB which are important to solve \eqref{marge} in the presence of the ${\cal O}({\rm M}_p, g_s)$ corrections. (Detail of these corrections are not necessary, so we will not elaborate anymore here and leave them for \cite{hete8}.)
There is also another subtlety now: the type IIB coupling is no longer at the constant coupling point of F-theory. Due to choice of \eqref{ripley}, and consequently \eqref{marge}, the IIB coupling $g_s^{(b)}$ takes the form:
\bg\label{sherwood}
\left(g_s^{(b)}\right)^2 = \sum_{k = 0}^\infty h_k \left({g_s\over {\rm H}(y) {\rm H}_o({\bf x})}\right)^{2k\vert\check{f}_k(t)\vert\over 3}, \nd
with $h_0 = 1$, and is therefore slowly varying with time. Depending on the signs of $h_k$, the coupling could be vary with time slightly above or below the constant coupling point of F-theory. This temporal variation, albeit small, is precisely due to the temporal variation of the M-theory torus as one of the direction of the torus determines the dual IIA coupling and the other direction fixes the dual IIB coupling. Once we know the IIB string coupling, and after combining \eqref{ripley}, \eqref{marge} and \eqref{sherwood} together, the emergent metric $\langle {\bf g}^{(b)}_{\rm AB}\rangle_\omega$, with $\vert\omega\rangle$ being the Glauber-Sudarshan state in the IIB side, now takes the following form:

{\footnotesize
 \bg\label{engrose}
 \begin{split}
 ds^2 & \equiv  \langle {\bf g}^{(b)}_{\rm AB}\rangle_\omega d{\rm Y}^{\rm A} d{\rm Y}^{\rm B} = 
 \sum_{k=0}^\infty h_k \left({g_s\over {\rm H}(y){\rm H}_o({\bf x})}\right)^{-2 + {2k\vert {\check{f}_k(t)}\vert\over 3}}\left(-dt^2 + \sum_{i = 1}^3 dx_i^2\right) \\
 &+  \sum_{k=0}^\infty {\bf g}^{(k)}_{mn}({\bf x}, y) 
\left({g_s\over {\rm H}(y){\rm H}_o({\bf x})}\right)^{ \check{\alpha}(t) + {2k\vert{\check{\alpha}_k(t)}\vert\over 3}} + \sum_{k=0}^\infty {\bf g}^{(k)}_{\rho\sigma}({\bf x}, y) 
\left({g_s\over {\rm H}(y){\rm H}_o({\bf x})}\right)^{\check{\beta}(t) + {2k\vert{\check{\beta}_k(t)}\vert\over 3}},\\
\end{split}
\nd}
which is clearly different from what we had in \eqref{metansatze}. The split of the base ${\cal M}_6$ to ${\cal M}_4 \times {\cal M}_2$ is useful because we can easily make the volume of the internal manifold, and consequently the emergent four-dimensional Newton's constant, time-independent by relating $\check{\alpha}(t)$ with $\check{\beta}(t)$ and also with $\check{\alpha}_k(t)$ and $\check{\beta}_k(t)$, although this may be relaxed now. (In any case any temporal dependence will be sub-dominant as we shall see below.) A special case is when $\check{\alpha}(t) = \check{\beta}(t) = 0$ with a certain relation between $\check{\alpha}_k(t)$ and $\check{\beta}_k(t)$ to keep the volume time-independent. 

The above emergent metric configuration \eqref{engrose} in type IIB would make sense if the original metric components from \eqref{ripley} {\it solve} the Schwinger-Dyson equations \eqref{tranishonal3}. This is a non-trivial exercise and in section \ref{sec7.3} we will argue that this is indeed possible. Assuming this to be the case, the pertinent question is the following: how does our little exercise above show that the four-dimensional cosmological constant is slowly varying with time? There are in fact two ways to show this. The first one is relatively straightforward and may be easily ascertained by looking at the form of the IIB metric \eqref{engrose}. We demand:
\bg\label{dakuta}
\sum_{k=0}^\infty h_k \left({g_s\over {\rm H}(y){\rm H}_o({\bf x})}\right)^{-2 + {2k\vert {\check{f}_k(t)}\vert\over 3}} = {1\over \Lambda(t)\vert t\vert^2} = {\left(1 + {\check{\Lambda}(t)\over \Lambda}\right)^{-1}\over \Lambda \vert t\vert^2}, \nd
where $\Lambda(t) \equiv \Lambda + \check{\Lambda}(t)$, with $\check{\Lambda}(t)$ is a very slowly varying function. (In fact because of this very slow variation we could express the RHS of \eqref{dakuta} in the way shown there as an almost flat-slicing but with varying $\Lambda(t)$.) It should be clear from \eqref{dakuta} that, for $\check{f}_k(t) = 0$, the correction $\check{\Lambda}(t) = 0$ and $\left({g_s\over {\rm H}(y){\rm H}_o({\bf x})}\right) = \sqrt{\Lambda} \vert t\vert$. Switching on $\check{f}_k(t)$ is equivalent to switching on $\check{\Lambda}(t)$, and since $\left({g_s\over {\rm H}(y){\rm H}_o({\bf x})}\right) < 1$, this would make the {\it effective} cosmological constant to monotonically decrease with time. An interesting solution to \eqref{dakuta} is to take $\left({g_s\over {\rm H}(y){\rm H}_o({\bf x})}\right) = \sqrt{\Lambda} \vert t\vert$ and to solve the following equation:

{\footnotesize
\bg\label{chatme}
\begin{split}
\sum_{k = 0}^\infty h_k \left(\sqrt{\Lambda}\vert t \vert\right)^{2k\vert \check{f}_k(t)\vert\over 3}  & = 1 + h_1 \left(\sqrt{\Lambda}\vert t \vert\right)^{2\vert \check{f}_1(t)\vert\over 3} + h_2 \left(\sqrt{\Lambda}\vert t \vert\right)^{4\vert \check{f}_2(t)\vert\over 3} + ... =\left(1 + {\check{\Lambda}(t)\over \Lambda}\right)^{-1}\\
& = 1 - {\check{\Lambda}(t)\over \Lambda} + \left( {\check{\Lambda}(t)\over \Lambda}\right)^2 - \left( {\check{\Lambda}(t)\over \Lambda}\right)^3 + 
\left( {\check{\Lambda}(t)\over \Lambda}\right)^4 + ... = \sum_{n = 0}^\infty (-1)^n\left( {\check{\Lambda}(t)\over \Lambda}\right)^n
\end{split}
\nd}
where the Taylor expansion in the second line is only justified if 
$\check{\Lambda}(t) << \Lambda$ within the temporal domain \eqref{tcc}.  This is also the case where $\left({g_s\over {\rm H}(y){\rm H}_o({\bf x})}\right) < 1$ with the aforementioned choice for $g_s$. The equation \eqref{chatme} is easily solved once we choose $h_k = (-1)^k$ which fixes the following functional form for $\check{f}_k(t)$ in terms of $\check{\Lambda}(t)$:
\bg\label{elzbet}
\vert \check{f}_k(t)\vert = \vert \check{f}_1(t)\vert = {3\over 2} \left[{{\rm log}\left({\check{\Lambda}(t)\over \Lambda}\right) \over {\rm log}\left(\sqrt{\Lambda}\vert t\vert\right)}\right], ~~~ \forall k \ge 1. \nd
Since the conformal time is bounded by \eqref{tcc} and $\check{\Lambda}(t) << \Lambda$ within the bound \eqref{tcc}, the RHS is a positive definite quantity consistent with the modulus of $\check{f}_k(t)$. In retrospect, the series $\left({g_s\over {\rm H}(y){\rm H}_o({\bf x})}\right)^{2\vert\check{f}_1(t)\vert\over 3}$ got by imposing \eqref{elzbet} shouldn't appear that surprising in the light of similar series, but with $\check{f}_1(t) = 1$, studied in \cite{desitter2, coherbeta, coherbeta2} for the internal manifold. The latter was shown to be consistent with the Schwinger-Dyson equations so the present case, with slight modification from the introduction of $\check{f}_1(t)$  along the space-time directions, should not pose any insurmountable difficulties. A possible solution, directly from M-theory, would be to define:
\bg\label{natasbhalo}
f_k(t) = \widetilde{f}_k(t) = \alpha_k(t) = \beta_k(t) = g_k = f_1(t), ~~ 
\alpha(t) = \beta(t) = 0, \nd
in \eqref{ripley}. Such a choice is clearly consistent with \eqref{elzbet} and, since $\vert \check{f}_1(t) - f_1(t)\vert = {\cal O}(g_s, {\rm M}_p)$, we expect $\check{f}_1(0) = f_1(0)$ at late time as $t \to 0$ (or $g_s \to 0$). Comparing it with \eqref{elzbet}, $\hat{\Lambda}(t)$ also needs to decrease monotonically for this limit to make sense so that the effective cosmological constant varies from $\Lambda + \check{\Lambda}\left(-{1\over \sqrt{\Lambda}}\right)$ to $\Lambda$ in the temporal domain \eqref{tcc}. This is clearly consistent with what we predicted earlier, so now we need to figure out the limiting value on the RHS of \eqref{elzbet} at late time. This may be determined in the following way. At late time if the parameters in \eqref{elzbet} approach zero as $\sqrt{\Lambda} \vert t\vert \to \epsilon$ and $\hat{\Lambda}(t) \to \Lambda \epsilon^a$ for $\epsilon \to 0^+$ and $a > 0$, then $\check{f}_1(0) \to {3a\over 2}$ implying $f_1(0) = {3a\over 2}$. Additionally, 
to see whether \eqref{natasbhalo} solves the Schwinger-Dyson equations, we can first define an effective coupling 
$ \check{\bf g}_s \equiv  \left({g_s\over {\rm H}(y){\rm H}_o({\bf x})}\right)^{\vert f_1(t)\vert}$. Plugging this in \eqref{kaittami4}, one may check that every term of the equation there scales with respect to $g_s$ as a dominant and a sub-dominant piece in the following way:
\bg\label{matpluslip}
\sum_{k = 0}^\infty {\bf a}_k({\bf x}, y) \left({g_s\over {\rm H}(y){\rm H}_o({\bf x})}\right)^{\rm dominant} \times \check{\bf g}^{{2k\over 3}~ +~ {\rm log~corrections}}, \nd
with a functional form ${\bf a}_k({\bf x}, y)$ that may be easily specified for each term. There are also log corrections that are a bit non-trivial to derive but are nevertheless tractable. These corrections tend to be very small, so we won't need them here, and the full derivation of them will be presented in \cite{hete8}. Therefore ignoring these corrections and equating the dominant scalings over all the terms 
in \eqref{kaittami4} would specify the dominant quantum corrections needed to solve the Schwinger-Dyson equation \eqref{kaittami4}\footnote{There is a bit of a subtlety here because the dominant scalings may themselves be functions of $t$ $-$ and consequently of $g_s$ $-$ so this matching needs to be done carefully. We will discuss this briefly in section \ref{sec7.3.3} and leave a more detailed study for \cite{hete8}.}. After which one can equate the subdominant scalings over all the terms to solve the system to all orders in $\check{\bf g}_s$. The strategy to solve the EOM with the log corrections will be discussed in \cite{hete8}.
In fact in sections \ref{sec7.3.1} and \ref{sec7.3.2} we will take a slightly simplified version with $f_k(t) = \widetilde{f}_k(t) = \alpha(t) = \beta(t) = 0$ and $\alpha_k(t) = \beta_k(t) = g_k(t) = 1$ to show the consistency of the aforementioned construction.

One of the advantage of the solution \eqref{elzbet} is that the both the definition of the dual IIA coupling $g_s$ and the temporal domain of validity \eqref{tcc} do not change. One could envision a different solution of $g_s$ in \eqref{dakuta} that would depend on the conformal time $t$, as well as the functional forms for $\check{\Lambda}(t)$ and 
$\check{f}_k(t)$. One example would be to keep $f_1(t) = 1$ in \eqref{natasbhalo} where we cannot always demand $\left({g_s\over {\rm H}(y){\rm H}_o({\bf x})}\right) = \sqrt{\Lambda}\vert t\vert$ to solve 
\eqref{dakuta}.
Needless to say, this is a slightly difficult exercise and is not clear whether a closed-form expression for $g_s$ exists. Such a solution will also change the temporal domain \eqref{tcc}, albeit to sub-leading order, but we will not address this any further here. 

The second way to justify the slow variation of the four-dimensional cosmological constant is to go back to the derivation of the expectation value using the Borel-$\grave{\rm E}$calle resummation as shown for the scalar field case in the second relation of \eqref{recelcards}. For our metric configuration this will take the following form:
\bg\label{carolgenius}
\begin{split}
\langle {\bf g}_{\mu\nu}\rangle_{\sigma} & = {1\over g^{1/\bar\alpha}}\left[\int_0^\infty d{\rm S} ~{\rm exp}\left(-{{\rm S}\over g^{1/\bar\alpha}}\right) {1\over 1 - \check{f}_{\rm max} {\rm S}^{\bar\alpha}}\right]_{\rm P.V} \int_{k_{\rm IR}}^\mu d^{11}k ~{\alpha(k) \eta_{\mu\nu}\over a(k)}~\psi_k(x, y)\\
& =  \sum_{k = 0}^\infty h_k \eta_{\mu\nu}\left({g_s\over {\rm H}(y){\rm H}_o({\bf x})}\right)^{-{8\over 3} + {2k\vert f_k(t)\vert\over 3}} = 
\sum_{k = 0}^\infty h_k \eta_{\mu\nu} \left(\sqrt{\Lambda}t\right)^{-{8\over 3} + {2k\vert\check{f}_1(t)\vert\over 3}},\\
\end{split}
\nd
where all the parameters in the first line above have already been defined in section \ref{patho}, and in the second line we have used \eqref{elzbet}. It is not too hard to see that now $\Lambda$ may continue to appear from the Borel resummed integral as in \eqref{tinmey}, or the modified one with the introduction of the Borel boxes from \eqref{palazzome}, and the Glauber-Sudarshan state may be determined from the following Fourier integral:
\bg\label{verotak}
\int_{k_{\rm IR}}^\mu d^{11}k ~{\alpha(k) \eta_{\mu\nu}\over a(k)}~\psi_k(x, y) = {\bf FT} \left[{\left(1 + {\check{\Lambda}(t)\over \Lambda}\right)^{-1} \over ({\rm M}_p \vert t\vert)^{8\over 3}}\right]_{\rm Riemann-Lebesgue}, \nd
where on the RHS we first take the Fourier transform, and then rewrite this using the Riemann-Lebesgue lemma to express it within the energy domain $k_{\rm IR} < k < \mu$. On the LHS, we can express $\alpha(k)$ as a series in Hermite polynomials as done in \cite{borel}. One may then impose a delta function constraint via $\alpha(k)$ or via the wave-function $\psi_k(x, y)$ ($y \in {\cal M}_4 \times {\cal M}_2$), to restrict the integral to be a function of $t$ only thus matching with the series on the RHS. More details on this will be presented elsewhere.

Therefore to summarize, it appears that we can easily accommodate small temporal variation of the cosmological constant in our model by slightly changing the Glauber-Sudarshan state as per the prescription \eqref{dakuta}, \eqref{elzbet} and \eqref{verotak}\footnote{Despite having a possible way to explain the slow temporal variation of the cosmological constant in our set-up, we should point out that this is {\it not} a model of quintessence as we are reinterpreting the one-point function from the Glauber-Sudarshan state in a slightly different way. As such the issues pointed out for a model of quintessence do not plague us. Of course the success of this scenario will depend on the success of the DESI-BAO result \cite{desibao}. More details on this will be presented in \cite{hete8}.}. A discussion of how 
a simpler version of \eqref{ripley} solves the Schwinger-Dyson equation \eqref{tranishonal3} will be presented in section \ref{sec7.3}.

\section{Schwinger-Dyson's equations, EOMs and the excited states \label{sec4}}

The readers who have skipped section \ref{smallcc} would be pleased to know that one can define a small and positive four-dimensional cosmological constant using the Glauber-Sudarshan states. However two issues prohibit us to declare this as a solution to the cosmological constant problem.

\vskip.1in

\noindent ${\bf 1.}$ The analysis is presented completely using four set of scalar degrees of freedom instead of the actual gravitational, fluxes and other on-shell degrees of freedom.

\vskip.1in

\noindent ${\bf 2.}$ Even with scalar degrees of freedom, it is not clear whether the procedure to reduce the value of the cosmological constant by introducing multiple Borel boxes would produce a value small enough to actually reproduce the oft-quoted $10^{-120}{\rm M}_p^2$ result.

\vskip.1in

\noindent Despite this our analysis, probably for the first time, has provided a reason for the smallness of the positive cosmological constant that only relies on non-perturbative effects. We will not elaborate further on this interesting direction $-$ which we leave it for our upcoming work \cite{ccpaper} $-$ instead we will answer a pertinent question related to how these Glauber-Sudarshan states overcome the so-called {\it swampland criteria} \cite{swampland}. Before that however we need to know the EOMs governing these states which in turn requires us to determine the resolution of identity that we briefly touched upon in section \ref{sec2.2.3}. These will be the contents of the following section.

\subsection{Resolution of identity and EOMs for the Glauber-Sudarshan states \label{sec4.1}}

To study the EOM we will first go back to the form of the Glauber-Sudarshan state $ \vert\sigma\rangle$ that we discussed in \eqref{saw1}. 
These states are not normalized, so $\langle\sigma\vert\sigma\rangle \ne 1$.  We can express this as the following path-integral:

{\scriptsize
\bg\label{straokua}
\begin{split}
{\langle \sigma\vert \sigma\rangle\over \langle \Omega\vert \Omega\rangle} &= {\langle\Omega\vert\mathbb{D}^\dagger\mathbb{D}\vert\Omega\rangle\over \langle\Omega\vert\Omega\rangle} = \int\prod_{i = 1}^4 [{\cal D}\varphi_i]
~{\rm exp}\left(-{\bf S}_{\rm tot}(\varphi_1,..,\varphi_4)\right) \mathbb{D}^\dagger(\sigma, \varphi_1,..,\varphi_4)\mathbb{D}(\sigma,\varphi_1,..,\varphi_4)\\
& = \int\prod_{i = 1}^4 [{\cal D}\varphi_i]
~{\rm exp}\left(-{\bf S}_{\rm tot}(\varphi_1+ \epsilon(x, y), \varphi_2,..,\varphi_4)\right) \mathbb{D}^\dagger(\sigma, \varphi_1+\epsilon(x, y), \varphi_2,..,\varphi_4)\mathbb{D}(\sigma,\varphi_1 + \epsilon(x, y), \varphi_2,..,\varphi_4)\\
& = \int\prod_{i = 1}^4 [{\cal D}\varphi_i]
~{\rm exp}\left(-{\bf S}_{\rm tot}(\varphi_1,.., \varphi_4)- \epsilon(x, y) {\delta{\bf S}_{\rm tot}\over \delta\varphi_1}\right) \mathbb{D}^\dagger(\sigma, \varphi_1,..,\varphi_4)\mathbb{D}(\sigma,\varphi_1,..,\varphi_4)\left(1 + \epsilon(x, y) {\delta\over \delta\varphi_1}\log\left(\mathbb{D}^\dagger \mathbb{D}\right)\right)\\
\end{split}
\nd}
where in the second line we have taken a dummy variable $\varphi'_1(x, y) \equiv \varphi_1(x, y) + \epsilon(x, y)$, and in the third line we have Taylor expanded to first order in $\epsilon(x, y)$, where $\epsilon(x, y)$ is a function satisfying 
$0 <\epsilon(x, y) << 1$ everywhere in eleven-dimensional space-time. It is also easy to see that ${\cal D}\varphi'_1 = {\cal D}\varphi_1$. Now comparing the first line with the third line, we easily get the following equation of motion for the on-shell field $\varphi_1$:

{\footnotesize
\bg\label{harryreidtag}
\epsilon(x, y)\int\prod_{i = 1}^4 [{\cal D}\varphi_i]
~{\rm exp}\left(-{\bf S}_{\rm tot}(\varphi_1,..,\varphi_4)\right) \mathbb{D}^\dagger(\sigma, \varphi_1,..,\varphi_4)\mathbb{D}(\sigma,\varphi_1,\varphi_4)\left({\delta{\bf S}_{\rm tot}\over \delta\varphi_1} -
{\delta\over \delta\varphi_1}\log\left(\mathbb{D}^\dagger \mathbb{D}\right)\right) = 0, \nd}
which would imply the vanishing of the integral because $\epsilon(x, y)$ is a non-zero function everywhere. Note the relative minus sign between the two terms above coming entirely from the fact that we took an Euclidean action. Had we taken a Lorentzian action, there would have been a relative $\sqrt{-1}$ factor. We will avoid factors of $\sqrt{-1}$ in our analysis. The vanishing of \eqref{harryreidtag} leads to the following EOM:
\bg\label{elenacigu}
\left\langle{\delta{\bf S}_{\rm tot}\over \delta\varphi_1}\right\rangle_\sigma = \left\langle
{\delta\over \delta\varphi_1}\log\left(\mathbb{D}^\dagger \mathbb{D}\right)\right\rangle_\sigma, \nd
which is precisely the EOM satisfied by the Glauber-Sudarshan state. This is a Schwinger-Dyson (SD) type EOM, but differs from the usual SD equation in at least two respects.

\vskip.1in

\noindent ${\bf 1.}$ It is defined over the Glauber-Sudarshan state $\vert\sigma\rangle$ instead over the interacting vacuum $\vert\Omega\rangle$.

\vskip.1in

\noindent ${\bf 2.}$ It differs from the expected EOM of the form 
$\left\langle{\delta{\bf S}_{\rm tot}\over \delta\varphi_1}\right\rangle_\sigma = 0$, because of the presence of a non-unitary\footnote{Had we chosen a unitary $\mathbb{D}(\sigma)$, then no EOM of the form \eqref{elenacigu} would have appeared because $\langle\sigma\vert\sigma\rangle = \langle\Omega\vert\Omega\rangle = 1$. This is another crucial way the Glauber-Sudarshan states differ from the coherent states.} $\mathbb{D}(\sigma, \varphi_1,..,\varphi_4)$. 

\vskip.1in

\noindent The above EOM is important to solve first to get the form of $\rho^\ast(k)$ in \eqref{traderjoeme}, which will eventually fix the functional form for $\alpha_1(k)$ in \eqref{robinright}. Unfortunately the form of the EOM in \eqref{elenacigu} is pretty unwieldy to yield the value \eqref{traderjoeme} for $\rho^\ast(k)$ so easily. Part of the difficulty of course lies in the form of the EOM itself, but more importantly, since:
\bg\label{elrancho}
\langle\varphi_i^n\rangle_\sigma \ne \langle\varphi_i\rangle^n_\sigma, ~~~~
\langle\partial^m\varphi_i^n\rangle_\sigma \ne \partial^m\langle\varphi_i\rangle^n_\sigma,\nd
the expectation values for $\varphi_i^n$, {\it i.e.} $\langle\varphi_i^n\rangle_\sigma$, or the ones with derivatives, {\it i.e.} $\langle\partial^m\varphi_i^n\rangle_\sigma$, need to be computed for all values of $(n, m)$ in \eqref{elrancho}. Recalling the fact that even for a simple computation like $\langle\varphi_1\rangle_\sigma$ involves Borel resummations of Gevrey series as well as summations over Borel boxes, going to $n > 1$ in \eqref{elrancho} will be a formidable exercise. Is there a way we can simplify \eqref{elenacigu} to efficiently solve the EOMs and get the desired functional form for $\rho^\ast{k}$ as in \eqref{traderjoeme}? The answer is {\it yes} provided we can have a well defined notion of the {\it resolution of identity using the Glauber-Sudarshan states}. How does that help us?

Before going into the benefits of {\it resolution of identity}, let us discuss whether it is even possible to have such resolution with Glauber-Sudarshan states. We already discussed one possible way in \eqref{metmedicmey} using the displacement operator \eqref{28rms}. The suppression factor in the path-integral using \eqref{metmedicmey} is given by \eqref{wombet}. However the aforementioned analysis only works well when ${\rm Im}~\sigma > {\rm Im}~\sigma'$ in $\langle\sigma\vert\sigma'\rangle$. This is an unsatisfactory feature, although for many practical purposes the choice of ${\cal O}$ in \eqref{metmedicmey} suffices. Clearly for generic application, a better formalism for the resolution of identity is called for. How can we derive this? 

One possibility would be to use the complete set of states for the full interacting Hamiltonian ${\bf H}(t) \equiv \int d^2{\bf x}~d^6y~d^2w ~{\bf H}_{\rm tot}(x,y, w)$, where ${\bf H}_{\rm tot}$ may be determined from the {\it full non-perturbative completion of} ${\bf S}_{\rm tot}$ (see section \ref{sec4.2})\footnote{This is also the Hamiltonian that entered in \eqref{saw1}.}. We will make two assumptions to simplify the ensuing analysis: \textcolor{blue}{one}, the interacting vacuum $\vert\Omega\rangle$ will be assumed to be non-degenerate, and \textcolor{blue}{two}, there is an energy gap between the first excited state and the vacuum. Keeping these in mind we express the completeness condition as:
\bg\label{sfotegre}
{\bf 1} ~ = ~ \vert\Omega\rangle \langle\Omega\vert + \sum_{n = 1}^\infty c_n\vert n\rangle\langle n \vert, \nd
where $c_n$ is the degeneracy of the higher excited states $\vert n \rangle$. $\vert\Omega\rangle$ is the interacting vacuum that we have been using so far $-$ which is an eigenstate of ${\bf H}$ $-$ and whose properties may be easily discerned from the non-perturbative (and non-local) completion of ${\bf S}_{\rm tot}$ in the path-integral formalism. However the properties of $\vert n\rangle$ are much harder because they are no longer harmonic oscillator states. Is there a way to avoid this subtlety and get the completeness relation for the Glauber-Sudarshan states? The answer is surprisingly {\it yes} with the aid of a few notations. We define an {\it ordered} operator $\Sigma_{{\rm L}{\rm R}}$ whose action on any operator ${\cal O}$ is:

{\footnotesize
\bg\label{enmillheno}
\Sigma_{{\rm L}{\rm R}} {\cal O} \equiv 
\Bigg[\mathbb{D}(\sigma;\varphi_i) ~{\rm exp}\Big(-i {\bf H} {\rm T}\Big)\Big]_{\overrightarrow{\rm L}} \otimes\Big[\mathbb{D}^\dagger(\sigma; \varphi_i)\Big]_{\overleftarrow{\rm R}}~ {\cal O} = \mathbb{D}(\sigma; \varphi_i) ~{\rm exp}\Big(-i {\bf H} {\rm T}\Big)~ {\cal O}~ \mathbb{D}^\dagger(\sigma; \varphi_i), \nd}
where the first equality provides the form of the operator. The ordering of the operator action means the following: the subscript $\overrightarrow{\rm L}$ denotes action from the left and the subscript $\overleftarrow{\rm R}$ denotes action from the right on the operator ${\cal O}$ (the arrows signify the directions of the operators actions). The action of $\Sigma_{\rm LR}$ on \eqref{sfotegre} then provides the following relation:

{\scriptsize
\bg\label{charhart}
\sum_{\sigma, \sigma^\ast}~ {\Bigg[\mathbb{D}(\sigma;\varphi_i) ~{\rm exp}\Big(-i ({\bf H} - {\rm E}_0) {\rm T}\Big)\Big]_{\overrightarrow{\rm L}} \otimes\Big[\mathbb{D}^\dagger(\sigma;\varphi_i)\Big]_{\overleftarrow{\rm R}} \over \langle\sigma\vert\sigma\rangle} ~=~ \sum_{\sigma, \sigma^\ast} {\vert\sigma\rangle \langle\sigma\vert\over \langle\sigma\vert\sigma\rangle} ~+~ \sum_{\sigma,\sigma^\ast}\sum_{n = 1}^\infty c_n~{\rm exp}\Big(-i({\rm E}_n - {\rm E}_0){\rm T}\Big)~ {\mathbb{D}(\sigma; \varphi_i)\vert n\rangle \langle n\vert \mathbb{D}^\dagger(\sigma; \varphi_i) \over \langle\sigma\vert\sigma\rangle}, \nd}
where ${\bf H}$ measured over the Minkowski minimum is taken to be independent of time, and ${\rm E}_0$ and ${\rm E}_n$ denote the energies of the vacuum $\vert\Omega\rangle$ and the higher excited states $\vert n \rangle$ respectively. 
It should also be clear that ${\rm E}_n > {\rm E}_0$, $\forall n$, which means $\vert \Omega \rangle$ to be the lowest energy state ({\it i.e.} the vacuum). This also means that if we take ${\rm T}$ to infinity along slightly imaginary direction, we can decouple all the higher excited states. This would immediately imply:
\bg\label{hubeimach}
\sum_{\sigma, \sigma^\ast} {\vert\sigma\rangle \langle\sigma\vert\over \langle\sigma\vert\sigma\rangle} ~= ~ \lim_{{\rm T} \to \infty(1-i\epsilon)}\sum_{\sigma, \sigma^\ast}~ {\left[\mathbb{D}(\sigma;\varphi_i) ~{\rm exp}\Big(-i ({\bf H} - {\rm E}_0) {\rm T}\Big)\right]_{\overrightarrow{\rm L}} \otimes\Big[\mathbb{D}^\dagger(\sigma;\varphi_i)\Big]_{\overleftarrow{\rm R}} \over \langle\sigma\vert\sigma\rangle}, \nd
with $\vert \sigma\rangle \langle\sigma\vert \equiv\underset{k}{\bigotimes}\vert\sigma(k)\rangle \langle \sigma(k)\vert$, where the operation $\underset{k}{\bigotimes}$ collects all the kets with different momenta on one side and the bras, again with different momenta, on the other side without contractions; and $\langle\sigma\vert\sigma\rangle = \prod\limits_k \langle\sigma(k)\vert \sigma(k)\rangle$, where $\prod\limits_k$ is the usual product over the momenta. The relation \eqref{hubeimach} produces something akin to a resolution of identity, but not quite. One immediate issue is the choice of the {\it ordered} operator \eqref{enmillheno}. This is a bit unsatisfactory, so the question is whether the ordering is important or not. To see this, let us act the {\it unordered} operator on \eqref{sfotegre}. The result we get is:

{\scriptsize
\bg\label{labusx}
\begin{split}
\mathbb{D}(\sigma;\varphi_i) ~{\rm exp}\Big(-i {\bf H} {\rm T}\Big) \mathbb{D}^\dagger(\sigma;\varphi_i) &= ~\mathbb{D}(\sigma; \varphi_i) ~{\rm exp}\Big(-i {\bf H} {\rm T}\Big) \mathbb{D}^\dagger(\sigma; \varphi_i) ~\vert\Omega\rangle \langle\Omega\vert + \mathbb{D}(\sigma; \varphi_i) ~{\rm exp}\Big(-i {\bf H} {\rm T}\Big) \mathbb{D}^\dagger(\sigma; \varphi_i) \sum_{n=1}^\infty c_n \vert n \rangle \langle n\vert \\
 & = ~\mathbb{D}(\sigma; \varphi_i) ~{\rm exp}\Big(-i {\bf H} {\rm T}\Big)\Big(\underbrace{\vert\Omega\rangle \langle\Omega\vert + \sum_{n = 1}^\infty c_n\vert n\rangle\langle n \vert}_{\bf 1}\Big) \mathbb{D}^\dagger(\sigma; \varphi_i) ~\vert\Omega\rangle \langle\Omega\vert\\
 & + ~ \mathbb{D}(\sigma; \varphi_i) ~{\rm exp}\Big(-i {\bf H} {\rm T}\Big) \Big(\underbrace{\vert\Omega\rangle \langle\Omega\vert + \sum_{n = 1}^\infty c_n\vert n\rangle\langle n \vert}_{\bf 1}\Big) \mathbb{D}^\dagger(\sigma; \varphi_i) \sum_n c_n \vert n \rangle \langle n\vert \\
 & =~ {\rm exp}\Big(-i{\rm E}_0{\rm T}\Big)~{\vert\sigma\rangle \langle\sigma\vert} ~+~ \sum_{n = 1}^\infty c_n~{\rm exp}\Big(-i{\rm E}_n{\rm T}\Big)~ {\mathbb{D}(\sigma; \varphi_i)\vert n\rangle \langle n\vert \mathbb{D}^\dagger(\sigma; \varphi_i)}\\
\end{split}
\nd}
which is exactly what we had on the RHS of \eqref{charhart}, implying that the ordering of the operator is not necessary, at least for the above case. Summing over $\sigma$ and $\sigma^\ast$, and taking the limit ${\rm T} \to \infty(1 - i\epsilon)$, then leads back to \eqref{hubeimach} but without any ordering. The RHS of \eqref{hubeimach} deviates from the identity because both the operators $\mathbb{D}(\sigma)\mathbb{D}^\dagger(\sigma)$ and ${\rm exp}\Big(-i ({\bf H} - {\rm E}_0) {\rm T}\Big)$ as well as the expectation value $\langle\sigma\vert\sigma\rangle$ deviate from the identity (in fact this is exactly how the Glauber-Sudarshan states differ from the coherent states). This means:

{\scriptsize
\bg\label{hubeimach2}
\sum_{\sigma, \sigma^\ast} {\vert\sigma\rangle \langle\sigma\vert\over \langle\sigma\vert\sigma\rangle} ~= ~ \lim_{{\rm T} \to \infty(1-i\epsilon)}\sum_{\sigma, \sigma^\ast}~ {\mathbb{D}(\sigma; \varphi_1,..,\varphi_4) ~{\rm exp}\Big(-i ({\bf H} - {\rm E}_0) {\rm T}\Big) \mathbb{D}^\dagger(\sigma; \varphi_1,..,\varphi_4) \over \langle\sigma\vert\sigma\rangle} ~ = ~{\bf 1} + {\bf Q}(\varphi_1,..,\varphi_4), \nd}
where ${\bf Q}(\varphi_1,..,\varphi_4)$, which is a function of only the on-shell fields, is another operator that may be derived from expanding the middle operator in \eqref{hubeimach2}. The identity ${\bf 1}$ should be understood as the identity operator in the Hilbert space of the over-complete set of $\vert\sigma\rangle$. This way no extra normalization factor needs to be introduced in \eqref{hubeimach2}. After the dust settles, the resolution of identity 
for the Glauber-Sudarshan states takes the following form:
\bg\label{mariesored}
{\bf 1} =  \sum_{\sigma, \sigma^\ast} {\vert\sigma\rangle \langle\sigma\vert\over \langle\sigma\vert\sigma\rangle} ~ - ~ {\bf Q}(\varphi_1,..,\varphi_4) ~ = ~ \int d^2\sigma~{\vert\sigma\rangle \langle\sigma\vert\over \langle\sigma\vert\sigma\rangle} ~ - ~ {\bf Q}(\varphi_1,..,\varphi_4), \nd
which elucidates the precise difference from the coherent states. For the latter ${\bf Q}(\varphi_1,..,\varphi_4)$ vanishes and $\langle\sigma\vert\sigma\rangle = 1$. In the second equality of \eqref{mariesored} we have converted the sum to an integral over the continuous parameters $\sigma$ and $\sigma^\ast$. (The way we have presented \eqref{mariesored}, the integral or the summation over the Glauber-Sudarshan states appearing therein do not involve any extra function of $\sigma$ and $\sigma^\ast$.)
Looking at \eqref{hubeimach2} however raises an immediate question: in the limit 
${\rm T}\to \infty(1-i\epsilon)$, the higher states $\vert n\rangle$ decouple faster compared to the vacuum state $\vert\Omega\rangle$, so shouldn't this also decouple the Glauber-Sudarshan state $\vert\sigma\rangle$ since it's energy is higher than the vacuum state? The answer may be presented in various ways. If the energy of the Glauber-Sudarshan state is only {\it very slightly above the vacuum state} then these states will not decouple. Such a consideration should be consistent with the fact that we expect the cosmological constant, which is the energy associated with the Glauber-Sudarshan state, to be {\it very} small. Note however that this is not a proof for the smallness of the cosmological constant because the completeness condition \eqref{sfotegre} doesn't exactly involve the $\vert\sigma\rangle$ states directly, so the energy argument is only an indirect one. More appropriately, the Glauber-Sudarshan states are not eigenstates of the total Hamiltomian ${\bf H}_{\rm tot}$, and therefore they do not appear in the completeness condition \eqref{sfotegre}. These states are only consistently described for a very short temporal domain of 
$-{1\over \sqrt{\Lambda}} < t < 0$, and as such are not decoupled by the asymptotic temporal condition of ${\rm T} \to \infty(1 - i\epsilon)$.

With this, the resolution of identity takes a consistent form that doesn't depend upon any constraints on ${\rm Im}~\sigma$. Any constraints that might have resulted earlier from using \eqref{metmedicmey}, are removed by the inclusion of the operator ${\bf Q}(\varphi_1,..,\varphi_4)$. Two questions arise now.

\vskip.1in

\noindent ${\bf 1.}$ Can we insert \eqref{mariesored} inside the expectation-value form of the EOM \eqref{elenacigu} to bring it to a more manageable form? 

\vskip.1in

\noindent ${\bf 2.}$ The expectation-value form \eqref{elenacigu} could in principle give rise to the expectation value of the action itself, {\it i.e.} $\langle {\rm S}_{\rm tot}\rangle_\sigma$. Since the latter involves Borel resummation, is the Wilsonian effective action (WEA) compatible with Borel resummation?

\vskip.1in

\noindent These two questions may seem somewhat unrelated, but we will argue that they are not. This is what we turn to next.

\subsection{Compatibility of the Wilsonian action with Borel resummations and EOMs \label{sec4.20}}

Let us start with a simple toy model to illustrate the underlying story. We will take a single real scalar field $\varphi$, which we shall call the ``on-shell'' degree of freedom, and another scalar field $\phi$, that we shall denote as the ``off-shell'' degree of freedom. (As an example of this, consider the metric \eqref{mmetric}. The on-shell degrees of freedom are the metric components ${\bf g}_{\mu\nu}, {\bf g}_{mn}$ and ${\bf g}_{ab}$, whereas the off-shell degrees of freedom are ${\bf g}_{0m}, {\bf g}_{mb}$ {\it et cetera}.  The former set is denoted by $\varphi$ and the latter set by $\phi$.) We will also not worry about momentum conservation or about other off-shell degrees of freedom. The latter will be very important once we take the actual the metric components, the fluxes and the fermionic degrees of freedom. The action we shall consider will be:

{\footnotesize
\bg\label{enmilltriangle}
{\bf S}_{\rm tot}(\varphi, \phi) \equiv {\bf S}_1(\varphi) + {\bf S}_2(\varphi, \phi) = {{\rm M}_p^9}\int d^{11} x\left(-\varphi \square \varphi - \phi \square \phi + {1\over {\rm M}_p^{2k-2}}\sum_{n, m, k} c_{nmk} \{\partial^{2k}\}\varphi^m \phi^n\right), \nd}
with $c_{200} = c_{020} = 0$ now\footnote{This should be compared to what we have been taking so far, namely $c_{200} \ne 0$ (see discussion after \eqref{mmtarfox}). Now allowing the off-shell degrees of freedom to be massless will have important implications.} and $\{\partial^k\}$ is the distribution of the derivatives among the two scalar fields\footnote{For example 
$c_{nm2}\{\partial^2\}\varphi^m\phi^n \equiv c_{nm20} \partial^2\varphi^m \phi^n + c_{nm02}\varphi^m \partial^2 \phi^n + c_{nm11} \partial_\mu \varphi^m \partial^\mu\phi^n$. We can make further finer distribution with $(n, m)$ but we don't do so here to keep the story simple. \label{cchoice}}. Both the fields $\varphi$ and $\phi$ are dimensionless, and so are the coupling constants $c_{nmk}$. The action is defined at the scale ${\rm M}_p$, and between this and the IR scale $k_{\rm IR}$ we will assume no other degrees of freedom. The raising and lowering of the indices are done using the flat eleven-dimensional metric, which is the Minkowski minimum that we alluded to earlier. Supersymmetry however will not be important for the following discussion so we will ignore it. 

\vskip.3in

\begin{figure}[h]
    \centering
\begin{tikzpicture}
\draw (0,0) circle [radius =3];
\draw[blue] (0,0) --(2.12,2.12); 
\filldraw (0,0) circle (1pt) (2.12,2.12) circle (1pt);
\node (x) at (0,0) [label=left:$x$] {};
\node (x) at (0.7,0.7) [label=right:$\mathbb{F}(x-y')$] {};
\node (x) at (3,0) [label=right:$y'\in\mathcal{M}_6$] {};
\node (x) at (2.12,2.12) [label=right:$ {\bf Q}_{\rm pert} \lc c\,; \varphi \lc x;y'\rc \rc$] {};
\end{tikzpicture}
    \caption{Diagrammatic representation of the non-local terms, with the circle denoting the internal manifold $\mathcal{M}_6$, and $\mathbb{F}(x-y')$ denoting their associated weights in the action. The effect of the quantum terms at a point $x$ may be determined by integrating the perturbative series ${\bf Q}_{\rm pert}(c, \varphi(x, y'))$ over all $y' \in {\cal M}_6$ as in \eqref{fepoladom}.}
    \label{nonpertdiag}
\end{figure}
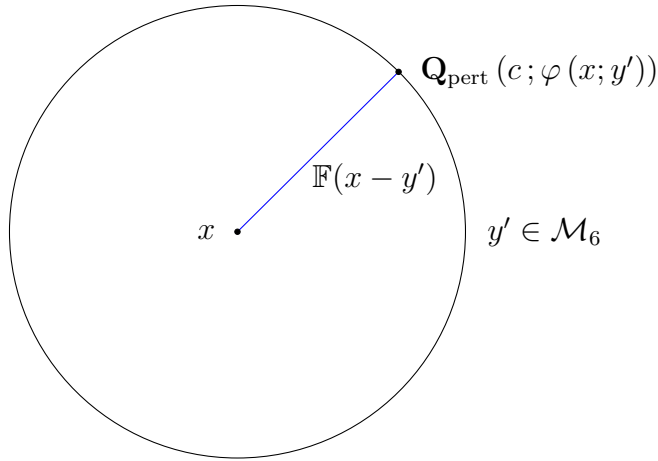

Let us now find out how the off-shell degree of freedom influences the Schwinger-Dyson's equations for both the on-shell and the off-shell degrees of freedom. This may sound a bit puzzling and counter-intuitive path to follow, but we shall see that such a procedure leads to the EOM for the {\it emergent} on-shell degree of freedom $\langle\varphi\rangle_\sigma$ along-with another EOM from varying $\phi$ even though $\langle\phi\rangle_\sigma = 0$. This criterion also justifies what we mean by the emergent on-shell and off-shell degrees of freedom: the on-shell degree of freedom will have a non-zero expectation value over the Glauber-Sudarshan state $\vert\sigma\rangle$, whereas the off-shell degree of freedom will have zero expecation value. The reason for the latter is simple. If we denote the on-shell degrees of freedom using a set $\Xi = \{\varphi\}$, then the displacement operator is exclusively given in terms of $\sigma$ and $\Xi$, {\it i.e.} $\mathbb{D} = \mathbb{D}(\sigma, \Xi)$. This means, even though we allow non-zero $\phi$, the one-point function $\langle\phi\rangle_\sigma = 0$. The EOM for $\varphi$ from \eqref{elenacigu} now takes the following form:
\bg\label{emergencefeb9}
\begin{split}
0 & = \int {\cal D}\varphi ~{\cal D}\phi~e^{-{\bf S}_1(\varphi) - {\bf S}_2(\varphi, \phi)}\left(\square\varphi + f_1(\varphi) - {\delta {\bf S}_2(\varphi, \phi)\over \delta \varphi}\right) \mathbb{D}^\dagger(\sigma, \varphi) \mathbb{D}(\sigma, \varphi)\\
 & + \int {\cal D}\varphi ~{\cal D}\phi~e^{-{\bf S}_1(\varphi) - {\bf S}_2(\varphi, \phi)} ~{\delta\over \delta\varphi}\Big( \mathbb{D}^\dagger(\sigma, \varphi) \mathbb{D}(\sigma, \varphi)\Big)\\ 
 & = \int {\cal D}\varphi ~ e^{-{\bf S}_1(\varphi)}\left(\square\varphi + f_1(\varphi)\right)~  \mathbb{D}^\dagger(\sigma, \varphi) \mathbb{D}(\sigma, \varphi) \int {\cal D}\phi~ e^{- {\bf S}_2(\varphi, \phi)} \\
 & + \int {\cal D}\varphi ~ e^{-{\bf S}_1(\varphi)}\mathbb{D}^\dagger(\sigma, \varphi) \mathbb{D}(\sigma, \varphi)~{\delta\over \delta\varphi} \int {\cal D}\phi~ e^{- {\bf S}_2(\varphi, \phi)} \\
  & +  \int {\cal D}\varphi ~e^{-{\bf S}_1(\varphi)} ~{\delta\over \delta\varphi}\Big(\log  
  \mathbb{D}^\dagger(\sigma, \varphi) \mathbb{D}(\sigma, \varphi)\Big)  
  \mathbb{D}^\dagger(\sigma, \varphi) \mathbb{D}(\sigma, \varphi) 
  \int {\cal D}\phi~e^{- {\bf S}_2(\varphi, \phi)}  
  \\  \end{split}
\nd
where $f_1(\varphi)$ depends only on $\varphi$ (and it's derivative) and may be derived from the part of the action given by ${\bf S}_1(\varphi)$. The splitting of the path integral in terms of $\varphi$ and $\phi$ now instructs us to integrate away the massless off-shell field $\phi$ leading to a non-local contribution of the form:
\bg\label{TIogyan}
\int {\cal D}\phi~ e^{-{\bf S}_2(\varphi, \phi)} ~ = ~ e^{-{\bf S}_{\rm nloc}(\varphi)}, \nd
which is expressed completely in terms of the on-shell field component $\varphi$. (Recall that at the Minkowski minimum $\langle\phi\rangle_0 = \langle\phi\rangle_\sigma = 0$.) As the readers may have noticed, $f_1(\varphi)$ will be related to the energy-momentum tensor once we replace $\varphi$ with the on-shell metric components. The integral result in \eqref{TIogyan} is now important: it tells us how the non-local energy-momentum tensor could enter the EOM. Combining everything together, the result we get for the EOM is:

{\footnotesize
\bg\label{2emergenmey}
\begin{split}
0 & = \int {\cal D}\varphi~e^{-{\bf S}_1(\varphi) - {\bf S}_{\rm nloc}(\varphi)} \left(\square\varphi + f_1(\varphi) - {\delta {\bf S}_{\rm nloc}(\varphi)\over \delta\varphi} - {\delta\over \delta\varphi}\Big(\log  
  \mathbb{D}^\dagger(\sigma, \varphi) \mathbb{D}(\sigma, \varphi)\Big) + ...\right) 
  \mathbb{D}^\dagger(\sigma, \varphi) \mathbb{D}(\sigma, \varphi)\\
  & = \left\langle  \square\varphi + f_1(\varphi) - {\delta {\bf S}_{\rm nloc}(\varphi)\over \delta\varphi} - {\delta\over \delta\varphi}\Big(\log  
  \mathbb{D}^\dagger(\sigma, \varphi) \mathbb{D}(\sigma, \varphi)\Big) + ...\right\rangle_\sigma\\
  \end{split}
\nd}
where the dotted terms would be related to the non-perturbative contributions (that we still haven't determined). A similar flow of arguments work for the off-shell field $\phi$, whose EOM now takes the following form:
\bg\label{leeswift}
\begin{split}
0 & = \int {\cal D}\varphi ~{\cal D}\phi~e^{-{\bf S}_1(\varphi) - {\bf S}_2(\varphi, \phi)}\left(\square\phi + \widetilde{f}_1(\phi) + \widetilde{f}_2(\varphi) + \widetilde{f}_3(\varphi, \phi)\right) \mathbb{D}^\dagger(\sigma, \varphi) \mathbb{D}(\sigma, \varphi)\\
& = \int {\cal D}\varphi~  e^{-{\bf S}_1(\varphi)} \widetilde{f}_2(\varphi)
\mathbb{D}^\dagger(\sigma, \varphi) \mathbb{D}(\sigma, \varphi)
\int {\cal D}\phi~ e^{-{\bf S}_2(\varphi, \phi)}\\
& + \int {\cal D}\varphi~  e^{-{\bf S}_1(\varphi)} \mathbb{D}^\dagger(\sigma, \varphi) \mathbb{D}(\sigma, \varphi)
\int {\cal D}\phi~ e^{-{\bf S}_2(\varphi, \phi)}\left(\square\phi + \widetilde{f}_1(\phi) + \widetilde{f}_3(\varphi, \phi)\right)\\
\end{split}
\nd
where all four functions, namely $\square\phi, \widetilde{f}_1(\phi), \widetilde{f}_2(\varphi)$ and $\widetilde{f}_3(\varphi, \phi)$ now appear from ${\bf S}_2(\varphi, \phi)$ with no contributions from 
${\bf S}_1(\varphi)$. This is unlike the previous case wherein both parts of the total action contributed to the EOM. To proceed, let us define:
\bg\label{TIogyan2}
\int {\cal D}\phi~ e^{-{\bf S}_2(\varphi, \phi)}\left(\square\phi + \widetilde{f}_1(\phi) + \widetilde{f}_3(\varphi, \phi)\right) \equiv ~ e^{-{\bf S}_{\rm nloc}(\varphi)}
 {\rm T}_{\rm nloc}(\varphi), \nd
which may be compared to \eqref{TIogyan}, and more generally to the energy-momentum tensor appearing from the non-local action once we take metric degrees of freedom despite the fact that they are not directly related by a field derivative. Plugging \eqref{TIogyan} and \eqref{TIogyan2} in \eqref{leeswift}, the EOM we get is the following:

{\footnotesize
\bg\label{kinkisole}
\int {\cal D}\varphi~  e^{-{\bf S}_1(\varphi) - {\bf S}_{\rm nloc}(\varphi)} \mathbb{D}^\dagger(\sigma, \varphi) \mathbb{D}(\sigma, \varphi)
\left(\widetilde{f}_2(\varphi) + {\rm T}_{\rm nloc}(\varphi) + ..\right) 
\equiv \left\langle \widetilde{f}_2(\varphi) + {\rm T}_{\rm nloc}(\varphi) + ..\right\rangle_\sigma ~ = ~ 0, \nd}
where the dotted terms are again the non-perturbative contributions that we will discuss soon. Note that the EOM does not have any contribution from the $\log \left(\mathbb{D}^\dagger(\sigma, \varphi) \mathbb{D}(\sigma, \varphi)\right)$ term unlike the previous case in \eqref{2emergenmey}. 

Our analysis above has raised one subtlety regarding the expectation value computation $\langle\varphi_1\rangle_\sigma$ itself once we consider only massless on-shell and off-shell degrees of freedom. The subtlety has to do with the presence of the non-local action ${\bf S}_{\rm nloc}$ that would now appear to contribute to, and therefore could influence, the factorial growth in $\langle\varphi_1\rangle_\sigma$ that we computed earlier. This should then affect the analysis for the cosmological constant. To quantify this let us consider a generic non-local interaction of the form:

{\footnotesize
\bg\label{jacksharp}
{\bf S}_{\rm nloc} = {\rm M}_p^{11}\int d^{11}x \sqrt{-{\bf g}_{11}(x)}~ \sum_{n = 0}^\infty d_n \bigg(\underbrace{{\rm M}_p^6\int d^6y \sqrt{{\bf g}_6(y)}~\mathbb{F}(x - y) \mathbb{Q}_{\rm pert}(y, \{l_i\}, \{n_j\})}_{\mathbb{Q}_{\rm nloc}}\bigg)^n, \nd}
where $\mathbb{Q}_{\rm pert}$ is the set of dimensionless perturbative quantum series expressed completely in terms of the on-shell degrees of freedom as in \eqref{QT2}\footnote{There is however a key difference between \eqref{QT2} and $\mathbb{Q}_{\rm pert}$. The latter is perturbative and is defined over a supersymmetric background (see details below), whereas the former is defined over the eleven-dimensional uplift of a type IIB de Sitter {\it vacuum}. The matching points are that (a) both use the same set of curvature and flux tensors, and (b) both are expressed over the same topology. The actual values of the corresponding tensors are however quite different because in the former case there are temporal dependence with zero supersymmetries whereas in the latter case there are no temporal dependence with non-zero supersymmetries.}, and $\mathbb{F}(x-y)$ is the dimensionless non-locality factor. Such a factor would appear naturally when we take, for example, non-localities of the form inverse derivatives et cetera. All the metric components are defined over a spacetime with a topology ${\bf R}^{2, 1} \times {\cal M}_4 \times {\cal M}_2 \times {\mathbb{T}^2\over {\cal G}}$, with a flat metric over ${\bf R}^{2, 1} \times {\mathbb{T}^2\over {\cal G}}$ and a non-K\"ahler metric over ${\cal M}_4 \times {\cal M}_2$. For us, since we are only considering scalar fields $\varphi$ and $\phi$, the aforementioned details are not important. However it suffices to say that, the Glauber-Sudarshan state would change the metric appropriately (by introducing temporal dependence as in \eqref{mmetric}) keeping the topology unchanged\footnote{For simplicity here we will avoid discussing topology changing effects.}. 

The non-local action presented in \eqref{jacksharp} is not useful because by itself it is not necessarily convergent, which means expanding ${\rm exp}\left(-{\bf S}_{\rm nloc}\right)$, would bring in a more complicated asymptotic series. However we can rewrite the series \eqref{jacksharp} in a slightly different way by expressing $d_n$ itself as the following series:
\bg\label{dnseries}
d_n \equiv \sum_{p = 1}^\infty {c_p(-p)^n\over n!} = {c_1(-1)^n\over n!} + 
{c_2(-2)^n\over n!} + {c_3(-3)^n\over n!} + {c_4(-4)^n\over n!} + ....., \nd
without loss of generalities. If we define ${\bf d} = \left(\begin{matrix} d_1 & d_2 & d_3 & d_4 & ...\end{matrix}\right)^{\rm T}$ and similarly 
${\bf c} = \left(\begin{matrix} c_1 & c_2 & c_3 & c_4 & ...\end{matrix}\right)^{\rm T}$, then \eqref{dnseries} can be written as 
${\bf d} = \mathbb{M} {\bf c}$, where $\mathbb{M}$ is now the following matrix:
\bg\label{poladom}
\mathbb{M} =  \left(\begin{matrix} - 1 & -2 & - 3 & -4& - 5& - 6 & - 7 & ....\\
~ & ~ & ~ & ~ & ~ & ~ & ~ & ~ \\
~{1\over 2} & ~2 & ~{9\over 2} & ~8 & ~{25\over 2} & ~18 & ~{49\over 2} & .... \\
~ & ~ & ~ & ~ & ~ & ~ & ~ & ~ \\
- {1\over 6} & - {4\over 3} & -{9\over 2} & - {32\over 2} & -{125\over 2} & - 36 & - {343\over 6} & ....\\
~ & ~ & ~ & ~ & ~ & ~ & ~ & ~ \\
~{1\over 24} & ~{2\over 3} & ~{27\over 8} & ~{32\over 3} & ~{625\over 24}  &~ 54 & ~{2401\over 24} & .... \\
~ & ~ & ~ & ~ & ~ & ~ & ~ & ~ \\
- {1\over 120} & - {4\over 15} & - {81\over 40} & - {128\over 15} & - {625\over 24} & - {324\over 5} & 
- {16807\over 120} & ....\\
~ & ~ & ~ & ~ & ~ & ~ & ~ & ~ \\
.... & .... & .... & .... & .... & .... & .... & ....  \end{matrix}\right), \nd
which, despite being infinite dimensional, has a {\it finite} determinant\footnote{Somewhat surprisingly, ${\rm det}~\mathbb{M} = \pm 1$  depending on the dimension of the matrix $\mathbb{M}$ as we shall see in section \ref{mathy} below.} and an inverse (see section \ref{mathy}). This would imply that we can substitute \eqref{dnseries} in \eqref{jacksharp}, to express it alternatively as:

{\footnotesize
\bg\label{fepoladom}
{\bf S}_{\rm nloc} = {\rm M}_p^{11}\int d^{11}x \sqrt{-{\bf g}_{11}(x)}~ \sum_{p = 1}^\infty c_p\left[{\rm exp}\left({-p{\rm M}_p^6\int d^6y \sqrt{{\bf g}_6(y)}~\big\vert\mathbb{F}(x - y) \mathbb{Q}_{\rm pert}(y, \{l_i\}, \{n_j\})\big\vert}\right) - 1\right], \nd}
defined over the supersymmetric warped Minkowski background (with an internal non-K\"ahler space). The modulus sign in \eqref{fepoladom} is motivated from the fact that the series in \eqref{fepoladom} is always convergent no matter what overall sign we assign to $\mathbb{Q}_{\rm pert}$. This is borne out of our choice in \eqref{dnseries}. For example if the overall sign of $\mathbb{Q}_{\rm pert}$ is negative then we can resort to positive $p$ in \eqref{dnseries}, or alternatively change $p \to -p$ in \eqref{dnseries} (and consequently positive row elements in \eqref{poladom}), which would again make the series in \eqref{fepoladom} convergent. The inverse of the matrix $\mathbb{M}$ continues to exists for either signs of $p$ in \eqref{dnseries} because ${\rm det}~\mathbb{M} = +1$ now. See {\bf figure \ref{nonpertdiag}} for a diagrammatic view of the non-local interaction. \textcolor{blue}{In the following section we will provide a mathematical proof of how to go from \eqref{jacksharp} to \eqref{fepoladom}. Readers wanting to see the consequence of \eqref{fepoladom} may skip this section on their first reading and go directly to section \ref{sec6.2.2}.}

\subsubsection{A mathematical proof to convert \eqref{jacksharp} to \eqref{fepoladom} and convergence \label{mathy}}

Let us take a short mathematical detour to explain some of the steps used to get \eqref{fepoladom} from \eqref{jacksharp}. Our aim would also be to 
show that the aforementioned transformations are not related in any way to the Borel transformation. The latter will be used to further resum the series to bring it in a trans-series form. 

Our starting point would be \eqref{dnseries} with the coefficients $d_n$ expressed using the coefficients $c_p$.
The fact that we can define \eqref{dnseries} may be verified to any order (although as we shall see, there is no need to go beyond certain order in the series because of relative suppressions). Before moving ahead, let us verify that \eqref{dnseries} works to any order. It is easy to see that the coefficients $d_n$ are related to the coefficients $c_p$ via the matrix \eqref{poladom} which we express in following by showing more rows and columns:

{\footnotesize
\bg\label{crystals}
\left(\begin{matrix} d_1\\ ~\\ d_2\\ ~ \\d_3\\ ~\\d_4\\ ~\\ d_5\\ ~\\d_6\\ ~\\ d_7\\ ~\\ ... 
\end{matrix}\right) = \left(\begin{matrix} -1 & -2 & -3 & -4& -5& -6 & -7 & ....\\
~ & ~ & ~ & ~ & ~ & ~ & ~ & ~ \\
~{1\over 2} & ~2 & ~{9\over 2} & ~8 & ~{25\over 2} & ~18 & ~{49\over 2} & .... \\
~ & ~ & ~ & ~ & ~ & ~ & ~ & ~ \\
-{1\over 6} & -{4\over 3} & -{9\over 2} & -{32\over 2} & -{125\over 6} & -36 & -{343\over 6} & ....\\
~ & ~ & ~ & ~ & ~ & ~ & ~ & ~ \\
~{1\over 24} & ~{2\over 3} & ~{27\over 8} & ~{32\over 3} & ~{625\over 24}  &~ 54 & ~{2401\over 24} & .... \\
~ & ~ & ~ & ~ & ~ & ~ & ~ & ~ \\
-{1\over 120} & -{4\over 15} & -{81\over 40} & -{128\over 15} & -{625\over 24} & -{324\over 5} & 
-{16807\over 120} & ....\\
~ & ~ & ~ & ~ & ~ & ~ & ~ & ~ \\
~{1\over 720} & ~{4\over 45} & ~{81\over 80} & ~{256\over 45} & ~{3125\over 144} & ~{324\over 5} & ~{117649\over 720} & ....\\
~ & ~ & ~ & ~ & ~ & ~ & ~ & ~ \\
-{1\over 5040} & -{8\over 315} & -{243\over 560} & -{1024\over 315} & -{15625\over 1008} & 
-{1944\over 35} & -{117649\over 720} & ...\\
~ & ~ & ~ & ~ & ~ & ~ & ~ & ~ \\
.... & .... & .... & .... & .... & .... & .... & ....  \end{matrix}\right) \left(\begin{matrix}
c_1\\ ~\\ c_2\\~\\c_3\\ ~\\c_4\\ ~\\c_5\\ ~\\c_6\\ ~\\c_7\\~\\ ... 
\end{matrix}\right), \nd}
which we can write as $d_n = \mathbb{M}_{n l} c_l$ where $\mathbb{M}$ in general is 
a ${\rm N} \times {\rm N}$ matrix in the limit ${\rm N } \to \infty$. Solution would exist if the matrix $\mathbb{M}$ has an inverse. Since the matrix is infinite-dimensional, one might worry that it will be very hard to compute the determinant and show that it is non-zero. Somewhat surprisingly, taking ${\rm N} = 2, 3,..., 7$
tell us that the determinants of the corresponding matrices are either $+1$ or $-1$ arranged in alternate ways. This is shown in {\bf Table \ref{lilluth}} and thus implies:
\bg\label{benedetta}
{\rm det}~\mathbb{M}_{2{\rm N} \times 2{\rm N}} = {\rm cos}~{\rm N}\pi, ~~~~~
{\rm det}~\mathbb{M}_{(2{\rm N} + 1) \times (2{\rm N} + 1)} = {\rm sin}~\left({\rm N} + {1\over 2}\right)\pi, \nd 
which would at least provide a chance that the inverse of matrix $\mathbb{M}$ could exist. To justify whether the inverse really exists, we can take various dimensions of $\mathbb{M}_{{\rm N} \times {\rm N}}$ and determine the inverses\footnote{An actual proof of \eqref{benedetta} appears to be quite non-trivial so, in the absence of this, we will perform various tests to justify the aforementioned result. It will be interesting to find a formal proof of the result explicitly and then compare the subsequent result to the Borel resummation, as both provides convergent answers.}. This gives: 

{\footnotesize
\bg\label{reshish}
\mathbb{M}^{-1}_{7 \times 7} = \left(\begin{matrix}
\textcolor{red}{-7} & -{223\over 10} & -{638\over 15} & -{111\over 2}& -{295\over 6} & -27 &\textcolor{red}{-7}\\
~ & ~ & ~ & ~ & ~ & ~ & ~  \\
~{21 \over 2} & ~{879\over 20} & ~{3929\over 40} & ~142 & ~{135} & ~78 & ~{21} \\
~ & ~ & ~ & ~ & ~ & ~ & ~ \\
-{35 \over 3} & -{949\over 18} & -{389\over 3} & -{1219\over 6} & -{1235\over 6} & -125 & -{35}\\
~ & ~ & ~ & ~ & ~ & ~ & ~ \\
~{35\over 4} & ~{41} & ~{2545\over 24} & ~{176} & ~{565\over 3}  &~ 120 & ~{35} \\
~ & ~ & ~ & ~ & ~ & ~ & ~ \\
-{21\over 5} & -{201\over 10} & -{268\over 5} & -{185\over 2} & -{207\over 2} & -{69} & 
-{21}\\
~ & ~ & ~ & ~ & ~ & ~ & ~ \\
~\textcolor{red}{{7\over 6}} & ~{1019\over 180} & ~{1849\over 120} & ~{82\over 3} & ~{95\over 3} & ~{22} & ~\textcolor{red}{7}\\
~ & ~ & ~ & ~ & ~ & ~ & ~\\
\textcolor{red}{-{1\over 7}} & -{7\over 10} & -{29\over 15} & -{7\over 2} & -{25\over 6} & 
-{3} &\textcolor{red}{-{1}}\\\end{matrix}\right), ~~~~~
\mathbb{M}^{-1}_{5 \times 5} = 
\left(\begin{matrix}
\textcolor{red}{-5} & -{77\over 6} & -{71\over 4} & -{14}&\textcolor{red}{-5}\\
~ & ~ & ~ & ~ & ~   \\
~{5} & ~{107\over 6} & ~{59\over 2} & ~26 & ~{10} \\
~ & ~ & ~ & ~ & ~ \\
-{10 \over 3} & -{13} & -{49\over 2} & -{24} & -{10} \\
~ & ~ & ~ & ~ & ~ \\
~\textcolor{red}{{5\over 4}} & ~{61\over 12} & ~{41\over 4} & ~{11} & ~ 
\textcolor{red}{5}  \\
~ & ~ & ~ & ~ & ~  \\
\textcolor{red}{-{1\over 5}} & -{5 \over 6} & -{7\over 4} & -{ 2} & 
\textcolor{red}{-{1}}\\
\end{matrix}\right) \nd}
for the two cases where we take $7 \times 7$ and $5 \times 5$ matrices. The good thing about 
\eqref{reshish} is that the inverse matrices are non-singular, but we notice something more. There are certain patterns to these inverses which we denote above in \textcolor{red}{red}. (There are also other pattens in the 
last columns of the two inverse matrices, and maybe more if we go to higher rank.) This means in the rank ${\rm N}$ case we at least expect:
\bg\label{kettleman}
\mathbb{M}^{-1}_{{\rm N} \times {\rm N}} = 
\left(\begin{matrix}
{-{\rm N}} &  ... &  ... &  ... &  ... & -{\rm N}\\
~ & ~ & ~ & ~ & ~ &  ~  \\
... & ...& ... & ...& ...&  ... \\
~ & ~ & ~ & ~ & ~ & ~ &   ~\\
... & ...& ... & ...&  ...& ... \\
 ~ & ~ & ~ & ~ & ~ &  ~ \\
~ {{\rm N}\over {\rm N - 1}} &  ... &  ... & ... &  ... &  ~{\rm N} \\
  ~ & ~ & ~ & ~ & ~ & ~ & ~ \\
  -{1\over {\rm N}} &  ... &   ... &  ... & ... &  -1\\
 \end{matrix}\right), \nd
implying that in the limit ${\rm N} \to \infty$ many of the terms of the matrix \eqref{kettleman} blow up. This is exactly what we would have expected if we try to express a {\it perturbative} series $-$ for example every terms of \eqref{jacksharp} $-$ as an exponential series. An easy way to justify this would be to note that 
${\rm exp}\left(p \mathbb{X}\right) - 1$ vanishes when $\mathbb{X} \to 0$, where $\mathbb{X}$ is the integral in \eqref{fepoladom}. Thus the coefficients have to be arbitrarily large for the exponential series to give the same finite answer as the original perturbative series. This is of course the reason why we generally do not express a perturbative series in terms of an exponential one. On the other hand if $\mathbb{X} >> 1$, then the only way a {\it perturbative} series would make sense if the coefficients are arbitrarily small.  This means $d_n \to 0$  in \eqref{jacksharp} and:
\bg\label{apr24}
c_l = \mathbb{M}^{-1}_{ln} d_n, \nd
which would give finite coefficients for the exponential series even if ${\rm N} \to \infty$ in \eqref{kettleman}. This is a bit of a subtle issue so let us clarify a few points starting with the choice \eqref{dnseries} itself. First of all, despite certain similarities, \eqref{dnseries} is {\it not} a Borel resummation. We will discuss Borel resummation soon, but it is instructive here to point out the key differences. First, the expansion in \eqref{dnseries} is {\it not} a weak-coupling expansion and secondly, the coefficients $d_n$ will have to have zero radii of convergences for the series \eqref{jacksharp} to make sense. Thus the procedure here is simply a resummation but not of the Borel kind. The above statements need further clarifications, so let us elaborate the story a bit more. The function in the exponential of \eqref{fepoladom} typically scales as $\left({g_s\over {\rm HH}_o}\right)^{-(2 - \theta)}$ where $\theta$ is the scaling of $\mathbb{Q}_{\rm pert}$. For $\theta < 2$ it is non-perturbative in 
${g_s\over {\rm HH}_o}$ and therefore the series \eqref{jacksharp} doesn't make sense unless we can control every term. On the other hand, for $\theta > 2$ the series becomes perturbative in ${g_s\over {\rm HH}_o}$. Also, and as we saw above, \eqref{apr24} is possible if the coefficients $d_n$ can be made arbitrarily small. Let us then take the following limits:
\bg\label{ivana22}
b_n ~ \to ~ \epsilon^{\bar\alpha}, ~~~~~~ {g_s\over {\rm HH}_o} ~ \to ~
\epsilon^{1/\bar\beta}, ~~~~~~ {\rm N} ~ \to ~ \epsilon^{-\bar\alpha},
~~~~~~ \epsilon \to 0, \nd
where $\bar\beta >> 1$ which would typically make ${g_s\over {\rm HH}_o} \le 1$ but {\it not} ${g_s\over {\rm HH}_o} \to 0$, indicating early times. Such a choice of ${g_s \over {\rm HH}_o}$ allows us to take any values of $\theta$ (meaning, larger the value of $\bar\beta$, closer is ${g_s\over {\rm HH}_o}$ to identity). On the other hand the choice of ${\rm N}$ in \eqref{ivana22} would keep 
$c_l$ finite in \eqref{apr24} (whose exact values may then be fixed by other means, like supersymmetry constraints {\it et cetera}). This means the maximum value of $\bar\alpha$, for a given value of $\bar\beta$, may be determined from the following transcendental equation:
\bg\label{ghoulash}
\bar\alpha ~{\rm log}~\epsilon + {\rm log}~\bar\alpha + {\rm log}~\bar\beta = 0, \nd
which could be solved using the Lambert ${\rm W}$-function (sometimes also called the omega function or the ProductLog function), and has real solutions as long as $\bar\beta >> 1$ consistent with the early-time picture\footnote{For example, if we take $\epsilon = 10^{-40}$, then $\bar\alpha = 10$ and $\bar\beta = 10^{399}$ would consistently solve \eqref{ghoulash} for ${g_s\over {\rm HH}_o} = 0.9999$, $d_{n} = 10^{-400}$ and ${\rm N} = 10^{400}$. Plugging this in \eqref{apr24} would provide ${\cal O}(1)$ values for $c_l$ exactly as we wanted (whose precise values may be fixed by supersymmetry).}.
Of course if $\mathbb{X}$ is large then we are not required to write a perturbative series {\it per se}, and we can simply express the series in terms of decaying exponential functions. However if we want to go via the perturbative route, then the coefficients of the perturbative series have to be arbitrarily small. Once the series \eqref{fepoladom} is resummed to the trans-series form, we can relax all the aforementioned conditions and take finite values for $c_p$ for all 
${g_s\over {\rm HH}_o} < 1$. We will discuss the second resummation process in section \ref{sec7.3.2}.

\begin{table}[tb]  
 \begin{center}
\renewcommand{\arraystretch}{1.5}
\begin{tabular}{|c||c||c||c||c||c||c||c|| c|}\hline ${\rm N}$  & 2 & 3 & 4 & 5 & 6 & 7 & ... \\ \hline\hline
${\rm det}$ & $-$ & $+$ & $+$ & $-$ & $-$ & $+$ & ... \\ \hline 
 \end{tabular}
\renewcommand{\arraystretch}{1}
\end{center}
 \caption[]{The determinants of $\mathbb{M}_{{\rm N} \times {\rm N}}$ which only take values $\pm 1$.} 
  \label{lilluth}
 \end{table}

\subsubsection{Non-local and non-perturbative terms and compatibility issues \label{sec6.2.2}}

Our little exercise above, and for the readers who have skipped section \ref{mathy},  suggests that we can regard \eqref{fepoladom} to be the action for the non-local interactions (got from integrating out the massless off-shell states) without worrying too much about the connection between $c_p$ and $d_n$. However we should keep in mind of the following three points.
\vskip.1in

\noindent ${\bf 1.}$ The conversion of \eqref{jacksharp} to \eqref{fepoladom} is {\it not} a Borel resummation because the procedure of resummation does not involve any factorial growths. 

\vskip.1in

\noindent ${\bf 2.}$ The relation ${\bf c} = \mathbb{M}^{-1} {\bf d}$ makes sense for arbitrarily small $d_n$, so one may simply view the procedure of getting \eqref{fepoladom} as a trick to convert the non-local interactions into a convergent series.

\vskip.1in

\noindent ${\bf 3.}$ The action \eqref{fepoladom} is still {\it not} the full story, as there would be additional corrections to it. How and why such corrections are necessary require more preparation than what we have developed so far. This will be clarified in section 
\ref{sec7.3.2}. 

\vskip.1in

\noindent With this, we can now answer the question raised earlier above \eqref{jacksharp} regarding ${\bf S}_{\rm nloc}$: because of it's convergence property, it will no longer contribute additional factorial growth when we compute $\langle\varphi\rangle_\sigma$. All the factorial growth leading to a Gevrey series continues to come from ${\bf S}_1(\varphi)$ {\it only}, thus keeping our earlier conclusions unchanged.
Despite this however it still doesn't answer the origin of the non-perturbative contributions. Where are the non-perturbative contributions to the action coming from?

The answer is not hard to see. Recall that non-perturbative effects enter the system through Borel resummation of an asymptotic series. This means when we computed $\langle\varphi\rangle_\sigma$ the non-perturbative effects entering the computation from Borel resummation should also dictate the non-perturbative contribution to the action itself. To see this let us revisit the expectation value computation by rewriting 
${\bf S}_1(\varphi) \equiv {\bf S}_{\rm kin}(\varphi) + {\bf S}_{\rm pert}(\varphi)$. The expectation value then becomes:
\bg\label{posnerdead}
\langle\varphi\rangle_\sigma = {\int {\cal D}\varphi~e^{-{\bf S}_{\rm kin}(\varphi) - {\bf S}_{\rm pert}(\varphi) - {\bf S}_{\rm nloc}(\varphi) +  \log~\mathbb{D}^\dagger(\sigma,\varphi)\mathbb{D}(\sigma,\varphi)} ~\varphi(x, y) \over\int {\cal D}\varphi~e^{-{\bf S}_{\rm kin}(\varphi) - {\bf S}_{\rm pert}(\varphi) - {\bf S}_{\rm nloc}(\varphi) + \log~\mathbb{D}^\dagger(\sigma,\varphi)\mathbb{D}(\sigma,\varphi)}}, \nd
with ${\bf S}_{\rm pert}(\varphi)$ capturing the perturbative interactions, and the logarithmic piece with the displacement operators would not only shift the kinetic term ${\bf S}_{\rm kin}(\varphi)$, but will also renormalize the coefficients in the interacting perturbative 
action ${\bf S}_{\rm pert}(\varphi)$ (see \eqref{monwhip} and \eqref{indydial} for details). Now there is the two-step process:
\vskip.1in
\noindent ${\bf 1.}$ The asymptotic nature of the perturbative series with and without the presence of a source $\varphi(x, y)$ leading to a Gevrey growth.
\vskip.1in
\noindent ${\bf 2.}$ Subsequent Borel resumming the asymptotic series and introducing the non-perturbative contribution as in \eqref{rubiem} or as in the second relation in \eqref{recelcards}.

\vskip.1in

\noindent The non-pertubative contribution to the action would simply {\it reduce the two-step process to just one-step process}! In other words, replacing the perturbative action by a non-perturbative one would give us the {\it same} answer for $\langle\varphi\rangle_\sigma$ that we got by the aforementioned two-step process\footnote{Turning the argument around, we can ask what non-perturbative action we can add that gives us the same answer as in say \eqref{rubiem} or \eqref{capquebcmey} below.}. To appreciate the last point, let us look again at the wave-function renormalization factor from \eqref{rubiem} which appears from the two-step process mentioned above. This takes the following form:

{\footnotesize
\bg\label{capquebcmey}
\int_0^\infty d{\rm S}~ {{\rm exp}\left({-{\rm S}/g^{1/\bar\alpha}}\right)\over 1 - \check{f}_{\rm max} {\rm S}^{\bar\alpha}} = \sum_{{\rm N} = 0}^\infty 
~~\underset{\rm saddle}{\underbrace{{\rm exp}\left(-{{\rm N}\over g^{1/\bar\alpha}}\right)}_{\rm instanton}} ~~\Bigg[\underbrace{{1\over 1 - {\rm N}^{\bar\alpha} \check{f}_{\rm max}} + \sum_{n = 1}^\infty {\bar\alpha^n \check{f}^n_{\rm max} {\rm N}^{n(\bar\alpha-1)} \over \left(1- {\rm N}^{\bar\alpha} \check{f}_{\rm max}\right)^{n+1}} \left({\rm s} - {\rm N}\right)^n}_{\rm fluctuation~determinant}\Bigg], 
\nd}
where $\check{f}_{\rm max}$ and $\bar\alpha$ may be read from \eqref{runikate} and \eqref{halfprice} respectively; and ${\rm s - N} \equiv \Delta {\rm s}_{\rm N}$ is the fluctuation over a given instanton specified by the instanton number ${\rm N}$. From \eqref{capquebcmey} we see that the wave-function renormalization factor is exactly the sum over instanton saddles and the corresponding fluctuations over them, confirming the fact that the two-step process may be replaced by a one-step process if we incorporate the non-perturbative effects from the instanton sums. Moreover,
the non-perturbative action being {\it convergent}\footnote{This may be verified from \eqref{halfprice}. Since $g^{1/\bar\alpha} = \check{g}^{\check{\alpha}}$ and $\check{g} = {1\over {\rm M}_p} << 1$ with $\check{\alpha} > 1$, the instanton series is convergent even for increasing $\check{\alpha}$. However once we take the fluctuation determinants into account, the series becomes a {\it resurgent} trans-series where the convergence may be realized from the resurgence theory \cite{resurgence}. We will elaborate more on this in section \ref{sec7.1}.} will not grow factorially, so the path-integral computation will become much simpler. This means we can replace \eqref{posnerdead} by the following:
\bg\label{dunbar}
\langle\varphi\rangle_\sigma = {\int {\cal D}\varphi~e^{-{\bf S}_{\rm kin}(\varphi) - \textcolor{red}{{\bf S}_{\rm NP}(\varphi)} - {\bf S}_{\rm nloc}(\varphi) +  \textcolor{red}{\log~\mathbb{D}_o^\dagger(\sigma,\varphi)\mathbb{D}_o(\sigma,\varphi)}} ~\varphi(x, y) \over\int {\cal D}\varphi~e^{-{\bf S}_{\rm kin}(\varphi) - \textcolor{red}{{\bf S}_{\rm NP}(\varphi)} - {\bf S}_{\rm nloc}(\varphi) + \textcolor{red}{\log~\mathbb{D}_o^\dagger(\sigma,\varphi)\mathbb{D}_o(\sigma,\varphi)}}}, \nd
where ${\bf S}_{\rm NP}(\varphi)$ is the non-perturbative action which is by construction convergent, and $\mathbb{D}_o(\sigma, \varphi)$ is only the {\it linear} part of $\varphi$ in say \eqref{28rms}. Both ${\bf S}_{\rm pert}(\varphi)$ and the non-linear terms in \eqref{28rms} are replaced by the non-perturbative action ${\bf S}_{\rm NP}(\varphi)$. Note that we make no changes to ${\bf S}_{\rm nloc}(\varphi)$ because it is already resummed to take a convergent form as in \eqref{fepoladom}. The above rewriting justifies what we have been emphasizing all along: the perturbative interactions are red herrings in the problem. In fact there are no perturbative interactions at all! The total action:
\bg\label{katuli}
\hat{\bf S}_{\rm tot}(\varphi) \equiv {\bf S}_{\rm kin}(\varphi) + {\bf S}_{\rm NP}(\varphi) + {\bf S}_{\rm nloc}(\varphi), \nd
differs from ${\bf S}_{\rm tot}(\varphi)$ in \eqref{enmilltriangle} in many respects. It not only replaces the perturbative series therein by non-perturbative and non-local actions, but also replaces the perturbative series from the displacement operator \eqref{28rms}. Question can be raised whether this action is compatible with the Wilsonian Effective Action (WEA), which eventually hinges down to the deeper question of whether WEA is compatible with Borel resummation. Our simple analysis presented above suggests otherwise, although the opinion seems to be varied in the literature \cite{WEA?}. We will however avoid taking any sides here, leaving this delicate topic for later works, and instead concentrate on the consequence of the action \eqref{katuli} on the Schwinger-Dyson's equations. 

\subsection{Revisiting the Schwinger-Dyson's equation \eqref{elenacigu} and background EOMs \label{sec6.3}}

Our analysis above, wherein we integrate out the off-shell degree of freedom and replace the perturbative interactions by non-perturbative effects, suggests that we should revisit the computation leading to the Schwinger-Dyson's equation \eqref{elenacigu} starting from \eqref{straokua} itself. We will continue using the toy set-up with one massless on-shell field $\varphi$ and one massless off-shell field $\phi$ 
controlled by the perturbative action \eqref{enmilltriangle}. The relation \eqref{straokua} gets replaced by:
\bg\label{kinmarie}
{\langle\sigma\vert\sigma\rangle\over \langle\Omega\vert\Omega\rangle} = 
\int {\cal D}\varphi~e^{-{\bf S}_{\rm kin}(\varphi) - {{\bf S}_{\rm NP}(\varphi)} - {\bf S}_{\rm nloc}(\varphi) + {\log~\mathbb{D}_o^\dagger(\sigma,\varphi)\mathbb{D}_o(\sigma,\varphi)}}, \nd
where, as noted earlier, we only have $\mathbb{D}_o^\dagger(\sigma,\varphi)\mathbb{D}_o(\sigma,\varphi)$ {\it i.e.} the linear terms in \eqref{28rms}. These terms are expressed in terms of four on-shell fields $\varphi_i$ in \eqref{28rms}, but here we only take one on-shell field $\varphi$. For this specific case, \eqref{28rms} may be rewritten as:
\bg\label{enmillboro}
\log~\mathbb{D}_o(\sigma, \varphi) = \sum_{n, p}\int_{k_{\rm IR}}^\mu d^{11}k ~{\rm z}_{n1p} ~k^{2n} \sigma^{(1, p)} ~\widetilde\varphi^\ast(k), \nd
where ${\rm z}_{n1p}$ are dimensionful constants ({\it i.e.} suppressed by powers of 
${\rm M}_p$), and $\widetilde\varphi(k)$ is the Fourier transform of the on-shell field $\varphi(x, y)$. The series in ${\rm z}_{n1p}$ is important, because it changes \eqref{monwhip} to  the following transformation:
\bg\label{monwhip2}
k^2\vert \widetilde\varphi(k)\vert^2 ~ \longrightarrow ~ k^2\Big\vert \widetilde\varphi(k) - 
\sum_{n, p} {\rm z}_{n1p}~k^{2n-2} \sigma^{(1, p)}\Big\vert^2, \nd
thus providing a series in \eqref{rubiem} which is suppressed by an increasing powers of ${\rm M}_p$. Such an argument is important when we want to compare two theories that are related by T-dualities, as is known that T-duality transformations include higher order $\alpha'$ corrections. For example, going from type IIB theory to M-theory include one T-duality. Such a duality will relate the metric components of M-theory to various metric and flux components in the IIB side, expressed  by increasing powers of $\alpha'$. Our simple toy model only takes a scalar field, so the duality transformation is not relevant for the present discussion, but once we go to realistic metric and flux degrees of freedom, we can in fact incorporate subtle nuances like the $\alpha'$ corrections to the duality rules. Even more so, we can incorporate $g_s$ corrections to the duality rules as evidenced by our upcoming work \cite{hete8}.

After this small detour, let us get back to the question whose answer we seek here, namely, what replaces the Schwinger-Dyson's equation \eqref{elenacigu}? To answer this we will follow the same set of manipulations on \eqref{kinmarie} that we performed starting with \eqref{straokua} to get to \eqref{elenacigu}. After the dust settles, the answer we get is:
\bg\label{elenacigu2}
\left\langle{\delta\hat{\bf S}_{\bf tot}\over \delta\varphi}\right\rangle_\sigma = \left\langle
{\delta\over \delta\varphi}\log\left(\mathbb{D}_o^\dagger \mathbb{D}_o\right)\right\rangle_\sigma, \nd
where $\hat{\bf S}_{\rm tot}$ is given by \eqref{katuli}. If there are Faddeev-Popov ghosts, then the action $\hat{\bf S}_{\rm tot}$ will get additional contribution (see \eqref{palpucki} and also \cite{hete8}), but the scalar field model that we take here allows no such ghosts, so this subtlety does not enter here. For more realistic situations, ghosts are essential, and the story will be  elaborated in \cite{hete8} (see also \cite{coherbeta, coherbeta2}). 

The EOM, as also noticed in \eqref{elenacigu}, is not ready for synthesis yet. To make it closer to something we know, we will have to use the resolution of identity that we derived in section \ref{sec4.1} culminating in the formula \eqref{mariesored}. To see how this helps, we will go back to the action \eqref{enmilltriangle} but take the sector with $\vert c_{0mk}\vert > 0$ only to avoid incorporating the off-shell states. Let us now use the following manipulations:
\bg\label{katmarzoe}
\begin{split}
\langle\varphi^2\rangle_\sigma = {\langle\sigma\vert \varphi^2\vert\sigma\rangle\over \langle\sigma\vert \sigma\rangle} & = 
{1\over \langle\sigma\vert\sigma\rangle} \langle\sigma\vert \varphi \Big(\sum_{\sigma'} {\vert\sigma'\rangle \langle\sigma'\vert\over \langle\sigma'\vert \sigma'\rangle} - {\bf Q}(\varphi)\Big)\varphi \vert\sigma\rangle\\
& = {\langle\sigma\vert\varphi\vert\sigma\rangle \langle\sigma\vert\varphi\vert\sigma\rangle\over \langle\sigma\vert\sigma\rangle \langle\sigma\vert\sigma\rangle} + \sum_{\sigma' \ne \sigma}   {\langle\sigma\vert\varphi\vert\sigma'\rangle \langle\sigma'\vert\varphi\vert\sigma\rangle\over \langle\sigma'\vert\sigma'\rangle \langle\sigma\vert\sigma\rangle} - 
{\langle \sigma\vert \varphi {\bf Q}(\varphi)\varphi\vert\sigma\rangle\over \langle\sigma\vert\sigma\rangle}\\
& = \langle\varphi\rangle^2_\sigma + \sum_{\sigma'\ne\sigma} \langle\varphi^2\rangle_{(\sigma'|\sigma)} - \langle \varphi {\bf Q}(\varphi)\varphi\rangle_\sigma \\
\end{split}
\nd
where ${\bf Q}(\varphi)$ is the operator appearing in \eqref{hubeimach} and in the resolution of identity \eqref{mariesored}. This is a function of the on-shell field $\varphi$ because the Hamiltonian ${\bf H}$ entering it's definition may be expressed completely in terms of the on-shell fields by incorporating the non-perturbative and the non-local contributions. However ${\bf H}$ doesn't exactly appear from $\hat{\bf S}_{\rm tot}$ because $\hat{\bf S}_{\rm tot}$ involves the non-perturbative completion in the presence of $\mathbb{D}^\dagger(\sigma, \varphi) \mathbb{D}(\sigma, \varphi)$, whereas ${\bf H}$ is defined over the Minkowski minimum. However over the Glauber-Sudarshan state $\vert\sigma\rangle$, it is easy to infer that:
\bg\label{capqubec2mey}
\langle \varphi ~{\bf Q}({\bf H}(\varphi), \varphi)~\varphi\rangle_\sigma = 
\langle \varphi ~\hat{\bf Q}(\hat{\bf H}(\varphi), \varphi)~\varphi\rangle_\sigma, \nd
where $\hat{\bf H}(\varphi)$ can now be derived from $\hat{\bf S}_{\rm tot}(\varphi)$, and $\hat{\bf Q}$ involves $\hat{\bf S}_{\rm tot}(\varphi)$ and $\hat{\bf H}(\varphi)$. The result \eqref{capqubec2mey} can be easily justified by opening up the path integral and replacing all perturbative terms by the non-perturbative terms in the same vein as \eqref{dunbar}. Once we go to ${\rm N}$-th order, it is not hard to justify the following decomposition:
\bg\label{meysmelme}
\langle\varphi^{\rm N}\rangle_\sigma = \langle\varphi\rangle^{\rm N}_\sigma + \sum_{\{\sigma_i\}\ne \sigma} \langle\varphi^{\rm N}\rangle_{(\sigma_1, \sigma_2,.., \sigma_{{\rm N} -1}\vert\sigma)} - 
\langle\varphi~ \underbrace{\hat{\bf Q}~ \varphi~ \hat{\bf Q}.... \hat{\bf Q}}_{({\rm N} - 1) ~\hat{\bf Q}~ {\rm terms}}~\varphi\rangle_\sigma, \nd
where $\vert\sigma_i\rangle$ are the intermediate Glauber-Sudarshan states which we are summing over. In fact \eqref{meysmelme} extends to the derivatives of $\varphi^{\rm N}$ because 
$\langle \partial^{2{\rm M}} \varphi^{\rm N}\rangle_\sigma = \partial^{2{\rm M}}\langle\varphi^{\rm N}\rangle_\sigma$ which in-turn could be easily verified from the following path integral structure:
\bg\label{rambagan}
\begin{split}
\langle\partial^{2{\rm M}}\varphi^{\rm N}\rangle_\sigma  &= 
{\int {\cal D}\varphi ~e^{-\hat{\bf S}_{\rm tot}(\varphi) + \log~\mathbb{D}_o^\dagger(\sigma, \varphi) \mathbb{D}_o(\sigma, \varphi)}
~\partial^{2{\rm M}}\varphi^{\rm N}(x, y) \over \int {\cal D}\varphi ~e^{-\hat{\bf S}_{\rm tot}(\varphi) + \log~\mathbb{D}_o^\dagger(\sigma, \varphi) \mathbb{D}_o(\sigma, \varphi)}}\\
&= \partial^{2{\rm M}}\left({\int {\cal D}\varphi ~e^{-\hat{\bf S}_{\rm tot}(\varphi) + \log~\mathbb{D}_o^\dagger(\sigma, \varphi) \mathbb{D}_o(\sigma, \varphi)}
~\varphi^{\rm N}(x, y) \over \int {\cal D}\varphi ~e^{-\hat{\bf S}_{\rm tot}(\varphi) + \log~\mathbb{D}_o^\dagger(\sigma, \varphi) \mathbb{D}_o(\sigma, \varphi)}}\right) = \partial^{2{\rm M}} \langle\varphi^{\rm N}\rangle_\sigma,\\
\end{split}
\nd 
where $\hat{\bf S}_{\rm tot}(\varphi)$ is given by \eqref{katuli}, and one may now plug in \eqref{meysmelme} to get the required decomposition. Looking at \eqref{meysmelme} and \eqref{rambagan}, it is not too hard to convince oneself of the following decomposition:
\bg\label{palfasa}
\langle f(\varphi) \rangle_\sigma = f(\langle \varphi\rangle_\sigma) + 
\sum_{\{\sigma_i\} \ne \sigma} \langle f(\varphi)\rangle_{(\{\sigma_i\}\vert \sigma)} + \langle f(\varphi)\rangle_{(\hat{\bf Q}\vert \sigma)}, \nd
where $f(\varphi)$ is a generic function of $\varphi$ and it's derivatives. The first term in \eqref{palfasa} simply resonates the first term from \eqref{meysmelme}, and the corresponding one with derivatives. More importantly it suggests that we can replace the field $\varphi$ in $f(\varphi)$ by its {\it expectation value} $\langle\varphi\rangle_\sigma$. The second and the third terms are additional corrections, coming from the presence of the intermediate Glauber-Sudarshan states $\vert \sigma_i\rangle$ and the $\hat{\bf Q}$ operators respectively, needed to compute the full expectation value $\langle f(\varphi)\rangle_\sigma$. Taking cues from \eqref{meysmelme}, \eqref{rambagan} and \eqref{palfasa}, the EOM \eqref{elenacigu2} can be decomposed in the following way:
\bg\label{palpenti}
\begin{split}
\left\langle {\delta\hat{\bf S}_{\rm tot}(\varphi)\over \delta\varphi}\right\rangle_\sigma & = {\delta\hat{\bf S}_{\rm tot}(\langle\varphi\rangle_\sigma)\over \delta \langle\varphi\rangle_\sigma} + \sum_{\{\sigma_i\} \ne \sigma}\left\langle {\delta\hat{\bf S}_{\rm tot}(\varphi)\over \delta\varphi}\right\rangle_{(\{\sigma_i\}\vert\sigma)} + 
~\left\langle {\delta\hat{\bf S}_{\rm tot}(\varphi)\over \delta\varphi}\right\rangle_{(\hat{\bf Q}\vert\sigma)}\\
& =  \left\langle
{\delta\over \delta\varphi}\log\left(\mathbb{D}_o^\dagger \mathbb{D}_o\right)\right\rangle_\sigma, \\
\end{split}
\nd
where $\hat{\bf S}_{\rm tot}(\varphi)$ is given by \eqref{katuli} and 
$\mathbb{D}_o(\sigma, \varphi)$ by \eqref{enmillboro}. In the presence of Faddeev-Popov ghosts $\psi$ we can replace $\hat{\bf S}_{\rm tot}(\varphi)$
by $\hat{\bf S}_{\rm full}(\varphi, \psi) \equiv \hat{\bf S}_{\rm tot}(\varphi) + {\bf S}_{1}^{({\rm g})}(\varphi, \psi) + {\bf S}^{(g)}_{\rm nloc}(\varphi, \psi)$, and in this case \eqref{palpenti} gets replaced by the following EOM:
\bg\label{palpucki}
\begin{split}
\left\langle {\delta\hat{\bf S}_{\rm full}(\varphi, \psi)\over \delta\varphi}\right\rangle_\sigma & = {\delta\hat{\bf S}_{\rm tot}(\langle\varphi\rangle_\sigma)\over \delta \langle\varphi\rangle_\sigma} + \sum_{\{\sigma_i\} \ne \sigma}\left\langle {\delta\hat{\bf S}_{\rm tot}(\varphi)\over \delta\varphi}\right\rangle_{(\{\sigma_i\}\vert\sigma)} + 
~\left\langle {\delta\hat{\bf S}_{\rm tot}(\varphi)\over \delta\varphi}\right\rangle_{(\hat{\bf Q}\vert\sigma)}\\
& + \left\langle {\delta{\bf S}_{1}^{({\rm g})}(\varphi, \psi)\over \delta\varphi}\right\rangle_\sigma + \left\langle {\delta{\bf S}_{\rm nloc}^{({\rm g})}(\varphi, \psi)\over \delta\varphi}\right\rangle_\sigma 
=  \left\langle
{\delta\over \delta\varphi}\log\left(\mathbb{D}_o^\dagger \mathbb{D}_o\right)\right\rangle_\sigma, \\
\end{split}
\nd
where the ghost contributions may be explained in the following way. Consider the original action ${\bf S}_{\rm tot}(\varphi, \phi)$ from \eqref{enmilltriangle}. Add to it the ghost action ${\bf S}_{\rm ghost}(\varphi, \phi, \psi)$, where $\psi$ is the set of ghost fields, and then split ${\bf S}_{\rm ghost}(\varphi, \phi, \psi)$ as ${\bf S}_{\rm ghost}(\varphi, \phi, \psi) = {\bf S}^{(1)}_{\rm ghost}(\varphi, \psi) + {\bf S}^{(2)}_{\rm ghost}(\varphi, \phi, \psi)$. Following the same strategy that dictated the non-local interactions from \eqref{TIogyan}, we now claim:
\bg\label{TIogyan4}
\int {\cal D}\phi {\cal D}\psi~e^{-{\bf S}_2(\varphi, \phi) - {\bf S}^{(1)}_{\rm ghost}(\varphi, \psi) - {\bf S}^{(2)}_{\rm ghost}(\varphi, \phi, \psi)} ~ = ~ e^{-{\bf S}_{\rm nloc}(\varphi)} \int{\cal D}\psi~ e^{- {\bf S}_1^{(g)}(\varphi, \psi) - {\bf S}_{\rm nloc}^{({\rm g})}(\varphi, \psi)}, \nd
which would keep the form of $\hat{\bf S}_{\rm tot}(\varphi)$ from \eqref{katuli} unchanged\footnote{It is clear that this is only possible if both ${\bf S}^{({\rm g})}_{1}(\varphi, \psi)$ and ${\bf S}_{\rm nloc}^{(g)}(\varphi, \psi)$ are expressed using convergent series. Failure to maintain this would result in changing ${\bf S}_{\rm NP}(\varphi)$.}, but introduce two new ghost contributions: ${\bf S}_{1}^{({\rm g})}(\varphi, \psi)$ and ${\bf S}_{\rm nloc}^{({\rm g})}(\varphi, \psi)$, both defined in terms of on-shell field component $\varphi$. These changes \eqref{palpenti} to \eqref{palpucki}. The EOM \eqref{palpucki}, while much more manageable than \eqref{elenacigu2}, still requires some adjustments to bring it in a form where we can identify it to a more realistic scenario. By {\it realistic} we mean a scenario close to the vacuum solution that we studied earlier in section \ref{sec2}. Looking at {\bf Table \ref{firoksut}}, we see that the terms in the energy-momentum tensors that would {\it almost} reproduce a de Sitter {\it vacuum} solution are the ones coming from the non-perturbative and the non-local parts of the action as elucidated in section \ref{sec2.1.4}. Unfortunately, and as discussed in section \ref{sec3}, the construction fails because of the absence of a well-defined Wilsonian effective action over a temporally varying de Sitter vacuum among other issues. On the other hand, if we demand that the equation in \eqref{palpucki} splits into the following two equations:

{\scriptsize
\bg\label{eventhorizon}
\boxed{\begin{split}
& ~~~ {\delta\hat{\bf S}_{\rm tot}(\langle\varphi\rangle_\sigma)\over \delta \langle\varphi\rangle_\sigma} ~ = ~ 0\\
& \sum_{\{\sigma_i\} \ne \sigma}\left\langle {\delta\hat{\bf S}_{\rm tot}(\varphi)\over \delta\varphi}\right\rangle_{(\{\sigma_i\}\vert\sigma)} + 
~\left\langle {\delta\hat{\bf S}_{\rm tot}(\varphi)\over \delta\varphi}\right\rangle_{(\hat{\bf Q}\vert\sigma)} +  \left\langle {\delta{\bf S}_1^{({\rm g})}(\varphi, \psi)\over \delta\varphi}\right\rangle_\sigma + \left\langle {\delta{\bf S}_{\rm nloc}^{({\rm g})}(\varphi, \psi)\over \delta\varphi}\right\rangle_\sigma = ~ \left\langle
{\delta\over \delta\varphi}\log\left(\mathbb{D}_o^\dagger \mathbb{D}_o\right)\right\rangle_\sigma \\
\end{split}}
\nd}
then we are very close to the same scenario that we studied in section \ref{sec2} and especially in section \ref{sec2.1.4}! However now there are a few key differences. 

\vskip.1in

\noindent ${\bf 1.}$ The first EOM in \eqref{eventhorizon} is now for the {\it emergent} on-shell degree of freedom $\langle\varphi\rangle_\sigma$ and {\it not} for the vacuum configuration. Resorting to actual metric, flux and fermionic degrees of freedom, an equivalent picture\footnote{For example if $\langle\Xi\rangle_\sigma = (\langle{\bf g}_{\rm AB}\rangle_\sigma, \langle{\bf C}_{\rm ABC}\rangle_\sigma, \langle\Psi_{\rm A}\rangle_\sigma, \langle \overline{\Psi}_{\rm A}\rangle_\sigma)$, then we are effectively looking at EOMs of the form ${\delta\hat{\bf S}_{\rm tot}(\langle\Xi\rangle_\sigma)\over \delta \langle\Xi\rangle_\sigma} ~ = ~ 0$. This would lead to EOMs as in {\bf Table \ref{firoksut}} but for the emergent on-shell degrees of freedom $\langle {\bf g}_{\rm AB}\rangle_\sigma, \langle {\bf C}_{\rm ABC}\rangle_\sigma$ and $\langle \Psi_{\rm A}\rangle_\sigma$.} would lead to the EOMs for the emergent on-shell degrees of freedom $\langle {\bf g}_{\rm AB}\rangle_\sigma, \langle {\bf C}_{\rm ABC}\rangle_\sigma$ and $\langle \Psi_{\rm A}\rangle_\sigma$, where $({\rm A, B, C}) \in {\bf R}^{2, 1} \times {\cal M}_4 \times {\cal M}_2\times {\mathbb{T}^2\over {\cal G}}$.

\vskip.1in

\noindent {\bf 2.} The second equation in \eqref{eventhorizon} is something that we did not encounter earlier. This equation weaves the three crucial elements in the construction together: the intermediate Glauber-Sudarshan states, the Faddeev-Popov ghosts (if any) and the linear term in the displacement operator. Note that all the three contributions cannot be visible at the ``classical'' level, {\it i.e.} at the level of the first equation in \eqref{eventhorizon}, thus explaining the natural split of the EOM \eqref{palpucki} into a set of two EOMs. 

\vskip.1in

\noindent ${\bf 3.}$ The issues which prohibited us to declare the results from {\bf Table \ref{firoksut}} as evidence for the existence of a de Sitter {\it vacuum} solution now no longer plagues us. In fact the first equation in \eqref{eventhorizon} clearly suggests that de Sitter space-time should be an {\it emergent} solution appearing from the Glauber-Sudarshan state $\vert\sigma\rangle$. The emergent background is now backed by a well defined effective field theory description that is  consistent with the trans-Planckian bound and other constraints.

\vskip.1in

\noindent There is however a bit more to the above story. The second equation in \eqref{eventhorizon} is not the only {\it non-classical} equation that we encounter here. There is yet another set of equations that could appear from the EOM associated from the off-shell field $\phi$. This is \eqref{kinkisole}, and as before we have three possibilities.

\vskip.1in

\noindent ${\bf 1.}$ Rewrite it using the non-perturbative completion that we did for the on-shell field $\varphi$ in \eqref{elenacigu2}.

\vskip.1in

\noindent ${\bf 2.}$ Add in the Faddeev-Popov ghosts and see how the EOM changes to compare with \eqref{palpucki}.

\vskip.1in

\noindent ${\bf 3.}$ Try to see if the equation could be split into two equations as in \eqref{eventhorizon} for the on-shell case.

\vskip.1in

\noindent All the three steps above are now more involved than their on-shell counterparts. To see this, let us start by answering the first two points above, namely the non-perturbative completion of the action and the insertion of the Faddeev-Popov ghosts. For that we can express the Schwinger-Dyson's equation for $\phi$ field in the following way:

{\scriptsize
\bg\label{marbello}
\int {\cal D}\varphi {\cal D}\phi {\cal D}\psi~e^{-{\bf S}_1(\varphi) - {\bf S}_2(\varphi, \phi) - {\bf S}^{(1)}_{\rm ghost}(\varphi, \psi) - {\bf S}^{(2)}_{\rm ghost}(\varphi, \phi, \psi)}~\mathbb{D}^\dagger(\sigma, \varphi) \mathbb{D}(\sigma, \varphi)
\left({\delta {\bf S}_2(\varphi, \phi)\over \delta\phi} + 
{\delta {\bf S}^{(2)}_{\rm ghost}(\varphi, \phi, \psi)\over \delta\phi}\right) = 0, \nd}
where we have followed the same decomposition for ${\bf S}_{\rm tot}(\varphi, \phi)$ and ${\bf S}_{\rm ghost}(\varphi, \phi, \psi)$ as before. One may compare this equation with the one we got earlier, namely \eqref{leeswift}, and it is clear that all the $\widetilde{f}_i$ functions appearing therein may be derived from ${\delta{\bf S}_2\over \delta\phi}$ above. A more elaborate comparison could be done, but we will leave this for our diligent readers to work them out. The above equation has the following non-perturbative completion:
\bg\label{lulupupu}
\int {\cal D}\varphi {\cal D}\psi ~  \mathbb{D}_o^\dagger(\sigma, \varphi) \mathbb{D}_o(\sigma, \varphi)~e^{-{\bf S}_{\rm kin}(\varphi) - {\bf S}_{\rm NP}(\varphi)- {\bf S}_{1}^{({\rm g})}(\varphi, \psi)} \int {\cal D}\phi~{\delta\over \delta\phi}\left(e^{- {\bf S}_2(\varphi, \phi)  - {\bf S}_{2}^{({\rm g})}(\varphi, \phi, \psi)}\right) = 0, \nd
where ${\bf S}_{1}^{({\rm g})}(\varphi, \psi)$ and ${\bf S}_{2}^{({\rm g})}(\varphi, \phi, \psi)$ appear from ${\bf S}^{(1)}_{\rm ghost}(\varphi, \psi)$ and ${\bf S}^{(2)}_{\rm ghost}(\varphi, \phi, \psi)$ respectively by integrating over the off-shell field $\phi$, and $\mathbb{D}_o(\sigma,\varphi)$ is from \eqref{enmillboro}. ${\bf S}_1^{(g)}(\varphi, \psi)$ as before is given by a convergent series, so that it doesn't effect the non-perturbative action. The second term in \eqref{lulupupu} can be integrated over to give us the following EOM:

{\scriptsize
\bg\label{belloliu}
\int {\cal D}\varphi {\cal D}\psi ~e^{-\hat{\bf S}_{\rm tot}(\varphi) - {\bf S}_{1}^{({\rm g})}(\varphi, \psi) - {\bf S}^{(g)}_{\rm nloc}(\varphi, \psi)}
\left(\mathbb{T}_{\rm nloc}(\varphi) + \mathbb{T}^{(g)}_{\rm nloc}(\varphi, \psi)\right)
~  \mathbb{D}_o^\dagger(\sigma, \varphi) \mathbb{D}_o(\sigma, \varphi) = 
\langle \mathbb{T}_{\rm nloc}(\varphi) + \mathbb{T}^{(g)}_{\rm nloc}(\varphi, \psi)\rangle_\sigma = 0, \nd}
where both ${\bf S}_{\rm nloc}(\varphi)$ and $\mathbb{T}_{\rm nloc}(\varphi)$ are related and so are ${\bf S}^{(g)}_{\rm nloc}(\varphi, \psi)$ and $\mathbb{T}^{(g)}_{\rm nloc}(\varphi, \psi)$. Moreover, $\mathbb{T}_{\rm nloc}(\varphi) = \widetilde{f}_2(\varphi) + {\rm T}_{\rm nloc}(\varphi)$ from \eqref{kinkisole}, and $\hat{\bf S}_{\rm tot}(\varphi)$ is the same one that appeared in \eqref{katuli}. Using the resolution of identity now splits \eqref{belloliu} into the following two equations:
\bg\label{necampbel}
\boxed{
\begin{split}
&\mathbb{T}_{\rm nloc}(\langle\varphi\rangle_\sigma) = 0\\
&\langle \mathbb{T}_{\rm nloc}(\varphi)\rangle_{(\{\sigma_i\}|\sigma)} + 
\langle \mathbb{T}_{\rm nloc}(\varphi)\rangle_{(\hat{\bf Q}|\sigma)} + 
\langle  \mathbb{T}^{(g)}_{\rm nloc}(\varphi, \psi)\rangle_\sigma = 0\\
\end{split}} \nd
where combining the first equations from \eqref{eventhorizon} and \eqref{necampbel} respectively would provide the necessary EOMs that would control the dynamics of the emergent de Sitter space-time from the Glauber-Sudarshan state. These are the equations that would replace the vacuum EOMs from {\bf Table \ref{firoksut}}. The remain two equations from \eqref{eventhorizon} and \eqref{necampbel} respectively would relate the dynamics of ghosts and the intermediate Glauber-Sudarshan states with the displacement operator $\mathbb{D}_o(\sigma, \varphi)$ from \eqref{enmillboro}.

The set of equations in \eqref{necampbel} signal the {\it non-classical} nature of the system and would not appear if we had taken coherent states. One may think of them as the first concrete sign of four-dimensional de Sitter space being an excited state coming from an expectation value of the graviton operator over a Glauber-Sudarshan state.
A specific consequence of these equations appear from the cross-term EOM of ${\bf g}_{0n}$ where $y^n \in {\cal M}_6$ despite the fact that ${\bf g}_{0n}$ is an off-shell state. A more detailed discussion of this will appear elsewhere \cite{hete8}.

Finally, in the third point after \eqref{eventhorizon}, we argued that the Glauber-Sudarshan states are well within the temporal bound advocated by the trans-Planckian censorship, and they always satisfy the EFT criteria\footnote{It is important to note that the Glauber-Sudarshan states are {\it coherent} only within the temporal regime governed by \eqref{tcc}, beyond which we expect the system to go to a supersymmetric Minkowski state. Thus in the Penrose diagrams, both in sections \ref{sec2} and \ref{sec3}, the relevant domains are the interval $-{1\over \sqrt{\Lambda}} < t < 0$. The system differs from KKLT \cite{kklt} in the sense that we no longer have a de Sitter vacuum configuration but a de Sitter excited state. Despite this certain similarities between the two pictures can be found. For example,
as in KKLT \cite{kklt}, the actual Penrose diagram beyond the TCC temporal regime will be hard to represent pictorially, but the mathematical analysis is much simpler. Interestingly, while in KKLT the system tunnels from a de Sitter minima to Minkowski, in our case the system transitions from being an excited state to a Minkowski vacuum state beyond the temporal domain \cite{tcc}.}.  In fact satisfying the trans-Planckian bound is enough to justify the latter \cite{vafareview}. However we would like to dwell a bit on this topic with the aim to dispel any doubts that reader might have regarding the existence of these states {\it without violating any {\rm EFT} criteria}. This will be the content of the following section.

\section{Glauber-Sudarshan states and Effective Field Theory criteria \label{sec7}}

After having got all the EOMs for the {\it emergent} on-shell degrees of freedom, it is time to ask how the Glauber-Sudarshan states themselves satisfy the EFT criteria. Of course the way we constructed these states over supersymmetric time-independent background clearly follow the necessary EFT constraints $-$ coming from the ERG procedure and from the restriction we imposed earlier to remain withing the energy range $k_{\rm IR} < k < \mu << \hat\mu$ as depicted in {\bf figure \ref{scales}} $-$ so it is necessary to explain what we mean by ``satisfying the EFT criteria''. Somewhat surprisingly, even after following the aforementioned procedure to construct the Glauber-Sudarshan states, and from there determine the dynamics of the emergent on-shell degrees of freedom, it is still necessary to satisfy the following {\it two} criteria \cite{coherbeta2}:
\bg\label{atryan}
\boxed{g_s ~ < ~ 1, ~~~~~~~~~ {\partial g_s\over \partial t} ~ \propto ~ g_s^{\rm +ive}} \nd 
where $g_s$ is the one that we encountered in \eqref{mmetric}, and is the dual type IIA coupling with $t$ being the conformal time. The first criterion leads to \eqref{tcc} which restricts the Glauber-Sudarshan states to lie within the temporal domain \eqref{tcc}. The second criterion boils down to the statement that the Null Energy Condition (NEC) should {\it not} be violated: any violation of the NEC is a violation of the underlying EFT \cite{coherbeta2}. One might wonder how do \eqref{atryan} appear from the two set of equations \eqref{eventhorizon} and \eqref{necampbel}. In the following sections we will first give a warm-up example and then, in section \ref{sec7.2} onwards, we will take a more realistic example and argue that many of the EFT criteria alluded to earlier are in fact tied up to \eqref{atryan}. \textcolor{blue}{Readers wishing to delve directly to the application of \eqref{atryan} on an actual de Sitter background, may skip the following section and go to section \ref{sec7.2}.}

\subsection{A warm-up example on the EFT criteria for the emergent states \label{sec7.1}}

The two sets of equations \eqref{eventhorizon} and \eqref{necampbel} contain all the necessary information to determine not only the evolution of the emergent on-shell degrees of freedom, but also to determine $\rho^\ast(k)$ in \eqref{robinright} necessary to construct the Glauber-Sudarshan states. Moreover, it is only the first equations in each of the two sets, that are actually needed to fix the form of $\rho^\ast(k)$. The other two equations aren't directly involved but are essential for the consistency of the underlying picture. The two relevant equations from \eqref{eventhorizon} and \eqref{necampbel} are:
\bg\label{ryanmukh}
{\delta \hat{\bf S}_{\rm tot}(\langle\varphi\rangle_\sigma)\over \delta\langle\varphi\rangle_\sigma} ~ = ~ 0 ~ = ~ \mathbb{T}_{\rm nloc}(\langle\varphi\rangle_\sigma), \nd
with $\hat{\bf S}_{\rm tot}(\varphi)$ is given by \eqref{katuli}. Despite appearance, \eqref{ryanmukh} is not exactly related to the classical EOM. Three key differences are as follows.

\vskip.1in

\noindent ${\bf 1.}$ The equations are for the {\it emergent} on-shell degree of freedom $\langle\varphi\rangle_\sigma$ and {\it not} for the on-shell field $\varphi$. 

\vskip.1in

\noindent ${\bf 2.}$ The action governing the emergent degree of freedom is not the classical action with field $\varphi$ replaced by $\langle\varphi\rangle_\sigma$, but differs from it by having non-local and non-perturbative contributions that are usually absent in a given classical action (or even with an action with perturbative corrections). 

\vskip.1in

\noindent ${\bf 3.}$ The second equation with $\mathbb{T}_{\rm nloc}$ is purely a consequence of having a Glauber-Sudarshan state, and therefore completely invisible from either the vacuum or the coherent state point of view. In a similar vein the other two equations, from the two sets \eqref{eventhorizon} and \eqref{necampbel} respectively, are also invisible from either the vacuum or the coherent state point of view.  

\vskip.1in

\noindent To see the effect of \eqref{ryanmukh} on the EFT, it would be best to first start with a toy model and then go to the original de Sitter space-time from the Glauber-Sudarshan state. For that let us specify the perturbative action as:
\bg\label{ladinpink}
\begin{split}
{\bf S}_{\rm pert}(c; \varphi) & = {\rm M}_p^{11}\int d^{11} x \left(-\varphi(x)~ {\square\over {\rm M}_p^2} ~\varphi(x) +~{\bf Q}_{\rm pert}(c; \varphi(x))\right)\\
&= {\rm M}_p^{11}\int d^{11} x \left(-\varphi(x, y)~ {\square\over {\rm M}_p^2} ~\varphi(x, y) + {1\over {\rm M}_p^{2n}}\sum_{n, p} c_{np} ~\partial^{2n} \cdot \varphi^p(x)\right),\\
\end{split}
\nd
where the action of the derivative follows the procedure outlined in footnote \ref{cchoice}, and $n \ge 0, p > 2$ to avoid generating mass term in the energy scale $k_{\rm IR} < k < \mu << \hat\mu$. The relevant action\footnote{This is an {\it intermediate} action because the actual action governing the evolution of the emergent state $\langle\varphi(x, y)\rangle_\sigma$ is $\hat{\bf S}_{\rm tot}(\langle\varphi\rangle_\sigma)$ which is clearly not a non-perturbative completion of ${\bf S}_{\rm pert}(c; \varphi)$ with $\varphi$ replaced by $\langle\varphi\rangle_\sigma$ from \eqref{ladinpink} as cautioned earlier.} is $\hat{\bf S}_{\rm tot}$, which doesn't have any perturbative pieces, and from \eqref{katuli} it may be expressed as:

{\scriptsize
\bg\label{kaittami}
\begin{split}
\hat{\bf S}_{\rm tot}(\varphi) & = {\rm M}_p^{11}\mathbb{V}_2
\int d^{3} x ~d^6y \Bigg\{-\varphi(x, y)~ {\square\over {\rm M}_p^2} ~\varphi(x, y) +
\sum_{s = 0}^\infty d_s~{\bf Q}_{\rm pert}\left(\bar{c}; \varphi(x, y)\right)~{\rm exp}\left(-s {\rm M}_p^6 \int_0^y d^6y' \big\vert {\bf Q}_{\rm pert}(\hat{c}; \varphi(y', x))\big\vert \right)\\
&~~~~~~ + \sum_{p = 1}^\infty 
b_p \left[{\rm exp}\left({-p{\rm M}_p^6\int_{{\cal M}_6} d^6y' \big\vert \mathbb{F}(y -y') {\bf Q}_{\rm pert}(\tilde{c}; \varphi(y', x))\big\vert}\right) - 1\right]\Bigg\},
\end{split}
\nd}
where $x \in {\bf R}^{2, 1}, y \in {\cal M}_6$, $w \in {\mathbb{T}^2\over {\cal G}}$ and we have restricted the on-shell field to be independent of the coordinates of the toroidal direction (whose volume is $\mathbb{V}_2$).
In \eqref{kaittami}, the first term is the kinetic term, the second  and the third are the non-perturbative and the non-local contributions respectively. Note that the latter two are expressed using ${\bf Q}_{\rm pert}$, but the coefficients differ. The non-perturbative term takes a trans-series form as one would have expected from the instanton saddles and the fluctuation determinants of \eqref{capquebcmey}. This may also be justified from the growth of the non-perturbative and the non-local terms in the path integral. Fixing a particular value of $p$ in \eqref{ladinpink}, these two terms grow respectively as:
\bg\label{claudannie}
{\left(p{\rm N}\right)!\over {\rm N}!}~{\rm exp}(-{\rm N}), ~~~~~~~ 
{1\over {\rm N}!}~{\rm exp}(-{\rm N}!), \nd
at ${\cal O}({\rm N})$. It is easy to see from {\bf figure \ref{growth}} that the first series is asymptotic\footnote{Since $p \ge 3$ in \eqref{ladinpink}, we have $\lim\limits_{{\rm N}\to \infty} ~{\left(p{\rm N}\right)!\over {\rm N}!}~{\rm exp}(-{\rm N}) = {\rm exp}\left[{\rm N}(p-1){\rm log}~{\rm N} + p{\rm N}({\rm log}~p - 1)\right]_{{\rm N} \to \infty} ~ \to ~ \infty$.} while the second one is convergent.
The asymptotic nature shows that the perturbative series from the fluctuation determinants {\it knows} about the next order non-perturbative effect. This is of course the precise requirement to apply the {\it resurgence theory} \cite{resurgence}, implying that the non-perturbative trans-series from \eqref{kaittami} itself may be studied using the resurgent trans-series when inserted in the path integral. Additionally, the generic formulation\footnote{We have avoided using eight-dimensional integral $d^8y$ because of the assumption that all fields are independent of the toroidal direction ${\mathbb{T}^2\over {\cal G}}$ in M-theory. Thus {\it generic} here is up to this assumption. We will relax this condition when we study a more realistic scenario later.} of the two contributions is important as it would specify the structure of these effects which, with appropriate choices of the $(d, b, \bar{c}, \tilde{c}, \hat{c})$ coefficients, may be tuned to any specific construction. In fact a more appropriate way of representing the $c$-parameters would be to demand:
\bg\label{dusamey}
\bar{c} \equiv \bar{c}(s), ~~~~ \hat{c} \equiv \hat{c}(s), ~~~~ \tilde{c} \equiv \tilde{c}(p), \nd
this way the non-perturbative series would spell out exactly what we had in \eqref{capquebcmey}, namely, for every instanton saddle, the corresponding fluctuation determinant would be expectedly different. Similar story would resonate for the non-local terms too: at every order of the expansion, the corresponding quantum terms would be distributed differently. On the other hand, the absence of any fluctuation determinants in the non-local quantum series may strike as odd, although so far we have seen no inconsistencies with it. However there is an issue that remains invisible with scalar degrees of freedom till we study the EOMs with actual metric and flux components. We will come back to it soon. Meanwhile
the Schwinger-Dyson equation, from the first equation in \eqref{ryanmukh}
in the scalar field set-up, becomes:

\begin{figure}
    \centering
    \includegraphics[scale=0.8]{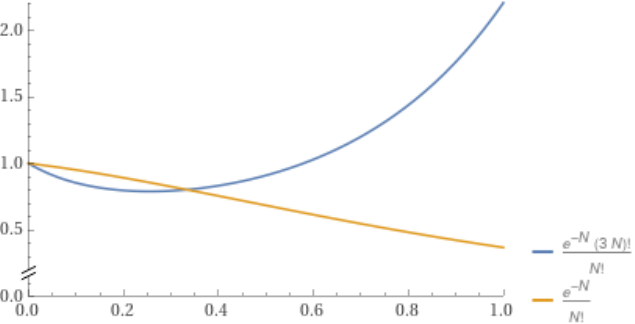} 
    \caption{The growths of the non-perturbative  series (in \textcolor{blue}{blue}) and the non-local series (\textcolor{orange}{orange}) at ${\cal O}({\rm N})$, with $c_{np} = 0$ except $c_{n3}$ in \eqref{ladinpink}. The first one is asymptotic while the second one is convergent. The trans-series form in \eqref{kaittami} leading to the asymptotic growth can be precisely tackled by the resurgence theory.}
    \label{growth}
\end{figure}
{\scriptsize
\bg\label{leahcall}
\begin{split}
 & ~ \square \textcolor{red}{\langle\varphi(x, y)\rangle_\sigma} ~
 - ~\sum_{s= 0}^\infty d_s~{\delta {\bf Q}_{\rm pert}(\bar{c}(s); \textcolor{red}{\langle\varphi(x, y)\rangle_\sigma})\over \delta\textcolor{red}{\langle\varphi(x, y)\rangle_\sigma}}~ ~{\rm exp} \left(-s{\rm M}_p^6 \int_0^{y} d^6y' \big\vert {\bf Q}_{\rm pert}(\hat{c}(s);\textcolor{red}{\langle\varphi(y', x)\rangle_\sigma})\big\vert\right)\\
 & + ~{\rm M}_p^6 \sum_{s = 1}^\infty s ~d_s \int_{{\cal M}_6} d^6 y'~
{\bf Q}_{\rm pert}(\bar{c}(s); \textcolor{red}{\langle\varphi(y', x)\rangle_\sigma})~{\delta 
 \big\vert{\bf Q}_{\rm pert}(\hat{c}(s);\textcolor{red}{\langle\varphi(y, x)\rangle_\sigma})\big\vert\over \delta \textcolor{red}{\langle\varphi(y, x)\rangle_\sigma}}~ \Theta(y' - y) ~{\rm exp} \left(-s{\rm M}_p^6 \int_0^{y'} d^6y'' \big\vert {\bf Q}_{\rm pert}(\hat{c}(s);\textcolor{red}{\langle\varphi(y'', x)\rangle_\sigma})\big\vert\right)\\
&+ ~ {\rm M}_p^6\sum_{p = 1}^\infty p ~b_p \int_{{\cal M}_6} d^{6} y'~{\delta\big\vert \mathbb{F}(y'-y){\bf Q}_{\rm pert}(\tilde{c}(p); \textcolor{red}{\langle\varphi(x, y)\rangle_\sigma})\big\vert\over \delta \textcolor{red}{\langle\varphi(x, y)\rangle_\sigma}}~  
{\rm exp}\left({-p{\rm M}_p^6\int_{{\cal M}_6} d^6 {y''} \big\vert \mathbb{F}(y'-{y''}) {\bf Q}_{\rm pert}(\tilde{c}(p);\textcolor{red}{\langle\varphi(x, {y''})\rangle_\sigma)\big\vert}}\right) ~ = ~ 0,\\
\end{split}
\nd}
where the perturbative series in \eqref{ladinpink} is expressed using 
${\textcolor{red}{\langle\varphi(x, y)\rangle_\sigma}}$. Note the conditional integral over $y'$ in the first line governed by the Heaviside step-function $\Theta(y' - y)$. The non-locality factor $\mathbb{F}(y'-y)$ is a bit of a concern now as one might worry that the EOM becomes non-local because of it's presence. This is actually {\it not} the case because the non-locality factor integrates out in the following way:
\bg\label{virperri}
{\rm M}_p^6\int d^{6} y'~\mathbb{F}(y'-y) ~  {\rm exp}\left({-p{\rm M}_p^6\int d^6 {y''} \big\vert \mathbb{F}(y'- {y''}) {\bf Q}_{\rm pert}(\tilde{c}(p); \langle\varphi(x, {y''})\rangle_\sigma)\big\vert}\right) ~\equiv ~ {\cal F}_p(x, y), \nd
thus leading to a local function! In doing the computation we have assumed that 
$\mathbb{F}(y_1- y_2) > 0$ for simplicity. If $\mathbb{F}(y_1- y_2) < 0$, then we can simply take $\left\vert\mathbb{F}(y_1 - y_2)\right\vert$ or 
$-\mathbb{F}(y_1 - y_2)$
in \eqref{virperri}. Combining \eqref{virperri} with \eqref{leahcall}, then provides the evolution dynamics for the emergent state $\langle\varphi(x, y)\rangle_\sigma$ in a potential given by:

{\scriptsize
\bg\label{masole2tois}
\begin{split}
{\bf V}(\langle\varphi(x, y)\rangle_\sigma) = & ~\langle\varphi(x, y)\rangle_\sigma~ {{\grad}^2\over {\rm M}_p^2} ~\langle\varphi(x, y)\rangle_\sigma-\sum_{s = 0}^\infty d_s~ {\bf Q}_{\rm pert}(\bar{c}(s); \langle\varphi(x, y)\rangle_\sigma) ~{\rm exp}\left(-s {\rm M}_p^6 \int_0^y d^6y' \big\vert {\bf Q}_{\rm pert}(\hat{c}(s); \langle\varphi(y', x)\rangle_\sigma)\big\vert \right) \\
& ~~~ - \sum_{p = 1}^\infty 
b_p \left[{\rm exp}\left({-p{\rm M}_p^6\int_{{\cal M}_6} d^6y' \big\vert \mathbb{F}(y -y') {\bf Q}_{\rm pert}(\tilde{c}(p); \langle\varphi(y', x)\rangle_\sigma)\big\vert}\right) - 1\right], \\
\end{split}
\nd}
where $\grad^2 = \eta^{ij}\partial_i\partial_j$ with mostly plus signature, and 
the overall signs for second and the third terms are controlled by the signs of $d_s$ and $b_p$ $\forall ~(s, b)$. In fact for large values of $s$ and $p$, the terms in the two exponential series die off, which is of course the reason for their convergence properties (recall the trans-series structure discussed earlier). Taking a derivative with respect to $\langle\varphi(x, y)\rangle_\sigma$ gives us the following functional form:

{\scriptsize
\bg\label{leahcall2}
\begin{split}
& {\delta {\bf V}(\langle\varphi(x, y)\rangle_\sigma)\over \delta \langle\varphi(x, y)\rangle_\sigma}  =  
~{\grad^2\over {\rm M}_p^2}~ \langle\varphi(x, y)\rangle_\sigma 
- ~\sum_{s= 0}^\infty d_s~{\delta {\bf Q}_{\rm pert}(\bar{c}(s); {\langle\varphi(x, y)\rangle_\sigma})\over \delta{\langle\varphi(x, y)\rangle_\sigma}}~ ~{\rm exp} \left(-s{\rm M}_p^6 \int_0^{y} d^6y' \big\vert {\bf Q}_{\rm pert}(\hat{c}(s);{\langle\varphi(y', x)\rangle_\sigma})\big\vert\right)\\
&+ ~ {\rm M}_p^6\sum_{p = 1}^\infty p ~b_p \int_{{\cal M}_6} d^{6} y' ~{\delta \big\vert \mathbb{F}(y'-y) {\bf Q}_{\rm pert}(\tilde{c}(p); \langle\varphi(x, y)\rangle_\sigma)\big\vert\over \delta \langle\varphi(x, y)\rangle_\sigma}~ 
{\rm exp}\left({-p{\rm M}_p^6\int_{{\cal M}_6} d^6 {y''} \big\vert \mathbb{F}(y'-{y''}) {\bf Q}_{\rm pert}(\tilde{c}(p); \langle\varphi(x, {y''})\rangle_\sigma)\big\vert}\right), \\
&+  ~{\rm M}_p^6 \sum_{s = 1}^\infty s ~d_s \int_{{\cal M}_6} d^6 y'~
{\bf Q}_{\rm pert}(\bar{c}(s); {\langle\varphi(y', x)\rangle_\sigma})~{\delta \big\vert{\bf Q}_{\rm pert}(\hat{c}(s); \langle\varphi(y, x)\rangle_\sigma)\big\vert\over \delta\langle\varphi(y, x)\rangle_\sigma}~ \Theta(y' - y) ~{\rm exp} \left(-s{\rm M}_p^6 \int_0^{y'} d^6y'' \big\vert {\bf Q}_{\rm pert}(\hat{c}(s); \langle\varphi(y'', x)\rangle_\sigma)\big\vert\right)\\
\end{split}
\nd}
where the overall sign of \eqref{leahcall2} now depends on the signs of 
$b_p$ and $d_s$ as well as the signs of ${\delta \left\vert{\bf Q}_{\rm pert}(\hat{c}(s); \langle\varphi(y, x)\rangle_\sigma)\right\vert\over \delta\langle\varphi(y, x)\rangle_\sigma}, {\delta \big\vert \mathbb{F}(y'-y) {\bf Q}_{\rm pert}(\tilde{c}(p); \langle\varphi(x, y)\rangle_\sigma)\big\vert\over \delta \langle\varphi(x, y)\rangle_\sigma}$ and 
${\delta {\bf Q}_{\rm pert}(\bar{c}(s); \langle\varphi(y, x)\rangle_\sigma)\over \delta\langle\varphi(y, x)\rangle_\sigma}$; and we have assumed that 
$\mathbb{F}(y'-y) > 0$. In a more {\it realistic} case, which we are going to discuss soon, the signs of ${\delta \left\vert{\bf Q}_{\rm pert}(\hat{c}(s); \langle\varphi(y, x)\rangle_\sigma)\right\vert\over \delta\langle\varphi(y, x)\rangle_\sigma}, {\delta \big\vert \mathbb{F}(y'-y) {\bf Q}_{\rm pert}(\tilde{c}(p); \langle\varphi(x, y)\rangle_\sigma)\big\vert\over \delta \langle\varphi(x, y)\rangle_\sigma}$ 
and  ${\delta {\bf Q}_{\rm pert}(\bar{c}(s); \langle\varphi(y, x)\rangle_\sigma)\over \delta\langle\varphi(y, x)\rangle_\sigma}$
will depend on the specific configurations of local and global fluxes as well as the warp-factor and curvature components. Question is whether it is necessary for us to recover a condition like:
\bg\label{mcakasi}
{\delta {\bf V}(\langle\varphi(x, y)\rangle_\sigma)\over \delta \langle\varphi(x, y)\rangle_\sigma} ~ \ge ~ c_1 {\bf V}(\langle\varphi(x, y)\rangle_\sigma), \nd
for the effective emergent potential ${\bf V}(\langle\varphi(x, y)\rangle_\sigma)$. To answer this, first note that the vacuum Minkowski space (or more appropriately a warped Minkowski spacetime with an internal compact non-K\"ahler space) does not allow the potential to be ``uplifted'' beyond the Minkowski minima. Secondly, the equation of motion from \eqref{leahcall}, that fixes a form of $\langle\varphi(x, y)\rangle_\sigma$ as well as that of the Glauber-Sudarshan state $\vert\sigma\rangle$ will always be:
\bg\label{alto2mei}
{\delta {\bf V}(\langle\varphi(x, y)\rangle_\sigma)\over \delta \langle\varphi(x, y)\rangle_\sigma} ~ = ~ {\partial^2\over \partial t^2} \langle\varphi(x, y)\rangle_\sigma, \nd
and {\it never} ${\delta{\bf V}(\langle\varphi\rangle_\sigma)\over \delta\langle\varphi\rangle_\sigma} = 0$,
implying that the emergent degrees of freedom $-$ from metric, fluxes and fermions $-$ should always have temporal dependence. This fits well with the EFT criteria analysed in \cite{desitter2, coherbeta, coherbeta2}, namely that if the emergent degrees of freedom become time-independent then there is a loss of both $g_s$ and ${\rm M}_p$ hierarchies implying a breakdown of EFT\footnote{One might question as to what happens in the static patch. First of course static patch doesn't mean that there are no temporal dynamics. As shown in {\bf figure \ref{staticpatch3}} the hidden temporal behavior actually requires us to keep the degrees of freedom with certain amount of temporal dependence. Moreover, the dual IIA string coupling $g_s$, albeit looks time independent in \eqref{polter2}, has other issues that we will discuss in section \ref{sec7.2}. These arguments confirm once again that \eqref{alto2mei} is the correct EOM for the emergent states. We will discuss more on this as we go along. \label{polkchabik}}. This is captured by the second criterion in \eqref{atryan}. 

\begin{figure}
    \centering
    \includegraphics[scale=0.8]{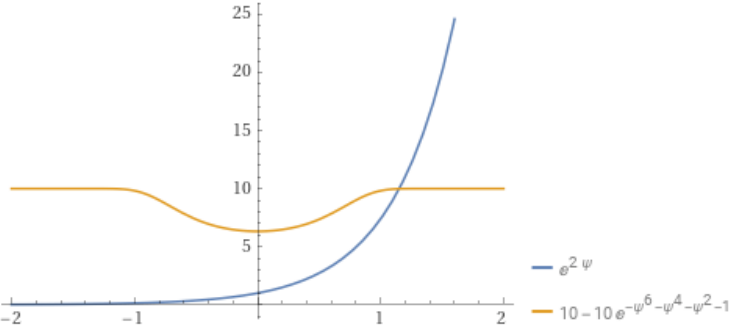} 
    \caption{Plot of two functions ${\bf V}_1 = e^{2\psi}$ (in \textcolor{blue}{blue}) and 
    ${\bf V}_2 = 10 - 10 e^{-\psi^6 - \psi^4 - \psi^2 - 1}$ (in \textcolor{orange}{orange}), with $\psi \equiv \langle\varphi\rangle_\sigma$. ${\bf V}_1$ appears from the first equation in \eqref{milenjon} with $c_1 = 2$, and ${\bf V}_2$ appears from \eqref{masole2tois} with $d_s = \bar{c}(s) = 0, b_1 = 10$,  $\grad^2\psi = \partial_0^2\psi = 0$ and $b_p = 0$ for $p > 1$. Such a configuration doesn't satisfy \eqref{alto2mei}, and therefore we expect the EFT criteria outlined in \eqref{atryan} to not be satisfied either.}
    \label{swamp1}
\end{figure}

The above discussion means that \eqref{mcakasi} is {\it unnecessary}. The derivative of the emergent potential, {\it i.e.} ${\delta{\bf V}(\langle\varphi\rangle_\sigma)\over \delta\langle\varphi\rangle_\sigma}$ never hits zero, even at the so-called ``classical'' level because of 
\eqref{alto2mei}. Thus there appears to be no strong reason why the EFT criteria of \cite{swampland} should hold for the emergent potential \eqref{masole2tois} although \eqref{atryan} should continue to hold. The worry that de Sitter space may not exist due to vanishing minima of the potential is now unfounded. For historical reasons however we note that 
the coefficient $c_1$ is positive, and it's value may be easily extracted from the integral form in \eqref{leahcall2}, but we will not do so here. In the presence of multiple fields, we can try to find the second derivative and again ask which of the following two conditions:
\bg\label{milenjon}
\begin{split}
&{\delta {\bf V}(\{\langle\varphi_l(x, y)\rangle_\sigma\})\over \delta \langle\varphi_j(x, y)\rangle_\sigma} ~ \ge ~ c_1 {\bf V}(\{\langle\varphi_l(x, y)\rangle_\sigma\})\\
{\rm min}&\left({\delta^2 {\bf V}(\{\langle\varphi_l(x, y)\rangle_\sigma\})\over \delta \langle\varphi_i(x, y)\rangle_\sigma \delta\langle\varphi_j(x, y)\rangle_\sigma}\right) ~ \le ~ -c_2 {\bf V}(\{\langle\varphi_l(x, y)\rangle_\sigma\}),\\ 
\end{split}
\nd
are necessary to capture the EFT criteria alluded to earlier
for the set of emergent fields $\{\langle\varphi_l(x, y)\rangle_\sigma\}$. Following \cite{swampland} one might expect the answer to be the first of the two in \eqref{milenjon}, because the Glauber-Sudarshan states are defined in the temporal domain \eqref{tcc} $-$ which appears from the first criterion in \eqref{atryan} $-$ and so first condition in \eqref{milenjon} should be automatically satisfied. One might even take an emergent field 
that satisfies $\grad^2 \langle\varphi(x, y)\rangle_\sigma = \partial_0^2 \langle\varphi(x, y)\rangle_\sigma = 0$ (a plot of such a scenario is presented in {\bf figure \ref{swamp1}}) and compare the first equation in \eqref{milenjon} with the potential in \eqref{masole2tois}.

Needless to say, since the emergent potential has a minimum, the condition \eqref{mcakasi}, or the first criterion in \eqref{milenjon}, is not satisfied. As mentioned earlier, this is of no concern because the emergent potential is {\it not} required to satisfy either of the two conditions in \eqref{milenjon}. Additionally, the derivative of the 
 emergent potential in \eqref{leahcall2} cannot vanish. We do however need to satisfy the EFT criteria outlined in \eqref{atryan}. For this  it is necessary to not only include the consequence from the temporal derivative of the emergent field, but to also keep $\bar{c}(s) \ne 0$ in \eqref{leahcall2}. The latter would imply the inclusion of the fluctuation determinants over each instanton saddles. 
 As an example, let us consider a toy example motivated from the scalar field theory that we discussed here. We will only consider the non-perturbative sector in the potential \eqref{masole2tois} coming from the instanton saddles and their corresponding fluctuation determinants. To make this precise, we can choose $\hat{c}(s)$ and $\bar{c}(s)$ such that:
\bg\label{mbbrown}
\begin{split}
& {\bf Q}_{\rm pert}(\bar{c}(s); \langle\varphi(x, y)\rangle_\sigma) = \sum_{q = 0}^\infty {c_{sq}\over \sqrt{2^q q!}} ~\mathbb{H}_q\left(\langle\varphi(x, y)\rangle_\sigma\right)\\
& {\rm M}_p^6\int_0^y d^6y'~ \big\vert {\bf Q}_{\rm pert}(\hat{c}(s); \langle\varphi(y', x)\rangle_\sigma)\big\vert = {1\over 2s} \sum\limits_{p = 0}^\infty {\vert h_{sp}~\mathbb{H}_p(\langle\varphi(x, y)\rangle_\sigma)\vert \over \sqrt{2^p p!}},\\
\end{split}
\nd
where $\mathbb{H}_q\left(\langle\varphi(x, y)\rangle_\sigma\right)$ are the dimensionless Hermite polynomials $\forall q \in \mathbb{Z}$, and $h_{sp}, c_{sq}$ are constants. The way we have expressed the above form of the perturbative corrections, one could make this fit with any given theory by appropriately taking a linear combination of the Hermite polynomials using $c_{sq}$ and $h_{sp}$ without compromising any generalities. Using these the non-perturbative potential may be expressed in the following way:
\bg\label{claunikind}
{\bf V}_{\rm NP}(\langle\varphi(x, y)\rangle_\sigma) = -\sum\limits_{s, q=0}^\infty {d_s c_{sq} \over \sqrt{2^q q!}} ~\mathbb{H}_q\left(\langle\varphi(x, y)\rangle_\sigma\right)~{\rm exp}\left(-{1\over 2}\sum\limits_{p = 0}^\infty {\vert h_{sp}~\mathbb{H}_p(\langle\varphi(x, y)\rangle_\sigma)\vert \over \sqrt{2^p p!}}\right), \nd
with $d_s$ appearing from the second term in \eqref{masole2tois}. With appropriate choice of the coefficients $d_s$, the non-perturbative potential could either be a normalizable or a non-normalizable function (or a combination thereof) in the field space governed by $\langle\varphi(x, y)\rangle_\sigma$. (Demanding a potential whose derivative is always bigger than the potential itself in the field space can be easily constructed but, as mentioned earlier, is not necessary here.) The total package with the fluctuation determinants inserted as in \eqref{claunikind} contributes in the right way to precisely restore the EFT criteria outlined in \eqref{atryan}. This again justifies what we said earlier: the EFT criteria are simply the ones from \eqref{atryan}. 

Therefore to summarize: The original motivation to impose \eqref{mcakasi} or \eqref{milenjon} was to eliminate four-dimensional de Sitter spacetime which would appear, for example, in the static patch as a {\it minimum} of the potential. Unfortunately both these motivations are on the wrong footing: de Sitter spacetime is never a vacuum solution, nor is the static patch a good coordinate choice to study de Sitter. The correct picture is that the de Sitter spacetime is generated from a Glauber-Sudarshan state with an emergent potential \eqref{masole2tois} satisfying \eqref{alto2mei}, implying that even in this language the de Sitter excited state is {\it not} a minimum of the emergent potential. This makes the two criteria in \eqref{milenjon} unnecessary but does lead to two other criteria \eqref{atryan} as the necessary requirements for the validity of EFT. 

\subsection{Glauber-Sudarshan states and the first EFT criterion from \eqref{atryan} \label{sec7.2}}

The lessons that we learnt from the warm-up example in the previous section with one on-shell field $\varphi(x, y)$ $-$ along-with the assumption that we have integrated out at least one off-shell field $\phi(x, y)$ $-$ are the following.
\vskip.1in

\noindent ${\bf 1.}$ The action that controls the dynamics of the emergent field $\langle\varphi(x, y)\rangle_\sigma$ is neither ${\bf S}_{\rm pert}(c; \varphi)$ from \eqref{ladinpink} nor $\hat{\bf S}_{\rm tot}(\varphi)$ from \eqref{kaittami}. Rather it is given by 
$\hat{\bf S}_{\rm tot}(\langle\varphi\rangle_\sigma)$. 

\vskip.1in

\noindent ${\bf 2.}$ The effective field theory criteria that control the dynamics of the emergent degree of freedom, in the M-theory uplift of the type IIB scenario, are completely given in terms of the type IIA dual string coupling $g_s$ as in \eqref{atryan}. 

\vskip.1in

\noindent ${\bf 3.}$ The effective field theory criteria proposed in \cite{swampland} are {\it designed} to rule out four-dimensional de Sitter spacetime that is a {\it minimum} of a vacuum potential. Since the emergent de Sitter spacetime that we get (from an equally emergent potential) is neither a minimum of a potential nor controlled by a vacuum potential, the EFT criteria of \cite{swampland} play no role here and may thus be ignored. The only useful aspect that survives from the conjectures in \cite{swampland} is the trans-Planckian censorship conjecture which, here, appears from the first criterion of \eqref{atryan}. 

\vskip.1in

The last point regarding (a) the trans-Planckian {\it censorship} conjecture, and (b) the de Sitter spacetime not being the minimum of the emergent potential, deserves some discussion in light of the EFT criteria proposed in \eqref{atryan}. We will start with the trans-Planckian censorship conjecture and then discuss the issues surrounding \eqref{alto2mei}.

 The idea behind the trans-Planckian censorship conjecture is simple. It states that a reverse time evolution of excitations with ${\cal O}(1)$ energies living on a late time slice leads to trans-Planckian energies for the same at a finite time in the past. In terms of future time evolution, trans-Planckian modes in the past become increasingly long wavelength and subsequently freeze upon exiting the Hubble horizon. Thus at late times, one can classically detect trans-Planckian physics at ${\cal O}(1)$ scales, for instance, in the CMB data.

   If we are {\it only} considering a generic higher-derivative gravity theory then these trans-Planckian modes will be problematic because the short-distance behavior of such a theory itself is problematic. The non-normalizability issue, appearing at a scale ${\rm M}_p$ and discussed right at the start of section \ref{sec3}, is the root cause of the problem. Of course now we know that this isn't really an issue in string or M- theory because new degrees of freedom appear at that scale to render all UV amplitudes finite. In the original formulation \cite{tcc}, where the theory was restricted to obey Einstein gravity only, 
    the {\it trans-Planckian censorship conjecture} prohibits the existence of such a scenario by simply bounding the temporal domain for these UV modes to not exit the Hubble horizon. More precisely, the conjecture states that any solution where we can see a classical imprint of trans-Planckian features upon exiting the de Sitter horizon is not possible since the time scale is bounded by:
    \bg\label{tcccrit}
\frac{a_f}{a_i} ~ < ~ \frac{\rm M_{pl}}{\cal{H}}, 
    \nd
    where the labels $i,f$ denote the initial and final times, $a$ is the expansion parameter, ${\rm M_{pl}}$ is the four-dimensional Planck mass, and $\cal{H}$ is the Hubble parameter (not to be confused with the warp-factor ${\rm H}(y)$ used earlier). If the Hubble parameter is constant over time, then the time scale where the de Sitter state can be valid is given by:
    \bg\label{tempo}
\Delta t_{\text{TCC}} ~=~ \frac{1}{\cal{H}} ~\log \left(\frac{\rm M_{pl}}{\cal{H}}\right) ~=~ 
\frac{1}{\sqrt{\Lambda}} ~\log \left(\frac{\rm M_{pl}}{\sqrt{\Lambda}}\right),
    \nd
where $\Lambda$ is the four-dimensional cosmological constant. Assuming 
$\Lambda = {10}^{-120} {\rm M}_p^2$, we see that $\Delta t_{\rm TCC} = 
{138.55\over \sqrt{\Lambda}} ~\log\left({{\rm M_{pl}} \over {\rm M}_p}\right)$. In the flat-slicing, this temporal domain is precisely related to the temporal domain \eqref{tcc} got by demanding the first EFT criterion from \eqref{atryan}, namely $g_s < 1$. 

In light of what we said above, the similarity between the two temporal domain is a bit surprising. The original trans-Planckian censorship conjecture deals with Einstein gravity (or more appropriately, generic higher derivative gravity) only whereas our analysis is in full string theory with the UV completeness inherently built in. (In fact the UV finiteness manifests as finite Wilsonian coefficients of the interactions in the low energy EFT. These are exactly the finite coefficients $(d_s, b_p, \bar{c}(s), \hat{c}(s), \tilde{c}(p))$ that we encountered in the toy set-up of \eqref{kaittami}.) There are no pathologies for the far UV modes in our set-up, so the natural question is why would the two temporal domain match-up so nicely. Moreover, {\it nothing really bad happens} if we go beyond the temporal domain \eqref{tcc}. We simply lose control of the non-perturbative and non-local interactions that go as ${\rm exp}\left(-{1\over g_s^p}\right)$, for $g_s > 1$ in the M-theory side, implying that the system no longer has a Glauber-Sudarshan state beyond the temporal domain \eqref{tcc}. Four-dimensional de Sitter space would then exist only in this limited temporal domain and beyond that the theory might simply go back to the warped-Minkowski description. This conclusion matches somewhat with the {\it quantum breaking time} of \cite{dvali} although the specifics of the detail differ.

There is yet another matter of contention here.
It can be shown that imposing the trans-Planckian censorship conjecture for a scalar field with say, a monotonically increasing potential, leads to satisfying the first EFT condition of \eqref{milenjon} $-$ which we will call as the ``de Sitter potential conjecture'' \cite{vafareview, swampland}. However in contrast to the de Sitter potential conjecture which does not allow for a de Sitter vacuum solution, the trans-Planckian censorship conjecture does allow for vacuum solution provided they satisfy the temporal bound \eqref{tempo}. Our analysis has ruled out any possibility of a de Sitter vacuum solution, so the trans-Planckian conjecture should be re-interpreted appropriately for the case with  Glauber-Sudarshan states. Putting things together, it appears that there are at least {\it three} different computations that lead to similar temporal domain of validity.
\vskip.1in

\noindent ${\bf 1.}$ The weak coupling $g_s < 1$ limit for the Glauber-Sudarshan states.

\vskip.1in

\noindent ${\bf 2.}$ The quantum breaking time within the standard formulation of the coherent states \cite{dvali}.

\vskip.1in

\noindent ${\bf 3.}$ The trans-Planckian censorship conjecture with Einstein, or higher derivative, gravity \cite{robert1}.

\vskip.1in

\noindent All three computations are different, and in particular the first and the second computations differ because the Glauber-Sudarshan states by construction are different from the coherent states. Why would then the temporal domain of validity match up in the aforementioned three computations? First, of course this match-up is not precise. For example \eqref{tcc} doesn't exactly equate to say 
\eqref{tempo}: there is some difference of factors, although if we construct de Sitter Glauber-Sudarshan state in global coordinates we can come very close to matching with \eqref{tempo}. But this is besides the point, and it doesn't answer the pertinent question of {\it why} is there any equivalence at all?

The answer may lie on the fact that, a subset of these cases (especially the first and the third ones), rely on a specific ingredient of the computation: the temporal variation of the fluctuating frequencies. This temporal variation appears to source both the trans-Planckian issue with Einstein, or higher derivative, gravity and the breakdown of a Wilsonian integrating-out procedure within string theory. The latter of course led to the formulation of the Glauber-Sudarshan states as discussed in much detail in section \ref{sec3}. While the ingredient is similar in both cases, the consequences are quite different. In the third case, the UV finiteness becomes the main issue, whereas in the first case, while the UV finiteness is no longer an issue, the convergence property of the non-perturbative and the non-local quantum series now becomes problematic. (The second issue with quantum break time for the coherent states \cite{dvali} should also come under similar scrutiny but we will not discuss it here.) Question is whether there is any quantity in string theory that can succinctly capture all the aforementioned issues. The answer turns out to be the dual type IIA coupling $g_s$. By definition it can be expressed in terms of the Hubble parameter $\cal{H}$ $-$ which appears from the T-duality rules $-$ and therefore keeping $g_s < 1$
takes care of both the convergence property of the instanton and non-local series, and the UV problems of Einstein gravity. \textcolor{red}{Thus this is one of the main reason for choosing $g_s < 1$ as our first EFT criterion in \eqref{atryan}}.

\begin{figure}
    \centering
    \includegraphics[scale=0.8]{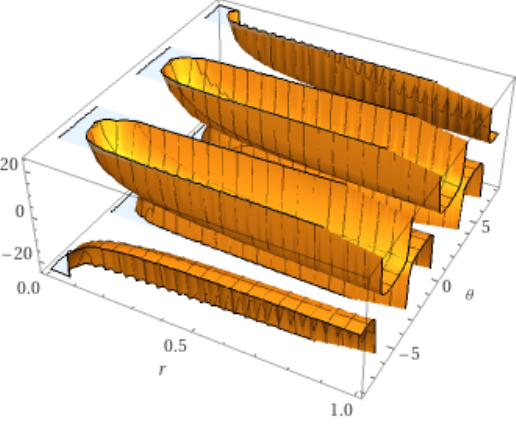} 
    \caption{Plot $g_s = {g_b~{\rm H}(y)\over r}~{\rm cosec}~\theta$ for $0 \le r \le 1$, and $-8 \le \theta \le 8$. We have also taken $g_b = {\rm H}(y) \equiv 1$. The former equality is to remain at the constant coupling point of F-theory, and the latter is to simplify the analysis.}
    \label{gsplot}
\end{figure}

Another more subtle reason for choosing $g_s < 1$ appears when we try to 
study a de Sitter state in a {\it static patch} (see also footnote \ref{polkchabik}). If $\vert\alpha\rangle$ denotes the Glauber-Sudarshan state in type IIB theory, then the emergent type IIB metric in a static patch takes the following form:

{\footnotesize
\bg\label{jobeth}
\begin{split}
ds^2  &= \langle {\bf g}_{\rm MN}\rangle_\alpha d {\rm Y}^{\rm M} d{\rm Y}^{\rm N} \\
& = {1\over {\rm H}^2(y)}\left[-(1- \Lambda r_s^2) dt_s^2 + 
{dr_s^2\over 1- \Lambda r_s^2 } + r_s^2\left(d\theta_s^2 + \sin^2\theta_s~d\varphi_s^2\right)\right] + {\rm H}^2(y) ~g_{mn}(y) dy^m dy^n\\
\end{split}
\nd}
where ${\rm Y}^{\rm M} = \left(t_s, r_s, \theta_s, \varphi_s, y^m\right)$ and $y^m \in {\cal M}_6$, with ${\rm H}(y)$ and $\Lambda$ being the warp-factor and the four-dimensional cosmological constant. The radial coordinate $r_s$ is bounded by $r_s < {1\over \sqrt{\Lambda}}$ as shown as the middle figure in {\bf figure \ref{staticpatch3}}. One would expect some emergent {\it static} flux configuration to support a background like \eqref{jobeth}. Unfortunately problem lies when we try to explore region $r_s > {1\over \sqrt{\Lambda}}$. The absence of a globally time-like Killing vector suggests that outside the static patch the Killing vector could become space-like or/and even change orientations. Interchanging the time-like and space-like coordinates would mean that outside the static patch a static flux configuration would develop temporal dependence leading to fluctuating frequencies shown on the extreme right of {\bf figure \ref{staticpatch3}}. The consequence of such a behavior has already been pointed out in section \ref{sec3.1}, so we will not elaborate the story further and instead discuss yet another problem  that may be tied up to the first EFT criterion of \eqref{atryan}. This appears when we uplift \eqref{jobeth} to M-theory by defining the type IIA dual coupling $g_s$ as:
\bg\label{polter2}
g_s ~ = ~ {1\over r_s}\left(g_b ~{\rm H}(y) ~ {\rm cosec}~\theta_s\right), \nd
where $g_b$ is the type IIB coupling that may be kept fixed at the constant coupling point of F-theory \cite{senkd}. Since ${\rm H}(y)$ is a smooth function, and $r_s$ is bounded by $0 \le r_s < {1\over \sqrt{\Lambda}}$, we see that $g_s$ is typically {\it bigger} than 1 along specific paths, {\it i.e.} $g_s > 1$, because $1 \le {\rm cosec}~\theta_s \le \infty$; implying that the non-perturbative and the non-local series that go as ${\rm exp}\left(-{1\over g_s^p}\right)$ with $p > 0$ cannot be made convergent (see also {\bf figure \ref{gsplot}})\footnote{More appropriately, at any point in $r_s$, there is at least one point in the angular direction $\theta_s$ for which $g_s$ blows up. What really matters is not that $g_s < 1$ at other points, but the fact that the instanton and non-local corrections are {\it not} convergent along certain directions in ${\bf R}^{2, 1}$ in M-theory. Since the action involves an integral over space-time, this naturally becomes a problem. Alternatively, if we absorb the spatial factors in the definition of ${\rm H}_o({\bf x})$, then in the static patch ${g_s\over {\rm H}(y) {\rm H}_o({\bf x})} = 1$.}. Thus the behavior in a static patch appears to be what we discussed earlier {\it outside the trans-Planckian bound}. Due to the UV completeness of string theory, one wouldn't expect things to go bad in this regime, but we do expect the Glauber-Sudarshan state to not exist in this regime.  

One might however wonder why is this the case if we compare it to say the flat-slicing. Going from type IIB to M-theory involves one T-duality, so in both flat-slicing and the static patch should involve one T-duality in each case respectively. Why is then the physics different? The answer has to do with the duality directions. For the case with a flat-slicing, the duality direction is $x_3$, whereas in the case with the static patch, the duality direction is $\varphi_s$. Comparing the two set of coordinate systems, it is easy to convince oneself that $\varphi_s$ is at least a function of {\it both} $x_3$ and the conformal time $t$ of the flat-slicing coordinates. This means, while in the flat-slicing case we are making a T-duality along $x_3$ direction, in the static patch we are actually making a T-duality along a {\it combination} of $x_3$ and $t$ directions. Since the duality directions are different, we don't expect the physics to be the same in the two cases. This should also explain why we see the variation of the frequencies {\it outside} the static-patch in {\bf figure \ref{staticpatch3}}.

The above arguments should convince the readers that the static patch construction clashes with the first criterion of \eqref{atryan}, and therefore we expect only those patches of emergent de Sitter space to exist that allow explicit temporal dependence of the underlying on-shell degrees of freedom. This is consistent with the EOM \eqref{alto2mei}, justifying what we said earlier: a de Sitter emergent state is never a minimum of the emergent potential, rather it satisfies \eqref{alto2mei}. To summarize, it appears that:

\vskip.1in

\noindent ${\bf 1.}$ The Minkowski minimum may not be uplifted further leading us to construct a four-dimensional de Sitter spacetime in type IIB theory as an emergent excited state over the Minkowski minimum.

\vskip.1in

\noindent ${\bf 2.}$ The four-dimensional emergent de Sitter state may not be well defined using a static patch leading us to consider only those patches that allow some inherent temporal dependence implying \eqref{alto2mei} as the correct EOM to study the dynamics of such a state.

\vskip.1in

\noindent Both the aforementioned results should also embody the second EFT criterion of \eqref{atryan}. How far this is realized in an actual set-up with metric and fluxes will be the topic of the following section.

\subsection{Glauber-Sudarshan states and the second EFT criterion from \eqref{atryan} \label{sec7.3}}

The second EFT criterion from \eqref{atryan} deals with the derivative of $g_s$ with respect to the conformal time $t$. In \cite{coherbeta2} we have already connected this to the non-violation of the Null Energy Condition (NEC), so we will not discuss this connection any further here. Instead we will see how, in a more realistic case with metric and fluxes, 
the second criterion from \eqref{atryan} is embedded in the EOMs from \eqref{ryanmukh}.

As mentioned earlier, the action governing the EOMs of the emergent space-time is neither the perturbative action (say \eqref{ladinpink}), nor the one with non-perturbative and the non-local completion (say \eqref{kaittami}), rather it is the second one with the metric and the fluxes replaced by their expectation values over any arbitrary Glauber-Sudarshan state. Both the state and the dynamics of the emergent space-time are fixed by the Schwinger-Dyson equations:
\bg\label{libhitftk0}
{\delta\hat{\bf S}_{\rm tot}(\langle {\bf \Xi}\rangle_\sigma)\over \delta \langle {\bf \Xi}\rangle_\sigma} ~ = ~ 0, \nd
where  ${\bf \Xi}$ is the set of on-shell fields 
${\bf \Xi} = ({\bf g}_{\rm AB}, {\bf C}_{\rm ABD}, {\bf \Psi}_{\rm A}, \overline{\bf \Psi}_{\rm A})$ with $({\rm A, B, D}) \in {\bf R}^{2, 1} \times {\cal M}_6 \times {\mathbb{T}^2\over {\cal G}}$; and $\hat{\bf S}_{\rm tot}$ is from \eqref{katuli}. There are also other equations accompanying \eqref{libhitftk0}, but we will not consider them here. A more complete analysis of the EOMs, including additional EOMs from the {\it off-shell} degrees of freedom (like the second constraint equation in \eqref{ryanmukh}), will be dealt in our upcoming work \cite{hete8}.

To get the form of $\hat{\bf S}_{\rm tot}$ we will follow the procedure as outlined earlier going from the perturbative action \eqref{ladinpink} to the non-perturbative and the non-local completion as in \eqref{kaittami}, and then finally replace the on-shell fields by their expectation value over any Glauber-Sudarshan state. The perturbative action takes the form:

{\scriptsize
\bg\label{rian432avn}
{\bf S}_{\rm pert}(c; {\bf \Xi}) = {\rm M}_p^9\int d^{11}x \sqrt{-{\bf g}_{11}}\left[{\bf R}_{11} + {\bf G}_4 \wedge \ast_{11}{\bf G}_4  + {\bf Q}_{\rm pert}(c; {\bf \Xi})\right] + 
{\rm M}_p^9\int {\bf C}_3 \wedge {\bf G}_4 \wedge {\bf G}_4 + {\rm M}_p^3 \int {\bf C}_3 \wedge \mathbb{X}_8, \nd}
where we have included all the kinetic terms for the metric and the three-form field, the topological terms, and the perturbative interactions. These have already been discussed earlier after \eqref{einstein2}. What we have ignored are the fermionic Rarita-Schwinger terms and the M2 and the M5-brane terms. (The latter should directly appear from the asymptotic nature of the perturbative series itself, so we will not worry too much about them now.) These contributions, and especially the fermionic and the world-volume {\it perturbative} terms, are pretty non-trivial and we will discuss them in our upcoming paper \cite{hete8}. This also means that henceforth the set of on-shell fields will be
${\bf \Xi} = ({\bf g}_{\rm AB}, {\bf C}_{\rm ABD})$ without the fermionic fields. In addition to what we see from \eqref{einstein2}, we have allowed an exhaustive list of perturbative terms denoted by ${\bf Q}_{\rm pert}(c; {\bf \Xi})$. This is similar to what we had in \eqref{QT2}, but without the non-perturbative contributions. For an easy access, we quote it again in the following way:
\bg\label{QT3}
{\bf Q}_{\rm pert}(c; {\bf \Xi}({\bf x}, y, w)) = \sum_{\{l_i\}, \{n_j\}} {{c}_{nl} \over {\rm M}_p^{\sigma_{nl}}} \left[{\bf g}^{-1}\right] \prod_{j = 0}^3 \left[\partial\right]^{n_j} \prod_{k = 1}^{60} \left({\bf R}_{\rm A_k B_k C_k D_k}\right)^{l_k} \prod_{p = 61}^{100} \left({\bf G}_{\rm A_p B_p C_p D_p}\right)^{l_p}, \nd
where ${c}_{nl}$ are dimensionless constants for a given values in the set $(n, l) \equiv (\{n_i\}, \{l_j\})$; $(n_i, l_j) \in (+\mathbb{Z}, +\mathbb{Z})$;
$(n_0, n_1, n_2, n_3)$ are the number of derivatives along temporal, ${\bf R}^2, {\cal M}_6$ and ${\mathbb{T}^2\over {\cal G}}$ respectively; and 
$\sigma_{nl} \equiv \sum\limits_{j = 0}^3 n_j + 2\sum\limits_{k = 1}^{60} l_k + \sum\limits_{p = 61}^{100} l_p$ fixes the ${\rm M}_p$ dimension of the operators for a given choice of $(n, l)$ \cite{desitter2, coherbeta, coherbeta2}. All the curvature and the flux tensors, including the inverse metric components are expressed in terms of their warped form, and therefore once we take their expectation values over the appropriate Glauber-Sudarshan state they would include the $g_s$ factors. This in particular implies that \eqref{mmetric} may be re-expressed simply as:
\bg\label{mmetric2}
\begin{split}
ds^2 & = \langle {\bf g}_{\rm AB}\rangle_\sigma dx^{\rm A} dx^{\rm B}\\
& = g_s^{-8/3}\left(-dt^2 + dx_1^2 + dx_2^2\right) + g_s^{-2/3}{\rm H}^2(y) g_{mn}(y) dy^m dy^n + g_s^{4/3} \eta_{ab} dw^a dw^b,
\end{split}
\nd
with $g_s \equiv {\rm H}(y) \sqrt{\Lambda} t$, $t$ being the conformal time, and ${\rm H}(y)$ being the smooth warp-factor. The coordinate choice is as follows. $(w^a, w^b) = (x_3, x_{11}) \in {\mathbb{T}^2\over {\cal G}}$ with ${\cal G}$ being an orbifold action without a fixed point; and $(y^m, y^n) \in {\cal M}_6$ such that the internal eight-manifold is locally ${\cal M}_6 \times {\mathbb{T}^2\over {\cal G}}$.
We have also generically taken the on-shell fields to depend on all the eleven-dimensional coordinates which are distributed as above, {\it i.e.} $({\bf x}, y, w) \in ({\bf R}^2, {\cal M}_6, {\mathbb{T}^2\over {\cal G}})$. One may check that ${\partial g_s\over \partial t} \propto g_s^0$, consistent with the second EFT criterion from \eqref{atryan}.

The quantum series ${\bf Q}_{\rm pert}(c; {\bf \Xi})$ from \eqref{QT3} doesn't appear in the total effective action $\hat{\bf S}_{\rm tot}({\bf \Xi})$ directly as discussed earlier, but does appear in the construction of the non-perturbative and the non-local quantum terms. 

{\scriptsize
\bg\label{kaittami2}
\begin{split}
& \hat{\bf S}_{\rm tot}({\bf \Xi}) = {\bf S}_{\rm kin}({\bf \Xi}) + {\bf S}_{\rm NP}({\bf \Xi}) + {\bf S}_{\rm nloc}({\bf \Xi})\\ 
& = {\rm M}_p^{9}
\int d^{3} x ~d^6y~d^2w \sqrt{-{\bf g}_{11}({\rm X})} 
\left[{\bf R}_{11}({\rm X}) + {\bf G}_4({\rm X}) \wedge \ast_{11}{\bf G}_4({\rm X})\right] + {\rm M}_p^9\int  
{\bf C}_3({\rm X}) \wedge {\bf G}_4({\rm X}) \wedge {\bf G}_4({\rm X)} + {\rm M}_p^3\int {\bf C}_3({\rm X}) \wedge \mathbb{X}_8({\rm X})\\
& + {\rm M}_p^{11}\int d^3x~d^6y~d^2w \sqrt{-{\bf g}_{11}({\rm X})}~
\sum_{s = 0}^\infty d_s~{\bf Q}_{\rm pert}(\bar{c}(s); {\bf \Xi}({\rm X}))~{\rm exp}\left(-s {\rm M}_p^8 \int_0^y \int_0^w d^6y'~d^2w' \sqrt{{\bf g}_8({\rm Y}', x)} \big\vert {\bf Q}_{\rm pert}(\hat{c}(s); {\bf \Xi}({\rm Y}', x))\big\vert \right)\\
&+{\rm M}_p^{11}\int d^3x~d^6y~d^2w \sqrt{-{\bf g}_{11}({\rm X})}~\sum_{p = 1}^\infty 
b_p \left[{\rm exp}\left({-p{\rm M}_p^8\int_{{\cal M}_8} d^6y'~d^2w' 
\sqrt{{\bf g}_8({\rm Y}', x)}\big\vert \mathbb{F}(y -y'; w -w') {\bf Q}_{\rm pert}(\tilde{c}(p); {\bf \Xi}({\rm Y}', x))\big\vert}\right) - 1\right], \\
\end{split}
\nd}
where ${\rm X} \equiv (x, y, w)$, ${\rm Y}' \equiv (y', w')$, ${\cal M}_8 \equiv {\cal M}_6 \times {\mathbb{T}^2\over {\cal G}}$, and we have generically taken all the fields to be functions of $(x, y, w)$. The second line is the set of kinetic and the topological terms, the third line is the non-perturbative series expressed using sum over all the instanton saddles and their corresponding fluctuation determinants; 
and the last line is
the non-local quantum series expressed using the non-locality function $\mathbb{F}(y-y', w-w')$. Finally, all the quantum series are expressed using 
\eqref{QT3}, but with different values of the coefficients $c_{nl}$.

The action \eqref{kaittami2} is a more elaborate version of \eqref{kaittami} because the former includes all the real on-shell degrees of freedom like metric and fluxes. The non-perturbative term has the standard form of instanton saddles accompanied by the corresponding fluctuation determinants, but the non-local terms (as derived in \eqref{fepoladom}) appear with no such pre-factors. Are these pre-factors necessary, or we can simply suffice with what we have now? Clearly it appears that such pre-factors are not necessary at least in \eqref{kaittami}: the EOM in \eqref{leahcall} shows no untoward pathologies. We will however show that, in a more realistic case, the $g_s$ scaling forces us to introduce additional fluctuation determinants around each of the non-local saddles. To show this we will have to develop the story a bit more, and in the following we elaborate on this. 

The action controlling the dynamics of the evolution of the Glauber-Sudarshan states is not \eqref{kaittami2}, rather it is $\hat{\bf S}_{\rm tot}(\langle {\bf \Xi}\rangle_\sigma)$. The typical Schwinger-Dyson equations governing the evolutions of the emergent on-shell states are now:
\bg\label{trancom}
{\delta \hat{\bf S}_{\rm tot}(\langle{\bf \Xi}\rangle_\sigma)\over \delta\langle {\bf g}_{\rm AB}\rangle_\sigma} ~ = ~  {\delta \hat{\bf S}_{\rm tot}(\langle{\bf \Xi}\rangle_\sigma)\over \delta \langle {\bf C}_{\rm ABD}\rangle_\sigma} ~ = ~ {\delta \hat{\bf S}_{\rm tot}(\langle{\bf \Xi}\rangle_\sigma)\over \delta \langle \overline{\bf \Psi}_{\rm A}\rangle_\sigma} ~ = ~ 0, \nd
where to derive the fermionic EOMs one will have to introduce the fermionic perturbative series in \eqref{QT3}. This will be discussed in \cite{hete8}. There are also other equations coming from integrating out the off-shell degrees of freedom. Those will also be discussed in \cite{hete8}. 

With this we are ready to tackle the Schwinger-Dyson equations for the emergent on-shell degrees of freedom. Here however we will only discuss the dynamics of the metric components $\langle {\bf g}_{\rm AB}\rangle_\sigma$ and leave the discussion on other components for a future work (see also \cite{hete8}). Using the first equation in \eqref{trancom}, the EOM takes the following form:

{\scriptsize
\bg\label{tranishonal}
\begin{split}
& {\bf R}_{\rm AB}(\langle {\bf \Xi}({\rm X})\rangle_\sigma) - 
{1\over 2} \langle {\bf g}_{\rm AB}({\rm X})\rangle_\sigma {\bf R}(\langle {\bf \Xi}({\rm X})\rangle_\sigma)\\ 
& = ~
{2 \over \sqrt{{\bf g}_{11}(\langle{\bf \Xi}({\rm X})\rangle_\sigma)}} {\delta\over \delta \langle{\bf g}^{\rm AB}({\rm X})\rangle_\sigma} \left(\sqrt{{\bf g}_{11}(\langle{\bf \Xi}({\rm X})\rangle_\sigma)} {\bf G}_4( \langle{\bf \Xi}({\rm X})\rangle_\sigma) \wedge \ast_{11}{\bf G}_4( \langle{\bf \Xi}({\rm X})\rangle_\sigma)\right)\\
& - ~{2{\rm M}_p^2\over \sqrt{{\bf g}_{11}(\langle{\bf \Xi}({\rm X})\rangle_\sigma)}}
\sum_{s= 0}^\infty d_s {\delta\over \delta \langle{\bf g}^{\rm AB}({\rm X})\rangle_\sigma}
\left[\sqrt{{\bf g}_{11}(\langle{\bf \Xi}({\rm X})\rangle_\sigma)}{\bf Q}_{\rm pert}(\bar{c}(s); \langle{\bf \Xi}({\rm X})\rangle_\sigma)\right] \\
& \times ~{\rm exp}\left(-s {\rm M}_p^8 \int_0^y \int_0^w d^6y'~d^2w' \sqrt{{\bf g}_8(\langle {\bf \Xi}({\rm Y}', x)\rangle_\sigma)} \big\vert {\bf Q}_{\rm pert}(\hat{c}(s); \langle{\bf \Xi}({\rm Y}', x)\rangle_\sigma)\big\vert \right)\\
& + ~{\rm M}_p^{10} \int d^6y' ~d^2w' \sqrt{ {\bf g}_{11}(\langle{\bf \Xi}({\rm Y}', x)\rangle_\sigma) \over{\bf g}_{11}(\langle{\bf \Xi}({\rm X})\rangle_\sigma)}~\sum_{s = 0}^\infty ~s d_s ~
{\bf Q}_{\rm pert}(\bar{c}(s); \langle{\bf \Xi}({\rm Y'}, x)\rangle_\sigma)\\
& \times ~ {\delta\over \delta \langle{\bf g}^{\rm AB}({\rm X})\rangle_\sigma}
\left[\sqrt{{\bf g}_{8}(\langle{\bf \Xi}({\rm X})\rangle_\sigma)}{\bf Q}_{\rm pert}(\hat{c}(s); \langle{\bf \Xi}({\rm X})\rangle_\sigma)\right]
~\Theta(y'-y)~\Theta(w'-z)\\
& \times ~{\rm exp}\left(-s {\rm M}_p^8 \int_0^{y'} \int_0^{w'} d^6y''~d^2w'' \sqrt{{\bf g}_8(\langle {\bf \Xi}({\rm Y}'', x)\rangle_\sigma)} \big\vert {\bf Q}_{\rm pert}(\hat{c}(s); \langle{\bf \Xi}({\rm Y}'', x)\rangle_\sigma)\big\vert \right)\\
& -~{\rm M}_p^2\langle {\bf g}_{\rm AB}({\rm X})\rangle_\sigma 
\sum_{p = 1}^\infty 
b_p \left[{\rm exp}\left({-p{\rm M}_p^8\int_{{\cal M}_8} d^6y'~d^2w' 
\sqrt{{\bf g}_8(\langle{\bf \Xi}({\rm Y}', x)\rangle_\sigma)}\big\vert \mathbb{F}({\rm Y - Y'}) {\bf Q}_{\rm pert}(\tilde{c}(p); \langle{\bf \Xi}({\rm Y}', x)\rangle_\sigma)\big\vert}\right) - 1\right]\\
& + ~ {\rm M}_p^{10} \int d^6y' ~d^2w' \sqrt{ {\bf g}_{11}(\langle{\bf \Xi}({\rm Y}', x)\rangle_\sigma) \over{\bf g}_{11}(\langle{\bf \Xi}({\rm X})\rangle_\sigma)}~\mathbb{F}({\rm Y'-Y})\sum_{p = 1}^\infty pb_p~
{\delta\over \delta \langle{\bf g}^{\rm AB}({\rm X})\rangle_\sigma}
\left[\sqrt{{\bf g}_{8}(\langle{\bf \Xi}({\rm X})\rangle_\sigma)}{\bf Q}_{\rm pert}(\tilde{c}(p); \langle{\bf \Xi}({\rm X})\rangle_\sigma)\right] \\
& \times ~ {\rm exp}\left({-p{\rm M}_p^8\int_{{\cal M}_8} d^6y''~d^2w'' 
\sqrt{{\bf g}_8(\langle{\bf \Xi}({\rm Y}'', x)\rangle_\sigma)}\big\vert \mathbb{F}(y' -y''; w' -w'') {\bf Q}_{\rm pert}(\tilde{c}(p); {\bf \Xi}({\rm Y}'', x))\big\vert}\right)\\
\end{split}
\nd}
where ${\rm Y} = (y, w), {\rm X} = (x, {\rm Y}) = (x, y, w)$, ${\rm Y}' = (y', w')$ and $\mathbb{F}({\rm Y' - Y}) \equiv \mathbb{F}(y' - y; w' - w) > 0$. We will consider the EOM along the space-time direction so that 
$({\rm A, B}) \equiv (\mu, \nu) \in {\bf R}^{2, 1}$ in M-theory. As a warm-up let us consider the following metric ans\"atze:
\bg\label{duipasea}
\langle{\bf g}_{\mu\nu}\rangle_\sigma = g_s^{-{8\over 3}}\eta_{\mu\nu}, ~~~~ 
\langle{\bf g}_{mn}\rangle_\sigma = g_s^{-{2\over 3}} g_{mn}(y), ~~~~ 
\langle{\bf g}_{ab}\rangle_\sigma = g_s^{4\over 3} \delta_{ab}, \nd
which is similar to the scalings that we discussed in sections \ref{analysis} and \ref{sec2.1.3}, with ${\rm H}(y) = {\rm H}_o({\bf x}) \equiv 1$ taken to simplify the ensuing analysis. A detailed analysis will be presented in our upcoming paper \cite{hete8}.

\begin{table}[tb]  
 \begin{center}
\renewcommand{\arraystretch}{1.5}
{\begin{tabular}{|c|c|}
\hline
Terms in the EOM \eqref{tranishonal} & $g_s$ scalings  \\
\hline\hline
$ \langle {\bf g}_{\mu\nu}\rangle_\sigma$ & $g_s^{-{8\over 3}}$\\ \hline
${\bf R}_{\mu\nu}(\langle {\bf \Xi}\rangle_\sigma) - 
{1\over 2} \langle {\bf g}_{\mu\nu}\rangle_\sigma {\bf R}(\langle {\bf \Xi}\rangle_\sigma)$ & ${1\over g_s^2}$ \\
\hline
${1 \over \sqrt{{\bf g}_{11}(\langle{\bf \Xi}\rangle_\sigma)}} {\delta\over \delta \langle{\bf g}^{\mu\nu}\rangle_\sigma} \left(\sqrt{{\bf g}_{11}(\langle{\bf \Xi}\rangle_\sigma)} {\bf G}_4( \langle{\bf \Xi}\rangle_\sigma) \wedge \ast_{11}{\bf G}_4( \langle{\bf \Xi}\rangle_\sigma)\right)$ & $g_s^{\theta_{\rm class} - {8\over 3}}$\\
\hline
${1 \over \sqrt{{\bf g}_{11}(\langle{\bf \Xi}\rangle_\sigma)}}
~{\delta\over \delta \langle{\bf g}^{\mu\nu}\rangle_\sigma}
\left[\sqrt{{\bf g}_{11}(\langle{\bf \Xi}\rangle_\sigma)}~{\bf Q}_{\rm pert}(\bar{c}; \langle{\bf \Xi}\rangle_\sigma)\right]$ & 
$g_s^{\theta_{\bar{c}} - {8\over 3}}$ \\ 
\hline
${\rm exp}\left(-{\rm M}_p^6 \int_0^y d^6y'\sqrt{{\bf g}_6(\langle {\bf \Xi}(y')\rangle_\sigma)} \big\vert {\bf Q}_{\rm pert}(\hat{c}; \langle{\bf \Xi}(y')\rangle_\sigma\big\vert \right)$ & ${\rm exp}\left(-{g_s^{\theta_{\hat{c}}}\over g_s^2}\right)$\\
\hline
${\bf Q}_{\rm pert}(\bar{c}; \langle{\bf \Xi}\rangle_\sigma)~
{\delta\over \delta \langle{\bf g}^{\mu\nu}\rangle_\sigma}
\left[\sqrt{{\bf g}_{6}(\langle{\bf \Xi}\rangle_\sigma)}{\bf Q}_{\rm pert}(\hat{c}; \langle{\bf \Xi}\rangle_\sigma)\right]$ & ${g_s^{\theta_{\bar{c}} + \theta_{\hat{c}} -{8\over 3}}\over g_s^2}$\\
\hline
${\rm exp}\left({-{\rm M}_p^6\int_{{\cal M}_6} d^6y' 
\sqrt{{\bf g}_6(\langle{\bf \Xi}(y')\rangle_\sigma)}\big\vert \mathbb{F}({y - y'}) {\bf Q}_{\rm pert}(\tilde{c}; \langle{\bf \Xi}({ y}')\rangle_\sigma)\big\vert}\right)$ & ${\rm exp}\left(-{g_s^{\theta_{\tilde{c}}}\over g_s^2}\right)$\\
\hline
${\delta\over \delta \langle{\bf g}^{\mu\nu}\rangle_\sigma}
\left[\sqrt{{\bf g}_{6}(\langle{\bf \Xi}\rangle_\sigma)}{\bf Q}_{\rm pert}(\tilde{c}; \langle{\bf \Xi}\rangle_\sigma)\right]$ & ${g_s^{\theta_{\tilde{c}} -{8\over 3}}\over g_s^2}$\\
\hline
\end{tabular}}
\renewcommand{\arraystretch}{1}
\end{center}
 \caption[]{The {\it dominant} $g_s$ scalings of various terms appearing the EOMs \eqref{tranishonal} in the presence of BBS like M5-brane instantons. Details about how these $g_s$ scalings help us to solve EOM are described in the text.} 
  \label{firozwalkup}
 \end{table}

\subsubsection{Five-brane instantons wrapped on six-manifold ${\cal M}_6$ \label{sec7.3.1}}

To proceed with \eqref{tranishonal}, we need to quantify the non-perturbative and the non-local interactions carefully right at the level of the action \eqref{kaittami2}. We will start by dealing with BBS like instanton effects which are related to M5-brane instantons wrapped on 
${\cal M}_6$ \cite{bbs}. For this, let us define the perturbative series in the following way:
\bg\label{elishkile}
{\bf Q}_{\rm pert}(c; {\bf \Xi}({\rm Y}', x)) \equiv {\bf Q}_{\rm pert}^{\rm BBS}(c; {\bf \Xi}(y', x)) ~{\delta^2(w'-w_o)\over {\rm M}_p^2 \sqrt{{\bf g}_2(w', x)}}, \nd
where $c = (\hat{c}(s), \bar{c}(s), \tilde{c}(p))$ and ${\bf g}_8({\rm Y}', x) = {\bf g}_6(y', x) {\bf g}_2(w', x)$. The delta function localizes these M5-brane instantons to be at a point $w_o$ on the toroidal manifold ${\mathbb{T}^2\over {\cal G}}$ parametrized by $w'$ and wrapped on ${\cal M}_6$. In the language of \cite{coherbeta}, these are the {\it localized BBS instantons}. (One could also allow {\it delocalized} instantons, but we will not deal with them here to avoid complicating the story.) Plugging \eqref{elishkile} in \eqref{kaittami2}, would convert all the terms in the EOM \eqref{tranishonal} in a way shown on the LHS of 
{\bf Table \ref{firozwalkup}}.

There is still a couple more adjustments need to be done before we can use the EOM \eqref{tranishonal}. One is to make all the emergent degrees of freedom to be independent of the toroidal direction $w \in {\mathbb{T}^2\over {\cal G}}$, and two, is to convert the metric ans\"atze \eqref{duipasea} to a slightly more refined form as in \eqref{ripley} but with $\alpha(t) = \beta(t) = f_k(t) = \widetilde{f}_k(t) = 0, \alpha_k(t) = \beta_k(t) = g_k(t) = 1$. This 
is a simplified version that doesn't take into account the temporal variation of the cosmological constant. In section \ref{sec5.3} we discussed how one might solve the EOMs order by order in $\check{\bf g}_s$. It is quite straightforward to extend the present set-up to a more involved one from section \ref{sec5.3} but we will not study this any further here. Instead we will take the following emergent metric and flux components: 

{\scriptsize
\bg\label{duipasea2}
\langle{\bf g}_{\mu\nu}\rangle_\sigma = g_s^{-{8\over 3}}\eta_{\mu\nu}, ~~ 
\langle{\bf g}_{mn}\rangle_\sigma = \sum_{k = 0}^\infty g^{(k)}_{mn}(y) ~g_s^{-{2\over 3} + {2k\over 3}}, ~~ 
\langle{\bf g}_{ab}\rangle_\sigma = g_s^{4\over 3} \delta_{ab}, ~~ 
{\bf G}_{\rm ABCD}(\langle {\bf \Xi}\rangle_\sigma) = \sum_{k = 0}^\infty 
{\rm G}^{(k)}_{\rm ABCD}(y) ~g_s^{{2k\over 3} + l^{\rm CD}_{\rm AB}},
\nd}
where $g^{(k)}_{mn}(y)$ are chosen such that the volume of the internal six-manifold ${\cal M}_6$ is time-independent keeping, in turn, the four-dimensional Newton's constant time independent\footnote{Although this is not necessary if we want to realize the configuration with temporally varying cosmological constant in section \ref{sec5.3}. For genericity, we will not impose this constraint in solving the Schwinger-Dyson equations. }; and $({\rm A, B}) \in {\bf R}^{2, 1} \times {\cal M}_6 \times {\mathbb{T}^2\over {\cal G}}$. 
$l^{\rm AB}_{\rm CD} > 0$ is the dominant scaling for the emergent G-flux degrees of freedom, and is non-zero to keep in tune with the conditions 
studied in \cite{desitter2, coherbeta, coherbeta2}. Thus the emergent flux components are never time-independent.
Note that we have kept the definition of the emergent four-form flux, compared to \eqref{gfluxes}, generic to include fermionic terms. This is not important here but will become important once we study the full EOMs in \cite{hete8}.

With these at hand, we are ready to tackle the EOM \eqref{tranishonal} in the presence of M5-brane instantons using \eqref{elishkile}. Because of the complexity of the form of \eqref{tranishonal}, we will follow the strategy of comparing the EOM order-by-order in $g_s$. The dominant $g_s$ scalings of the various terms of \eqref{tranishonal} is shown in {\bf Table \ref{firozwalkup}}. The first two set of terms are easy to quantify, and their 
full $g_s$ scaling take the following form:

{\scriptsize
\bg\label{liblilsekrt}
\begin{split}
& ~ {\bf R}_{\rm AB}(\langle {\bf \Xi}({\rm X})\rangle_\sigma) - 
{1\over 2} \langle {\bf g}_{\rm AB}({\rm X})\rangle_\sigma {\bf R}(\langle {\bf \Xi}({\rm X})\rangle_\sigma) = ~ {1\over g_s^2} \sum_{k = 0}^\infty {\bf G}^{(k)}_{\mu\nu}(y)~g_s^{2k\over 3} \\ 
& \\
& ~ {1 \over \sqrt{{\bf g}_{11}(\langle{\bf \Xi}({\rm X})\rangle_\sigma)}} {\delta\over \delta \langle{\bf g}^{\rm AB}({\rm X})\rangle_\sigma} \left(\sqrt{{\bf g}_{11}(\langle{\bf \Xi}({\rm X})\rangle_\sigma)} {\bf G}_4( \langle{\bf \Xi}({\rm X})\rangle_\sigma) \wedge \ast_{11}{\bf G}_4( \langle{\bf \Xi}({\rm X})\rangle_\sigma)\right) = ~ g_s^{\theta_{\rm class} - {8\over 3}} \sum_{k = 0}^\infty {\bf H}^{(k)}_{\mu\nu}(y) g_s^{2k\over 3}\\
\end{split}
\nd}
where ${\bf G}_{\mu\nu}^{(k)}(y)$ appears from the higher order series in $g_s$ for the metric components $\langle {\bf g}_{mn}\rangle_\sigma$, that enter via the emergent Ricci scalar ${\bf R}(\langle{\bf \Xi}\rangle_\sigma)$. In a similar vein we define the Hodge star using the emergent metric components from \eqref{duipasea2} and use the flux definition from there to construct the emergent energy-momentum tensor from the G-fluxes. All these information appear in the tensor 
${\bf H}^{(k)}_{\mu\nu}(y)$ with the $g_s$ scaling shown above. The dominant $g_s$ scaling from the fluxes, coming from a combination of $l^{\rm AB}_{\rm CD}$ and the metric scalings, is given by $\theta_{\rm class}$. For $k = 0$, we can compare the $g_s$ scaling of the two set of terms above to get:
\bg\label{kathtuch}
\theta_{\rm class} ~ = ~ {2\over 3}, \nd
which, while allowing the fluxes to contribute to the EOM, does not help in solving it. (The contributing flux components at this order are only ${\bf G}_{mnab}(\langle{\bf \Xi}\rangle_\sigma)$ with $(m, n) \in {\cal M}_6, (a, b) \in {\mathbb{T}^2\over {\cal G}}$ and 
$l^{ab}_{mn} = 1$.)
The underlying argument is similar to what we discussed in section \ref{sec2.1.4}, so we will not repeat it here\footnote{Recall that we need a positive cosmological constant solution. The fluxes can only solve the EOM with negative cosmological constant.}.

\begin{figure}
    \centering
    \includegraphics[scale=0.8]{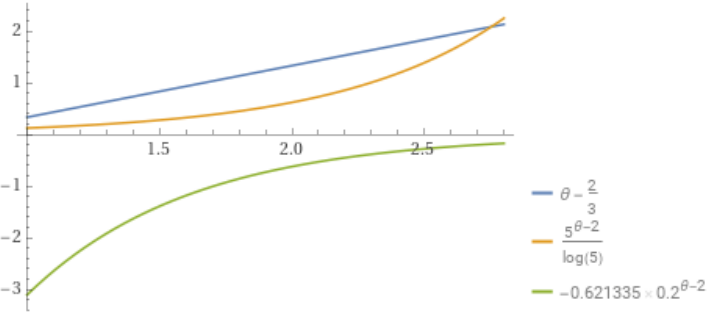} 
    \caption{Plot of the functions $y = \theta -{2\over 3}$ in \textcolor{blue}{blue} and 
    $y = {g_s^{\theta-2}\over {\rm log}~g_s}$ for $g_s = 5$ in \textcolor{orange}{orange} and $g_s = 0.2$ in \textcolor{green}{green}. We see that there are points of intersection between the 
     \textcolor{blue}{blue} curve and the \textcolor{orange}{orange} 
     curve only when $g_s > 1$. When $g_s < 1$, 
     there are no points of intersection between the \textcolor{blue}{blue} curve and the \textcolor{green}{green} curve.}
    \label{plottheta1}
\end{figure}

\begin{figure}
    \centering
    \includegraphics[scale=0.8]{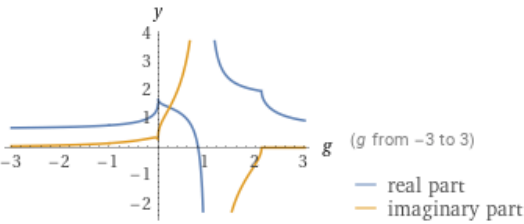} 
    \caption{Plot of the real and imaginary parts of the function $y = {2\over 3} - {\mathbb{W}(-g^{-4/3})\over {\rm log}~g}$ in \textcolor{blue}{blue} and 
     \textcolor{orange}{orange} respectively. Note that solution exists only for $g > 2$ here because ${\bf Im}~y = 0$. For $g < 1$, both 
     ${\bf Im}~y$ and ${\bf Re}~y$ are non-zero.}
    \label{prodlog}
\end{figure}

The third set of terms is more non-trivial, because this is where the non-perturbative corrections first enter. The non-perturbative corrections in 
\eqref{tranishonal} are given in terms of the instanton saddles and their corresponding fluctuation determinants. Including them, the contribution to the EOM may be quantified in the following way:

{\scriptsize
\bg\label{libsekrt2}
\begin{split}
& ~{1 \over \sqrt{{\bf g}_{11}(\langle{\bf \Xi}({\rm X})\rangle_\sigma)}}
\sum_{s= 0}^\infty d_s {\delta\over \delta \langle{\bf g}^{\rm AB}({\rm X})\rangle_\sigma}
\left[\sqrt{{\bf g}_{11}(\langle{\bf \Xi}({\rm X})\rangle_\sigma)}{\bf Q}_{\rm pert}(\bar{c}(s); \langle{\bf \Xi}({\rm X})\rangle_\sigma)\right]
 \\
& \times ~{\rm exp}\left(-s {\rm M}_p^8 \int_0^y \int_0^w d^6y'~d^2w' \sqrt{{\bf g}_8(\langle {\bf \Xi}({\rm Y}', x)\rangle_\sigma)} \big\vert {\bf Q}_{\rm pert}(\hat{c}(s); \langle{\bf \Xi}({\rm Y}', x)\rangle_\sigma)\big\vert \right)\\
& = ~ \sum_{r = 0}^\infty d_r g_s^{\theta_{{\bar{c}(r)}} -{8\over 3}} \sum_{k = 0}^\infty {\bf I}^{(k)}_{\mu\nu}(\bar{c}(r); y) g_s^{2k\over 3} ~{\rm exp}\left(-r {\rm M}_p^6 g_s^{\theta_{\hat{c}(r)} - 2} \int_0^y d^6y'\sum_{l = 0}^\infty\vert {\bf I}^{(l)}_1(\hat{c}(r); y) \vert g_s^{2l\over 3} \Theta(w_{1}-w_{2})\right),\\
\end{split}
\nd}
where ${\bf Q}_{\rm pert}(c; \langle {\bf \Xi}\rangle_\sigma)$ is from \eqref{QT3} and 
$c = \bar{c}(s)$ and $c = \hat{c}(s)$ with appropriate choices for these parameters. The LHS is over the full eight dimensional internal 
manifold ${\cal M}_6 \times {\mathbb{T}^2\over {\cal G}}$, and the RHS is over the six-dimensional base ${\cal M}_6$ where we used \eqref{elishkile}. The Heaviside step-function $\Theta(w_1-w_2)$ with 
$w_i \in {\mathbb{T}^2\over {\cal G}}$ is used here to localize the integrals over the six-dimensional base. The $g_s$ scalings of the two aforementioned perturbative series appear as $\theta_{\bar{c}(r)}$ and 
$\theta_{\hat{c}(r)}$ for $r \in \mathbb{Z}$ alongwith the corresponding functions ${\bf I}_{\mu\nu}^{(k)}(\bar{c}(r); y)$ and ${\bf I}_1^{(l)}(\hat{c}(r); y)$ respectively. As before, the dominant scaling will appear when $k = l = 0$. Ignoring the functional dependence, there is now a puzzle when we try to compare the $g_s$ scalings on the RHS of \eqref{libsekrt2} with the dominant $g_s$ scaling from the emergent Einstein tensor of \eqref{liblilsekrt}. A naive comparison gives us the following:
\bg\label{dotdreslib}
\theta_{\bar{c}(r)} - {2\over 3} = {g_s^{\theta_{\hat{c}(r)} - 2} \over 
{\rm log}~g_s}, \nd
$\forall r \in \mathbb{Z}$. There is already an issue here as seen from 
{\bf figure \ref{plottheta1}}: there is no solution to the system for $g_s < 1$. (Only the \textcolor{orange}{orange} curve intersects with the \textcolor{blue}{blue} curve in {\bf figure \ref{plottheta1}}.) If we try to solve \eqref{dotdreslib}, we find the following solution for 
$\theta_{\bar{c}(r)} =\theta_{\hat{c}(r)} \equiv\theta_{{c}(r)}$:
\bg\label{eliish2me}
\theta_{{c}(r)} = {2\over 3} - {\mathbb{W}(-g_s^{-{4/3}})\over {\rm log}~g_s}, \nd
where $\mathbb{W}(z)$ is the Lambert W-function (also known as the Product Log function). Generically $\mathbb{W}(z)$ has both real and complex parts, and therefore $\theta_{c(r)}$ will have both real and complex pieces as shown in {\bf figure \ref{prodlog}}. Solution would exist in the regime where the imaginary part vanishes, and this happens 
for $g_s > 1$ as shown in {\bf figure \ref{prodlog}}. Unfortunately, even in the regime where solutions exist (ignoring the fact that $g_s > 1$ there), this is unacceptable because $\theta_{c(r)}$ is not a constant but is a function of $g_s$. More so, it will also be a function of $y \in {\cal M}_6$ because of the functions ${\bf I}_{\mu\nu}^{(k)}(y)$ and 
${\bf I}^{(l)}(y)$ appearing in \eqref{libsekrt2}. 

All in all it appears that equating the $g_s$ scalings in the aforementioned way is not the correct way to go. How should we then proceed to solve the EOM \eqref{tranishonal}? 
The hint appears from the $g_s$ scaling itself: Since the $g_s$ dependence in the exponential factor is the main source of contention,  
we cannot presume that $\theta_{\bar{c}(r)} =\theta_{\hat{c}(r)}$ for 
$k \ge 0, l\ge 0$ in \eqref{libsekrt2}. Instead we can equate ${1\over g_s^2}$ from the emergent Einstein tensor in \eqref{liblilsekrt} with the dominant scaling $g_s^{\theta_{\bar{c}(r)} - {8\over 3}}$ to demand the following scaling of the quantum terms from the fluctuation determinants:
\bg\label{carrolb}
\theta_{\bar{c}(r)} ~ = ~ {2\over 3}, \nd
and take $\theta_{\hat{c}(r)} \ge {2\over 3}$. Interestingly, for 
${2\over 3} \le \theta_{\hat{c}(r)} < 2$, and $l = 0$, the instanton contributions are heavily suppressed (recall that ${\rm M}_p >> 1$ from the point of view of the low energy scale that we have been using). On the other hand, if $g_s$ and ${\rm M}_p$ go to zero and infinity respectively as $g_s = \epsilon$ and ${\rm M}_p = \epsilon^{-b}$ for 
$b > 0$, then the exponential part can be perturbatively expanded if the quantum terms from the instanton saddles scale as:
\bg\label{carbaker}
\theta_{\hat{c}(r)} ~ > ~ 2 + 6b, \nd
implying that even in the range $2 \le \theta_{\hat{c}(r)} < 2 + 6b$, we expect suppressions of the exponential parts. Unfortunately, for 
$\theta_{\bar{c}(r)} ~ = ~ {2\over 3}$ in \eqref{carrolb}, there are {\it no} quantum terms that could contribute! In fact this is the same issue that we encountered with the perturbative quantum series, implying that to this order {\it there are no non-trivial contributions} from \eqref{libsekrt2}.

What would then contribute? Contributions can come from the two set of quantum terms with $\theta_{\bar{c}(r)} > {2\over 3}$ and $\theta_{\hat{c}(r)} > {2\over 3}$ which are {\it not} suppressed by the exponential terms. This is possible for contributions to the EOM \eqref{tranishonal} that take the following form:

{\scriptsize
\bg\label{libsekrt3}
\begin{split}
& \int d^6y' ~d^2w' \sqrt{ {\bf g}_{11}(\langle{\bf \Xi}({\rm Y}', x)\rangle_\sigma) \over{\bf g}_{11}(\langle{\bf \Xi}({\rm X})\rangle_\sigma)}~\sum_{s = 0}^\infty ~s d_s ~
{\bf Q}_{\rm pert}(\bar{c}(s); \langle{\bf \Xi}({\rm Y'}, x)\rangle_\sigma)\\
& \times ~ {\delta\over \delta \langle{\bf g}^{\rm AB}({\rm X})\rangle_\sigma}
\left[\sqrt{{\bf g}_{8}(\langle{\bf \Xi}({\rm X})\rangle_\sigma)}{\bf Q}_{\rm pert}(\hat{c}(s); \langle{\bf \Xi}({\rm X})\rangle_\sigma)\right]
~\Theta(y'-y)~\Theta(w'-z)\\
& \times ~{\rm exp}\left(-s {\rm M}_p^8 \int_0^{y'} \int_0^{w'} d^6y''~d^2w'' \sqrt{{\bf g}_8(\langle {\bf \Xi}({\rm Y}'', x)\rangle_\sigma)} \big\vert {\bf Q}_{\rm pert}(\hat{c}(s); \langle{\bf \Xi}({\rm Y}'', x)\rangle_\sigma)\big\vert \right)\\
& = ~\sum_{r = 0}^\infty rd_r \int d^6y'\sum_{k = 0}^\infty \sum_{l = 0}^\infty {\bf I}_2^{(k)}(\bar{c}(r); y') {\bf J}_{\mu\nu}^{(l)}(\hat{c}(r); y) g_s^{\theta_{\bar{c}(r)} + \theta_{\hat{c}(r)} - {14\over 3} + {2\over 3}(k + l)} ~\Theta(y'-y) \Theta(w_{1}-w_{2}) \\
& ~~~ \times ~ {\rm exp}\left(-r {\rm M}_p^6 g_s^{\theta_{\hat{c}(r)} - 2} \int_0^{y'} d^6y''\sum_{n = 0}^\infty\vert {\bf I}^{(n)}_1(\hat{c}(r); y'') \vert g_s^{2n\over 3} \Theta(w_{1}-w_{2})\right)\\
\end{split}
\nd}
where ${\bf I}_2^{(k)}(\bar{c}(r); y')$ is similar to the quantum series 
${\bf I}_1^{(l)}(\hat{c}(r); y'')$ but defined with different coefficients $\bar{c}(r)$ and $\hat{c}(r)$ respectively. (The latter is also similar to one in \eqref{libsekrt2}.) ${\bf J}_{\mu\nu}^{(l)}(\hat{c}(r); y)$ appears from the instanton contributions, and as before the dominant contribution comes from $k = l = n = 0$ in \eqref{libsekrt3}. Since now the $g_s$ scalings have changed, we can try to equate them with the dominant ${1\over g_s^2}$ scaling of the emergent Einstein tensor from \eqref{liblilsekrt}. As before we will assume that 
$\theta_{\bar{c}(r)} = \theta_{\hat{c}(r)} \equiv \theta_{c(r)}$, and ignore the terms that depend on $y \in {\cal M}_6$ in the exponential part. Doing this leads to:
\bg\label{conermey}
\theta_{c(r)} = {4\over 3} - {\mathbb{W}\left(-{1\over 2g_s^{2/3}}\right)\over {\rm log}~g_s}, \nd
showing that this cannot be the right answer because of the $g_s$ dependence. Worse, solutions exist only for $g_s > 1$; and $\mathbb{W}(z)$ is the Lambert W-function. The plots are very similar to {\bf figure \ref{plottheta1}} and {\bf figure \ref{prodlog}}. Of course now we know how to proceed. We start by not identifying $\theta_{\bar{c}(r)}$ to $\theta_{\hat{c}(r)}$, and then by comparing the $g_s$ scaling with the emergent Einstein tensor from \eqref{liblilsekrt} we get the following relation:
\bg\label{rianchatu}
\theta_{\bar{c}(r)} + \theta_{\hat{c}(r)} ~ = ~ {8\over 3}, \nd
which is a significant improvement over \eqref{carrolb} because solutions would now exist where none existed before. For example consider 
$\theta_{\bar{c}(r)} = 0$, and $\theta_{\hat{c}(r)} = {8\over 3}$. Since the emergent Riemann curvature tensors generically scale as $g_s^{2/3}$ (see for example \cite{desitter2}), it would imply a {\it quartic} order curvature contribution among other stuff. They are all coming from the instanton sectors and interestingly it was predicted earlier \cite{issues} that such terms would be necessary to allow for a de Sitter solution to exist. We now see that for a de Sitter excited state to occur, at least quartic order curvature corrections are necessary. A similar conclusion can be made for the fluctuation determinants if we take $\theta_{\bar{c}(r)} = {8\over 3}$, and $\theta_{\hat{c}(r)} = 0$. However now two questions arise.

\vskip.1in

\noindent ${\bf 1.}$ What happens to the exponential pieces with the choice \eqref{rianchatu}?

\vskip.1in

\noindent ${\bf 2.}$ How and where does the second EFT criterion from \eqref{atryan} fits inside \eqref{rianchatu}?

\begin{table}[tb]  
 \begin{center}
\renewcommand{\arraystretch}{1.5}
{\begin{tabular}{|c|c|c|}
\hline
$\theta_{\bar{c}(r)}$  &  $\theta_{\hat{c}(r)}$  & $\theta_{\hat{c}(r)}-2$ \\
\hline\hline
${8\over 3}$ & 0 & $-2$ \\
\hline
${5\over 3}$ & 1 & $-1$ \\
\hline
${4\over 3}$ & ${4\over 3}$ & $-{2\over 3}$ \\
\hline
${1}$ & ${5\over 3}$ & $-{1\over 3}$ \\
\hline
${2\over 3}$ & ${2}$ & $~~0$ \\
\hline
${0}$ & ${8\over 3}$ & $~~{2\over 3}$ \\
\hline
\end{tabular}}
\renewcommand{\arraystretch}{1}
\end{center}
 \caption[]{The $g_s$ scalings $\theta_{\bar{c}(r)}$ and $\theta_{\hat{c}(r)}$ satisfying \eqref{rianchatu}.} 
  \label{firpashachum}
 \end{table}

\vskip.1in

\noindent We will start with the first question and to answer this we refer the readers to {\bf Table \ref{firpashachum}} where the various distributions of $\theta_{\bar{c}(r)}$ and $\theta_{\hat{c}(r)}$ satisfying \eqref{rianchatu} are laid out. Let ${\rm M}_p = \epsilon^{-b}$ and $g_s = \epsilon^a$, for $(a, b) > (0, 0)$ in the limit $\epsilon \to 0$.  The exponential piece in \eqref{libsekrt3}, behaves as:
\bg\label{lorenzo}
\lim_{\epsilon \to 0}~{\rm exp}\left[-r \epsilon^{a\theta_{\hat{c}(r)} - 2a -6b}\right], \nd
for $r \in \mathbb{Z}$ and $n = k = l = 0$. The choice of $g_s = \epsilon^a$ with $a > 0$ is the late time, but we can keep $a \to 0$ to cover the whole range in the temporal domain 
$-{1\over \sqrt{\Lambda}} < t \le 0$. Even with this, the first four entries in {\bf Table \ref{firpashachum}} suggests that the exponential piece dies-off pretty fast, implying that the contributions from the instanton sector may be made arbitrarily small. Note that the choice 
$\theta_{\bar{c}(r)} = {2\over 3}$ makes the exponential piece to be independent of $g_s$. This is not a classical contribution because $\theta_{\hat{c}(r)} = 2$ adds in the necessary quantum corrections both to the fluctuation determinants and to the exponential factors.  From 
\eqref{lorenzo}, the exponential piece would still appear to die-off, but if the value of the integral coming from ${\bf I}_1^{(0)}(\hat{c}(r), y'')$ is kept small then we can have finite answers for small $r$. 
${\bf I}_1^{(0)}(\hat{c}(r), y'')$ can have contributions from cubic order curvature terms, but also from the emergent Rarita-Schwinger fermionic condensates.
Although we don't discuss the fermionic story here (see \cite{hete8} for details on this), it is known for sometime now \cite{kklt} that type IIB gaugino condensates on seven-branes play a crucial role in generating a positive cosmological constant solution. 
The series constructed out of them is clearly convergent as one would have expected from it's trans-series form, thus quantifying the precise contributions from the instanton sectors for this scenario. For the choice $\theta_{\bar{c}(r)} = 0$, we have $\theta_{\hat{c}(r)} = {8\over 3}$ from \eqref{rianchatu}. On the other hand from {\bf Table \ref{firpashachum}} and \eqref{lorenzo}, 
we see that as long as $a > 9b$ $-$ which is late time $-$ the exponential piece may be perturbatively expanded, otherwise it will die-off pretty fast. Finally the choice of $\theta_{\hat{c}(r)} = {2\over 3}$ doesn't produce anything quantum so we will ignore it. Putting everything together would tell us how much do the instantons contribute to the EOM \eqref{tranishonal}.

This brings us to the final question: where does the second EFT criterion
from \eqref{atryan} fits in the aforementioned discussion? The answer lies in the form of the two scalings $\theta_{\bar{c}(r)}$ and $\theta_{\hat{c}(r)}$. We can rewrite them in the following way:
\bg\label{secretl}
\theta_{\bar{c}(r)} = \theta^{(1)}_{\bar{c}(r)} + \vert \theta^{(2)}_{\bar{c}(r)}\vert ~{\rm sign}~ \theta^{(2)}_{\bar{c}(r)}, ~~~~~
\theta_{\hat{c}(r)} = \theta^{(1)}_{\hat{c}(r)} + \vert \theta^{(2)}_{\hat{c}(r)}\vert ~{\rm sign}~ \theta^{(2)}_{\hat{c}(r)}, \nd
where $\theta^{(i)}_{c(r)}$ for $i = 1, 2$ are used to split the scaling $\theta_{c(r)}$ into two pieces with the second one being sensitive to the second criterion of \eqref{atryan}. This is captured by ${\rm sign}~\theta_{c(r)}$ defined in the following way:
\bg\label{evaliker}
{\rm sign}~\theta_{c(r)} \equiv \begin{cases} ~~+1
~~~~~~ {\rm for}~~ {\partial g_s\over \partial t} \propto g_s^{+{\rm ive}} \\
~~~ \\
~~-1  ~~~~~~ {\rm for}~~ {\partial g_s\over \partial t} \propto g_s^{-{\rm ive}}
\end{cases}
\nd
Plugging this in either \eqref{carrolb} or \eqref{rianchatu} would create relative minus signs implying that there are now infinite possible solutions for $\theta^{(i)}_{c(r)}$. This would clearly be a breakdown of the $g_s$ hierarchy. The ${\rm M}_p$ hierarchy is also broken because of the presence of the emergent localized G-fluxes in a similar vein as in 
\cite{desitter2, coherbeta2}. Thus if the second criterion of \eqref{atryan} is not satisfied then EFT would break down because of the presence of an infinite number of operators at order ${g_s^a\over {\rm M}_p^b}$, $\forall (a, b)$ with $(a, b) \in (\mathbb{Z}, \mathbb{Z})$. 

Therefore to summarize: the second criterion in \eqref{atryan} is absolutely essential for the consistency of the computations presented in this section done by including the instanton corrections. What about the non-local interactions? It turns out that the story is bit more subtle than what we discussed earlier. In the following we elaborate on this.

\subsubsection{Revisiting the non-local interactions from \eqref{fepoladom} and \eqref{kaittami2} \label{sec7.3.2}}

In the three points mentioned just below \eqref{fepoladom}, we cautioned the readers that \eqref{fepoladom} {\it may not} be the full answer, as there could exist the possibility of additional corrections to \eqref{fepoladom}. However the analysis that we performed to reach to 
\eqref{fepoladom} did not suggest any such possibilities, so the question is how and why do we need additional corrections to \eqref{fepoladom}. Note that a similar form for the instanton terms did require additional contributions coming from the fluctuation determinants around the instantons, so therein the corrected action did have a solid reasoning for back up. On the other hand, the computation that we performed to convert \eqref{jacksharp} to \eqref{fepoladom} was aimed to get a {\it convergent} form for the non-local interactions without losing any information contained in \eqref{jacksharp}. This resulted in a specific choice for the coefficients in \eqref{fepoladom} related to the ones from
\eqref{jacksharp} via the matrix $\mathbb{M}$ in \eqref{poladom}. The only way to relax this, yet still keeping the convergence property intact, is to convert \eqref{fepoladom} to a trans-series much like what we had for the instanton terms. However this begs the question as to {\it why} should we do this. Is there anything in particular missing from
\eqref{fepoladom} that requires us to change the format of \eqref{fepoladom}? 

To see the short-comings of the non-local interactions \eqref{fepoladom} appearing in the action \eqref{kaittami2}, let us go back to the terms  on the eighth line of \eqref{tranishonal}. Our aim would be to check the $g_s$ scalings of the various terms therein using the metric scalings \eqref{duipasea2}. For simplicity, we can restrict ourselves to the space-time directions only. The $g_s$ scaling then becomes:

{\scriptsize
\bg\label{libsekrt4}
\begin{split}
&\langle {\bf g}_{\rm AB}({\rm X})\rangle_\sigma 
\sum_{p = 1}^\infty 
b_p \left[{\rm exp}\left({-p{\rm M}_p^8\int_{{\cal M}_8} d^6y'~d^2w' 
\sqrt{{\bf g}_8(\langle{\bf \Xi}({\rm Y}', x)\rangle_\sigma)}\big\vert \mathbb{F}({\rm Y - Y'}) {\bf Q}_{\rm pert}(\tilde{c}(p); \langle{\bf \Xi}({\rm Y}', x)\rangle_\sigma)\big\vert}\right) - 1\right]\\
& ~ = ~ g_s^{-{8\over 3}} \sum_{p = 1}^\infty b_p \left[{\rm exp}\left(-p{\rm M}_p^6 g_s^{\theta_{\tilde{c}(p)} - 2} \int_{{\cal M}_6} d^6y' \sum_{n = 0}^\infty \vert \mathbb{F}(y - y') {\bf I}_3^{(n)}(\tilde{c}(p); y')\vert g_s^{2n\over 3}\right) - 1\right]\\
\end{split}
\nd}
where ${\bf I}_3^{(n)}(\tilde{c}(p); y')$ is similar to ${\bf I}_1^{(n)}(\hat{c}(r); y'')$ appearing in say \eqref{libsekrt3} but expressed using different coefficients $\tilde{c}(p)$ in \eqref{elishkile}. Comparing the $g_s$ scalings of this term with the ones from \eqref{liblilsekrt}, we already see an insurmountable issue: the $g_s$ scaling of the last term above can never match with the $g_s$ scalings of \eqref{liblilsekrt}! However in the limit $g_s \to 0$ this terms is cancelled by the exponential piece whenever $\theta_{\tilde{c}(p)} > 2$, so doesn't give enough reason to discard it yet. On the other hand, comparing the $g_s$ scaling of the exponential piece with the ones from \eqref{liblilsekrt}, we get the following value for $\theta_{\tilde{c}(p)}$:
\bg\label{taschioni}
\theta_{\tilde{c}(p)} ~ = ~ 2 + {{\rm log}(2r~{\rm log}~g_s) - {\rm log}~3\over {\rm log}~g_s} , \nd
\begin{figure}
    \centering
    \includegraphics[scale=0.8]{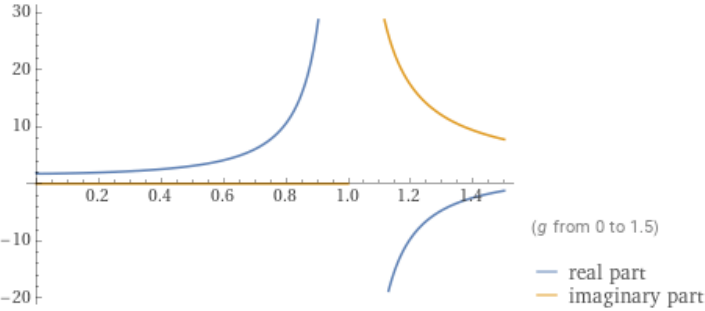} 
    \caption{Plot of the real and imaginary parts of the function $\theta_{\tilde{c}(p)} = 2 + {{\rm log}\left(-{2\over 3}~{\rm log}~g\right)\over {\rm log}~g}$ in \textcolor{blue}{blue} and 
     \textcolor{orange}{orange} respectively. Note that although solutions exist for $g < 1$, because ${\bf Im}~\theta_{\tilde{c}(p)} = 0$, they are functions of $g$. For $g > 1$, clearly both 
     ${\bf Im}~\theta_{\tilde{c}(p)}$ and ${\bf Re}~\theta_{\tilde{c}(p)}$ are non-zero, so no solutions exist.}
    \label{plotr=-1}
\end{figure}
with $r = -1$ and $r = 3$. For generic value of $g_s < 1$ this is problematic because it will make $\theta_{\tilde{c}(p)}$ to be $g_s$ dependent, which is unacceptable. In fact for $r = -1$ in \eqref{taschioni}, solutions for  $\theta_{\tilde{c}(p)}$  appear to exist for $g_s < 1$ but they are still slowly varying functions of $g_s$ as shown in {\bf figure \ref{plotr=-1}}. For $r = 3$ in \eqref{taschioni}, there are no solutions for $g_s < 1$ as shown in {\bf figure \ref{plotr=3}}. From this figure one may see that solutions exist, albeit as functions of $g_s$, only for $g_s \ge \left[\mathbb{W}(1)\right]^{-{1\over 2}} \approx 1.328$. ($\mathbb{W}(z)$ is the Lambert W-function.) This implies that $\theta_{\tilde{c}(p)}$ we got in \eqref{taschioni} for either values of $r$ is unacceptable. 

The above analysis suggests that the non-local interactions appearing in the fourth line of \eqref{kaittami2} cannot be the complete answer. Note that this mismatch is only visible once we used the Schwinger-Dyson equations. In fact the reason why we couldn't match the $g_s$ scalings was because of the absence of any operators accompanying the exponential piece. These operators would be similar to the fluctuation determinants that appear in a similar setting with the instanton terms, but now there is a difference: we can use both local and non-local operators as the fluctuation determinants now. This means we can quantify the ``fluctuation determinants$"$ by the following set of operators:

{\footnotesize
\bg\label{whiteang}
{\bf Q}_{\rm pert}(\check{c}(p); {\bf \Xi}({\rm X})) + 
{\rm M}_p^8\int_{{\cal M}_8} d^6y'' d^2w''\sqrt{{\bf g}_8({\rm Y}'', x)} ~{\bf Q}_{\rm pert}(\grave{c}(p); {\bf \Xi}({\rm Y}'', x)) \mathbb{F}(y-y''; w-w''), \nd}
where ${\bf Q}_{\rm pert}$ is defined as in \eqref{QT3} but now with coefficients $\check{c}(p) \equiv \check{c}_{nl}(p)$ and $\grave{c}(p) \equiv \grave{c}_{nl}(p)$. The other notations are: ${\rm X} = (x, y, w)$,
${\rm Y}'' = (y'', w'')$ and ${\bf \Xi}$ is the set of on-shell fields as before. In the Schwinger-Dyson equations the latter has to be replaced by its expectation value $\langle {\bf \Xi}\rangle_\sigma$, but there is still one more subtlety remaining related to the piece $-b_p \sqrt{-{\bf g}_{11}({\rm X})}$ that appears on the fourth line of \eqref{kaittami2} without an exponential factor. This is also the piece that led to an earlier mismatch with the $g_s$ scalings. In the presence of the operators \eqref{whiteang}, this particular piece will lead to two set of operator series: a perturbative quantum series and a non-local quantum series. The former is what we had as ${\bf S}_{\rm pert}$ in \eqref{posnerdead} and the latter as ${\bf S}_{\rm nloc}$ in \eqref{jacksharp} {\it before} we converted them to ${\bf S}_{\rm NP}$ and ${\bf S}_{\rm nloc}$  in \eqref{dunbar} and \eqref{fepoladom} respectively. The former simply renormalizes the coefficient $d_s$ in the third line of \eqref{kaittami2}, while the latter $-$ taking the form  \eqref{fepoladom} $-$ will again have the same issue as before because of the $-1$ factor. The correct way out of this is to rewrite the fourth line of \eqref{kaittami2} in the following way:

{\scriptsize
\bg\label{angeltrans}
\begin{split}
{\bf S}_{\rm nloc}({\bf \Xi}) & = 
{\rm M}_p^{11}\int d^3x~d^6y~d^2w \sqrt{-{\bf g}_{11}({\rm X})}\\
& \times \sum_{p = 0}^\infty 
b_p \left[{\bf Q}_{\rm pert}(\check{c}(p); {\bf \Xi}({\rm X})) + 
{\rm M}_p^8\int_{{\cal M}_8} d^6y'' d^2w''\sqrt{{\bf g}_8({\rm Y}'', x)} ~{\bf Q}_{\rm pert}(\grave{c}(p); {\bf \Xi}({\rm Y}'', x)) \mathbb{F}(y-y''; w-w'')\right]\\
& \times {\rm exp}\left({-p{\rm M}_p^8\int_{{\cal M}_8} d^6y'~d^2w' 
\sqrt{{\bf g}_8({\rm Y}', x)}\big\vert \mathbb{F}(y -y'; w -w') {\bf Q}_{\rm pert}(\tilde{c}(p); {\bf \Xi}({\rm Y}', x))\big\vert}\right)\\
\end{split}
\nd}
where the second line above may be interpreted as the fluctuation determinants and the third line as the saddles from the non-local effects. This will effect the EOM in \eqref{tranishonal}, and consequently the eighth, ninth and the tenth lines will get modified, including some of the entries in {\bf Table \ref{firozwalkup}}. These changes are easy to quantify and 
in the following we will elaborate on them.

Let us start with the simpler piece in \eqref{angeltrans} where the fluctuation determinant does not depend on the non-locality factor. The Schwinger-Dyson equation corresponding to this term may be expressed in the following way:

{\scriptsize
\bg\label{avrilgulb1}
\begin{split}
& -{1\over \sqrt{{\bf g}_{11}(\langle {\bf \Xi}({\rm X})\rangle_\sigma)}} \sum_{p = 0}^\infty b_p{\delta\over \delta \langle{\bf g}^{\rm AB}({\rm X})\rangle_\sigma}
\left(\sqrt{{\bf g}_{11}(\langle{\bf \Xi}({\rm X})\rangle_\sigma)}{\bf Q}_{\rm pert}(\check{c}(p); \langle{\bf \Xi}({\rm X})\rangle_\sigma)\right)\\
& \times ~{\rm exp}\left({-p{\rm M}_p^8\int_{{\cal M}_8} d^6y'~d^2w' 
\sqrt{{\bf g}_8(\langle{\bf \Xi}({\rm Y}', x)\rangle_\sigma)}\big\vert \mathbb{F}(y -y'; w -w') {\bf Q}_{\rm pert}(\tilde{c}(p); \langle{\bf \Xi}({\rm Y}', x)\rangle_\sigma)\big\vert}\right)\\
& ~ = ~  \sum_{p = 0}^\infty b_p g_s^{\theta_{\check{c}(p)}-{8\over 3}}  \sum_{r = 0}^\infty {\bf K}^{(r)}_{\mu\nu}(\check{c}(p); y) g_s^{2r\over 3} {\rm exp}\left(-p{\rm M}_p^6 g_s^{\theta_{\tilde{c}(p)} - 2} \int_{{\cal M}_6} d^6y' \sum_{n = 0}^\infty \vert \mathbb{F}(y - y') {\bf I}_3^{(n)}(\tilde{c}(p); y')\rangle_\sigma)\vert g_s^{2n\over 3}\right)\\
\end{split}
\nd}
where ${\bf I}_3^{(n)}(\tilde{c}(p); y')$ is the same operator that we encountered in \eqref{libsekrt4}, and ${\bf K}^{(r)}_{\mu\nu}(\check{c}(p); y)$ is constructed from the first line above. The dominant $g_s$ scaling, for $r = n = 0$ and $p =1$, when compared to \eqref{liblilsekrt} satisfies:
\bg\label{dotdreslib2}
\theta_{\check{c}(p)} - {2\over 3} = {g_s^{\theta_{\tilde{c}(p)} - 2} \over {\rm log}~g_s}, \nd
which is similar to what we had in \eqref{dotdreslib}, and motivated from there, as well as from {\bf figures \ref{plottheta1}} and {\bf \ref{prodlog}}, we cannot make $\theta_{\check{c}(p)} =\theta_{\tilde{c}(p)}$. One solution is that $\theta_{\check{c}(p)} = {2\over 3}$, and for 
${2\over 3} \le \theta_{\tilde{c}(p)} < 2$, the non-local contributions are heavily suppressed. Additionally the absence of quantum terms for the scaling $\theta_{\check{c}(p)} = {2\over 3}$, along with similar arguments following \eqref{carbaker}, shows that \eqref{avrilgulb1} {\it does not} contribute significantly to the Schwinger-Dyson equations.

The next contribution would appear from the non-local fluctuation determinants from \eqref{whiteang}. It contains more complicated set of terms, so one might wonder if they can contribute to the Schwinger-Dyson equations. To see this let us express the EOMs in the following way:

{\scriptsize
\bg\label{claugarphone}
\begin{split}
& ~ - {\rm M}_p^8\sum_{p = 0}^\infty \langle {\bf g}_{\rm AB}({\rm X})\rangle_\sigma 
\int_{{\cal M}_8} d^6y' d^2w'\sqrt{{\bf g}_8(\langle{\bf \Xi}({\rm Y}', x)\rangle_\sigma)} ~{\bf Q}_{\rm pert}(\grave{c}(p); \langle{\bf \Xi}({\rm Y}', x)\rangle_\sigma) \mathbb{F}(y-y'; w-w')\\
& ~\times ~{\rm exp}\left({-p{\rm M}_p^8\int_{{\cal M}_8} d^6y''~d^2w'' 
\sqrt{{\bf g}_8(\langle{\bf \Xi}({\rm Y}'', x)\rangle_\sigma)}\big\vert \mathbb{F}(y -y''; w -w'') {\bf Q}_{\rm pert}(\tilde{c}(p); \langle{\bf \Xi}({\rm Y}'', x)\rangle_\sigma)\big\vert}\right)\\
& ~ = ~  {\rm M}_p^6\sum_{p = 0}^\infty b_p g_s^{\theta_{\grave{c}(p)}-{14\over 3}}  \sum_{r = 0}^\infty \int_{{\cal M}_6} d^6y'{\bf L}^{(r)}_{\mu\nu}(\grave{c}(p); y) \mathbb{F}(y - y')g_s^{2r\over 3}
~ {\rm exp}\left(-p{\rm M}_p^6 g_s^{\theta_{\tilde{c}(p)} - 2} \int_{{\cal M}_6} d^6y' \sum_{n = 0}^\infty \vert \mathbb{F}(y - y') {\bf I}_3^{(n)}(\tilde{c}(p); y')\vert g_s^{2n\over 3}\right)\\
\end{split}
\nd}
where ${\bf I}_3^{(n)}(\tilde{c}(p); y')$ is the same operator that we encountered earlier, but ${\bf L}^{(r)}_{\mu\nu}(\grave{c}(p); y)$ is new and is constructed from the first line above. A naive $g_s$ scaling of the dominant term for $r = n = 0$ and $p = 1$ with the $g_s$ scalings from \eqref{liblilsekrt} gives us:
\bg\label{tarkapat}
\theta_{\grave{c}(p)} - {8\over 3} = {g_s^{\theta_{\tilde{c}(p)} - 2} \over {\rm log}~g_s}, \nd
which while doesn't have the same problem that we faced with \eqref{libsekrt4}, doesn't also provide a $g_s$ independent solution 
for $\theta_{\grave{c}(p)} =\theta_{\tilde{c}(p)}$. This is of course expected, and therefore we can keep $\theta_{\grave{c}(p)} \ne \theta_{\tilde{c}(p)}$ and assume the following solution for $\theta_{\grave{c}(p)}$:
\bg\label{kifulapat}
\theta_{\grave{c}(p)} = {8\over 3}, \nd
implying that for ${2\over 3} \le \theta_{\tilde{c}(p)} < 2$ the exponential piece will continue to be suppressed for $p \ge 1$ at late time. For $\theta_{\tilde{c}(p)} \ge 2$, one may perturbatively expand the exponential piece in powers of $g_s$. They will then contribute to the higher order Schwinger-Dyson equations. Again, we see that the contributing terms come typically from quartic order curvature terms satisfying \eqref{kifulapat}. In fact we expect similar contributions from another related term from the non-local action \eqref{angeltrans}, namely:

{\scriptsize
\bg\label{patricdal}
\begin{split}
& -{\rm M}_p^8 \int d^6y' d^2w' \sqrt{\bf g_{11}(\langle {\bf \Xi}({\rm Y}', x)\rangle_\sigma \over {\bf g}_{11}(\langle {\bf \Xi}({\rm X})\rangle_\sigma}\sum_{p = 0}^\infty {\delta \over \delta \langle {\bf g}_{\rm AB}({\rm X})\rangle_\sigma} \left(\sqrt{{\bf g}_8(\langle{\bf \Xi}({\rm X})\rangle_\sigma)} ~{\bf Q}_{\rm pert}(\grave{c}(p); \langle{\bf \Xi}({\rm X})\rangle_\sigma)\right) \mathbb{F}(y'-y; w'-w)\\
& ~\times ~{\rm exp}\left({-p{\rm M}_p^8\int_{{\cal M}_8} d^6y''~d^2w'' 
\sqrt{{\bf g}_8(\langle{\bf \Xi}({\rm Y}'', x)\rangle_\sigma)}\big\vert \mathbb{F}(y' -y''; w' -w'') {\bf Q}_{\rm pert}(\tilde{c}(p); \langle{\bf \Xi}({\rm Y}'', x)\rangle_\sigma)\big\vert}\right)\\
& ~ = ~  {\rm M}_p^6\sum_{p = 0}^\infty b_p g_s^{\theta_{\grave{c}(p)}-{14\over 3}}  \sum_{r = 0}^\infty \int_{{\cal M}_6} d^6y'{\bf M}^{(r)}_{\mu\nu}(\grave{c}(p); y) \mathbb{F}(y - y')g_s^{2r\over 3}
 ~ {\rm exp}\left(-p{\rm M}_p^6 g_s^{\theta_{\tilde{c}(p)} - 2} \int_{{\cal M}_6} d^6y'' \sum_{n = 0}^\infty \vert \mathbb{F}(y' - y'') {\bf I}_3^{(n)}(\tilde{c}(p); y'')\vert g_s^{2n\over 3}\right)\\
\end{split}
\nd}
which has similar behavior as \eqref{claugarphone} except with a different function ${\bf M}^{(r)}_{\mu\nu}(\grave{c}(p); y)$ and a different distribution of integrals in the exponential term. This also means that \eqref{kifulapat} continues to be the quantum scaling here too, and the term could contribute non-trivially as long as $\theta_{\tilde{c}(p)} \ge 2$. The dominant contribution would come from say a quartic order curvature piece from the fluctuation determinants and a $g_s$ independent $\theta_{\tilde{c}(p)} = 2$ piece from the exponential factor.

There is still one more contribution to the Schwinger-Dyson equation coming from a mixture of the terms from the fluctuation determinants and the exponential factors. This may be presented in the following way:

{\scriptsize
\bg\label{libsekrt5}
\begin{split}
& {\rm M}_p^{8} \int d^6y' ~d^2w' \sqrt{ {\bf g}_{11}(\langle{\bf \Xi}({\rm Y}', x)\rangle_\sigma) \over{\bf g}_{11}(\langle{\bf \Xi}({\rm X})\rangle_\sigma)}~\mathbb{F}({\rm Y'-Y})\sum_{p = 1}^\infty pb_p~
{\delta\over \delta \langle{\bf g}^{\rm AB}({\rm X})\rangle_\sigma}
\left[\sqrt{{\bf g}_{8}(\langle{\bf \Xi}({\rm X})\rangle_\sigma)}{\bf Q}_{\rm pert}(\tilde{c}(p); \langle{\bf \Xi}({\rm X})\rangle_\sigma)\right] \\
& ~\times \left[{\bf Q}_{\rm pert}(\check{c}(p); \langle{\bf \Xi}({\rm Y}', x)\rangle_\sigma) + 
{\rm M}_p^8\int_{{\cal M}_8} d^6y'' d^2w''\sqrt{{\bf g}_8(\langle {\bf \Xi}({\rm Y}'', x)\rangle_\sigma)} ~{\bf Q}_{\rm pert}(\grave{c}(p); \langle{\bf \Xi}({\rm Y}'', x)\rangle_\sigma) \mathbb{F}(y'-y''; w'-w'')\right]\\
& ~ \times ~ {\rm exp}\left({-p{\rm M}_p^8\int_{{\cal M}_8} d^6y'''~d^2w''' 
\sqrt{{\bf g}_8(\langle{\bf \Xi}({\rm Y}''', x)\rangle_\sigma)}\big\vert \mathbb{F}(y' -y'''; w' -w''') {\bf Q}_{\rm pert}(\tilde{c}(p); \langle{\bf \Xi}({\rm Y}''', x)\rangle_\sigma)\big\vert}\right)\\
& = {\rm M}_p^6 \sum_{p = 1}^\infty pb_p \int_{{\cal M}_6} d^6y' g_s^{\theta_{\tilde{c}(p)} - {14\over 3}}
~ {\rm exp}\left(-p{\rm M}_p^6 g_s^{\theta_{\tilde{c}(p)} - 2} \int_{{\cal M}_6} d^6y''' \sum_{n = 0}^\infty \vert \mathbb{F}(y' - y''') {\bf I}_3^{(n)}(\tilde{c}(p); y''')\vert g_s^{2n\over 3}\right)\\
& ~\times \left[\sum_{k = 0}^\infty {\bf P}_{\mu\nu}^{(k)}(\tilde{c}(p), \check{c}(p); y')~ g_s^{\theta_{\check{c}(p)} + {2k\over 3}} + {\rm M}_p^6 \int_{{\cal M}_6} d^6y''
\sum_{k = 0}^\infty {\bf Q}_{\mu\nu}^{(k)}(\tilde{c}(p), \grave{c}(p); y'')~ \mathbb{F}(y' - y'') ~ g_s^{\theta_{\grave{c}(p)} + {2k\over 3}}\right] \mathbb{F}(y' - y) \\
\end{split}
\nd}
where ${\bf P}_{\mu\nu}^{(k)}(\tilde{c}(p), \check{c}(p); y')$ and 
${\bf Q}_{\mu\nu}^{(k)}(\tilde{c}(p), \grave{c}(p); y'')$ are two new functions appearing from the local and non-local pieces from the fluctuation determinants; and ${\bf I}_3^{(n)}(\tilde{c}(p); y''')$ is the same function that appeared earlier. One may also easily see that the quantum scalings now satisfy the following set of relations:
\bg\label{bound}
\theta_{\tilde{c}(p)} +  \theta_{\check{c}(p)} = {8\over 3}, ~~~~
\theta_{\tilde{c}(p)} +  \theta_{\grave{c}(p)} = {8\over 3}, \nd
where each piece is similar to what we encountered in \eqref{rianchatu}, and therefore would be classified individually as in {\bf Table \ref{firpashachum}}. To summarize: including the fluctuation determinants to modify the non-local interactions as in \eqref{angeltrans}, all ambiguities seem to go away and consistent matching of the $g_s$ scalings may be easily performed. 

\subsubsection{Five-brane instantons wrapped on six-manifold ${\cal M}_4 \times {\mathbb{T}^2\over {\cal G}}$ \label{sec7.3.3}}

A detailed study done in the previous section instructed us how to deal with five-brane instantons wrapped on the six-dimensional base ${\cal M}_6$ of the eight-manifold ${\cal M}_8 = {\cal M}_6 \times {\mathbb{T}^2\over {\cal G}}$. We saw the contributions from both the non-perturbative as well as the non-local terms in the action \eqref{kaittami2}, although a more careful study revealed that \eqref{kaittami2} needs to be modified because the non-local interactions given in \eqref{fepoladom} are incomplete. The action that actually takes care of all the aforementioned subtleties and from where the correct Schwinger-Dyson equations can be derived is given by the following:  

{\scriptsize
\bg\label{kaittami4}
\begin{split}
& \hat{\bf S}_{\rm tot}({\bf \Xi}) = {\bf S}_{\rm kin}({\bf \Xi}) + {\bf S}_{\rm NP}({\bf \Xi}) + {\bf S}_{\rm nloc}({\bf \Xi})\\ 
& = {\rm M}_p^{9}
\int d^{3} x ~d^6y~d^2w \sqrt{-{\bf g}_{11}({\rm X})} 
\left[{\bf R}_{11}({\rm X}) + {\bf G}_4({\rm X}) \wedge \ast_{11}{\bf G}_4({\rm X})\right] + {\rm M}_p^9\int  
{\bf C}_3({\rm X}) \wedge {\bf G}_4({\rm X}) \wedge {\bf G}_4({\rm X)} + {\rm M}_p^3\int {\bf C}_3({\rm X}) \wedge \mathbb{X}_8({\rm X})\\
& + {\rm M}_p^{11}\int d^3x~d^6y~d^2w \sqrt{-{\bf g}_{11}({\rm X})}~
\sum_{s = 0}^\infty d_s~{\bf Q}_{\rm pert}(\bar{c}(s); {\bf \Xi}({\rm X}))~{\rm exp}\left(-s {\rm M}_p^8 \int_0^y \int_0^w d^6y'~d^2w' \sqrt{{\bf g}_8({\rm Y}', x)} \big\vert {\bf Q}_{\rm pert}(\hat{c}(s); {\bf \Xi}({\rm Y}', x))\big\vert \right)\\
& + 
{\rm M}_p^{11}\int d^3x~d^6y~d^2w \sqrt{-{\bf g}_{11}({\rm X})}
\sum_{p = 0}^\infty b_p~ {\rm exp}\left({-p{\rm M}_p^8\int_{{\cal M}_8} d^6y'~d^2w' 
\sqrt{{\bf g}_8({\rm Y}', x)}\big\vert \mathbb{F}(y -y'; w -w') {\bf Q}_{\rm pert}(\tilde{c}(p); {\bf \Xi}({\rm Y}', x))\big\vert}\right)\\
& \times \left[{\bf Q}_{\rm pert}(\check{c}(p); {\bf \Xi}({\rm X})) + 
{\rm M}_p^8\int_{{\cal M}_8} d^6y'' d^2w''\sqrt{{\bf g}_8({\rm Y}'', x)} ~{\bf Q}_{\rm pert}(\grave{c}(p); {\bf \Xi}({\rm Y}'', x)) \mathbb{F}(y-y''; w-w'')\right], \\
\end{split}
\nd}
where ${\bf Q}_{\rm pert}(c; {\bf \Xi}(x, y, w))$ is defined in terms of the on-shell degrees of freedom ${\bf \Xi}(x, y, w)$ in \eqref{QT3} with $c \equiv \{c_{nl}\}$ forming the set of coefficients for the perturbative series in \eqref{QT3}. These set of coefficients may be used to differentiate between the various contributions from the instanton and the non-local saddles and their corresponding fluctuation determinants as shown in \eqref{kaittami4}. For the five-brane instantons wrapped on the six-manifold base, the perturbative series may be expressed as \eqref{elishkile}. On the other hand, to study the five-brane instantons wrapped on the toroidal manifold and a four-cycle in the six-dimensional base one can take ${\cal M}_6$ locally as ${\cal M}_4 \times {\cal M}_2$, and study the instantons wrapped on ${\cal M}_4 \times {\mathbb{T}^2\over {\cal G}}$. These instantons are related to the three-branes instantons in the type IIB side, which we can call as the KKLT instantons \cite{kklt}. For such a case the perturbative series may be defined as:
\bg\label{elishkile2}
{\bf Q}_{\rm pert}(c; {\bf \Xi}(y^m, y^\alpha, w, x)) \equiv {\bf Q}_{\rm pert}^{\rm KKLT}(c; {\bf \Xi}(y^m, w, x)) ~{\delta^2(y^\alpha-y^\alpha_o)\over {\rm M}_p^2 \sqrt{\bar{\bf g}_2(y^\alpha, x)}}, \nd
where $y^m \in {\cal M}_4, y^\alpha \in {\cal M}_2, w \in {\mathbb{T}^2\over {\cal G}}$ and $x \in {\bf R}^{2, 1}$. We have also assume that ${\bf g}_8(y^m, y^\alpha, w, x) = {\bf g}_4(y^m, x) \bar{\bf g}_2(y^\alpha, x) {\bf g}_2(w, x)$, where the coordinate dependence specifies the corresponding sub-manifold for which we take the metric determinant. 

The analysis with the KKLT instantons is very similar to the ones we did with the BBS instantons, except that the determinant of the metric for ${\cal M}_4 \times {\mathbb{T}^2\over {\cal G}}$ has zero scaling with $g_s$, so the quantum terms do not contribute much to the EOMs. Therefore
we will not go in details here and leave the analysis for the diligent readers. The procedure follows the strategy we laid out in section \ref{sec7.3.2}, namely substitute \eqref{elishkile2} in the Schwinger-Dyson equations and then compute the contributions from all the terms\footnote{There would also be contributions from the two-brane instantons, as discussed briefly in \cite{coherbeta}, including other instantons that may not have simple brane interpretations. For the latter one could, for example, also expect contributions from the stable non-BPS states in the theory that we haven't touched here. Additionally all the aforementioned instantons are {\it real}, but the resurgence story also includes {\it complex} instantons coming from the complex turning points in a potential with Minkowski minima (with possible scale-separated AdS minima as the scale-separated ones do not appear to allow for an EFT description). All of these should come together to explain the various instantonic saddles appearing in the action \eqref{kaittami4}. A detailed analysis of these effects is clearly beyond the scope of this paper and will be discussed elsewhere. \label{shomithi}}. The Schwinger-Dyson equation coming from the action \eqref{kaittami4} may be succinctly presented as:

{\scriptsize
\bg\label{tranishonal3}
\begin{split}
& {\bf R}_{\rm AB}(\langle {\bf \Xi}({\rm X})\rangle_\sigma) - 
{1\over 2} \langle {\bf g}_{\rm AB}({\rm X})\rangle_\sigma {\bf R}(\langle {\bf \Xi}({\rm X})\rangle_\sigma)\\ 
& = ~
{2 \over \sqrt{{\bf g}_{11}(\langle{\bf \Xi}({\rm X})\rangle_\sigma)}} {\delta\over \delta \langle{\bf g}^{\rm AB}({\rm X})\rangle_\sigma} \left(\sqrt{{\bf g}_{11}(\langle{\bf \Xi}({\rm X})\rangle_\sigma)} {\bf G}_4( \langle{\bf \Xi}({\rm X})\rangle_\sigma) \wedge \ast_{11}{\bf G}_4( \langle{\bf \Xi}({\rm X})\rangle_\sigma)\right)\\
& - ~{2{\rm M}_p^2\over \sqrt{{\bf g}_{11}(\langle{\bf \Xi}({\rm X})\rangle_\sigma)}}
\sum_{s= 0}^\infty d_s {\delta\over \delta \langle{\bf g}^{\rm AB}({\rm X})\rangle_\sigma}
\left[\sqrt{{\bf g}_{11}(\langle{\bf \Xi}({\rm X})\rangle_\sigma)}{\bf Q}_{\rm pert}(\bar{c}(s); \langle{\bf \Xi}({\rm X})\rangle_\sigma)\right] \\
& \times ~{\rm exp}\left(-s {\rm M}_p^8 \int_0^y \int_0^w d^6y'~d^2w' \sqrt{{\bf g}_8(\langle {\bf \Xi}({\rm Y}', x)\rangle_\sigma)} \big\vert {\bf Q}_{\rm pert}(\hat{c}(s); \langle{\bf \Xi}({\rm Y}', x)\rangle_\sigma)\big\vert \right)\\
& + ~{\rm M}_p^{10} \int d^6y' ~d^2w' \sqrt{ {\bf g}_{11}(\langle{\bf \Xi}({\rm Y}', x)\rangle_\sigma) \over{\bf g}_{11}(\langle{\bf \Xi}({\rm X})\rangle_\sigma)}~\sum_{s = 0}^\infty ~s d_s ~
{\bf Q}_{\rm pert}(\bar{c}(s); \langle{\bf \Xi}({\rm Y'}, x)\rangle_\sigma)\\
& \times ~ {\delta\over \delta \langle{\bf g}^{\rm AB}({\rm X})\rangle_\sigma}
\left[\sqrt{{\bf g}_{8}(\langle{\bf \Xi}({\rm X})\rangle_\sigma)}{\bf Q}_{\rm pert}(\hat{c}(s); \langle{\bf \Xi}({\rm X})\rangle_\sigma)\right]
~\Theta(y'-y)~\Theta(w'-z)\\
& \times ~{\rm exp}\left(-s {\rm M}_p^8 \int_0^{y'} \int_0^{w'} d^6y''~d^2w'' \sqrt{{\bf g}_8(\langle {\bf \Xi}({\rm Y}'', x)\rangle_\sigma)} \big\vert {\bf Q}_{\rm pert}(\hat{c}(s); \langle{\bf \Xi}({\rm Y}'', x)\rangle_\sigma)\big\vert \right)\\
& -{1\over \sqrt{{\bf g}_{11}(\langle {\bf \Xi}({\rm X})\rangle_\sigma)}} \sum_{p = 0}^\infty b_p{\delta\over \delta \langle{\bf g}^{\rm AB}({\rm X})\rangle_\sigma}
\left(\sqrt{{\bf g}_{11}(\langle{\bf \Xi}({\rm X})\rangle_\sigma)}{\bf Q}_{\rm pert}(\check{c}(p); \langle{\bf \Xi}({\rm X})\rangle_\sigma)\right)\\
& \times ~{\rm exp}\left({-p{\rm M}_p^8\int_{{\cal M}_8} d^6y'~d^2w' 
\sqrt{{\bf g}_8(\langle{\bf \Xi}({\rm Y}', x)\rangle_\sigma)}\big\vert \mathbb{F}(y -y'; w -w') {\bf Q}_{\rm pert}(\tilde{c}(p); \langle{\bf \Xi}({\rm Y}', x)\rangle_\sigma)\big\vert}\right)\\
& ~ - {\rm M}_p^8\sum_{p = 0}^\infty \langle {\bf g}_{\rm AB}({\rm X})\rangle_\sigma 
\int_{{\cal M}_8} d^6y' d^2w'\sqrt{{\bf g}_8(\langle{\bf \Xi}({\rm Y}', x)\rangle_\sigma)} ~{\bf Q}_{\rm pert}(\grave{c}(p); \langle{\bf \Xi}({\rm Y}', x)\rangle_\sigma) \mathbb{F}(y-y'; w-w')\\
& ~\times ~{\rm exp}\left({-p{\rm M}_p^8\int_{{\cal M}_8} d^6y''~d^2w'' 
\sqrt{{\bf g}_8(\langle{\bf \Xi}({\rm Y}'', x)\rangle_\sigma)}\big\vert \mathbb{F}(y -y''; w -w'') {\bf Q}_{\rm pert}(\tilde{c}(p); \langle{\bf \Xi}({\rm Y}'', x)\rangle_\sigma)\big\vert}\right)\\
& -{\rm M}_p^8 \int d^6y' d^2w' \sqrt{\bf g_{11}(\langle {\bf \Xi}({\rm Y}', x)\rangle_\sigma \over {\bf g}_{11}(\langle {\bf \Xi}({\rm X})\rangle_\sigma}\sum_{p = 0}^\infty {\delta \over \delta \langle {\bf g}_{\rm AB}({\rm X})\rangle_\sigma} \left(\sqrt{{\bf g}_8(\langle{\bf \Xi}({\rm X})\rangle_\sigma)} ~{\bf Q}_{\rm pert}(\grave{c}(p); \langle{\bf \Xi}({\rm X})\rangle_\sigma)\right) \mathbb{F}(y'-y; w'-w)\\
& ~\times ~{\rm exp}\left({-p{\rm M}_p^8\int_{{\cal M}_8} d^6y''~d^2w'' 
\sqrt{{\bf g}_8(\langle{\bf \Xi}({\rm Y}'', x)\rangle_\sigma)}\big\vert \mathbb{F}(y' -y''; w' -w'') {\bf Q}_{\rm pert}(\tilde{c}(p); \langle{\bf \Xi}({\rm Y}'', x)\rangle_\sigma)\big\vert}\right)\\
& ~ + ~ {\rm M}_p^{8} \int d^6y' ~d^2w' \sqrt{ {\bf g}_{11}(\langle{\bf \Xi}({\rm Y}', x)\rangle_\sigma) \over{\bf g}_{11}(\langle{\bf \Xi}({\rm X})\rangle_\sigma)}~\mathbb{F}({\rm Y'-Y})\sum_{p = 1}^\infty pb_p~
{\delta\over \delta \langle{\bf g}^{\rm AB}({\rm X})\rangle_\sigma}
\left[\sqrt{{\bf g}_{8}(\langle{\bf \Xi}({\rm X})\rangle_\sigma)}{\bf Q}_{\rm pert}(\tilde{c}(p); \langle{\bf \Xi}({\rm X})\rangle_\sigma)\right] \\
& ~\times \left[{\bf Q}_{\rm pert}(\check{c}(p); \langle{\bf \Xi}({\rm Y}', x)\rangle_\sigma) + 
{\rm M}_p^8\int_{{\cal M}_8} d^6y'' d^2w''\sqrt{{\bf g}_8(\langle {\bf \Xi}({\rm Y}'', x)\rangle_\sigma)} ~{\bf Q}_{\rm pert}(\grave{c}(p); \langle{\bf \Xi}({\rm Y}'', x)\rangle_\sigma) \mathbb{F}(y'-y''; w'-w'')\right]\\
& ~ \times ~ {\rm exp}\left({-p{\rm M}_p^8\int_{{\cal M}_8} d^6y'''~d^2w''' 
\sqrt{{\bf g}_8(\langle{\bf \Xi}({\rm Y}''', x)\rangle_\sigma)}\big\vert \mathbb{F}(y' -y'''; w' -w''') {\bf Q}_{\rm pert}(\tilde{c}(p); \langle{\bf \Xi}({\rm Y}''', x)\rangle_\sigma)\big\vert}\right)\\
\end{split}
\nd}
which modifies \eqref{tranishonal} appropriately. Therefore instead of taking the configuration \eqref{duipasea2}, we can take the configuration from \eqref{ripley} with the simplifying conditions from \eqref{natasbhalo}, except now $\alpha(t) \ne 0$ and $\beta(t) \ne 0$. Allowing a non-zero $\alpha(t)$ and $\beta(t)$ could in principle lead to late time singularities, which we should avoid. Since we have chosen $f_1(t)$ as the defining function for all the flux and the metric components, it will be useful to also define $\alpha(t)$ and $\beta(t)$ in terms of $f_1(t)$ in the following way: 
\bg\label{metsunmey}
\alpha(t) = \sum_{n = 1}^\infty l_n\left(f_1(t) - {3a\over 2}\right)^n, ~~~ \beta(t) = \sum_{n = 1}^\infty p_n\left(f_1(t) - {3a\over 2}\right)^n, \nd
with appropriate choice for the coefficients $(l_n, p_n)$ such that 
$(\alpha(t), \beta(t)) < \left({2\over 3}, {2\over 3}\right)$. The latter requirement is to ensure the correct limit of type IIB theory from M-theory. Note that as $t \to 0$, both $\alpha(t)$ and $\beta(t)$ vanishes as can be inferred from section \ref{sec5.3}. The Einstein term in \eqref{tranishonal3} scales in the following way:
\bg\label{henoaleng}
{\bf R}_{\rm AB}(\langle {\bf \Xi}({\rm X})\rangle_\sigma) - 
{1\over 2} \langle {\bf g}_{\rm AB}({\rm X})\rangle_\sigma {\bf R}(\langle {\bf \Xi}({\rm X})\rangle_\sigma) = ~ {1\over g_s^2} \sum_{k = 0}^\infty {\bf G}^{(k)}_{\mu\nu}(y, w)~\check{g}_s^{{2k\over 3} {\rm L}_1(f_1(t)) + {\rm log~corrections}}, \nd
which is significantly different from the scaling in \eqref{liblilsekrt} because of the presence of $\hat{g}_s \equiv g_s^{f_1(t)}$, ${\rm L}_1(f_1(t))$ and log corrections. They provide sub-dominant contributions in 
\eqref{henoaleng} because at late time $f_1(t)$ approaches a constant value of ${3a\over 2}$, and log corrections are typically very small. The explicit form for ${\rm L}_1(f_1(t))$ {\it can} be derived but the analysis is quite technical and the readers may get the derivation from \cite{hete8}. One may now try to compare the $g_s$ scalings of the Einstein term with all the other terms from \eqref{tranishonal3}. We will however not do the detailed analysis here, but only discuss one such term and leave the others for future work. The term that we will like to study is the following:

{\scriptsize
\bg\label{libsekrt8}
\begin{split}
& {\rm M}_p^{8} \int d^6y' ~d^2w' \sqrt{ {\bf g}_{11}(\langle{\bf \Xi}({\rm Y}', x)\rangle_\sigma) \over{\bf g}_{11}(\langle{\bf \Xi}({\rm X})\rangle_\sigma)}~\mathbb{F}({\rm Y'-Y})\sum_{p = 1}^\infty pb_p~
{\delta\over \delta \langle{\bf g}^{\rm AB}({\rm X})\rangle_\sigma}
\left[\sqrt{{\bf g}_{8}(\langle{\bf \Xi}({\rm X})\rangle_\sigma)}{\bf Q}_{\rm pert}(\tilde{c}(p); \langle{\bf \Xi}({\rm X})\rangle_\sigma)\right] \\
& ~\times \left[{\bf Q}_{\rm pert}(\check{c}(p); \langle{\bf \Xi}({\rm Y}', x)\rangle_\sigma) + 
{\rm M}_p^8\int_{{\cal M}_8} d^6y'' d^2w''\sqrt{{\bf g}_8(\langle {\bf \Xi}({\rm Y}'', x)\rangle_\sigma)} ~{\bf Q}_{\rm pert}(\grave{c}(p); \langle{\bf \Xi}({\rm Y}'', x)\rangle_\sigma) \mathbb{F}(y'-y''; w'-w'')\right]\\
& ~ \times ~ {\rm exp}\left({-p{\rm M}_p^8\int_{{\cal M}_8} d^6y'''~d^2w''' 
\sqrt{{\bf g}_8(\langle{\bf \Xi}({\rm Y}''', x)\rangle_\sigma)}\big\vert \mathbb{F}(y' -y'''; w' -w''') {\bf Q}_{\rm pert}(\tilde{c}(p); \langle{\bf \Xi}({\rm Y}''', x)\rangle_\sigma)\big\vert}\right)\\
& = {\rm M}_p^6 \sum_{p = 1}^\infty pb_p \int_{\widetilde{\cal M}_6} d^4y' d^2w' g_s^{\theta_{\tilde{c}(p)}(g_s) - {14\over 3}} \check{g}_s^{{\rm log~corrections}}\\
& \times 
{\rm exp}\left(-p{\rm M}_p^6 g_s^{\theta_{\tilde{c}(p)}(g_s) - 2}\check{g}_s^{{\rm log~corrections}} \int_{\widetilde{\cal M}_6} d^4y''' d^2w''' \sum_{n = 0}^\infty \vert \mathbb{F}({\rm Y}' - {\rm Y}''') {\bf I}_3^{(n)}(\tilde{c}(p); {\rm Y}''')\vert \check{g}_s^{{2n\over 3} {\rm L}_2(f_1(t))}\right)\\
& \times \Bigg[\sum_{k = 0}^\infty {\bf P}_{\mu\nu}^{(k)}(\tilde{c}(p), \check{c}(p); y')~ g_s^{\theta_{\check{c}(p)}(g_s)} \check{g}_s^{{2k\over 3}{\rm L}_3(f_1(t)) + {\rm log~corrections}}\\
& + {\rm M}_p^6 \int_{\widetilde{\cal M}_6} d^4y''d^2w''
\sum_{k = 0}^\infty {\bf Q}_{\mu\nu}^{(k)}(\tilde{c}(p), \grave{c}(p); {\rm Y}'')~ \mathbb{F}({\rm Y}' - {\rm Y}'') ~ g_s^{\theta_{\grave{c}(p)}(g_s)} \check{g}_s^{{2k \over 3}{\rm L}_4(f_1(t)) + {\rm log~corrections}}\Bigg] \mathbb{F}({\rm Y}' - {\rm Y}) \\
\end{split}
\nd}
where ${\rm Y} = (y, w)$ and $\widetilde{\cal M}_6 = {\cal M}_4 \times {\mathbb{T}^2\over {\cal G}}$. The matching of the $g_s$ scalings with the ones from \eqref{henoaleng} is now more complicated because the quantum scaling $\theta_c$ of \eqref{QT3} is now a function of $g_s$ itself because of its temporal dependence. In addition to that we now have three new functions of $f_1(t)$, namely ${\rm L}_i(f_1(t))$ for $i = 2, 3, 4$, plus the expected log corrections. The dominant scaling appears from $n = k = 0$ and, with the log corrections being very small, we have a similar relation like \eqref{bound}:
\bg\label{bound2}
\theta_{\tilde{c}(p)}(g_s) +  \theta_{\check{c}(p)}(g_s) = {8\over 3}, ~~~~
\theta_{\tilde{c}(p)}(g_s) +  \theta_{\grave{c}(p)}(g_s) = {8\over 3}, \nd
except that the LHS are functions of $g_s$. As it stands, \eqref{bound2} doesn't make any sense because the RHS are $g_s$ independent factors of ${8\over 3}$ whereas the LHS are functions of $g_s$, and therefore functions of the conformal time $t$. On the other hand, if $\theta_{{c}}(g_s)$ are {\it very slowly} varying functions of time
then \eqref{bound2} could still make sense and using it we can construct the required quantum terms that would support a configuration like \eqref{ripley} with the simplifying conditions \eqref{natasbhalo}. Recall that, one of the requirement on the temporally varying cosmological constant in section \ref{sec5.3} is to take a very slow variation. We now see that this is a self-consistent assumption from various angles. What this means with respect to the DESI BAO result \cite{desibao} remains to be seen, although issues pointed out in \cite{donof} may no longer persist. We will discuss more on this in future publication. 

\begin{figure}
    \centering
    \includegraphics[scale=0.8]{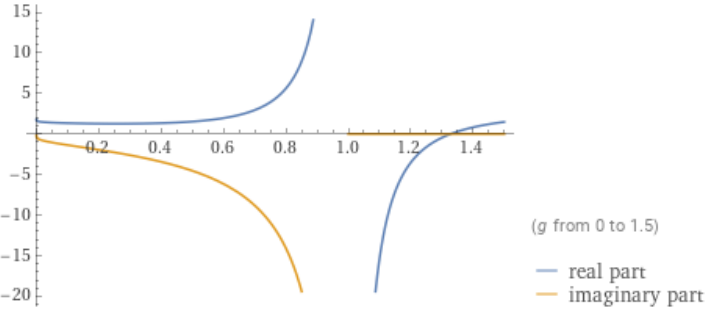} 
    \caption{Plot of the real and imaginary parts of the function $\theta_{\tilde{c}(p)}= {2} + {{\rm log}\left({2}~{\rm log}~g\right)\over {\rm log}~g}$ in \textcolor{blue}{blue} and 
     \textcolor{orange}{orange} respectively. Note that no solutions exist for $g < 1$ here because  both 
     ${\bf Im}~\theta_{\tilde{c}(p)}$ and ${\bf Re}~\theta_{\tilde{c}(p)}$ are non-zero. For $g \ge 1.33$ solutions exist, because ${\bf Im}~\theta_{\tilde{c}(p)} = 0$, but they are functions of $g$.}
    \label{plotr=3}
\end{figure}

\section{A non-technical summary on the construction of de Sitter GS states \label{sec8.1}}

We provide a non-technical summary of our work on the construction and the consistency of the de Sitter Glauber-Sudarshan (GS) states in a question-and-answer format. Our aim here is to elucidate the main points and the underlying physics of our construction without going into any technical details. A technical summary appears in section \ref{sec8}.

As unearthed\footnote{Also by G. Galilei and S. Coleman before.} by G. Moore and J. Harvey in \cite{mooreH} towards the end of last century, the discussion between Filippo Salviati, a scholar and a close friend of Galileo Galilei, and  Giovanni Francesco Sagredo, an intelligent layman, helped us to understand some basic principles behind the membrane instanton computations. Interestingly, unbeknownst to the general public, Salviati and Sagredo also had a detailed discussion on the two chief stringy systems related to positive cosmological constant solution in four spacetime dimensions, namely, the vacuum solution versus the excited ``Glauber-Sudarshan$"$ states. In the following we reproduce an unabridged version of the dialogue\footnote{Needless to say, any errors in translation leading to any subsequent misinterpretations of the text are the faults of the present authors of this paper.}. 

\vskip.1in

\subsection{The prologue: On the basics of the construction}

\vskip.1in

\noindent {\bf Sagredo:} I heard from Simplicio that you reproduced our four-dimensional de Sitter spacetime as a coherent state from M-theory...

\vskip.1in

\noindent {\bf Salviati:} Indeed but you got the details wrong: We reproduced four-dimensional de Sitter spacetime in string theory as a Glauber-Sudarshan state, and {\it not} as a coherent state. The difference between them is very important, but we are getting ahead of the story. And sorry, I interrupted you midway in your question. Please proceed.

\vskip.1in

\noindent {\bf Sagredo:} Yes, and of course that is one of my concern, but before I go into this I have some even simpler set of questions related to your construction. I understand that you realized de Sitter spacetime in type IIB theory, but I don't see why you have to do this from M-theory. Why not directly in type IIB theory?

\vskip.1in

\noindent {\bf Salviati:} That's a good question. You can of course do everything from type IIB, but the construction that we're looking at has vanishing axio-dilaton which is the constant coupling point of F-theory \cite{senkd}. This makes the type IIB string coupling to be exactly identity where even S-duality doesn't help. Plus the non-existence of a well-defined action in type IIB is yet another concern, although in recent times this has been surmounted \cite{senpaper}. 

\vskip.1in

\noindent {\bf Sagredo:} OK, but why do you need to be at a constant coupling point of F-theory? You can surely always go away from that point to be somewhere in the moduli space of F-theory where the coupling is weak..

\vskip.1in

\noindent {\bf Salviati:} Sure you can be at a generic point in the moduli space of F-theory, but this will make the axio-dilaton to become time-dependent which, in turn, will make the seven-branes dynamical. The fluxes are already time-dependent,  so the system will be highly non-trivial and we're not sure if it is easy to tackle the dynamics with the limited knowledge we have.

\vskip.1in

\noindent {\bf Sagredo:} I'm still a bit confused. We could always be in the so-called {\it static-patch} of de Sitter where there is no time-dependence. Couldn't we just study the dynamics from this point of view? 

\vskip.1in

\noindent {\bf Salviati:} You could in principle be any patch of a de Sitter space, but that doesn't mean that the system will be time-{\it independent}. Static patch in particular has a very deceptive dynamics. Since you asked me the question, let me clarify by first pointing you to 
{\bf figure \ref{staticpatch3}} in section \ref{sec3.1}. Please look at the middle and the right-most figures. In the middle figure the frequencies are not changing, however this doesn't mean that there is no time-dependence. Outside the static patch, the frequencies do change (the space-like vectors become time-like), and from Fourier decomposition we know that the frequencies are also changing with respect to time despite the metric appearing to be time-independent. (View this as though you are using global or Poincare patch temporal coordinate.) The difference is that we don't see these temporal variations from inside the static patch. Additionally, the dual IIA coupling \eqref{polter2} as discussed in section \ref{sec7.2} shows that it becomes strongly coupled along certain paths (see {\bf figure \ref{gsplot}}). 

\vskip.1in

\noindent {\bf Sagredo:} Thank you, but your answer has raised more questions for me. First, you said that the dual type IIA coupling becomes strong. But, aren't you supposed to allow strong IIA coupling since you are going to M-theory? 

\vskip.1in

\noindent {\bf Salviati:} Yes we are in M-theory, but we are in the opposite limit where the dual IIA coupling $g_s < 1$. Now you might ask: why is that? The reason is that we are not restricted to perturbative computations only. In fact, as will become clear soon, we really {\it do not} have any perturbative interactions at all. Due to an underlying resurgence structure (which I shall also clarify soon), the action that actually controls the dynamics of the de Sitter space only has non-perturbative and non-local interactions. Both these effects go as ${\rm exp}\left(-{1\over g_s^a}\right)$, and to allow for a convergent series of such terms, we need $g_s < 1$.  

\vskip.1in

\noindent {\bf Sagredo:} OK, this looks even more complicated than what I thought, as I do not understand many of the points that you referred to related to resurgence, convergence and the system not having any perturbative interactions at all. I also do not understand how and where are you getting the non-local interactions from. But that's for later. Meanwhile, even in field theory, don't we have perturbative interactions? For example $\lambda\varphi^4$ theory or even QED. They all have perturbative interactions as far as I know.

\vskip.1in

\noindent {\bf Salviati:} I see that you are going for even more basic questions. But I'm sorry to break it to you: perturbative interactions which were long thought to be useful techniques to study quantum field theory, are not very reliable as we go to higher orders in perturbation theory. This is in the light of more recent developments in the subject \cite{resurgence}. One needs to tread carefully here. The problem lies in the asymptotic nature of a perturbative series itself. The factorial growth of the Feynman diagrams calls for Borel resumming a perturbative series. Such a summation actually counts the number of instanton contributions to the system coming from {\it both} the real and the complex instantons. The net effect of this is to convert a perturbative series to a trans-series, or more appropriately to a {\it resurgence} trans-series. The meaning of such a trans-series should now be clear: in a path-integral representation this corresponds to taking the instanton saddles and the corresponding fluctuation determinants into account. This implies that your $\lambda\varphi^4$ theory and QED should be expressed in terms of such a trans-series, and {\it not} as perturbative theories, which is what Dyson \cite{dyson} and other mathematical physicists \cite{resurgence} have been emphasizing for some time now. Of course perturbation theory does give correct answer up to some order, for example QED up to say three-hundredth loop, beyond which the series stops being convergent.

\vskip.1in

\noindent {\bf Sagredo:} OK, I think I understand what you are saying, but what it has to do with your construction of a de Sitter spacetime? However before you answer that, maybe I want to take a few steps back and ask you an even more basic question. You are claiming to study de Sitter space from M-theory. But we know nothing of M-theory, except maybe the low energy dynamics. How are you going to tackle that?

\vskip.1in

\noindent {\bf Salviati:} Yes we know nothing of the UV behavior of either string theory or M-theory, but we do know one thing for sure. No matter how complicated the short distance behavior of these theories are, or what degrees of freedom reside there, the far UV dynamics are well-defined {\it without} any pathologies. This is important for the consistency of these theories. The point is that, no matter what degrees of freedom exist in the far UV, an Exact Renormalization Group (ERG) procedure \cite{erg} will tell us that once we integrate from $\Lambda_{\rm UV}$ till $\mu$ (see {\bf figure \ref{scales}}) all the information of the UV can be encoded in the {\it coefficients} of the infinite towers of marginal, relevant and irrelevant interactions of the massless states! Of course this means that we have to keep track of all possible operators constructed from the massless degrees of freedom. This way we lose no information of the UV, but the subtlety is that some of these massless states could be {\it off-shell}. Question then is how to deal with the off-shell degrees of freedom? I'll answer that question after I have motivated the construction of the Glauber-Sudarshan states. Meanwhile I'll let you ask other basic questions you might have in mind. 

\vskip.1in

\noindent {\bf Sagredo:} Thank you, that discussion is useful, and I'll ask about the off-shell states soon. As I understood, the infinite interactions of the massless states are important and truncating the series to only relevant and marginal operators is definitely not the right way to go. I suspected something along that direction, but I don't see why we cannot allow for de Sitter spacetime to be a vacuum solution to the aforementioned system. Of course the actual analysis could become technical, but is there a fundamental reason why such a solution {\it cannot} appear in string theory?   

\vskip.1in

\noindent {\bf Salviati:} Well, I'm guessing that you may have already heard of the no-go theorem \cite{GMN} that forbids a de Sitter spacetime with only classical sources. If not, the reasoning therein is rather simple and has to do with the strange behavior of gravity itself. Once we put matter or energy, the spacetime gets curved due to the back-reactions from these sources, and the resulting curvature creates a {\it negative} potential. This negative potential nullifies any positive energy that we put in the system. This is why a negative or a zero cosmological constant solution is ubiquitous in string theory (or in any theory of gravity), but a positive cosmological constant solution is rare. In string theory the no-go theorem states that any kind of matter, including fluxes, branes, anti-branes and orientifold planes, cannot give rise to a positive energy solution. We need something that does not back-react so strongly $-$ so that it is not nullified by the negative gravitational potential $-$ yet creates enough excess positive energy to give a positive cosmological constant solution. This is where the quantum terms become immensely useful. One of the surprise however is that the perturbative corrections {\it fail} to do this, so all responsibilities now lie on the shoulders of the non-perturbative and the non-local quantum terms... 

\vskip.1in

\noindent {\bf Sagredo:} OK, so the non-perturbative and the non-local quantum terms can give rise to a de Sitter vacuum solution. What prohibits us to use these corrections to get a de Sitter {\it vacuum} solution here?

\vskip.1in

\noindent {\bf Salviati:}  Mi dispiace, questa $\grave{\rm e}$ una conclusione affrettata. You are forgetting one thing that I emphasized earlier, namely the existence of a low energy effective action coming from implementing an Exact Renormalization Group procedure. What does such a procedure entail? 

\vskip.1in

\noindent {\bf Sagredo:} I'm guessing that this would entail integrating out the high frequency $-$ and consequently the high energy $-$ modes as well as integrating out the massive states above a certain energy scale.

\vskip.1in

\noindent {\bf Salviati:} Indeed! But what if we couldn't do such an integrating-out procedure? You may ask when and how could such a situation arise. Such a situation would appear if there is no well-defined UV completion or if the frequencies themselves are changing with respect to time. We can clearly rule out the former because, as I described a moment ago, no matter how complicated the short-distance behavior of string or M-theory is, the far UV is by definition a well-defined theory. However the latter is now an issue: one may easily check that the fluctuating frequencies over a four-dimensional de Sitter spacetime do become time-dependent. Such time-dependence would rule out the Wilsonian integrating-out procedure because  $-$ as the frequencies are constantly red-shifted  $-$ there is no meaning of the integrating out process now! You may immediately ask why we couldn't do this over a static patch. I already answered that: the dynamics over a static patch is highly deceptive. This should also be clear from {\bf figure \ref{staticpatch3}}: looking at the right-most diagram, the frequencies are actually {\it changing} even inside the static patch although they may not look like to the static patch observer. It would clearly be wrong to implement the ERG procedure and write an effective action at low energies. Moreover quantization of strings now becomes rather complicated because of this time-dependence. It is not at all clear that the same massless modes $-$ that we got from quantizing a string over a flat background $-$ continues to provide the dynamics over the static patch. Plus of course having no supersymmetry at the vacuum level simply adds more issues to an already problematic edifice.

\vskip.1in

\noindent {\bf Sagredo:} But if we are inside the static patch, we are confined to be within the boundary of the patch. Does it matter what happens {\it outside} the boundary as we have no access to it?

\vskip.1in

\noindent {\bf Salviati:} This is a wrong conception of the dynamics in the static patch. You have chosen a bad coordinate system that appears to disallow you to go outside, but that doesn't mean you have no access beyond the confines of the static patch. Outside the static patch what really happens is that the relevant time-like vector becomes space-like and vice versa. Thus all degrees of freedom that are not changing with respect to time inside the static patch will start having temporal dependence outside the static patch. This is of course the content of {\bf figure \ref{staticpatch3}} that I referred to you earlier.

\vskip.1in

\noindent {\bf Sagredo:} I'm still a bit puzzled by this. For the Schwarzschild black hole the static patch argument still works, isn't it? Since de Sitter spacetime is very similar to a Schwarzschild black hole, shouldn't the argument go through here too? 

\vskip.1in

\noindent {\bf Salviati:} Thank you for asking this, but it is a misconception that the physics for the two cases are similar. I don't want to go through the details of the black-hole physics here, but let me just point out the following. A drawback of analysing the static patch alone is that the cosmological horizon, and consequently the static patch, are both observer-dependent concepts. What is typically assumed when one draws the Penrose diagram of a de Sitter space is that there is a static observer sitting at the North Pole, and the static patch is the causal region of the de Sitter space accessible to the observer since this is the region to which he can both send and receive information from. However, for any two physical observers, their respective static patches would be different. This is in contrast to the 
case of the Schwarzschild black hole, whose exterior Schwarzschild patch is the same for all asymptotic observers.

\vskip.1in

\noindent {\bf Sagredo:} Thank you for the explanation. However I'll need some time to digest what you said as they are rather non-trivial concepts. Shall we break for lunch and we can resume after?

\vskip.1in

\noindent {\bf Salviati:}  Fare una pausa per il pranzo sembra un'ottima idea. Incontriamoci dopo.

\vskip.1in

\subsection{The intermezzo: On the subtleties of the construction}

\vskip.1in

\noindent {\bf Sagredo:} I had some time to ponder over the lunch. I think I'm beginning to understand the problem, but surely such a startling thing must have been noticed earlier? How did the physicists deal with such an outcome? 

\vskip.1in

\noindent {\bf Salviati:} For a long time static patch was considered to be the way out by somehow restricting the dynamics within the finite radius of the patch. However the study of de Sitter in string theory is relatively new and with it new surprises are coming forth. For example, there are various constraints that come with such a study that involve many things and not just reproducing the metric. Moreover in string theory we simply cannot restrict the fields to lie within a finite region of space and forget that there is something beyond the confines of the static patch! Additionally, ignoring the fact that we have no spacetime supersymmetry anymore,
a consistent compactification would imply quantized fluxes over the internal compact non-K\"ahler manifold. Outside the static patch, such a manifold itself will become time-dependent, implying subtleties with flux quantization, anomaly cancellation, dynamical motion of the branes, {\it e cos$\grave{i}$ via}. All these issues have to be solved simultaneously
before we can claim that we have a consistent solution. Putting it differently, in string theory there is no meaning of {\it one or the other}, rather {\it all} have to be solved together especially when the background has time-dependence. Going to 
other patches related to the static patch, like the Eddington-Finkelstein or the Kruskal slicings or the so-called ``nice-slices$"$, will unfortunately not alleviate the problem. They may hide it, but cannot remove the problem.

\vskip.1in

\noindent {\bf Sagredo:} Sorry, but I need to interrupt as I don't understand how the branes enter the picture. I thought you said that you have integrated out all massive degrees of freedom beyond some energy scale. Aren't branes included in those massive degrees of freedom?

\vskip.1in

\noindent {\bf Salviati:} First let me explain where the branes enter the story. Remember, I mentioned the asymptotic nature of the perturbative series, and how this is replaced by a resurgent trans-series? A part of the contribution comes from the brane-instantons as was first discussed in \cite{shenker}. However in the low energy limit that we study here, we don't see the branes, although we do see the instantons. The effects of the branes come from their {\it classical} back-reactions and from their massless world-volume interactions. The latter is again asymptotic and may be rewritten in terms of a resurgent trans-series. This is what I emphasised earlier: there are no perturbative interactions.  

\vskip.1in

\noindent {\bf Sagredo:} Thank you for the brane discussion, and I understand the seriousness of the problem regarding the Wilsonian integrating out procedure. But you haven't answered my previous question: How did the physicists deal with such an outcome? 

\vskip.1in

\noindent {\bf Salviati:} Yes, sorry I didn't answer that. There are at least two ways I know by which this problem was handled: one, using {\it open} quantum field theory, and two, using a certain {\it censorship} criterion called the trans-Planckian censorship. The first one, namely the 
open QFT \cite{vernon}, is traditionally described by isolating relevant degrees of freedom from an ``environment$"$ which allows them to gain or lose energies to the environment. Since neither energy is conserved, nor an EFT description exists, the dynamics of the theory typically follows some Markovian process which may be quantified in certain settings \cite{suddhomarkov}. The problem with this picture appears when we try to use it in string theory: in a UV complete theory, like string theory or M-theory, there appears to be no clear demarcation between the relevant degrees of freedom and the environment, and therefore a concrete realization over a temporally varying background is equally hard. So far there has not been much progress implementing this idea in string theory, although some recent works in field theories have shed some interesting light on the de Sitter problem \cite{burgess}.

\vskip.1in

\noindent {\bf Sagredo:} This means, 
if I understood correctly, since we are dealing with only the low-energy degrees of freedom in string or M-theory, there is no simple way to isolate the relevant degrees of freedom from an ``environment$"$. In fact ambiguities like whether we should view the massive $-$ or what fractions of the massless on-shell and off-shell $-$ degrees of freedom to be the ``environment$"$ now appear. 

\vskip.1in

\noindent {\bf Salviati:} Indeed that is the point. Because of these ambiguities there is no simple way to implement the idea of an open QFT in string theory.

\vskip.1in

\noindent {\bf Sagredo:} Thank you, although I still haven't understood how you are dealing with the massless off-shell degrees of freedom, and I'm guessing you'll explain that soon. But before we go into that, what about the trans-Planckian censorship criterion?

\vskip.1in

\noindent {\bf Salviati:} Yes, first let me explain what this entails. 
The idea behind the trans-Planckian censorship conjecture is simple. It states that a reverse time evolution of excitations with ${\cal O}(1)$ energies living on a late time slice leads to trans-Planckian energies for the same at a finite time in the past. In generic theories of higher derivative gravity, including Einstein gravity, this is problematic because the short-distance behaviors are not well-defined. In terms of future time evolution, trans-Planckian modes in the past become increasingly long wavelength and subsequently freeze upon exiting the Hubble horizon. Thus at late times, one can classically detect trans-Planckian physics at ${\cal O}(1)$ scales, for instance, in the CMB data.  In the original formulation \cite{tcc}, where the theory was restricted to obey Einstein gravity only, the {\it trans-Planckian censorship conjecture} prohibits the existence of such a scenario by simply bounding the temporal domain for these UV modes to not exit the Hubble horizon. More precisely, the conjecture states that any solution where we can see a classical imprint of trans-Planckian features upon exiting the de Sitter horizon is not possible since the time scale is bounded by \eqref{tcccrit}. Saying differently, as long as we are bounded by the temporal interval \eqref{tempo} we will not see the trans-Planckian problems.

\vskip.1in

\noindent {\bf Sagredo:} Thank you, but I'm puzzled by your explanation. I thought you said earlier that the short distance behaviors of string and M-theory have no pathologies. The UV dynamics therein are all well-defined, so why should we worry about the trans-Planckian modes? The new degrees of freedom that enter at scale ${\rm M}_p$ would render all UV amplitudes finite, so any modes frozen at late time should show no trans-Planckian issues.

\vskip.1in

\noindent {\bf Salviati:} I'm glad you asked this question. Indeed in string or M theories there are no such trans-Planckian issues because of the UV finiteness, although one might be concerned that the UV degrees of freedom are somehow getting classicalized due to the Hubble expansion. This is not a worrisome point but more like a technical challenge, so imposing the trans-Planckian bound here would simply amount to {\it hiding} it. However something else gets manifested here due to the temporal variation of the fluctuating frequencies: the convergence of the non-perturbative series. How do we see this? Remember I mentioned the fact that the dual IIA coupling has to satisfy $g_s < 1$? In a flat-slicing this actually leads to \eqref{tcc}. Now compare \eqref{tcc} to 
\eqref{tempo}: there appears to be an amazing match despite the fact that the two ways of reaching the temporal domain are quite different. This means, while for the generic higher derivative or Einstein gravities going beyond the temporal domain \eqref{tempo} classicalizes the trans-Planckian issues, for string or M theories it is the convergence of the non-perturbative series going as ${\rm exp}\left(-{1\over g_s^a}\right)$, $a > 0$,  
that becomes an issue beyond the temporal domain \eqref{tcc}.  

\vskip.1in

\noindent {\bf Sagredo:} I thought the instanton saddles go as ${\rm exp}\left(-{n\over g_s^a}\right)$ where $n \in \mathbb{Z}$ is the instanton number. So even if $g_s = {\cal O}(1)$ number, shouldn't the series be always convergent for large $n$?

\vskip.1in

\noindent {\bf Salviati:} Well, yes and no. If the instanton series were as simple as what you mentioned, then yes, the convergence is guaranteed for large instanton numbers. However the actual computation, with say M5-brane instantons wrapped on the six-dimensional non-K\"ahler base, is far more complicated. For example if you look at the third line in \eqref{kaittami2}, with ${\bf Q}_{\rm pert}$ therein given by the series \eqref{QT3}, it is not guaranteed that the series would converge even for large $s \in \mathbb{Z}$. However for $g_s << 1$, convergence can actually happen as discussed in the analysis presented in section \ref{sec7.3}.

\vskip.1in

\noindent {\bf Sagredo:} I see that the quantum series ${\bf Q}_{\rm pert}$ appears in both the fluctuation determinants as well as the instanton saddles, with different choices of the coefficients. This already looks pretty complicated, and I was wondering how do we even go about using it in the EOMs? But I think I'm getting ahead of myself. I need to first understand how and why you are constructing de Sitter spacetime as a Glauber-Sudarshan state. Could you please start by clarifying that?

\vskip.1in

\noindent {\bf Salviati:} Yes, it is high time we get into the main issue of de Sitter spacetime as a Glauber-Sudarshan state. Let me ask you something before we start. Do you agree that the kind of potentials we see in string or M theories only have Minkowski and AdS minima, but no minima with positive energies? In other words, in string/M theory we can see potentials like the ones in {\bf figure \ref{pot1}} $-$ with possible additional AdS minima $-$ but not a potential like the one on the right of {\bf figure \ref{multsaddle}} with unequal minima?

\vskip.1in

\noindent {\bf Sagredo:} Not really. Why do you say that?

\vskip.1in

\noindent {\bf Salviati:} Well, think of it in the following way. If the potential has a minimum like the one on the right of {\bf figure \ref{multsaddle}}, then we will have a positive energy solution in the {\it static patch}. (Or generically in any other patch.) From our discussion above, such a solution cannot allow a Wilsonian integrating out procedure resulting in an ill-defined EFT at low energies. Since string or M theories cannot allow this, the only way this can happen is by {\it modifying} the profile of the potential using the non-pertubative and the non-local quantum corrections so that the minimum is removed or smoothened out. In other words, the low energy EFT that I told you earlier must have come from potentials like the ones in {\bf figure \ref{pot1}} with additional AdS minima, but not the one on the right of {\bf figure \ref{multsaddle}} with unequal minima. You see that demanding Wilsonian integration procedure or an Exact Renormalization Group technique essentially rules out a four-dimensional de Sitter vacuum solution in string theory.

\vskip.1in

\noindent {\bf Sagredo:} This reminds be of the conjectures I read about in \cite{whatif} and \cite{swampland}. Your conclusion seems to coincide with what those papers said, but as far as I see, your methods are quite different. Is this correct? 

\vskip.1in

\noindent {\bf Salviati:} Yes, we do seem to agree with the final conclusion that a four-dimensional de Sitter space cannot exist as a vacuum solution in string theory. And you are also correct to say that the techniques we used to get to our conclusion are quite different from the ones used in the aforementioned papers. To be honest, I do not understand how the EFT criteria mentioned in \cite{swampland} were {\it derived}. I understand that they were {\it proposed} to rule out accelerating solutions in string theory, but I didn't see any derivations there. Our analysis, which has its roots in the ERG techniques, does not rule out de Sitter space in string theory. Instead, four-dimensional de Sitter space can exist in string theory $-$ and even within the temporal bound advocated by the so-called trans-Planckian censorship conjecture $-$ except not as a vacuum state but as an {\it excited state} in the low energy effective action of string theory.

\vskip.1in

\noindent {\bf Sagredo:} Actually I heard that the modified version of the criteria in \cite{swampland}, motivated from the trans-Planckian censorship conjecture \cite{tcc}, does allow a four-dimensional de Sitter space to exist.

\vskip.1in

\noindent {\bf Salviati:} Yes, but there is a difference: nothing really bad happens in our set-up beyond the temporal domain \eqref{tcc}. The convergence issue of the instanton series does come under scrutiny, but this simply means that the de Sitter excited state may no longer be sustained by the non-perturbative effects and may just go back to the vacuum Minkowski configuration. On the other hand, in the set-up of \cite{swampland}, the theory is strictly restricted to be within the temporal domain \eqref{tempo} because beyond which there are short distance pathologies.

\vskip.1in

\noindent {\bf Sagredo:} OK that explains things quite well. Now could you kindly explain how the de Sitter excited state is constructed? Plus you said at the begining that it is different from a coherent state. Could you elaborate on it?

\vskip.1in

\noindent {\bf Salviati:} Yes, after a long preparation I can finally explain to you our construction. However it is getting late, and we have been talking for quite a while. How about we resume our discussion tomorrow?

\vskip.1in

\noindent {\bf Sagredo:} S$\grave{\rm i}$ riprendiamo la discussione domani.

\subsection{The dialogue: On the construction of the GS states}

\vskip.1in

\noindent {\bf Sagredo:} Yesterday before breaking off, you said that you'll motivate the construction of four-dimensional de Sitter space as an excited state in the low energy effective field theory of type IIB string theory.

\vskip.1in

\noindent {\bf Salviati:} Yes indeed. The fact that string or M theory appears to  allow potentials that only have Minkowski and AdS minima, like the ones in {\bf figure \ref{pot1}}, but no potentials with unequal minima, like the one in right of {\bf figure \ref{multsaddle}}, convinced us to look for four-dimensional de Sitter space as an excited state over a supersymmetric Minkowski minimum in string theory. Now if you look around yourself, the surrounding spacetime is pretty {\it classical} in the sense that you can measure distances and do other experiments classically. So whatever excited state we get from the low energy effective action must be very close to the classical picture. Now what quantum state is closest to classical physics? The answer is clearly a {\it coherent state} so naively one would expect de Sitter space to be a coherent state. 

\vskip.1in

\noindent {\bf Sagredo:} Yes, that would also be my expectation. However you are going to say that it is incorrect. Why is that?

\vskip.1in

\noindent {\bf Salviati:} The reasoning is simple if you look at the definition of the coherent state in quantum mechanics. Coherent state is constructed by shifting the {\it free} vacuum by a unitary displacement operator. One immediate problem is that this definition is in quantum mechanics, but can be extended to field theory because quantization of modes does lead to an infinite collection of simple harmonic oscillators.
However the issue with the low energy effective field theory is that it is {\it not a free theory}. Recall our earlier discussion of the constant coupling limit of F-theory where I emphasised that the type IIB coupling is fixed to 1 and going away from that point entails complicated and possibly uncontrolled dynamics involving the seven-branes and whatnot. 

\vskip.1in

\noindent {\bf Sagredo:} Which means, if I understood correctly, construction of a coherent state doesn't make any sense here and probably doesn't even exist in a simple form that we usually see in quantum mechanics. Is that the only reason to discard this idea?

\vskip.1in

\noindent {\bf Salviati:} No, there are many other issues. Even if I grant you a weak coupling scenario, the construction of coherent state cannot be simple because of the numerous tunnelling effects from the nearby vacua in potentials  like the ones from {\bf figure \ref{pot1}}. These tunnelling effects do not just come from the {\it real} instantons, but they have sizable contributions from the {\it complex} instantons also. (These complex instantons come from the complex solutions of the turning points in an {\it inverted} potential.) Secondly, a coherent state over the lower minimum in (say) a potential like the one on the right in {\bf figure \ref{multsaddle}} is simply a Bogoliubov transformation of the state over the positive energy minimum. The second point is crucial. It simply says that a coherent state construction is nothing but a fancy way of saying that a vacuum state exists in a nearby minimum with positive energy. Plus the fact that the construction uses a unitary displacement operator means that the resulting analysis is just the tree-level answer in a free field theory. de Sitter space cannot appear from any tree-level computations, so coherent state {\it cannot} give rise to the kind of quantum state we want which overcomes the no-go conditions of \cite{GMN}.

\vskip.1in

\noindent {\bf Sagredo:} I see the issues with a coherent state. 
But what stops us to make simple modifications of the coherent state by adding in quantum corrections by hand? 
Are there problems with these kind of scenarios?

\vskip.1in

\noindent {\bf Salviati:} Yes because adding in pertubative quantum corrections cannot help simply because of the asymptotic nature of such quantum series. You may now say that we can allow only the non-perturbative corrections. However an unambiguous set of non-perturbative corrections 
could only appear once we fix the form of the perturbative correction unambiguously. The latter is only possible if we know what kind of quantum state we are looking for..

\vskip.1in

\noindent {\bf Sagredo:} I'm guessing that the state should be in the full interacting theory that produces a potential like the ones in {\bf figure \ref{pot1}} with possible additional AdS minima.

\vskip.1in

\noindent {\bf Salviati:} Yes, and in the {\it full} interacting theory we can choose the interacting vacuum $\vert\Omega\rangle$ and shift this by a displacement operator $\mathbb{D}(\sigma)$ to create the so-called Glauber-Sudarshan state $\vert\sigma\rangle$ associated to every {\it on-shell} degrees of freedom over a supersymmetric Minkowski minimum. Such a state is far from the Gaussian approximation we use in perturbative field theories. 

\vskip.1in

\noindent {\bf Sagredo:}  I understand the construction, but why shift the interacting vacuum? Why not just take the interacting vacuum $\vert\Omega\rangle$ itself?

\vskip.1in

\noindent {\bf Salviati:} This is simple to see once you realize that the {\it emergent} on-shell value would appear from an expectation value over the Glauber-Sudarshan state. For example the emergent spacetime metric configuration would now be $\langle \hat{\bf g}_{\mu\nu}\rangle_\sigma$ which takes the path-integral form as defined in \eqref{mmtarfox}. This is just a one-point function and clearly vanishes if we only take $\vert\Omega\rangle$. The displacement operators are responsible in shifting the vacuum, and now we see why we cannot allow unitary operators: they would satisfy $\mathbb{D}^\dagger \mathbb{D} = 1$ and therefore be inconsequential when inserted in the path-integral. 

\vskip.1in

\noindent {\bf Sagredo:} OK, but I could take any other state in the full theory and shift it by a displacement operator. What's so special about the interacting vacuum state $\vert\Omega\rangle$?

\vskip.1in

\noindent {\bf Salviati:} First of all, the vacuum state is the lowest energy state in the full interacting theory and it's wave-function in the configuration space will most likely be localized around the origin of the configuration space. A shift of the interacting vacuum does exactly what we want: it localizes over certain set of Fourier modes within the energy scale of our low energy effective action. However this localization is {\it not} over the classical Fourier modes that any set of coherent states would have done. This difference is crucial and the reason is because of what I emphasized earlier: the asymptotic nature and the underlying resurgence structure of the system. Your question is, why a preferential treatment to the interacting vacuum only? The answer is that we have no idea how the higher excited looks like in an interacting theory. In fact generically we don't even know the exact wave-function of the interacting vacuum in the configuration space, let alone the higher excited states!

\vskip.1in

\noindent {\bf Sagredo:} Thank you, but I'm confused: how do we then construct the Glauber-Sudarshan state $\vert\sigma\rangle \equiv \mathbb{D}(\sigma) \vert\Omega\rangle$ if we don't know the wave-function of the interacting vacuum state?

\vskip.1in

\noindent {\bf Salviati:} Well, the key word I used above is {\it generic}. In fact  {\it generically} we don't know the wave-functions in the Hilbert space of the full interacting theory, but there is one {\it specific} situation where we can at least determine the wave-function of the interacting vacuum, provided the following two conditions are met. One, the vacuum is non-degenerate, and two, there is an energy gap between the vacuum state and the first excited state. 

\vskip.1in

\noindent {\bf Sagredo:} How do the two conditions help us to know the ground state wave-function? And what is the {\it specific} situation?

\vskip.1in

\noindent {\bf Salviati:} Interestingly, the method that I'm going to propose solves {\it two} problems at the same time. To see this let us go back to the potentials appearing in {\bf figure \ref{pot1}}, and we will concentrate only on the Minkowski minima and not the AdS minima (if any).
Choose any one of the supersymmetric Minkowski minimum and let's call it $\vert 0 \rangle_{\rm min}$. This state can be expressed as a linear combination of all the eigenstates of the full interacting potential by appropriately choosing the coefficients. You may say that this is a fancy rewriting that is devoid of any content because we generically do not know the wave-functions of the eigenstates in the full Hilbert space of the interacting theory. This is of course what I emphasized earlier, but we can play the following computation trick: hit both left and right side of the series by the operator ${\rm exp}\left(-i{\bf H} {\rm T}\right)$ where ${\bf H}$ is the total interacting Hamiltonian, and ${\rm T}$ is the temporal coordinate over the Minkowski minimum. Now take ${\rm T}$ to infinity along a slightly imaginary direction, much like what is shown in footnote \ref{saddledistance}. Such a process will kill off the effects from all higher excited states precisely because of the aforementioned two conditions. The net result is \eqref{saw1}. Plugging this in the formula for the expectation value, say $\langle \hat{\bf g}_{\mu\nu}\rangle_\sigma$, immediately converts it into a path-integral form as shown in \eqref{mmtarfox}. Thus the two issues $-$ the form of the wave-function for the interacting vacuum and the choice of the Minkowski minimum around which we want to study the theory $-$ are resolved simultaneously by this technique. This is the {\it specific} situation alluded to earlier.

\vskip.1in

\noindent {\bf Sagredo:} Thank you, but I don't see how expanding over a supersymmetric Minkowski saddle answers all the questions related to the real and the complex instantons, non-local interactions, and the resurgence trans-series that you were telling me earlier. Also how is the supersymmetry broken now?

\vskip.1in

\noindent {\bf Salviati:} First, before I answer, I should tell you to be careful with terms like {\it saddles} and {\it minima}. We are expanding over a supersymmetric Minkowski {\it minimum}, and the action therein will have contributions from all the (real and the complex) instanton {\it saddles}, including their fluctuation determinants; plus the non-local interactions and their corresponding fluctuation determinants. The latter is quite tricky to see, but again I'm getting ahead of the story. I'll let you ask all the relevant questions here, and answer the supersymmery breaking scenario a little later because it requires an ingredient that we haven't discussed yet.

\vskip.1in

\noindent {\bf Sagredo:} Yes sorry, it's the instanton saddles over a given Minkowski minimum that captures the dynamics. And now I understand how one may choose any one of the Minkowski minimum in (say) {\bf figure \ref{pot1}}, and the theory therein will manifest the various tunnellings by allowing the instanton saddles plus their corresponding fluctuation determinants in the action as you showed me earlier in the third line of \eqref{kaittami2}. However I do not understand how you managed to get this result. What type of computations will reveal this kind of behavior?

\vskip.1in

\noindent {\bf Salviati:} Yes, now that we have reached the heart of the computation, let us fix some simple model to discuss the details of the analysis. We could take a set four real scalar fields $\varphi_i$ for $i = 1, .., 4$ with an interaction given by products of some arbitrary powers of these fields with arbitrary derivatives acting on them. We could also sum over all the powers, but we won't do it here just to keep the analysis simple. Our aim would be to use the path integral \eqref{mmtarfox} to determine the one-point function of the scalar fields, but we will concentrate only on one scalar field to determine 
$\langle\hat{\varphi}_1\rangle_\sigma$.

\vskip.1in

\noindent {\bf Sagredo:} Is there a reason why you are taking only four scalar fields? Also I guess the final goal would be to compute $\langle \hat{\bf g}_{\mu\nu}\rangle_\sigma$, and other related stuff. Is there an easy way to go from the scalar field results to the one with actual metric components?

\vskip.1in

\noindent {\bf Salviati:} You are right that the correct expectation value should be \eqref{russmameye}, but the actual analysis is hindered by our lack of  knowledge of the Faddeev-Popov ghosts and gauge fixing terms. In the actual model there are four set of fields from M-theory point of view: the metric components along the ${\bf R}^{2, 1}$ directions, the metric components along the internal eight-manifold (which is a toroidal fibration over a six-dimensional base), the G-flux components and the components of the Rarita-Schwinger fermions. The four scalar fields form the representative samples of the four set of the actual fields. (The last one being related to a condensate of the fermions so as to avoid doing Grassmanian integrals, but these details are not important here.) Once we use the scalar fields, we can define the displacement operator as in \eqref{28rms}, and plug everything in the path-integral 
\eqref{mmtarfox}. 

\vskip.1in

\noindent {\bf Sagredo:} And then?

\vskip.1in

\noindent {\bf Salviati:} And then we compute the one-point function. The analysis is pretty non-trivial as detailed in section \ref{patho} due to the non-trivial interaction and due to the fact that {\it we cannot use any approximation to terminate the perturbative series}. The reason is because of what I cautioned earlier: the asymptotic nature of the perturbative series coming from the factorial growth of the diagrams, which we termed as the {\it nodal diagrams}.

\vskip.1in

\noindent {\bf Sagredo:} Are these nodal diagrams the same as the Feynman diagrams? 

\vskip.1in

\noindent {\bf Salviati:} Unfortunately not. These new set of diagrams form a much bigger class of diagrams than the Feynman diagrams as recently studied in \cite{borel}. Typically one-point function in field theory vanishes, but the story here is different. Due to the shifted vacuum from the displacement operator \eqref{28rms}, the analysis takes on a completely new direction. It doesn't matter much whether we take real or complex Fourier components of the fields, the answer remains non-zero for either case. Recall that the complex parts of the Fourier components were absolutely essential to give meanings to the correlation functions in the Feynman path-integral. Here the story is different and consequently, richer.

\vskip.1in

\noindent {\bf Sagredo:} OK I understand that the asymptotic nature of the perturbative series would make our analysis harder. However is there a way to quantify the growth of the nodal diagrams? 

\vskip.1in

\noindent {\bf Salviati:} Fortunately it turns out that there is a way to {\it count} the diagrams to see how they grow. This is discussed in section \ref{factoria} where a careful counting of all the nodal diagrams is presented. After the dust settles, a simple way to decipher the results therein is to notice that the system shows a hidden {\it binomial} structure resulting in the so-called {\it Gevrey growth} of the nodal diagrams.

\vskip.1in

\noindent {\bf Sagredo:} I heard of the factorial growth, but what is a Gevrey growth? 

\vskip.1in

\noindent {\bf Salviati:} Gevrey growth is more generic then the simple factorial growth that we study within the realm of the Feynman diagrams. A detailed discussion of this is unfortunately not possible given the limited time we have, plus it will take us along a different trajectory which I'd prefer not to venture right now. I'd suggest you look up some of the details in section \ref{sec3.7}, and in the references suggested therein. The take-home message here is that we can {\it quantify} the growth of the nodal diagrams. 

\vskip.1in

\noindent {\bf Sagredo:} Thank you, this is way more than what I thought can be achieved from the path-integral \eqref{mmtarfox}. I'm OK with the explanation you provided and will look up the references you suggested. My question is, once we know that the system has a Gevrey growth, how do we proceed?

\vskip.1in

\noindent {\bf Salviati:} Here is where something rather remarkable happens. The Gevrey growth of the system implies that we can now sum over all the diagrams using the so-called {\it Borel-$\grave{E}$calle} summation procedure. This summation procedure is explained in \cite{borel, gevrey2, dorigoni, ecalle} and in section \ref{sec3.7}. The net result of such a summation is that now we can provide a {\it closed form} expression for the four-dimensional cosmological constant $\Lambda$ as shown in \eqref{tinmey}.

\vskip.1in

\noindent {\bf Sagredo:} Sorry, it was a bit too fast for me. How did you go from the {Borel-$\grave{\rm E}$calle} resummation of a Gevrey series to a closed form expression of the cosmological constant?  Plus how do you know that the expression for $\Lambda$ you got is {\it small} and {\it positive definite}? Could you explain this in simple terms?

\vskip.1in

\noindent {\bf Salviati:} Sure I could, and sorry that I went too fast over the crucial details of our construction. However I think, for your benefit, it might be better you have a quick look at the references I suggested, and we can meet after lunch. 

\vskip.1in 

\noindent {\bf Sagredo:} Penso che sia un'ottima idea. Mi 
dar$\grave{\rm a}$ anche un po' di tempo per digerire quello che hai detto.

\subsection{The epilogue: On the consequences of the construction}

\vskip.1in

\noindent {\bf Sagredo:} I did manage to go through a sizable portion of the references you suggested, but unfortunately I still don't see how a  {Borel-$\grave{\rm E}$calle} resummation could lead to a closed form expression for four-dimensional cosmological constant $\Lambda$. But before you answer that, could you explain how supersymmetry is broken here?

\vskip.1in

\noindent {\bf Salviati:} Yes, of course I should now tell you about the supersymmetry breaking. The vacuum, which is Minkowski, is supersymmetric and therefore the supersymmetry is broken by the Glauber-Sudarshan state. I'll call this as {\it spontaneously} broken supersymmetry. The first question is how is supersymmetry preserved at the Minkowski level in the presence of fluxes et cetera? The answer is simple. If we allow a configuration of {\it self-dual} four-form fluxes then supersymmetry remains unbroken \cite{DRS}. This is simply because the mass of the gravitino can be shown to be proportional to the difference between four-form flux and its Hodge-dual partner. (The Hodge-duality is taken with respect to the internal eight-manifold in M-theory.) You may now ask: how is supersymmetry broken? The answer is again simple. The action, or the potential, that determines the dynamics of the expectation values is an emergent one in which you basically replace all the on-shell fields by their expectation values. This in turn means that the gravitino mass will be proportional to the difference between the four-form flux {\it expectation} value and its Hodge-dual partner. Clearly there is no reason for the difference of the expectation values to vanish now! And indeed one may show that this is the case, implying that the supersymmetry is spontaneously broken by the Glauber-Sudarshan state.  

\vskip.1in

\noindent {\bf Sagredo:} In other words, due to the Glauber-Sudarshan state, it would appear that the fermionic partners, after dimensional reduction, are {\it effectively} massive compared to their bosonic counterparts. I understand the idea, and it looks very much like Higgs mechanism, right?

\vskip.1in

\noindent {\bf Salviati:} Almost, except that both the fluxes and the fermionic degrees of freedom are emergent quantities here. 

\vskip.1in

\noindent {\bf Sagredo:} How do we then understand the {\it fluctuations} that give rise to quanta now?

\vskip.1in

\noindent {\bf Salviati:} Well, the fluctuations are now controlled by yet another state called the Agarwal-Tara state \cite{agarwal, coherbeta, coherbeta2}. For example, graviton fluctuations over a de Sitter space would be replaced here by another state which is like a ``graviton-added$"$ Glauber-Sudarshan state. Similarly fermionic quanta will be a ``fermion-added$"$ Glauber-Sudarshan state. Unfortunately I will not have time to elaborate on this here anymore and I suggest you look up \cite{coherbeta, coherbeta2} for more details.

\vskip.1in

\noindent {\bf Sagredo:} Thank you for the clarification. I need to read up the references you suggested. 
Could you now kindly explain the connection between {Borel-$\grave{\rm E}$calle} resummation and the four-dimensional cosmological constant?

\vskip.1in

\noindent {\bf Salviati:} The analysis leading to the aforementioned connection is a bit tricky so I'll go in small steps. We will also take the scalar field model to simplify the discussion. The path integral \eqref{mmtarfox}, as I mentioned earlier, shows a Gevrey growth of the nodal diagrams which may be quantified as detailed in section \ref{patho}. However there is a subtlety. The most dominant of the diagrams, in which the momenta of the nodes {\it do not} match with that of the source, are eliminated. This is somewhat similar to the case in usual field theory where the external legs that do not share momenta with the interaction vertices are being eliminated as vacuum bubbles. In the case with nodal diagrams, somewhat similar story prevails with final answer being just the tree-level one. Therefore non-trivial result appears when at least one of the node has the same momentum as the source. This is a sub-dominant contribution and according to the rules of the path-integral, such diagrams are suppressed by a volume factor. If more than one nodes have the same momenta as the source, they are suppressed by even higher powers of the volume factor, so they are truly sub-subdominant. 

\vskip.1in

\noindent {\bf Sagredo:} This means in the limit where the volume goes to infinity, all these diagrams are eliminated. You already told me that the dominant diagrams are eliminated like the vacuum bubbles, and now it seems that the sub-dominant ones may also be eliminated in the infinite volume limit. I'm confused, how do you then get a non-zero answer here?

\vskip.1in

\noindent {\bf Salviati:} Ancora una volta stai saltando alla conclusione troppo in fretta! The sub-dominant diagrams are indeed suppressed by a volume factor, but there is a piece inside the subdominant contribution that is proportional to the volume itself! Such a piece, which is in fact also proportional to the inverse of the massless propagator, is responsible to provide a finite answer once we divide by the volume factor. 

\vskip.1in

\noindent {\bf Sagredo:} I see, and sorry that I missed the fact. So now we sum over all the next-to-leading-order (NLO) diagrams, right? 

\vskip.1in

\noindent {\bf Salviati:} Right, but the summation doesn't go very smoothly as before because of the difference between the tree-level and the higher order nodal diagrams. This was an unforeseen subtlety, but turned out to be quite useful. To proceed we will only keep the first order sub-leading diagrams (which are suppressed by one power of the volume factor). Because of this, the Borel-$\grave{\rm E}$calle summation procedure leads to two pieces: one, the summed-up result over all the quantum terms, and two, a tree-level piece.

\vskip.1in

\noindent {\bf Sagredo:} That's a bit surprising.  Shouldn't the tree-level piece be absent due to the no-go theorem that you mentioned earlier? 

\vskip.1in

\noindent {\bf Salviati:} Yes indeed! Identifying the tree-level piece to zero provides a relation between $\sigma_i$ appearing in the displacement operator \eqref{28rms}, the massless propagator and the suppression volume. This relation is in an integrated form so it is not directly useful in determining $\sigma_i$ in terms of the momenta, but it does become useful in a different way which I'll come to soon. 

\vskip.1in

\noindent {\bf Sagredo:} OK, and what about the piece that you get by performing the Borel-$\grave{\rm E}$calle summation over the Gevrey series?

\vskip.1in 

\noindent {\bf Salviati:} Interestingly the result that we get after performing the Borel-$\grave{\rm E}$calle resummation, goes in tandem with the result we get by imposing the vanishing of the tree-level contribution. To see this recall that $g_s = \sqrt{\Lambda} t$ with $t$ being the conformal time in the flat-slicing of de Sitter. Since the scalar field toy model represents the metric component ${\bf g}_{\mu\nu}$ along ${\bf R}^{2, 1}$ in M-theory, we expect $\langle \hat{\bf g}_{\mu\nu}\rangle_\sigma$ to be equal to 
$g_s^{-8/3}$ so that it reproduces a de Sitter emergent metric in type IIB theory. Now combining the result from the Borel-$\grave{\rm E}$calle resummation with the vanishing of the tree-level result gives us the final answer which is just a {\it wave-function renormalization} of the tree-level result as shown in \eqref{recelcards}! This means the whole instanton series add up in such a way as to give us the result that is exactly a renormalization of the tree-level answer. If we now identify the Fourier factor with $t^{-8/3}$, then the wave-function renormalization factor can be identified with an inverse power of the four-dimensional cosmological constant as in \eqref{tinmey}! 

\vskip.1in

\noindent {\bf Sagredo:} I understand that the whole resummation procedure could provide a closed form expression for the four-dimensional cosmological constant, but let me  be the devil's advocate here and ask you the following questions. How can we unambiguously fix the wave-function renormalization factor to $\Lambda$? What if I absorb some of it in the definition of the conformal temporal coordinate $t$? This would clearly lead to certain ambiguity here..

\vskip.1in

\noindent {\bf Salviati:} Thank you, these are indeed pertinent questions. The unambiguous determination of $\Lambda$ relies on yet another subtlety of the summation process that I haven't revealed to you yet. Previously I told you that the Borel-$\grave{\rm E}$calle resummation produces two pieces, one classical and the other quantum. This is not exactly right, because the quantum piece has a further sub-division: one {\it on-shell} piece and another {\it off-shell} piece. This off-shell piece should not be confused with the off-shell degrees of freedom whose consequences on the low energy effective action I promised to explain soon. The off-shell piece comes from zero momentum and zero frequency states and it is needed to counteract the Minkowski background so that we only get a de Sitter emergent metric. Of course since the off-shell piece has a wave-function, the counteraction should come from the highest amplitude of the wave-function. Alternatively you may think of this as a matching of the boundary condition, with the boundary being the Minkowski background, but the difference is that we are matching the modulus and not the sign. In fact it is this matching that actually identifies the wave-function renormalization factor with the inverse of the four-dimensional cosmological constant as given in \eqref{tinmey}. For more details on the computation you may refer to section \ref{sec3.8}.

\vskip.1in

\noindent {\bf Sagredo:} Thank you and it indeed clarifies a lot for me, but I'd still like to continue as the devil's advocate and ask you a few other questions. The expression that you got in \eqref{tinmey}, how do you know it is always positive definite? Also, how do you know that it is very small, {\it i.e.} as small as $10^{-120} {\rm M}_p^2$?  

\vskip.1in

\noindent {\bf Salviati:} I'll start by answering the first question which is easier. The expression for $\Lambda$ given in \eqref{tinmey} involves a principal value integral expressed using a parameter $\check{f}_{\rm max}$ defined via \eqref{stoneje}. If $\check{f}_{\rm max}$ is negative definite, then the principal value integral is always positive. Subtlety appears when $\check{f}_{\rm max}$ is positive. 
For this case, one may show that there is always {\it one} pole in the Borel plane, and the question is to justify the positivity of the principal value integral over all possible contours that go around that pole. For more on this I'll refer you to section 4.5 of our work \cite{borel} where a detailed mathematical proof of the positivity is demonstrated.

\vskip.1in

\noindent {\bf Sagredo:} Thank you, and I agree that the expression for $\Lambda$ in \eqref{tinmey} is always positive definite, but why is it small? Is that the denominator of the expression that gives us the small factor of say $10^{-120}$? 

\vskip.1in

\noindent {\bf Salviati:} If this was the case then it would have been really nice, but one can show that in the limit $g \to 0$, where $g$ is proportional the inverse of ${\rm M}_p$, the denominator simply goes to identity making $\Lambda$ to be of order ${\rm M}_p^2$ and not small. However there are many details that we did not carefully incorporate in our computation, and one of the crucial one is the contributions from those class of sub-leading diagrams that are still suppressed by one power of the volume but with set of two nodes having identical momenta that are not necessarily same as the node whose momentum match with that of the source. Again, these diagrams are not eliminated in the infinite volume limit because of hidden volume factors inside them. A Borel-$\grave{\rm E}$calle resummation of these diagrams can be done by introducing a new technical machinery called the {\it Borel Boxes}. (An example of this is shown in {\bf Table \ref{fiirzacutt2}} and \eqref{resorworlme}.) Including these contributions lead to a rather complicated structure as discussed in section \ref{sec3.9} which seems to suggest a reduction of the value of $\Lambda$. In section \ref{sec3.9} we included a subset of these diagrams and showed the reduction of the four-dimensional cosmological constant from the value quoted in \eqref{tinmey}. We believe an exhaustive inclusion of all the aforementioned effects would sizably reduce $\Lambda$, but that is a work in progress \cite{ccpaper}. I hope to tell you more on this in a month or so. 

\vskip.1in

\noindent {\bf Sagredo:} I understand that it's a work in progress, but why do you think incorporating these diagrams would reduce $\Lambda$ sizable? 

\vskip.1in

\noindent {\bf Salviati:} Remember when you asked me whether the denominator of an expression like \eqref{tinmey} is responsible in reducing $\Lambda$, I answered {\it no} for the case in question. This is true, but the actual answer is {\it yes} once we include the aforementioned diagrams. To see this, note that incorporating the effects of all the Borel Boxes actually contribute to the denominator thus reducing it. Since there are infinite possible contributions, and they all contribute to the denominator of \eqref{tinmey}, one would worry that this makes $\Lambda$ arbitrarily small. Fortunately this is not the case because the series sum of the Borel Boxes is a convergent one as discussed in section \ref{sec3.9}, implying that the contribution to the denominator of \eqref{tinmey} is very large but not infinite. (See for example \eqref{palazzome}.) Question is whether the reduction could be as small as $10^{-120} {\rm M}_p^2$. This is yet to be worked out. 

\vskip.1in

\noindent {\bf Sagredo:} I think I now understand the computational scheme. Also, since you are taking a supersymmetric Minkowski minimum, there are no contributions from the zero-point energies of the bosonic and the fermionic degrees of freedom. The cosmological constant thus has its root in the full non-perturbative corrections to the system. I have two related questions now. My first question is, how do we see the non-perturbative and the non-local corrections in \eqref{kaittami2}? My second question concerns the recent news I heard from \cite{desibao}: what if the cosmological constant is slowly varying with time? 

\vskip.1in

\noindent {\bf Salviati:} Thank you for your important questions, and I will address your query about the non-perturbative and the non-local corrections first, and leave the second question for discussion a bit later. To answer the first question, let me point out that getting the closed form expression for the cosmological constant, and fixing the form of the low energy effective action, go hand in hand. You may ask, why is that? Let me explain. The computation for the expectation value of the metric (here we took a representative scalar field) generically includes an exhaustive collection of the perturbative interactions. Needless to say, higher order nodal diagrams show Gevrey growths, which led us to resum the series using the Borel-$\grave{\rm E}$calle procedure. From there we got \eqref{tinmey}. We can turn this around, and ask the following question: what replacement should we do to the perturbative terms {\it that would produce the same effect in the path-integral analysis} as the one we got from the usage of the Borel-$\grave{\rm E}$calle resummation procedure? The generic answer is what I showed you on the third line of \eqref{kaittami2}: this is completely expressed using the instanton saddles alongwith their corresponding fluctuation determinants. Both use ${\bf Q}_{\rm pert}$ from \eqref{QT3} but with different set of coefficients $\{c_{nl}\}$. Such a procedure then allows us to replace the perturbative terms in the action by the non-perturbative series written as a resurgent trans-series, thus getting rid of any asymptotic behavior and keeping the action well-defined at all orders. 

\vskip.1in

\noindent {\bf Sagredo:} Thank you and that explains most of the terms in the action \eqref{kaittami2}, but what about the fourth line? I'm guessing that they are coming from the non-local interactions? How and where are the non-localities appearing here?  I thought the low energy limit of M-theory is a local theory? 

\vskip.1in

\noindent {\bf Salviati:} Indeed, the low energy limit of M-theory is typically a local theory as any higher order derivatives are suppressed by higher powers of ${\rm M}_p$. But these are not the non-localities that I was talking about earlier. The non-localities I alluded to earlier appear from integrating out the {\it massless} off-shell degrees of freedom in the theory. To see where they are coming from, let us go back to the path-integral analysis from 
\eqref{mmtarfox} and \eqref{russmameye}. The analysis therein was done by making one crucial assumption: the off-shell degrees of freedom are kept massive. In reality this cannot be the case, so one has to revisit the computations. A careful computation of the path-integral will incorporate both the on-shell and the off-shell degrees of freedom, both in the measure and in the action. To proceed, we'll take a simple model with one on-shell and one off-shell scalar fields like the one in \eqref{enmilltriangle}. (In the language of metric, the metric components along ${\bf R}^{2, 1}$ will be the on-shell ones whereas the metric components, with one leg along ${\bf R}^{2. 1}$ and the other leg along the internal eight-manifold, will be the off-shell ones.) An expectation value of the on-shell degree of freedom, using the path-integral formalism with the displacement operators will involve integrating over both the on-shell and the off-shell degrees of freedom. We can integrate away the massless off-shell degrees of freedom to give rise to a non-local action as shown in \eqref{TIogyan}. The massless on-shell degrees of freedom are then integrated over a shifted vacuum coming from the displacement operators using an action that now has both local and non-local pieces expressed using on the on-shell degrees of freedom.

\vskip.1in

\noindent {\bf Sagredo:} I see now how the off-shell piece in the action arises, but how do you know it will not give rise to additional asymptotic series? 

\vskip.1in

\noindent {\bf Salviati:} That's a good question and the answer is that if we choose an off-shell action of the form \eqref{jacksharp}, we are bound to get asymptotic divergence from factorial growths. However a little mathematical manipulations can make us rewrite \eqref{jacksharp} in a more convergent form as in \eqref{fepoladom}.

\vskip.1in

\noindent {\bf Sagredo:} This rewriting of \eqref{jacksharp} as \eqref{fepoladom} is {\it not} a Borel-$\grave{\rm E}$calle resummation right? 

\vskip.1in

\noindent {\bf Salviati:} Correct, because we are not going to higher orders in the path-integral expansion. You may view this as a mathematical trick to go from \eqref{jacksharp} to \eqref{fepoladom}, and in the end we can keep the coefficients $c_p$ in \eqref{fepoladom} unrelated to the coefficients $d_p$ in \eqref{jacksharp}. This rewriting is useful because it can convert the form of the expectation value of the on-shell field in \eqref{posnerdead} to the final form \eqref{dunbar} which has only the non-perturbative and the non-local terms. The relevant action then is \eqref{katuli} which, when we incorporate the actual on-shell metric and flux degrees of freedom, takes the form \eqref{kaittami2} that I showed you earlier.

\vskip.1in

\noindent {\bf Sagredo:} Thank you, and now I see how all are falling into places. However I'm still a bit confused about the fourth line in the action \eqref{kaittami2} that deals with the non-local interactions. This doesn't look like a trans-series because I don't see any fluctuation determinants appearing anywhere there.

\vskip.1in

\noindent {\bf Salviati:} Thank you for noticing this because the story therein is more subtle. A naive computation that converted \eqref{jacksharp} to \eqref{fepoladom} would suggest no such additional factors, but we will soon see that this leads to problems. However to appreciate this subtlety, I'll need to tell you about the equations of motion first.

\vskip.1in

\noindent {\bf Sagredo:} OK, in that case I'll wait for the explanation. I understand that, because you have an explicit form of the action, you can compute the equations of motion, right?

\vskip.1in

\noindent {\bf Salviati:} Right, but the EOMs story has an additional layer of subtleties that I haven't told you earlier. It is high time to bring forth these details now. The EOMs are not the usual EOMs, rather they take the form of the Schwinger-Dyson's equations simply because the on-shell degrees of freedom that we compute using the path-integral (with the displacement operators and the action \eqref{kaittami2}) are in the form of expectation values and are therefore {\it emergent} quantities. Schwinger-Dyson's equations are specifically designed to study the dynamics of the expectation values and are therefore the perfect tools to study the temporal and the spatial evolutions of these emergent on-shell degrees of freedom.

\vskip.1in

\noindent {\bf Sagredo:} One question about terminology before you proceed. Since the on-shell degrees of freedom appear as emergent quantities because of the path-integral analysis, they are actually {\it off-shell} states but give rise to {\it local} interactions in the action \eqref{fepoladom}. On the other hand, the off-shell states that you integrate away are also {\it off-shell} and give rise to {\it non-local} interactions in the action \eqref{fepoladom}. Is this the correct way to think about the system?

\vskip.1in

\noindent {\bf Salviati:} Indeed, everything is actually off-shell in this set-up. We simply distinguish between the {\it emergent} on-shell and the truly off-shell states. The latter gives us the non-local interactions in \eqref{fepoladom}. Coming back to the EOM $-$ the derivation of which is a little more non-trivial because of the various ingredients and subtleties $-$ is nevertheless easy to express in the form of expectation value as in \eqref{elenacigu2}. Notice the appearance of action from \eqref{fepoladom} as well as the {\it linear terms} in the displacement operator \eqref{28rms} in the Schwinger-Dyson's equation \eqref{elenacigu2}. 

\vskip.1in

\noindent {\bf Sagredo:} Why only linear terms from the displacement operator \eqref{28rms}? This is a bit surprising.

\vskip.1in

\noindent {\bf Salviati:} There is another layer of subtlety here that I don't think will have time to explain to you now. I'll let you read up 
section \ref{sec4} and especially section \ref{sec4.2}, because it opens up a Pandora's box of issues related to the compatibility of the Wilsonian effective action with Borel resummation, {\it et cetera}. 

\vskip.1in

\noindent {\bf Sagredo:} This is fine by me and I'll look up the sections that you mentioned. Please proceed with the discussion about the Schwinger-Dyson's equations.

\vskip.1in

\noindent {\bf Salviati:} Yes, as I was saying that the EOM takes the form \eqref{elenacigu2}, but unfortunately it is not very useful to extract the dynamics of the emergent quantities from it. We need to rewrite it in a form where it can be more useful and for this we need something called the {\it resolution of the identity} for the Glauber-Sudarshan states. This process turned out to be more complicated because the Glauber-Sudarshan states are not coherent states. Nevertheless it can be worked out with some effort $-$ see details in section \ref{sec4.1} $-$ and after the dust settles the final answer may be presented as \eqref{mariesored} using an operator ${\bf Q}$ which may be easily quantified. You may ask: why is this necessary? The answer may be ascertained from the series of manipulations starting from \eqref{katmarzoe} and culminating in \eqref{eventhorizon}. The first equation in \eqref{eventhorizon} is the answer that we have been looking for. This is exactly the EOM of the {\it emergent} on-shell metric and flux degrees of freedom but with two differences: one, the action has only the non-perturbative and the non-local terms as in \eqref{kaittami2} and two, all on-shell degrees of freedom in \eqref{kaittami2} are replaced by their expectation values. On the other hand, the second equation in \eqref{eventhorizon} is something that we did not encounter earlier. This equation weaves the three crucial elements in the construction together: the intermediate Glauber-Sudarshan states (that appear from \eqref{mariesored}), the Faddeev-Popov ghosts (if any) and the linear term in the displacement operator. Note that all the three contributions cannot be visible at the ``classical'' level, {\it i.e.} at the level of the first equation in \eqref{eventhorizon}, thus explaining the natural split of the EOM \eqref{palpucki} into a set of two EOMs. 

\vskip.1in

\noindent {\bf Sagredo:} This is fascinating! If I understood correctly, 
the issues which prohibited us to declare the results from {\bf Table \ref{firoksut}} as evidence for the existence of a de Sitter {\it vacuum} solution now no longer plagues us. In fact the first equation in \eqref{eventhorizon} clearly suggests that de Sitter space-time should be an {\it emergent} solution appearing from the Glauber-Sudarshan state $\vert\sigma\rangle$. The emergent background is now backed by a well defined effective field theory description that is  consistent with the trans-Planckian bound and other constraints. Right?

\vskip.1in

\noindent {\bf Salviati:} Right! There are also two more set of equations appearing from imposing non-localities. They are \eqref{necampbel} expressed using the local degrees of freedom. These set of equations signal the {\it non-classical} nature of the system and would not appear if we had taken coherent states. {\it You may think of them as the first concrete sign of being a Glauber-Sudarshan state}. They all seem to fit well, 
so the question is, where do we see the problems with the non-local terms in the fourth line of \eqref{kaittami2}? The answer appears rather surprisingly from analyzing the Schwinger-Dyson equation \eqref{tranishonal} itself. Observe that the equation is complicated, so one strategy to solve this equation would be to match the $g_s$ scalings of every term: start by matching the dominant scalings over all the terms, and then match the subsequent sub-dominant ones. This procedure works very well for the non-perturbative terms, as evident from the analysis presented in section \ref{sec7.3.1}. Problem appears when we compare the $g_s$ scalings with the eighth line of \eqref{tranishonal}: it doesn't match! The only way this matching would work correctly is if we allow a fluctuation determinant of the form \eqref{whiteang} accompanying the fourth line of \eqref{kaittami2}.

\vskip.1in

\noindent {\bf Sagredo:} Which means the action itself needs to be changed, right? This should also effect the Schwinger-Dyson equation, if I'm not mistaken.

\vskip.1in

\noindent {\bf Salviati:} Indeed, the action now becomes \eqref{kaittami4}, which modifies the Schwinger-Dyson equation to take the form \eqref{tranishonal3}. One needs to use this equation to compute the effects of the wrapped instantons as we discussed in section \ref{sec7.3}.

\vskip.1in

\noindent {\bf Sagredo:} Thank you and that answers most of my questions except the one on the temporally varying cosmological constant. How are you dealing with that? 

\vskip.1in

\noindent {\bf Salviati:} It was a bit surprising that with a slight change in the emergent field configuration from \eqref{duipasea2} to \eqref{ripley} allows us to construct a new Glauber-Sudarshan state that captures the slow variation of the cosmological constant. The analysis is a bit technical and I direct you to sections \ref{sec5.3} and \ref{sec7.3.3} for full details. Of course it is still a bit early to say what it all means with respect to the recent results from \cite{desibao}, although the issues pointed out in say \cite{donof} may no longer be worrisome. I can tell you more on this in near future. Meanwhile the take-home message is that the Glauber-Sudarshan state may provide a consistent framework to study positive cosmological constant solution whether or not it is constant or showing a very slow temporal variation. For your convenience I have summarised the results in {\bf Table \ref{firzacut3}} where you might get answers to other questions that we didn't have time to discuss.

\vskip.1in

\noindent {\bf Sagredo:} Thank you, and this means the Glauber-Sudarshan states are capable of capturing the late-time behavior of our universe. What about early times?

\vskip.1in

\noindent {\bf Salviati:} Questo lo lascio calcolare a voi. In altre parole, stai zitto e calcola!

\section{Conclusion, future directions and open questions \label{sec8}}

In this paper we have hopefully convinced the readers that the Glauber-Sudarshan states capture the late-time dynamics of our universe in a consistent way. Our analysis is presented in the M-theory uplift of type IIB background at far IR, and we have shown how one may compute the cosmological constant and argue its positivity, smallness and possible temporal variation. 

As mentioned, the consistency of such a state is important. This comes from variety of directions, namely consistency with respect to the quantum corrections, the Wilsonian integrating out method, the exact renormalization group procedure and other related effects. Somewhat surprisingly all of these boil down to {\it two} simple effective field criteria, namely \eqref{atryan} imposed on the dual type IIA coupling. Satisfying them is equivalent to satisfying all the aforementioned list of consistency conditions. 

In fact a careful study, keeping in mind the two criteria \eqref{atryan}, led us to find a number of interesting results which we listed in section \ref{newresults}. An interesting and somewhat surprising result appeared from analyzing the low energy effective action \eqref{kaittami4} which is expressed using non-perturbative and non-local interactions as a resurgent trans-series, but no perturbative terms $-$ beyond the fluctuation determinants over the zero-instanton sectors $-$ appear. Although this is consistent with the expectation from the underlying Borel resummations, it raises questions related to the compatibility with Wilsonian integration. This is a subtle topic and it will be interesting to pursue this further. Another interesting thing from \eqref{kaittami4} is the emergence of an effective potential that is very different from the vacuum potential, leading to the study of the dynamics of the {\it emergent} on-shell degrees of freedom as Schwinger-Dyson equations (see \eqref{alto2mei} and \eqref{tranishonal3}). These emergent on-shell degrees of freedom are ultimately responsible in realizing the de Sitter background in our model as an excited state and as such are not restricted by the EFT criteria of \cite{swampland}. 

The emergent action that we discussed here now raises questions regarding the precise set of non-perturbative corrections. In section \ref{sec7.3} we argued how the five-brane instantons contribute to the EOM
\eqref{tranishonal3} from the action \eqref{kaittami4}. However there are other instantons, like the M2-brane instantons or stable instantons that are non-BPS objects in the theory, that we did not touch upon (see also footnote \ref{shomithi}). Additionally we have not incorporated the effects of the fermionic sector fully, although we did present their contributions briefly. Interestingly, their contributions can be incorporated by replacing the emergent curvature and the G-flux tensors by generalized ones. The resulting analysis becomes very technical but fortunately the physics doesn't change. A complete study of these effects will be presented in \cite{hete8}.  

The potential that we discussed, say in {\bf figure \ref{pot1}} has Minkowski minima, but we also expect non scale-separated AdS minima as alluded to earlier. All these are supersymmetric minima so a natural question is how far are the AdS minima separated from the local Minkowski minima. It could be that there are natural potential wall between them so as to reduce the tunnelling effects, but to analyze the full landscape of all the supersymmetric minima of the theory will probably require string field theory. Interestingly however, the presence or absence of the non- scale-separated minima, does not effect any of the analysis that we presented here as the action \eqref{kaittami4} is sufficiently generic to capture all possible contributions. Nevertheless it would be interesting to study this as this will have consequence for AdS/CFT.

The emergent de Sitter space also raises an interesting question regarding the symmetry algebra. The vacuum state, which is Minkowski, has the full super-Poincare symmetry, and we expect the excited Glauber-Sudarshan state to break it down to a sub-algebra in tune with the underlying supersymmetry breaking.  How exactly the super-Poincare algebra (with appropriate internal symmetry group) breaks down to the full de Sitter symmetry algebra remains to be seen. Interestingly even for the case with slowly varying cosmological constant, it would be interesting to work out the symmetry breaking pattern. Moreover, the question of how probable is the Glauber-Sudarshan state in the configuration space of the full system, is very important. For example, whether the state of the
system is at least a local attractor in initial
condition space or how naturally it is 
generated in the early universe is a matter that needs to be  clarified\footnote{We thank Robert Brandenberger for discussion on this.}. We hope to return to these issues in near future.

Among other intriguing result is the study of the {\it smallness} of the four-dimensional positive cosmological constant, and its possible temporal variation. Concerning the smallness, we have argued in section \ref{smallcc} how using new mathematical tools we can lower the value of the cosmological constant. Whether the reduction is small enough to account for the expected answer is not clear yet. Moreover, the recent result from DESI \cite{desibao} may suggest that this small value also has a {\it very slow} temporal variation. In sections \ref{sec5.3} and \ref{sec7.3.3} we have argued quantitatively that such a configuration may be easily accommodated in our set-up by modifying the Glauber-Sudarshan state. However more work is needed to see whether this matches with the present dynamical evolution of our universe, and in particular whether the DESI results increase or decrease in statistical significance in the coming years (the last reference in \cite{desibao} seems to suggest the latter).

 Yet another important question is to extend the Glauber-Sudarshan state construction to describe Nariai black holes in order to address the Festina-Lente (FL) bound \cite{festina}. Specifically, we want to understand whether the FL bound constrains the most general low energy M-theory action \eqref{kaittami4} or whether the bound arises as a natural implication of the excited state picture taking into account stringy principles. A related issue is to understand more on the behaviour of the Glauber-Sudarshan state in the far IR. Specifically the question involves determining whether the state takes the familiar form of the Hartle-Hawking state, or other states in the late time canonical Hilbert space \cite{joydeep}. 

Due to the time dependence of $g_s$, the early time state before the de Sitter description is strongly coupled, i.e. $g_s \gg 1$, and is not well understood. At very early time, the Glauber-Sudarshan state is expected to lose its coherent features. Since the Glauber-Sudarshan state is an excited state over the Minkowski vacuum, the state possibly transitions quickly to the Minkowski minimum or slowly to a thermal gas in Minkowski space. The latter can probably be understood as an out of equilibrium system en route to attaining thermal equilibrium. More on this will be presented elsewhere.

\section*{Acknowledgements}
We thank Robert Brandenburger, Simon Caron-Huot, Dileep Jatkar, Bohdan Kulinich, P. Ramadevi, Ashoke Sen and Radu Tatar for useful discussions. Special thanks goes to Archana Maji for verifying many of the equations carefully and to Suddhasattwa Brahma for explaining to us the subtleties with the static patch, the temporal variation of the cosmological constant and other related topics. The work of JC is supported by the Simons
Collaboration on Nonperturbative Bootstrap. The work
of KD is supported in part by a Discovery Grant from
the Natural Sciences and Engineering Research Council of
Canada (NSERC). KD would also like to thank Ori Ganor and the members of the department of Physics, University of Berkeley, during his sabbatical visits this year.

\newpage

\begin{table}[H]  
 \begin{center}
\renewcommand{\arraystretch}{1.5}
\begin{tabular}{|c||c||c|}\hline {\bf Vacuum configuration}  & {\bf Replaced by} \\ \hline\hline
Type II, I and Heterotic theories & M-theory and duality sequences \\ \hline
de Sitter space-time & Glauber-Sudarshan state $\vert\sigma\rangle = 
\mathbb{D}(\sigma)\vert\Omega\rangle, \mathbb{D}^\dagger(\sigma)\ne \mathbb{D}^{-1}(\sigma) $\\ \hline
Fluctuations over de Sitter space-time & Agarwal-Tara state \\ \hline
Trans-Planckian bound & $g_s < 1$ $\implies ~ -{1\over \sqrt{\Lambda}} < t < 0$ \\ \hline
Black-hole in de Sitter space & Another Glauber-Sudarshan state $\vert\sigma'\rangle$ \\ \hline
Equations of motion & Schwinger-Dyson equations \\ \hline
On-shell states & Off-shell states \\ \hline
Off-shell states & Off-shell states and non-local quantum terms \\ \hline
Feynman diagrams & Nodal diagrams \\ \hline
Space-time metric $g_{\mu\nu}(x, y)$ & 
$\langle {\bf g}_{\mu\nu} \rangle_{\sigma} = {\int [{\cal D} \Xi][{\cal D}\Upsilon_g]~e^{-{\bf S}_{\rm tot}(\Xi, \Upsilon_g)}~ \mathbb{D}^\dagger(\sigma; \Xi) g_{\mu\nu}(x, y)
\mathbb{D}(\sigma; \Xi) \over 
\int [{\cal D} \Xi] [{\cal D}\Upsilon_g]~e^{-{\bf S}_{\rm tot}(\Xi, \Upsilon_g)} ~\mathbb{D}^\dagger(\sigma; \Xi) 
\mathbb{D}(\sigma; \Xi)}$ \\ \hline
Conformal time $t$, $-\infty < t < 0$ & $ -{1\over \sqrt{\Lambda}} < t = t(g_s) < 0$ \\ \hline
Internal metric $g_{mn}(y, t)$ &
$\langle {\bf g}_{mn} \rangle_{\sigma} = {\int [{\cal D} \Xi][{\cal D}\Upsilon_g]~e^{-{\bf S}_{\rm tot}(\Xi, \Upsilon_g)}~ \mathbb{D}^\dagger(\sigma; \Xi) g_{mn}(y, t)
\mathbb{D}(\sigma; \Xi) \over 
\int [{\cal D} \Xi] [{\cal D}\Upsilon_g]~e^{-{\bf S}_{\rm tot}(\Xi, \Upsilon_g)} ~\mathbb{D}^\dagger(\sigma; \Xi) 
\mathbb{D}(\sigma; \Xi)}$ \\ \hline
Positive cosmological constant $\Lambda$ & 
$ 0 < \Lambda = \lim\limits_{c_{({\bf s})}\to 0} {{\rm M}_p^2 \over \Big[\sum\limits_{\{{\bf s}\}}
{\mathbb{F}_{({\bf s})}\over c_{({\bf s})}}
\int_0^\infty du_{({\bf s})}~{{\rm exp}\left(-u_{({\bf s})}/c_{({\bf s})}\right) \over 1 - 
u_{({\bf s})}^l}\Big]^{3/4}_{{\rm P. V}}} << {\rm M}_p^2$ \\ \hline
Existence of EFT & ${\partial g_s\over \partial t} = g_s^{+ive} \implies$ non-violation of NEC\\ \hline
Quantum tunneling & Real and complex instantons \\ \hline
Non-perturbative effects & Borel resummation of Gevrey series \\ \hline
Open quantum field theories & Wilsonian EFT or Exact Renormalization Group\\ \hline
Contributions from zero-point energies & Cancelled Zero-point energies \\ \hline
Non-susy vacuum configuration & Spontaneously broken supersymmetry by 
GS state $\vert\sigma\rangle$\\\hline
Moduli stabilization & Dynamical moduli stabilization \\ \hline
Possibility of Boltzmann brains & No possibility of Boltzmann brains because $-{1\over \sqrt{\Lambda}} < t < 0$ \\ \hline
Background fluxes ${\bf G}_{\rm MNPQ}$ & 
$\langle {\bf G}_{\rm MNPQ} \rangle_{\sigma} = {\int [{\cal D} \Xi][{\cal D}\Upsilon_g]~e^{-{\bf S}_{\rm tot}(\Xi, \Upsilon_g)}\mathbb{D}^\dagger(\sigma; \Xi) 
{\rm G}_{\rm MNPQ}(x, y)
\mathbb{D}(\sigma; \Xi) \over 
\int [{\cal D} \Xi] [{\cal D}\Upsilon_g]~e^{-{\bf S}_{\rm tot}(\Xi, \Upsilon_g)} ~\mathbb{D}^\dagger(\sigma; \Xi) 
\mathbb{D}(\sigma; \Xi)}$ \\ \hline
D7-branes & Taub-NUT space as a Glauber-Sudarshan state $\vert\sigma_{\rm TN}\rangle$ \\ \hline
Gauge fields on D7-branes & Localized $\langle {\bf G}_{\rm MNPQ} \rangle_{\sigma}$ at Taub-NUT singularities \\ \hline
Orientifold 7-planes & Atiyah-Hitchin state $\vert \sigma_{\rm AH}\rangle$
\\ \hline
\end{tabular} 
\renewcommand{\arraystretch}{1}
\end{center}
 \caption[]{A comparison between the results from a vacuum configuration and an excited state.} 
  \label{firzacut3}
 \end{table}

\appendix
\section{Different coordinate patches of de Sitter \label{append}}

We briefly review coordinate patches of de Sitter spacetime which are used in our text. In four dimensions, the metric covering the global de Sitter is given by:
\beq \label{2.0}
ds^2 = -d\tau^2 + l^2 \cosh^2 \frac{\tau}{l} \lc d\chi^2 + \sin^2 \chi \, d\Omega^2\rc
\eeq
The global time lies in the range $\tau \in (-\infty, \infty)$ while $d\Omega^2$ denotes the standard two sphere metric:
\beq
d\Omega^2 = d \theta^2 + \sin^2 \theta d\phi^2.
\eeq
The global patch does not have any global timelike Killing vector. Plugging \eqref{2.0} into the Einstein equations, the cosmological constant takes the form $\Lambda = 3 l^{-2}$. Defining\footnote{A note on the notation: ${\cal H}$ is the Hubble constant and ${\rm H}(y)$ is the warp-factor.} ${\cal H} \equiv l^{-1}$, the flat slicing, which covers half of the global patch, takes the form:
\beq \label{2.00}
ds^2 = -dt'^2 + e^{2{\cal H} t'} d\vec{x}^2
\eeq
where $\vec{x}$ denotes a three-vector, while the time $t'$ ranges from $t' \in \lc -\infty, \infty \rc$. Another important coordinate system used in our work is the Poincare coordinates. We go to the Poincare coordinates by shifting to a conformal time ${\cal H} t = -e^{-t'{\cal H}}$.
For our purposes, it is convenient to work with the de Sitter metric in the Poincare patch as shown in {\bf figure \ref{planar}}. The metric takes the form:
\beq \label{2.1}
ds^2 = \frac{1}{{\cal H}^2 t^2} \lc -dt^2 + d\vec{x}^2\rc.
\eeq
where the time $t$ ranges from $t \in \lc -\infty,  0\rc$, with $t \to 0^-$ describing the late time asymptotic boundary. Note that the flat slicing and the Poincare denote the same patch, and our construction of the Glauber-Sudarshan state corresponds to the flat slicing. (We have absorbed a factor of 3 in the definition of the warp-factor ${\rm H}(y)$ and the internal metric $g_{mn}(y)$ in \eqref{metansatze} onwards.)

A single observer can only observe a part of the geometry. The metric inside this region is the static patch and is given by:
\beq\label{a5}
ds^2 = - f(r) dT^2 + \frac{dr^2}{f(r)} + r^2 d\Omega^2
\eeq
where $f(r) = \lc 1 - \frac{r^2}{l^2} \rc$, such that $r\in (0,l)$ and $T \in \lc -\infty, \infty \rc$. The patch is visually indicated in Regions ${\rm R}$ and ${\rm L}$ of {\bf figure \ref{staticpatch3}}. The patch has a local timelike Killing vector, and the generator of time translations in the static patch is simply the boost operator. 

All the aforementioned metric on the various patches can be studied in Type IIB by simply replacing the flat-slicing four-dimensional metric by 
\eqref{2.0}, \eqref{2.00} or \eqref{a5} respectively keeping the warp-factor and the internal metric unchanged. While the patches with temporal variations in \eqref{2.0}, \eqref{2.00} and \eqref{2.1} work well, the one with the static patch from \eqref{a5} has many problems as discussed in the text.


\end{document}